\documentclass[12pt]{book}
\usepackage[utf8]{inputenc}
\usepackage[english]{babel}
\usepackage{indentfirst}
\usepackage{amsmath}
\usepackage{amssymb}
\usepackage{physics}
\usepackage{braket}
\usepackage{fullpage}
\usepackage{bookmark}
\usepackage{multicol}
\usepackage{lipsum}
\usepackage{float}
\usepackage{lastpage}
\usepackage{calc}
\usepackage{graphicx}
\usepackage{cancel}
\usepackage[T1]{fontenc}
\usepackage{lmodern}
\usepackage{mathtools,amsfonts}
\usepackage{enumitem}
\usepackage[svgnames]{xcolor}
\usepackage[most]{tcolorbox}
\usepackage{babel}
\usepackage{blkarray}
\tcbset{
    exstyle/.style={enhanced, colback=Indigo!5, colframe=black, 
                   fonttitle=\bfseries,  
                   colbacktitle=white, coltitle=black,    
                   top=\tcboxedtitleheight,
                   boxed title style={},
                   attach boxed title to top left={yshift=-\tcboxedtitleheight/2,
                                                   xshift=4mm}%
                   },
    teostyle/.style={enhanced, colback=white, colframe=black, 
                   fonttitle=\bfseries, sharp corners, boxrule=0pt,
                   colbacktitle=DarkOrchid, coltitle=white,
                   drop fuzzy shadow,    
                   top=\tcboxedtitleheight,
                   boxed title style={sharp corners},
                   attach boxed title to top left={}%
                   },
}

\newtcbtheorem[number within=section]{exercise}{Exercise}{exstyle}{ex}
\newtcbtheorem[number within=section]{example}{Example}{exstyle}{ex}
\usepackage{tikz}
\usetikzlibrary{angles,quotes}
\usetikzlibrary{fit,shapes}
\usetikzlibrary{quantikz}
\begin{document}
\title{Lecture Notes on Quantum Algorithms}
\author{Muhammad Faryad\\~\\
Department of Physics,\\
Lahore University of Management Sciences, Lahore 54792, Pakistan.\\~\\
muhammad.faryad@lums.edu.pk}

\maketitle

\chapter*{Preface}
These lecture notes were developed for a semester-long course on quantum algorithms taught at the Lahore University of Management Sciences (LUMS) during Fall 2021 and Summer 2024. They are intended to help students—particularly those without a background in quantum physics—learn quantum algorithms with clarity and ease. The notes include detailed derivations of fundamental quantum algorithms to support self-guided learning.

Chapters 1–6 may be viewed as a companion to Quantum Computation and Quantum Information by Nielsen and Chuang, as they provide in-depth derivations and explanations of many algorithms covered in that text. Most of the material has been drawn from various sources for teaching purposes, and no claim to originality is made. However, I have made every effort to include detailed derivations and conceptual insights to facilitate understanding.

The notes begin in Chapter 1 with a review of linear algebra and the postulates of quantum mechanics, leading to an explanation of single- and multi-qubit gates. Chapter 2 explores the challenge of constructing arbitrary quantum states from given initial states, and introduces circuits for building oracles. Chapter 3 presents foundational algorithms such as entanglement creation, quantum teleportation, Deutsch-Jozsa, Bernstein-Vazirani, and Simon's algorithm.

Chapters 4 and 5 cover algorithms based on the quantum Fourier transform, including phase estimation, period finding, factoring, and logarithm computation. These chapters also include complexity analysis and detailed quantum circuits suitable for implementation in code. Chapter 6 introduces Grover's algorithm for quantum search and amplitude amplification, including its realization via Hamiltonian simulation and a method for derandomization.

Chapter 7 discusses basic techniques for Hamiltonian simulation, such as Lie-Trotter decomposition, sparse Hamiltonians, and the linear combination of unitaries. It also provides example circuits for simulating Hamiltonians expressed as linear combinations of Pauli operators. Chapter 8 introduces variational quantum algorithms, and Chapter 9 presents an algorithm for simulating fermionic many-particle systems, with an emphasis on molecular Hamiltonians. It also outlines the key transformations needed to map a molecular Hamiltonian to a form suitable for simulation on a quantum computer.

I am deeply grateful to Muhammad Faizan and Amber Riaz, both MS students in the Department of Physics at LUMS, for their invaluable assistance in typesetting these notes in LaTeX. Without their dedication and effort, these notes would likely not have taken their present form.

For corrections or inquiries, please feel free to contact me at: muhammad.faryad@lums.edu.pk.

\tableofcontents

\chapter{Fundamentals}\footnote{Copyrights: Muhammad Faryad, LUMS}

This chapter covers the fundamentals of quantum gates and qubits. We begin with the review of linear algebra and then discuss single and two-qubit quantum states and gates.
\section{Review of Linear Algebra}
A  vector space consists of two sets of elements and two algebraic rules:
\begin{enumerate}
\item a set of \emph{vectors} $V=\{\ket{v_1},\ket{v_2},\ket{v_3},\dotsc\}$ and set of scalars $\mathbb{C}=a,b,c,\dotsc$
\item a rule for vector \emph{addition} and a rule for scalar \emph{multiplication}.
\end{enumerate}
\begin{enumerate}[label=\alph*)]
\item \textbf{Addition Rule}\\
Addition rule has the following properties:\\
\begin{itemize}
\item {Closure}: if $\ket{v_1}$ and $\ket{v_2}$ are elements of vector space V, then their sum $\ket{v_1}+\ket{v_2}$ is also an element of the same space. Mathematically, $$\text{if}\ \ket{v_1},\ket{v_2}\in V \Rightarrow \ket{v_1}+\ket{v_2}\in V$$
\item {Commutativity}: $\ket{v_1}+\ket{v_2}=\ket{v_2}+\ket{v_1}$
\item {Associativity}: $(\ket{v_1}+\ket{v_2})+\ket{v_3}=\ket{v_1}+(\ket{v_2}+\ket{v_3})$
\item {Existence of Null Vector}: For each vector $\ket{v}$, there must exist a null vector $\ket{N}$ such that: $$\ket{N}+\ket{v}=\ket{v}+\ket{N}=\ket{v}$$
\item {Existence of Inverse}: Each vector $\ket{v}$ has its inverse vector $-\ket{v}$ such that $$\ket{v}+(-\ket{v})=(-\ket{v})+\ket{v}=\ket{N}$$
\end{itemize}
\item \textbf{Multiplication rule}\\
Multiplication of vectors by scalars(either $\mathbb{R}$ or $\mathbb{C}$) has the following properties:
\begin{itemize}
\item {Closure:} The product of a scalar with a vector gives another vector in the same space. Mathematically, $$\text{if}\ a\in \mathbb{C}\ \text{and}\ \ket{v}\in V \Rightarrow a\ket{v}\in V$$
\item Distributiviy with respect to addition:$$a(\ket{v_1}+\ket{v_2})=a\ket{v_1}+a\ket{v_2}\ \ , \ \ (a+b)\ket{v}=a\ket{v}+b\ket{v}$$
\item Associativity with respect to multiplication of scalars: $$a(b\ket{v})=(ab)\ket{v}$$
\item For each element $\ket{v}$, there must exists a unitary scalar $I$ and zero scalar $o$ such that: $$I\ket{v}=\ket{v}I=\ket{v}\ \text{and\ }o\ket{v}=\ket{v}o=o$$
\end{itemize}
\end{enumerate}
\section{Dimension and Basis of a Vector Space}
A set of $n$ nonzero vectors $\{\ket{v_1},\ket{v_2},\dotsc,\ket{v_n}\}$ is said to be \emph{linearly independent} iff the solution of the equation
\begin{align*}
\sum_{i=1}^nc_i\ket{v_i}=0
\end{align*} is $c_1=c_2=\dotsc c_n=0$. If $c_i$ are not all zero, so that one vectors (say $\ket{v_k}$) can be expressed as a linear combination of the others, then the set $\{\ket{v_i}\}$ is said to be \emph{linearly dependent}.\\
\textbf{Dimension: }The \emph{dimension} of a vector space is given by the number of linearly independent vectors in the basis set.\\
\textbf{Basis: }The \emph{basis} of a vector space consists of a set of the number of linearly independent vectors belonging to that space. Let say, a vector $\ket{w}$ can be written as a linear combination of $\ket{v_1},\ket{v_2},\dotsc \ket{v_n}$ i.e., $$\ket{w}=\sum_{i=1}^nc_i\ket{v_i}$$ Here, the set $\{\ket{v_i}\}$ is the basis of the vector space because it contains linearly independent vectors. It is convenient to choose the linearly independent vectors \emph{orthonormal}, that is, their inner product satisfy the relation $\braket{v_i|v_j}=\delta_{ij}$
\begin{equation*}
\text{where }\delta_{ij} =
    \begin{cases}
      0 &   i\neq j\\
      1 &   i=j
    \end{cases}       
\end{equation*}
\section{Dirac Notation}
\subsection{Kets}
It is the column vector in the Hilbert Space. For example, in $\mathbb{C}^2$, arbitrary ket can be written as: $$\ket{v}=\begin{pmatrix}
a\\ b
\end{pmatrix}\ ;\ a,b\in \mathbb{C}$$
\subsection{Bras}
For every ket $\ket{\ }$, there is a unique bra $\bra{\ }$ that belongs to the dual Hilbert space such that if: $$\ket{v}=\begin{pmatrix}
a\\b
\end{pmatrix} \text{\ then\ } \bra{v}=(\ket{v})^{\dagger}=[\ket{v}^*]^T=\begin{pmatrix}
a^* & b^*
\end{pmatrix}$$
where $a^*$ and $b^*$ are the complex conjugates of $a$ and $b$ respectively.
\subsection{Inner Product}
Inner product in Dirac Notation is denoted by $\braket{\ |\ }$, which is called \emph{bra-ket}. Let say we have two vectors $\ket{v}$ and $\ket{w}$, their inner product is given by:
\begin{align*}
\braket{v|w}& =\begin{pmatrix}a^* & b^*\end{pmatrix}\begin{pmatrix}c\\d\end{pmatrix}\\
& = a^*c + b^*d
\end{align*}
Also, note that $$\braket{v|w}=\left(\braket{w|v}\right)^*$$
Note: In the Hilber space, inner product of a vector $\ket{v}$ with itself is a positive real number: $$\braket{v|v}=a^*a+b^*b=\abs{a}^2+\abs{b}^2\geq 0$$
\textbf{What are orthogonal and orhonormal vectors?}\\
Those vectors whose inner product is equal to $0$ are called orthogonal. For example, $\ket{v}$ and $\ket{w}$ are orthogonal if $\braket{v|w}=\braket{w|v}=0$. Moreover if the magnitude of $\ket{v}$ and $\ket{w}$ are $1$, then it means vectors are normalized and both forms an orthonormal basis.\\
\textbf{What is Hilbert Space?}\\
A  vector space in which all vectors are normalized with an inner product defined.

\subsection{Linear operator}
An operator $\hat{A}$ is said to be \emph{linear} if it obeys the distributive law and like all operators, it commutes with constants. That is, an operator $\hat{A}$ is linear if, for any vectors $\ket{v_1}$ and $\ket{v_2}$ and any complex numbers $c_1$ and $c_2$, we have:
\begin{align*}
\hat{A}\left(c_1\ket{v_1}+c_2\ket{v_2}\right)=c_1\hat{A}\ket{v_1}+c_2\hat{A}\ket{v_2}
\end{align*}
For example, \\
\begin{align*}
\hat{X}& =\begin{pmatrix}0 & 1\\1 & 0\end{pmatrix}\\
\hat{X}\begin{pmatrix}1\\0\end{pmatrix}& =\begin{pmatrix}0 & 1\\1 & 0\end{pmatrix}\begin{pmatrix}1\\0\end{pmatrix}=\begin{pmatrix}0\\1\end{pmatrix}
\end{align*}
Similarly,
\begin{align*}
\hat{X}\begin{pmatrix}
0\\1
\end{pmatrix}&=\begin{pmatrix}
1\\0
\end{pmatrix}
\end{align*}
\subsection{Outer product}
Suppose we have two vectors $\ket{v}=\begin{pmatrix}
a\\b
\end{pmatrix}$ and $\ket{w}=\begin{pmatrix}
c\\d
\end{pmatrix}$, Then the outer product is simply:
\begin{align*}
\ketbra{v}{w} & = \begin{pmatrix}
a\\b
\end{pmatrix}\begin{pmatrix}
c^* & d^*
\end{pmatrix}\\
& = \begin{pmatrix}
ac^* & ad^* \\
bc^* & bd^*
\end{pmatrix}
\end{align*}
For example, \\
\begin{align*}
\ketbra{0}{0}& =\begin{pmatrix}
1\\0
\end{pmatrix}
\begin{pmatrix}
1 & 0
\end{pmatrix}= \begin{pmatrix}
1 & 0\\
0 & 0
\end{pmatrix}
\end{align*}
\begin{align*}
\ketbra{0}{1}& =\begin{pmatrix}
1\\0
\end{pmatrix}
\begin{pmatrix}
0 & 1
\end{pmatrix}= \begin{pmatrix}
0 & 1\\
0 & 0
\end{pmatrix}
\end{align*}
\begin{align*}
\ketbra{1}{0}& =\begin{pmatrix}
0\\1
\end{pmatrix}
\begin{pmatrix}
1 & 0
\end{pmatrix}= \begin{pmatrix}
0 & 0\\
1 & 0
\end{pmatrix}
\end{align*}
\begin{align*}
\ketbra{1}{1}& =\begin{pmatrix}
0\\1
\end{pmatrix}
\begin{pmatrix}
0 & 1
\end{pmatrix}= \begin{pmatrix}
0 & 0\\
0 & 1
\end{pmatrix}
\end{align*}
Therefore any arbitrary $2\times2$ matrix $\displaystyle{A=\begin{pmatrix}
a & b\\c & d
\end{pmatrix}}$ can be written as:
\begin{align*}
A & = \begin{pmatrix}
a & b\\c & d
\end{pmatrix} = a \ketbra{0}{0} + b\ketbra{0}{1} + c\ketbra{1}{0} + d\ketbra{1}{1}
\end{align*}\newpage

\subsection{Tensor Product}
Suppose that $H_1$ and $H_2$ are two Hilbert spaces of dimension $N_1$ and $N_2$.We can put these two Hilbert spaces together to construct a larger Hilbert space. We denote this larger space by $H$ and use the tensor product operation symbol $\otimes$. So we write,
\begin{align*}
H=H_1\otimes H_2
\end{align*}
And the $$dim(H)=N_1N_2$$
Suppose $\ket{\phi}=\begin{pmatrix}
a\\b
\end{pmatrix}\in H_1$ and $\ket{\chi}=\begin{pmatrix}
c\\d
\end{pmatrix}\in H_2$, then the composite state $\ket{\psi}$ that belongs to the big Hilbert space $H=H_1\otimes H_2$ is given by the tensor product:
\begin{align*}
\ket{\psi}& =\ket{\phi}\otimes\ket{\chi}=\begin{pmatrix}
a\\b
\end{pmatrix}\otimes\begin{pmatrix}
c\\d
\end{pmatrix}\\
& = \begin{pmatrix}
a\begin{pmatrix}c\\d\end{pmatrix}\\
b\begin{pmatrix}c\\d\end{pmatrix}\\
\end{pmatrix}=\begin{pmatrix}
ac\\ad\\bc\\bd
\end{pmatrix}
\end{align*}
Similarly, if there is an operator $\hat{A_1}\in H_1$ and $\hat{A_2}\in H_2$, then the composite operator $\hat{A}$ belonging to the large Hilbert space $H=H_1\otimes H_2$ is: 
\begin{align*}
\hat{A}& =\hat{A_1}\otimes\hat{A_2}=\begin{pmatrix}
a_1 & b_1\\c_1 & d_1
\end{pmatrix}\otimes\begin{pmatrix}
a_2 & b_2\\c_2 & d_2
\end{pmatrix}\\
\renewcommand*{\arraystretch}{2.5}
& = \begin{pmatrix}
a_1\begin{pmatrix}
a_2 & b_2\\c_2 & d_2
\end{pmatrix} & b_1\begin{pmatrix}
a_2 & b_2\\c_2 & d_2
\end{pmatrix} \\
c_1\begin{pmatrix}
a_2 & b_2\\c_2 & d_2
\end{pmatrix} & d_1\begin{pmatrix}
a_2 & b_2\\c_2 & d_2
\end{pmatrix}
\end{pmatrix}\\
& = \begin{pmatrix}
a_1a_2 & a_1b_2 & b_1a_2 & b_1b_2\\
a_1c_2 & a_1d_2 & b_1c_2 & b_1d_2\\
c_1a_2 & c_1b_2 & d_1a_2 & d_1b_2\\
c_1a_2 & c_1b_2 & d_1a_2 & d_1b_2
\end{pmatrix}
\end{align*}

\section{Postulates of Quantum Mechanics}
\subsection{State Space}
Associated to any isolated physical system is a complex vector space
with inner product (that is, a Hilbert space) known as the \emph{state space} of the system. The system is completely described by its \emph{state vector}, which is a unit vector in the system’s state space.\\
\indent The simplest quantum mechanical system is the \emph{qubit}. A qubit has two-dimensional state space. Suppose $\ket{0}$ and $\ket{1}$ form an orthonormal basis for that state space. Then an arbitraty state vector in the state space can be written as:
\begin{align}\label{equ1}
\ket{\psi}& =a \ket{0} + b\ket{1}
\end{align}
where $a$ and $b$ are complex numbers. Since $\ket{\psi}$ belongs to Hilbert space, it means that $\braket{\psi|\psi}=1$ which is equivalent to $\abs{a}^2+\abs{b}^2=1$. This is often known as the \emph{normalization condition} for the state vector.
\subsection*{Geometrical representation of state vector}
\begin{figure}[H]
\centering
\begin{tikzpicture}[line cap=round, line join=round, >=Triangle]
  \clip(-2.19,-2.49) rectangle (2.66,2.58);
  \draw [shift={(0,0)}, lightgray, fill, fill opacity=0.1] (0,0) -- (56.7:0.4) arc (56.7:90.:0.4) -- cycle;
  \draw [shift={(0,0)}, lightgray, fill, fill opacity=0.1] (0,0) -- (-135.7:0.4) arc (-135.7:-33.2:0.4) -- cycle;
  \draw(0,0) circle (2cm);
  \draw [rotate around={0.:(0.,0.)},dash pattern=on 3pt off 3pt] (0,0) ellipse (2cm and 0.9cm);
  \draw (0,0)-- (0.70,1.07);
  \draw [->] (0,0) -- (0,2);
  \draw [->] (0,0) -- (-0.81,-0.79);
  \draw [->] (0,0) -- (2,0);
  \draw [dotted] (0.7,1)-- (0.7,-0.46);
  \draw [dotted] (0,0)-- (0.7,-0.46);
  \draw (-0.08,-0.3) node[anchor=north west] {$\varphi$};
  \draw (0.01,0.9) node[anchor=north west] {$\theta$};
  \draw (-1.01,-0.72) node[anchor=north west] {$\mathbf {\hat{x}=\ket{+}}$};
  \draw (2.07,0.3) node[anchor=north west] {$\mathbf {\hat{y}}$};
  \draw (-0.5,2.6) node[anchor=north west] {$\mathbf {\hat{z}=\ket{0}}$};
  \draw (-0.4,-2) node[anchor=north west] {$-\mathbf {\hat{z}=\ket{1}}$};
  \draw (0.4,1.65) node[anchor=north west] {$|\psi\rangle$};
  \scriptsize
  \draw [fill] (0,0) circle (1.5pt);
  \draw [fill] (0.7,1.1) circle (0.5pt);
\end{tikzpicture}
\caption{Bloch sphere representation of a State Vector}\label{bloch}
\end{figure}
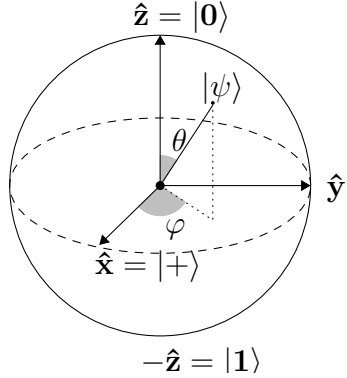

Two-level quantum mechanical system (qubit) can be represented geometrically on a unit sphere called \emph{Bloch Sphere} which is named after the physicist Felix Bloch. Since the state $\ket{\psi}$ in Eq.\ref{equ1} is written as a superposition of a basis vector where the coefficient of each basis vector is a complex number. It means that state $\ket{\psi}$ is described by four real numbers. However, only the relative phase\footnote[1]{the phase of the quantum system is not directly measurable} between the two coefficient has physical meaning. So we can take the coefficient of $\ket{0}$ to be real and positive which allow the state to be described by only three real numbers giving rise to three dimension of bloch sphere. Mathematically, 
\begin{align*}
\ket{\psi}& =a \ket{0} + b\ket{1}
\end{align*}
Since $a$ and $b$ are complex numbers. So we can write them in polar form.
\begin{align*}
\ket{\psi} &= \abs{a}e^{i\phi_1}\ket{0}+\abs{b}e^{i\phi_2}\ket{1}\\
& = e^{i\phi_1}\left(\abs{a}\ket{0}+\abs{b}e^{i(\phi_2-\phi_1)}\ket{1}\right)
\end{align*}
Normalization condition, $\displaystyle{\abs{a}^2+\abs{b}^2=1}$, allows me to take $\displaystyle{\abs{a}=\cos\left(\frac{\theta}{2}\right)}$ and $\abs{b}=\sin\left(\frac{\theta}{2}\right)$. Also take $\phi_2-\phi_1=\varphi$. Therefore, the state vector can be written as:
\begin{align}\label{eq2}
\ket{\psi}& = e^{i\phi_1}\left[\cos\left(\frac{\theta}{2}\right)\ket{0}+e^{i\varphi}\sin\left(\frac{\theta}{2}\right)\ket{1} \right]
\end{align}
Here $e^{i\phi_1}$ is global phase. Since we are concerned with only information which we can extract by doing measurement, presence of global phase has absolutely no effect on measurement. For instance, probability of $i\ket{0}$ and $\ket{0}$ are the same, because upon taking the modulus square, global phase vanishes.\newpage
So, finally our state vector can now be represented as a function of only two parameters $(\theta,\varphi)$ where $0\leq\varphi\leq2\pi$ and $0\leq\theta\leq\pi$.
\begin{align}\label{eq3ch1}
\ket{\psi}& = \cos\left(\frac{\theta}{2}\right)\ket{0}+e^{i\varphi}\sin\left(\frac{\theta}{2}\right)\ket{1}
\end{align}
Representation of state vector in Eq.\ref{eq3ch1} is illustrated in Fig.\ref{bloch}.
Here is a list of few examples.
\begin{enumerate}[label=\roman*)]
\item For $\theta=0$, $\ket{\psi}=\ket{0}$. Point on bloch sphere where $\theta=0$ represent $\ket{0}$
\item Similarly, for $\theta=\pi$ and $\varphi=0$, $\ket{\psi}=\ket{1}$
\item For $\displaystyle{\theta=\frac{\pi}{2}}$ and $\varphi=0$, $\displaystyle{\ket{\psi}=\frac{1}{\sqrt{2}}\ket{0}+\frac{1}{\sqrt{2}}\ket{1}=\ket{+}}$
\item At last, for $\displaystyle{\theta=\frac{\pi}{2}}$ and $\varphi=\pi$, $\displaystyle{\ket{\psi}=\frac{1}{\sqrt{2}}\ket{0}-\frac{1}{\sqrt{2}}\ket{1}=\ket{-}}$\footnote[1]{$\{\ket{+},\ket{-}\}$ also form a basis, because both the vectors are normalized and orthogonal to each other i.e., $\braket{+|-}=0$}
\end{enumerate}
Some of these states are represented on bloch sphere in Fig.\ref{bloch}.\\
Let's have a look an action of few gates on the state in Eq.\ref{eq3} on bloch sphere.
\begin{enumerate}[label=(\roman*)]
\item $X$ gate can be thought of as a rotation by $\pi~$radians around $x$-axis of bloch sphere.
Representation of action of $X$ gate on $\ket{\psi}$ is illustrated on bloch sphere in Fig.\ref{Xbloch}.
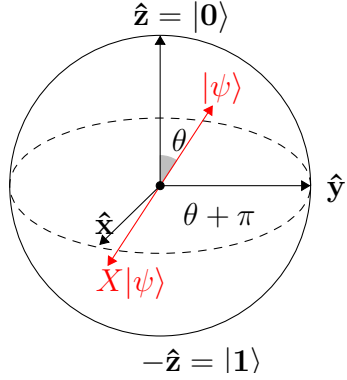
\begin{figure}[H]
\centering
\begin{tikzpicture}[line cap=round, line join=round, >=Triangle]
  \clip(-2.19,-2.49) rectangle (2.66,2.58);
     
     \draw [shift={(0,0)}, lightgray, fill, fill opacity=1] (0,0) -- (56.7:0.4) arc (56.7:90.:0.4) -- cycle;
 
  \draw(0,0) circle (2cm);
  \draw [rotate around={0.:(0.,0.)},dash pattern=on 3pt off 3pt] (0,0) ellipse (2cm and 0.9cm);
  \draw[color=red] [->] (0,0)-- (0.70,1.07); 
  \draw[color=red] [->] (0,0)-- (-0.70,-1.07); 
  \draw [->] (0,0) -- (0,2); 
  \draw [->] (0,0) -- (-0.81,-0.79); 
  \draw [->] (0,0) -- (2,0); 
  \draw (0.01,0.9) node[anchor=north west] {$\theta$}; 
  \draw (0.16,-0.09) node[anchor=north west] {$\theta+\pi$}; 
  \draw (-1.01,-0.2) node[anchor=north west] {$\mathbf {\hat{x}}$}; 
  \draw (2.07,0.3) node[anchor=north west] {$\mathbf {\hat{y}}$}; 
  \draw (-0.5,2.6) node[anchor=north west] {$\mathbf {\hat{z}=|0\rangle}$}; 
  \draw (-0.4,-2) node[anchor=north west] {$-\mathbf {\hat{z}=|1\rangle}$}; 
  \draw (0.4,1.65) node[color=red][anchor=north west] {$|\psi\rangle$}; 
  \draw (-1,-1.65) node[color=red][anchor=south west] {$X|\psi\rangle$}; 
  \scriptsize
  \draw [fill] (0,0) circle (1.5pt);
\end{tikzpicture}
\caption{State is rotated around $x$-axis by $\pi~$radians after applying $X$ gate}\label{Xbloch}
\end{figure}
\item $Y$ gate is rotation by $\pi~$radians around y-axis of bloch sphere.
Representation of action of $Y$ gate on $\ket{\psi}$ is illustrated on bloch sphere in Fig.\ref{Y-bloch}.
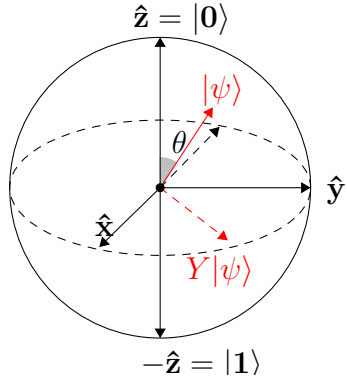
\begin{figure}[H]
\centering
\begin{tikzpicture}[line cap=round, line join=round, >=Triangle]
  \clip(-2.19,-2.49) rectangle (2.66,2.58);

  \draw [shift={(0,0)}, lightgray, fill, fill opacity=1] (0,0) -- (56.7:0.4) arc (56.7:90.:0.4) -- cycle;
  \draw(0,0) circle (2cm);
  \draw [rotate around={0.:(0.,0.)},dash pattern=on 3pt off 3pt] (0,0) ellipse (2cm and 0.9cm);
  \draw[color=red] [->] (0,0)-- (0.70,1.07); 
    \draw[color=red,dashed] [->] (0,0)-- (0.90,-0.7); 
  \draw [->] (0,0) -- (0,2); 
  \draw [->] (0,0) -- (0,-2); 
  \draw [->] (0,0) -- (-0.81,-0.79); 
  \draw [->,dashed] (0,0) -- (0.79,0.81); 
  \draw [->] (0,0) -- (2,0); 
  \draw (0.01,0.9) node[anchor=north west] {$\theta$}; 
  \draw (-1.01,-0.2) node[anchor=north west] {$\mathbf {\hat{x}}$}; 
  \draw (2.07,0.3) node[anchor=north west] {$\mathbf {\hat{y}}$}; 
  \draw (-0.5,2.6) node[anchor=north west] {$\mathbf {\hat{z}=|0\rangle}$}; 
  \draw (-0.4,-2) node[anchor=north west] {$-\mathbf {\hat{z}=|1\rangle}$}; 
  \draw (0.4,1.65) node[color=red][anchor=north west] {$|\psi\rangle$}; 
  \draw (0.2,-0.75) node[color=red][anchor=north west] {$Y|\psi\rangle$}; 
  \scriptsize
  \draw [fill] (0,0) circle (1.5pt);
\end{tikzpicture}
\caption{State is rotated around $y$-axis by $\pi~$radians after applying $Y$ gate}\label{Y-bloch}
\end{figure}
\item $Z$ gate is rotation by $\pi~$radians around $z$-axis of bloch sphere:
\begin{align*}
Z\ket{\psi} & = \cos\left(\frac{\theta}{2}\right)Z\ket{0}+e^{i\varphi}\sin\left(\frac{\theta}{2}\right)Z\ket{1}\\
& = \cos\left(\frac{\theta}{2}\right)\ket{0}-e^{i\varphi}\sin\left(\frac{\theta}{2}\right)\ket{1}\\
& = \cos\left(\frac{\theta}{2}\right)\ket{0}+e^{i(\pi+\varphi)}\sin\left(\frac{\theta}{2}\right)\ket{1}
\end{align*}
Representation of action of $Z$ gate on $\ket{\psi}$ is illustrated on bloch sphere in Fig.\ref{Z-bloch}.
\begin{figure}[H]
\centering
\begin{tikzpicture}[line cap=round, line join=round, >=Triangle]
  \clip(-2.19,-2.49) rectangle (2.66,2.58);

  \draw [shift={(0,0)}, lightgray, fill, fill opacity=1] (0,0) -- (56.7:0.4) arc (56.7:90.:0.4) -- cycle;
  \draw(0,0) circle (2cm);
  \draw [rotate around={0.:(0.,0.)},dash pattern=on 3pt off 3pt] (0,0) ellipse (2cm and 0.9cm);
  \draw[color=red] [->] (0,0)-- (0.70,1.07); 
    \draw[color=red] [->] (0,0)-- (-0.70,1.07); 
  \draw [->] (0,0) -- (0,2); 
  \draw [->] (0,0) -- (-0.81,-0.79); 
  \draw [->] (0,0) -- (2,0); 
  \draw (0.01,0.9) node[anchor=north west] {$\theta$}; 
  \draw (-1.01,-0.2) node[anchor=north west] {$\mathbf {\hat{x}}$}; 
  \draw (2.07,0.3) node[anchor=north west] {$\mathbf {\hat{y}}$}; 
  \draw (-0.5,2.6) node[anchor=north west] {$\mathbf {\hat{z}=|0\rangle}$}; 
  \draw (-0.4,-2) node[anchor=north west] {$-\mathbf {\hat{z}=|1\rangle}$}; 
  \draw (0.4,1.65) node[color=red][anchor=north west] {$|\psi\rangle$}; 
  \draw (-1.2,1.65) node[color=red][anchor=north west] {$Z|\psi\rangle$}; 
  \scriptsize
  \draw [fill] (0,0) circle (1.5pt);
\end{tikzpicture}
\caption{State is rotated around $z$-axis by $\pi~$radians after applying $Z$ gate}\label{Z-bloch}
\end{figure}
\item $H$ gate can be expressed as a $\pi/2~$radians rotation around $y$-axis, followed by $\pi~$radians rotation around $x$-axis.
\begin{align*}
H\ket{\psi} & = \cos\left(\frac{\theta}{2}\right)H\ket{0} + e^{i\varphi}\sin\left(\frac{\theta}{2}\right)H\ket{1} \\
& = \cos\left(\frac{\theta}{2}\right)\frac{\ket{0}+\ket{1}}{\sqrt{2}} + e^{i\varphi}\sin\left(\frac{\theta}{2}\right)\frac{\ket{0}-\ket{1}}{\sqrt{2}}\\
& = \frac{1}{\sqrt{2}}\left[\cos\left(\frac{\theta}{2}\right) + e^{i\varphi}\sin\left(\frac{\theta}{2}\right) \right]\ket{0} + \frac{1}{\sqrt{2}}\left[\cos\left(\frac{\theta}{2}\right) - e^{i\varphi}\sin\left(\frac{\theta}{2}\right) \right]\ket{1}
\end{align*}
Representation of action of $H$ gate on $\ket{\psi}$ is illustrated on bloch sphere in Fig.\ref{H-bloch}.
\begin{figure}[H]
\centering
\begin{tikzpicture}[line cap=round, line join=round, >=Triangle]
  \clip(-2.19,-2.49) rectangle (2.66,2.58);

  \draw [shift={(0,0)}, lightgray, fill, fill opacity=1] (0,0) -- (56.7:0.4) arc (56.7:90.:0.4) -- cycle;
  \draw(0,0) circle (2cm);
  \draw [rotate around={0.:(0.,0.)},dash pattern=on 3pt off 3pt] (0,0) ellipse (2cm and 0.9cm);
  \draw[color=red] [->] (0,0)-- (0.70,1.07); 
  \draw[color=blue,dashed] [->] (0,0)-- (0.60,-1.34); 
      \draw[color=red] [->] (0,0)-- (-0.70,1.54); 
  \draw [->] (0,0) -- (0,2); 
  \draw [->] (0,0) -- (-0.81,-0.79); 
  \draw [->] (0,0) -- (2,0); 
  \draw (-0.08,-0.3) node[anchor=north west] {$\varphi$}; 
  \draw (0.01,0.9) node[anchor=north west] {$\theta$}; 
  \draw (-1.01,-0.2) node[anchor=north west] {$\mathbf {\hat{x}}$}; 
  \draw (2.07,0.3) node[anchor=north west] {$\mathbf {\hat{y}}$}; 
  \draw (-0.5,2.6) node[anchor=north west] {$\mathbf {\hat{z}=|0\rangle}$}; 
  \draw (-0.4,-2) node[anchor=north west] {$-\mathbf {\hat{z}=|1\rangle}$}; 
  \draw (0.4,1.65) node[color=red][anchor=north west] {$|\psi\rangle$}; 
  \draw (0.35,-1.25) node[color=blue][anchor=north west] {$R_y\left(\frac{\pi}{2}\right)|\psi\rangle$}; 
  \draw (-1.8,2.4) node[color=red][anchor=north west] {$H|\psi\rangle$}; 
  \scriptsize
  \draw [fill] (0,0) circle (1.5pt);
  \draw [fill] (0.7,1.1) circle (0.5pt);
\end{tikzpicture}
\caption{State is rotated around $y$-axis by $\pi/2~$radians which is represented as dotted line followed by $\pi~$radians around $x$-axis}\label{H-bloch}
\end{figure}
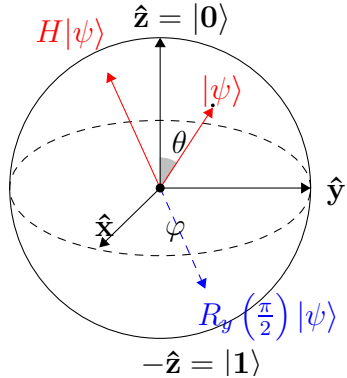
\end{enumerate}
\subsection{Evolution}
How does the state, $\ket{\psi}$, of a quantum mechanical system change with time? The following postulate gives a prescription for the description of such state changes.\\
\indent The evolution of a closed quantum system is described by a \emph{unitary transformation}. That is, the state $\ket{\psi}$ of the system at time $t_1$ is related to the state $\ket{\psi'}$ of the system at time $t_2$ by a unitary operator $U$ which depends only on times $t_1$ and $t_2$:
\begin{align}\label{pos2}
\ket{\psi'} & = U\ket{\psi}
\end{align}
\indent So, according to the postulate 2, the state vector transform in a such way that it remains normalized which is only possible if transformation occurs via \emph{unitary operators}.
\begin{example}{}{label}
An operator $U$ is unitary if $U^{-1}=U^{\dagger}$ where $U^{\dagger}$ is complex conjugate of $U$. Prove that norm is preserved under unitary transformation.
\textbf{Solution: }We know that 
\begin{align*}
\ket{\psi'}& = U\ket{\psi} \\
\bra{\psi'}& = (U\ket{\psi})^*=\bra{\psi}U^{\dagger}\\
\Rightarrow \braket{\psi'|\psi'}& = \bra{\psi}U^{\dagger}U\ket{\psi}\\
& = \bra{\psi}I\ket{\psi}\ \ \ \because U\ \text{is unitary}\\
& = \braket{\psi|\psi}\\
\Rightarrow \braket{\psi'|\psi'}& =1\ \ \Rightarrow \text{Norm is preserved.}
\end{align*}
\end{example}
Let's look at a few examples of unitary operators on a single qubit. The pauli matrices are all unitary operators.
\begin{enumerate}[label=\roman*)]
\item $X$ gate takes $\ket{0}$ to $\ket{1}$ and $\ket{1}$ to $\ket{0}$. This gate is classical analogue of NOT gate. It is sometimes referred to as the \emph{bitflip} gate. On bloch sphere, $X$ gate rotates the qubit about $x$-axis by $\pi~$radians.
\item $Y$ gates takes $\ket{0}$ to $i\ket{1}$ and $\ket{1}$ to $-i\ket{0}$. This gate adds phase and flips the qubit at the same time.
\item $Z$ gate leaves $\ket{0}$ invariant and transforms $\ket{1}$ to $-\ket{1}$. The negative sign is called phase factor. This gate is referred as \emph{phase flip}.
\item Another interesting unitary operator is \emph{Hadamard gate}. which is denoted by $H$. This gate takes qubits $\ket{0}$ and $\ket{1}$ to superposition state. Mathematically, 
\begin{align*}
H\ket{0}& =\frac{1}{\sqrt{2}}\big(\ket{0}+\ket{1})\\
H\ket{1}& =\frac{1}{\sqrt{2}}\big(\ket{0}-\ket{1})
\end{align*}
Matrix representation of Hadamard gate is:
\begin{align*}
H& = \frac{1}{\sqrt{2}}\begin{pmatrix}
1 & 1\\
1 & -1
\end{pmatrix}
\end{align*}
\end{enumerate}
\begin{example}{}{label}
Verify that Pauli matrices and Hadamard gates are unitary.\\\\
\textbf{Solution: }We know that an arbitrary operator $A$ is unitary if $A^{\dagger}A=AA^{\dagger}=I$. \begin{enumerate}[label=\roman*)]
\item \begin{align*}
X & = \begin{pmatrix}
0 & 1\\1 & 0
\end{pmatrix}\\
\Rightarrow X^{\dagger} & = (X^T)^* = \begin{pmatrix}
0 & 1\\1 & 0\end{pmatrix}\\
\therefore XX^{\dagger} & = X^{\dagger}X = \begin{pmatrix}
0 & 1\\1 & 0
\end{pmatrix}\begin{pmatrix}
0 & 1\\1 & 0
\end{pmatrix} = \begin{pmatrix}
1 & 0\\0 & 1
\end{pmatrix} = I
\end{align*}
\item \begin{align*}
Y & = \begin{pmatrix}
0 & -i\\i & 0
\end{pmatrix}\\
\Rightarrow Y^{\dagger} & = (Y^T)^*=\begin{pmatrix}
0 & i\\-i & 0
\end{pmatrix}^* = \begin{pmatrix}
0 & -i\\i & 0
\end{pmatrix}\\
\therefore YY^{\dagger} &=Y^{\dagger}Y=\begin{pmatrix}
0 & -i\\i & 0
\end{pmatrix}\begin{pmatrix}
0 & -i\\i & 0
\end{pmatrix}=\begin{pmatrix}
1 & 0\\0 & 1
\end{pmatrix}=I
\end{align*}
\item \begin{align*}
Z & = \begin{pmatrix}
1 & 0\\0 & -1
\end{pmatrix}\\
\Rightarrow Z^{\dagger} & = (Z^T)^*=\begin{pmatrix}
1 & 0\\0 & -1
\end{pmatrix}\\
\therefore ZZ^{\dagger} &=Z^{\dagger}Z=\begin{pmatrix}
1 & 0\\0 & -1
\end{pmatrix}\begin{pmatrix}
1 & 0\\0 & -1
\end{pmatrix}=\begin{pmatrix}
1 & 0\\0 & 1
\end{pmatrix}=I
\end{align*}
\item \begin{align*}
H & =\frac{1}{\sqrt{2}}\begin{pmatrix}
1 & 1\\1 & -1
\end{pmatrix} \\
\Rightarrow H^{\dagger} & = \frac{1}{\sqrt{2}}\begin{pmatrix}
1 & 1\\1 & -1
\end{pmatrix} \\
\therefore HH^{\dagger} & = H^{\dagger}H=\frac{1}{2}\begin{pmatrix}
1 & 1\\1 & -1
\end{pmatrix}\begin{pmatrix}
1 & 1\\1 & -1
\end{pmatrix} = \begin{pmatrix}
1 & 0\\0 & 1
\end{pmatrix} \\
\end{align*}
\end{enumerate}
\end{example}

\section{Gates in Outer-Product Form}
An operator $\hat{A}$ is a mathematical rule that when applied to ket $\ket{\psi}$ transforms $\ket{\psi}$ to another ket $\ket{\psi'}$ in the same space and when applied on bra $\bra{\psi}$ transform to another bra $\bra{\psi'}$ in the same space. Mathematically
\begin{align*}
\hat{A}\ket{\psi} & = \ket{\psi'} , \bra{\psi}\hat{A}^\dagger = \bra{\psi'}
\end{align*}
An operator is $n\times n$ matrix where $n$ is dimension of vector space. So for two level system: qubits, operators are $2\times2$ matrices. In quantum computing, operators are \emph{unitary} and are called \emph{gates}.
\begin{tcolorbox}
\begin{center}
\large\textbf{Unitary Operator}
\end{center}
An operator $\hat{U}$ will be unitary, if its inverse is equal to its hermitian conjugate. Mathematically,
\begin{align*}
\text{if }\hat{U}^{-1} & = \hat{U^{\dagger}}\\
\Rightarrow \hat{U}\hat{U^{\dagger}}  & = \hat{U^{\dagger}}\hat{U}=\hat{I}\\
\Rightarrow \hat{U}\ \text{is Unitary}
\end{align*}
\end{tcolorbox}
\noindent Let us take an arbitrary gate $G$:
\begin{align}\label{operatormatrix}
G & = \begin{pmatrix}
g_{11} & g_{12} \\\\
g_{21} & g_{22} 
\end{pmatrix}
\end{align}
Decomposing this gate gives us: 
\begin{align*}
G & = g_{11}\begin{pmatrix}
1 & 0\\
0 & 0
\end{pmatrix} + g_{12}\begin{pmatrix}
0 & 1\\
0 & 0
\end{pmatrix} + g_{21}\begin{pmatrix}
0 & 0\\
1 & 0
\end{pmatrix} + g_{22}\begin{pmatrix}
0 & 0\\
0 & 1
\end{pmatrix}
\end{align*}
This means that the set of matrices: $\displaystyle{\bigg\{\begin{pmatrix}
1 & 0\\0 & 0
\end{pmatrix},\begin{pmatrix}
0 & 1\\0 & 0
\end{pmatrix},\begin{pmatrix}
0 & 0\\1 & 0
\end{pmatrix},\begin{pmatrix}
0 & 0\\0 & 1
\end{pmatrix}\bigg\}}$ form a basis for gates since we witnessed that each gate can be written as a linear combination of such matrices. Most importantly, these matrices are nothing but the outer product\footnote{Outer product is simply the matrix multiplication of \emph{column vector} and \emph{row vector}. For example, if $\ket{u}=\begin{pmatrix}
a\\b
\end{pmatrix}$ and $\ket{v}=\begin{pmatrix}
c\\d
\end{pmatrix}$ then outer product $\ketbra{u}{v}=\begin{pmatrix}
a\\b
\end{pmatrix}\begin{pmatrix}
c^* & d^*
\end{pmatrix}=\begin{pmatrix}
ac^* & ad^* \\ bc* & bd^*
\end{pmatrix}$ } of basis vectors $\{\ket{0},\ket{1}\}$.
\begin{align*}
\ketbra{0}{0} & = \begin{pmatrix}
1\\0
\end{pmatrix}\begin{pmatrix}
1 & 0
\end{pmatrix} = \begin{pmatrix}
1 & 0\\
0 & 0
\end{pmatrix}\\
\ketbra{0}{1} & =\begin{pmatrix}
1\\0
\end{pmatrix}\begin{pmatrix}
0 & 1
\end{pmatrix} = \begin{pmatrix}
0 & 1\\
0 & 0
\end{pmatrix}\\
\ketbra{1}{0} & = \begin{pmatrix}
0\\1
\end{pmatrix}\begin{pmatrix}
1 & 0
\end{pmatrix} = \begin{pmatrix}
0 & 0\\
1 & 0
\end{pmatrix}\\
\ketbra{1}{1} & =\begin{pmatrix}
0\\1
\end{pmatrix}\begin{pmatrix}
0 & 1
\end{pmatrix} = \begin{pmatrix}
0 & 0 \\
0 & 1
\end{pmatrix}
\end{align*}
So, in Dirac notation any arbitrary gate can be written as: 
\begin{align}\label{operatorDirac}
G & = g_{11}\ketbra{0}{0} + g_{12}\ketbra{0}{1} + g_{21}\ketbra{1}{0} + g_{22}\ketbra{1}{1}
\end{align}
\begin{exercise}{}{label}
Express $\displaystyle{X=\begin{pmatrix}
0 & 1\\1 & 0
\end{pmatrix}}$ , $\displaystyle{Y=\begin{pmatrix}
0 & -i\\ i & 0
\end{pmatrix}}$ and $\displaystyle{Z = \begin{pmatrix}
1 & 0 \\ 0 & -1
\end{pmatrix}}$ in Dirac Notation and find $\displaystyle{X\ket{0},X\ket{1},Y\ket{0},Y\ket{1},Z\ket{0},Z\ket{1}}$.\\\\
\textbf{Hint: }Use Eq.\ref{operatormatrix} and Eq.\ref{operatorDirac} and use the fact that $\displaystyle{(\ketbra{a}{b})\ket{b}=\ket{a}}$ and $\displaystyle{(\ketbra{a}{b})\ket{a}=0}$, considering that $\ket{a}$ and $\ket{b}$ are orthonormal.
\end{exercise}

\subsection{Hadamard Gate}
Matrix representation of Hadamard gate is given by: 
\begin{align*}
H & = \frac{1}{\sqrt{2}} \begin{pmatrix}
1 & 1\\
1 & -1
\end{pmatrix}
\end{align*}
In Dirac notaion, it can be written as: 
\begin{align*}
H & = \frac{1}{\sqrt{2}}\big( \ketbra{0}{0} + \ketbra{0}{1} + \ketbra{1}{0} -\ketbra{1}{1}\big)
\end{align*}
Its action on the basis vectors is given by: 
\begin{align*}
H\ket{0} & = \frac{1}{\sqrt{2}}\big( \ketbra{0}{0} + \ketbra{0}{1} + \ketbra{1}{0} -\ketbra{1}{1}\big) \ket{0}\\
& = \frac{1}{\sqrt{2}}\left(\ket{0}+\ket{1}\right)\\
& = \ket{+} \\
H\ket{1} & = \frac{1}{\sqrt{2}}\big( \ketbra{0}{0} + \ketbra{0}{1} + \ketbra{1}{0} -\ketbra{1}{1}\big) \ket{1}\\
& = \frac{1}{\sqrt{2}}\left(\ket{0}-\ket{1}\right)\\
& = \ket{-} \\
\end{align*}
So we see that Hadamard gate actually transform the basis vectors into the superposition state.
\begin{exercise}{}{label}
Explain the action of Hadamard Gate on $\ket{+}$ and $\ket{-}$ states.
\end{exercise}
\subsection{Phase Gate}
Matrix representation of phase gate is given by: 
\begin{align*}
S & = \begin{pmatrix}
1 & 0\\
0 & i
\end{pmatrix} = \begin{pmatrix}
1 & 0\\
0 & e^{i\pi/2}
\end{pmatrix}
\end{align*}
In Dirac notation, it can be written as: 
\begin{align*}
S & = \ketbra{0}{0} + i \ketbra{1}{1}
\end{align*}
Its action on the basis vectors is given by:
\begin{align*}
S\ket{0} & = \left(\ketbra{0}{0} + i \ketbra{1}{1}\right)\ket{0}\\
& = \ket{0}\\
S\ket{1} & = \left(\ketbra{0}{0} + i \ketbra{1}{1}\right)\ket{1}\\
& = i\ket{1}
\end{align*}
So we see that, it does nothing to $\ket{0}$ but adds a phase to $\ket{1}$.

\begin{exercise}{}{label}
Express the rotation gates in Dirac notation.
\end{exercise}

\begin{example}{}{label}
For the single qubit gates, using the Dirac notation for the gates in outer product form, show that $Z=HXH$ and $X=HZH$.\\\\
\textbf{Solution: }Writing the gates in Dirac notation and computing the inner products:
\begin{equation*}\label{eq1}
\begin{split}
HXH & =\frac{1}{\sqrt{2}}\left(\ketbra{0}{0}+\ketbra{0}{1}+\ketbra{1}{0}-\ketbra{1}{1}\right)\left(\ketbra{0}{1}+\ketbra{1}{0}\right) H\\
& = \frac{1}{\sqrt{2}}\left(\ketbra{0}{1}+\ketbra{0}{0}+\ketbra{1}{1}-\ketbra{1}{0}\right)H \ \ \  \because \langle{i}|{j}\rangle =\delta_{ij} \\
& = \frac{1}{2}\left(\ketbra{0}{1}+\ketbra{0}{0}+\ketbra{1}{1}-\ketbra{1}{0}\right)\left(\ketbra{0}{0}+\ketbra{0}{1}+\ketbra{1}{0}-\ketbra{1}{1}\right)\\
& =  \frac{1}{2}\left(2\ketbra{0}{0}-\cancel{\ketbra{0}{1}}+\cancel{\ketbra{0}{1}}+\ketbra{1}{0}-2\ketbra{1}{1}\right)\\
& = \left(\ketbra{0}{0}-\ketbra{1}{1}\right)\\
& = Z
\end{split}
\end{equation*}
Hence Proved!!!\\
$X=HZH$ can be proved in a similar fashion.
\end{example}

\subsection*{Exponentiation}
\noindent Consider a matrix $U$ is hermitian and unitary as well, then we can write
\begin{align*}
e^{i\eta U} & = \cos \eta I - i \sin \eta U
\end{align*}
It can be proved in the following way.
\begin{align*}
e^{-i\eta U} & = I - i\eta U +(-1)^2\frac{\eta^2}{2!}U^2+(-i)^3\frac{\eta^3}{3!}U^3 + (-i)^4\frac{\eta^4}{4!}U^4 + \cdots
\end{align*}
Since, $U^2 = UU = UU^{\dagger} = I$, thus
\begin{align*}
e^{-i\eta U} & = \left(1 - \frac{\eta^2}{2!} + \frac{\eta^4}{4!} - \cdots \right)I - i\left(\eta - \frac{\eta^3}{3!}+ \frac{\eta^5}{5!} - \cdots \right)U \\
& = \cos \eta I - i \sin \eta U
\end{align*}
\subsection*{Rotation gates}
\noindent By exponentiating a given matrix, we can come up with more gates. In fact we can create
rotation operators to represent rotation about the x, y, and z axes on the Bloch sphere by
exponentiating the Pauli matrices. These are given by:
\begin{align*}
R_x(\eta) & = e^{-i\eta X/2} =\cos \left(\frac{\eta}{2}\right) I - i \sin \left(\frac{\eta}{2}\right) X = \begin{pmatrix}
\cos \left(\frac{\eta}{2}\right) & -i \sin\left(\frac{\eta}{2}\right) \\\\
-i \sin \left(\frac{\eta}{2}\right) & \cos\left(\frac{\eta}{2}\right)
\end{pmatrix} \\
R_y(\eta) & = e^{-i\eta Y/2} =\cos\left(\frac{\eta}{2}\right) I - i \sin\left(\frac{\eta}{2}\right) Y =  \begin{pmatrix}
\cos \left(\frac{\eta}{2}\right) & - \sin\left(\frac{\eta}{2}\right) \\\\
 \sin \left(\frac{\eta}{2}\right) & \cos\left(\frac{\eta}{2}\right)
\end{pmatrix} \\
R_z(\eta) & = e^{-i\eta Z/2} =\cos\left(\frac{\eta}{2}\right) I - i \sin \left(\frac{\eta}{2}\right) Z = \begin{pmatrix}
e^{-i\eta/2} & 0 \\
0 & e^{i\eta/2}
\end{pmatrix}
\end{align*}
\subsection*{$Z-Y$ decomposition for a single qubit}
\noindent An arbitrary unitary operator on a single qubit can be written in many ways as a combination of rotations together with global phase shifts on the qubit. Suppose $U$ is a unitary operation on a single qubit. Then there exist real numbers $\alpha$, $\beta$, $\gamma$ and $\delta$ such that
\begin{align}\label{zy-theorem}
U & = e^{i\alpha} R_z(\beta)R_y(\gamma)R_z(\delta)
\end{align}
\textbf{Proof: }Since $U$ is a unitary matrix, it means
\begin{align*}
UU^{\dagger} & = \begin{pmatrix}
a & b\\
c & d
\end{pmatrix}\begin{pmatrix}
a^* & c^* \\
b^* & d^*
\end{pmatrix} = I \\
\Rightarrow & \begin{pmatrix}
\abs{a}^2+\abs{b}^2 & ac^* + bd^* \\
ca^*+db^* & \abs{c}^2+\abs{d}^2
\end{pmatrix}  = \begin{pmatrix}
1 & 0 \\
0 & 1
\end{pmatrix}
\end{align*}
So, the unitary matrix should be such that
\begin{align*}
\abs{a}^2+\abs{b}^2 & = 1 \\
\abs{c}^2 + \abs{d}^2 & = 1 \\
ac^* + bd^* & = 0 \\
ca^*+db^* & = 0
\end{align*}
It means that rows and columns of $U$ are orthonormal, from which it follows that there exist real numbers $\alpha$, $\beta$, $\gamma$ and $\delta$ such that
\begin{align}\label{z-y-decomposition}
U & = e^{i\alpha} \begin{pmatrix}
e^{-i\beta/2}e^{-i\delta/2}\cos \frac{\gamma}{2} & -e^{-i\beta/2}e^{i\delta/2}\sin \frac{\gamma}{2} \\\\
e^{i\beta/2}e^{-i\delta/2}\sin \frac{\gamma}{2} & e^{i\beta/2}e^{i\delta/2}\cos \frac{\gamma}{2}
\end{pmatrix}
\end{align}
The Eq.(\ref{z-y-decomposition}) follows from Eq. (\ref{zy-theorem}).
\begin{exercise}{}{1}
Given the matrix representation of rotation gates, prove that Eq. (\ref{zy-theorem}) is indeed equal to Eq. (\ref{z-y-decomposition}).
\end{exercise}

\noindent\textbf{Corollary: }A general unitary operator can be written as:
\begin{align*}
U & = e^{i\alpha} AXBXC
\end{align*}
where $A$, $B$ and $C$ are unitary operators such that $ABC=I$. We can set
\begin{align*}
A & \equiv R_z(\beta)R_y(\gamma/2) \\
B & \equiv R_y(-\gamma/2)R_z\left(- \frac{\delta + \beta}{2} \right) \\
C & \equiv R_z\left(\frac{\delta - \beta}{2} \right)
\end{align*}
Clearly, $ABC=I$.
\begin{example}{}{label}
Prove that $AXBXC = R_z(\beta)R_y(\gamma)R_z(\delta) $ \\
\noindent\textbf{Solution: }\\
\begin{align*}
XBX & = X R_y(-\gamma/2)R_z\left(- \frac{\delta + \beta}{2} \right) X \\
& = X R_y(-\gamma/2) XX R_z\left(- \frac{\delta + \beta}{2} \right) X
\end{align*}
Since, $XX=I$ and $XR_y(\theta)X=R_n(-\theta)$, thus, 
\begin{align*}
XBX & = R_y(\gamma/2) R_z\left(\frac{\delta+\beta}{2}\right)
\end{align*}
Thus, 
\begin{align*}
AXBXC & = R_z(\beta)R_y(\gamma/2)R_y(\gamma/2)R_z\left(\frac{\delta+\beta}{2}\right)R_z\left(\frac{\delta-\beta}{2}\right) \\
& = R_z(\beta) R_y(\gamma) R_z(\delta)
\end{align*}
\end{example}
\subsection{Quantum Measurements}
Quantum measurements are described by a collection $\{M_m\}$ of \emph{measurement operators}. These are operators acting on the state space of the space the system being measured. The only outcome of such measurement is one of the eigenvalues $m$ of operator $M_m$. If the state of the quantum system prior measurement is $\ket{\psi}$ then probability that result $m$ occurs is:
\begin{align*}
p(m) & = \bra{\psi}M_m^{\dagger}M_m\ket{\psi}
\end{align*}
and the state of the system after the measurement is
\begin{align*}
\frac{M_m\ket{\psi}}{\sqrt{\bra{\psi}M_m^{\dagger}M_m\ket{\psi}}}
\end{align*}
The measurement operators satisfy \emph{completeness equation},
\begin{align*}
\sum_mM_m^{\dagger}M_m& =I
\end{align*}
The completeness relation states the fact that sum of probabilities is equal to one:
\begin{align*}
\sum_mp(m) & =\sum_m\bra{\psi}M_m^{\dagger}M_m\ket{\psi}=1
\end{align*}
For example, Let we have a state $\ket{\psi}=a\ket{0}+b\ket{1}$. The measurement operator that corresponds to the measurement of state being in $\ket{0}$ is $M_0=\ketbra{0}{0}$. Similarly, for the state being in $\ket{1}$ is $M_1=\ketbra{1}{1}$. Then the probability of obtaining measurement outcome 0 is:
\begin{align*}
p(0) & = \bra{\psi}M_0^{\dagger}M_0\ket{\psi}=\bra{\psi}M_0\ket{\psi}=\abs{a}^2
\end{align*}
$\because M^{\dagger}_0M_0=(\ketbra{0}{0})(\ketbra{0}{0})=\ketbra{0}{0}$\\
Similarly, the probability of obtaining the measurement outcome 1 is $p(1)=\abs{b}^2$.

\subsection{Representing composite systems and states}
Suppose we have Hilbert spaces $H_1,H_2,H_3,\dotsc,H_n$ that combine to form a big Hilbert space $H$ such that $H=H_1\otimes H_2\otimes\dotsc\otimes H_n$ and the state vectors $\ket{\psi_1},\ket{\psi_2},\ket{\psi_3},\dotsc,\ket{\psi_n}$ such that $\ket{\psi_1}\in H_1,\ \ket{\psi_2}\in H_2,\dotsc,\ket{\psi_n}\in H_n$. Then we can construct $\ket{\psi}$ such that $\ket{\psi}\in H$ using the tensor product in the following way:
\begin{align*}
\ket{\psi} & = \ket{\psi_1}\otimes\ket{\psi_2}\otimes\ket{\psi_3}\otimes\dotsc\otimes\ket{\psi_n}
\end{align*}
The tensor product of two vectors is linear that is:
\begin{align*}
\ket{v}\otimes\left[\ket{w_1} + \ket{w_2}\right] & = \ket{v}\otimes\ket{w_1} + \ket{v}\otimes\ket{w_2} \\
\left[\ket{w_1}+\ket{w_2}\right]\otimes\ket{v} & = \ket{w_1}\otimes\ket{v} + \ket{w_2}\otimes\ket{v}
\end{align*}
Tensor product is also linear with respect to scalar. Let $\alpha\in\mathbb{R}$, then
\begin{align*}
\ket{v}\otimes\left(\alpha\ket{w}\right)& = \alpha\ket{v}\ket{w}
\end{align*}
and vice versa. To construct basis for the larger Hilbert space, we simply form tensor product of basis vectors. Suppose the basis for the Hilbert space $H_1$ is denoted by $\ket{u_i}$ and that for $H_2$ is denoted by $\ket{v_i}$. Then the basis for the big Hilbert space $H=H_1\otimes H_2$ is given by:
\begin{align*}
\ket{w_i} & = \ket{u_i} \otimes \ket{v_i}
\end{align*}
Tensor product $\ket{\phi}\otimes \ket{\chi}$ is often written more simply as $\ket{\phi}\ket{\chi}$ or even $\ket{\phi\chi}$.\\
\begin{example}{}{label}
If the basis of $H_1$ and $H_2$ is $\{\ket{0},\ket{1}\}$. What will be the basis for $H=H_1\otimes H_2$?\\
\textbf{Solution: } The basis of $H$ can be found by writing all possible tensor products of the basis states for $H_1$ and $H_2$. Thus basis vectors are: 
\begin{align*}
\ket{w_1} & = \ket{0}\otimes\ket{0} = \ket{00} \\
\ket{w_2}& = \ket{0}\otimes\ket{1} = \ket{01} \\
\ket{w_3}& = \ket{1}\otimes\ket{0} = \ket{10} \\
\ket{w_4}& = \ket{1}\otimes\ket{1} = \ket{11}
\end{align*}
\end{example}

\subsubsection{Computing Inner Product}
Assume that $\{\ket{u_1},\ket{v_1}\}\in H_1$ and $\{\ket{u_2},\ket{v_2}\}\in H_2$. Then taking inner product of vectors belonging to big Hilbert space, $H$, is quite simple. We actually take inner product of vectors belonging to $H_1$ and $H_2$ and multiply them together. Let
\begin{align*}
\ket{w_1} & =\ket{u_1}\otimes\ket{v_1}\\
\ket{w_2} & =\ket{u_2}\otimes\ket{v_2}
\end{align*}
Then,
\begin{align*}
\braket{w_1|w_2} & = \left(\bra{u_1}\otimes\bra{v_1}\right) \left(\ket{u_2}\otimes\ket{v_2}\right)\\
& = \braket{u_1|u_2}\braket{v_1|v_2}
\end{align*}
\subsubsection{Operators and tensor products}
In language of quantum computing, we have gates that are actually unitary operators. An operator acts on tensor product in a following fashion. Suppose $\ket{u}\in H_1$ and $\ket{v}\in H_2$ be two state vectors that belong to the big Hilbert space $H=H_1\otimes H_2$. Now, suppose that an operator $\hat{A}$ acts on $\ket{u}$ and an operator $\hat{B}$ acts on $\ket{v}$. So, we create an operator $\hat{C}=\hat{A}\otimes\hat{B}$ that acts on $\ket{w}\in H$ such that $\ket{w}=\ket{u}\otimes\ket{v}$.
\begin{align*}
(\hat{A}\otimes\hat{B})\ket{w} & = (\hat{A}\otimes\hat{B})(\ket{u}\otimes\ket{v})\\
& = \hat{A}\ket{u}\otimes\hat{B}\ket{v}
\end{align*}
Let's take a look at an example. Suppose we have a gate $X$ that is acting on a state $\ket{u_1}$ and a gate $Y$ that is acting on the state $\ket{u_2}$. Then the composite gate(in matrix form) that acts on the composite state $\ket{u_1}\otimes\ket{u_2}$ is given by:
\begin{align*}
X \otimes Y & = \begin{pmatrix}
0 & 1\\
1 & 0
\end{pmatrix} \otimes \begin{pmatrix}
0 & -i\\
i & 0
\end{pmatrix} \\
& = \begin{pmatrix}
0 & 0 & 0 & -i \\
0 & 0 & i & 0 \\
0 & -i & 0 & 0\\
i & 0 & 0 & 0
\end{pmatrix}
\end{align*}
\subsubsection{Tensor Product of two qubit states}
Since we know how to calculate tensor product of two column vectors, we are going to show the matrix form of two qubits system that contains four possible states\footnote{These states are basis for two qubit system, because each state can be written as a linear combination of these states}, $\{\ket{00},\ket{01},\ket{10},\ket{11}\}$.
\begin{multicols}{2}
\begin{align*}
\ket{00}&=\ket{0}\otimes\ket{0}=\begin{pmatrix}
1\\0
\end{pmatrix}\otimes\begin{pmatrix}
1\\0
\end{pmatrix}=\begin{pmatrix}
1\\0\\0\\0
\end{pmatrix}
\end{align*}

\columnbreak

\begin{align*}
\ket{01}&=\ket{0}\otimes\ket{1}=\begin{pmatrix}
1\\0
\end{pmatrix}\otimes\begin{pmatrix}
0\\1
\end{pmatrix}=\begin{pmatrix}
0\\1\\0\\0
\end{pmatrix}
\end{align*}
\end{multicols}
Similarly, 
\begin{multicols}{2}
\begin{align*}
\ket{10}&=\ket{1}\otimes\ket{0}=\begin{pmatrix}
0\\1
\end{pmatrix}\otimes\begin{pmatrix}
1\\0
\end{pmatrix}=\begin{pmatrix}
0\\0\\1\\0
\end{pmatrix}
\end{align*}

\columnbreak

\begin{align*}
\ket{11}&=\ket{1}\otimes\ket{1}=\begin{pmatrix}
0\\1
\end{pmatrix}\otimes\begin{pmatrix}
0\\1
\end{pmatrix}=\begin{pmatrix}
0\\0\\0\\1
\end{pmatrix}
\end{align*}
\end{multicols}
\begin{example}{}{label}
Write the state $\displaystyle{\ket{\phi}=\frac{1}{\sqrt{3}}\begin{pmatrix}
i\\0\\i\\1
\end{pmatrix}}$ in terms of computational basis of bipartite system.\\
\textbf{Solution:}
\begin{align*}
\frac{1}{\sqrt{3}}\begin{pmatrix}
i\\0\\i\\1
\end{pmatrix} & = \frac{1}{\sqrt{3}}\left[i\begin{pmatrix}
1\\0\\0\\0
\end{pmatrix}+0\begin{pmatrix}
0\\1\\0\\0
\end{pmatrix}+i\begin{pmatrix}
0\\0\\1\\0
\end{pmatrix}+1\begin{pmatrix}
0\\0\\0\\1
\end{pmatrix}\right]\\
& = \frac{1}{\sqrt{3}}\left[i\ket{00}+0\ket{01}+i\ket{10}+1\ket{11}\right]\\
\end{align*}
\begin{align}\label{phi}
\Rightarrow\ket{\phi}& = \frac{1}{\sqrt{3}}\left[i\ket{00}+i\ket{10}+\ket{11}\right]
\end{align}
\end{example}

\subsection{Two-Qubit Gates in Outer-Product Form}
For the two state system, the gate has $4\times4$ dimension, because now the state vector resides in 4-dimensional Hilbert space. An arbitrary gate is given by:
\begin{align*}
G & = \begin{pmatrix}
g_{11} & g_{12} & g_{13} & g_{14} \\
g_{21} & g_{22} & g_{23} & g_{24} \\
g_{31} & g_{32} & g_{33} & g_{34} \\
g_{41} & g_{42} & g_{43} & g_{44} \\
\end{pmatrix}
\end{align*}
We can deocmpose this gate as: 
\begin{align}\label{4by4}
G & = g_{11}\begin{pmatrix}
1 & 0 & 0 & 0\\
0 & 0 & 0 & 0\\
0 & 0 & 0 & 0\\
0 & 0 & 0 & 0\\
\end{pmatrix} + \cdots +g_{44}\begin{pmatrix}
0 & 0 & 0 & 0\\
0 & 0 & 0 & 0\\
0 & 0 & 0 & 0\\
0 & 0 & 0 & 1\\
\end{pmatrix}
\end{align}
Matrix corresponding to each coefficient in Eq. \ref{4by4} forms a basis for gates in two qubit system. These matrices are nothing but just the outer product of basis vectors $\{\ket{00},\ket{01},\ket{10},\ket{11}\}$. There will be 16 possible outer products. One of them is given below:
\begin{align*}
\ketbra{00}{00} & = \begin{pmatrix}
1 & 0 & 0 & 0\\
0 & 0 & 0 & 0\\
0 & 0 & 0 & 0\\
0 & 0 & 0 & 0\\
\end{pmatrix}
\end{align*}
Thus in Dirac notation, the gate $G$ can be written as: 
\begin{align*}
G & = g_{11}\ketbra{00}{00} +g_{12}\ketbra{00}{01} + \cdots + g_{44}\ketbra{11}{11}
\end{align*}
\begin{exercise}{}{label}
\textbf{Exercise: }Write $X\otimes X$ in Dirac notation.\\\\
\textbf{Hint: } First, find the tensor product of the matrix and then use the matrix to write the composite gate in Dirac notation.
\end{exercise}

\section{Entangled states}
One of the most fascinating and unusual phenomenon of quantum mechanics is entanglement. For the simplest, if two quantum systems $A$ and $B$ are entangled, it means that the values of certain properties of $A$ are correlated with the values that those properties will assume for system $B$.\\
\indent In other words, entanglement explains the relationship that exist between two or more particles that interact in a way that makes it impossible to describe each particle independently.
\subsection{When is state entangled?}
Not all states are entangles. When two states are \emph{entangled}, the state of each system can only be described with reference to the other state i.e., state is not separable. On the contrary, if two states are \emph{not entangled}, the state of each system can be described independently. In other words, we can write the state in tensor product form.\\
\indent For example, the state $\displaystyle{\ket{\psi}=\frac{\ket{00}+\ket{01}}{\sqrt{2}}}$ is not entangled, because it can be written in the product form.
\begin{align*}
\ket{\psi}& =\frac{\ket{0}\otimes\left(\ket{0}+\ket{1}\right)}{\sqrt{2}}\\
& = \ket{0}\otimes \frac{\ket{0}+\ket{1}}{\sqrt{2}}
\end{align*}
Here $\displaystyle{\ket{0}\in H_1}$ and $\displaystyle{\frac{\ket{0}+\ket{1}}{\sqrt{2}}\in H_2}$. While the state in 
\begin{align*}
\ket{\phi}& =\frac{\ket{0}\otimes\ket{0}+\ket{1}\otimes\ket{1}}{\sqrt{2}}
\end{align*}
 cannot be written in the product form. Thus it can be concluded that $\ket{\phi}$ is entangled state.
\begin{tcolorbox}
\textbf{Quick test to check entanglement: }\\\\
\begin{align*}
\text{Suppose}\ \ \ket{\psi}& =\begin{pmatrix}
a\\b\\c\\d
\end{pmatrix} \in \mathbb{C}^4\\
\end{align*}
The state will be entangled if $ad\neq bc$
\end{tcolorbox}\bigskip
\begin{example}{}{label}
Express the state $\ket{++}$ in computational basis and determine whether state is entangled or not?\\
\textbf{Solution: }Clearly, state can be written as the product form implying that state is not entangled.
\begin{align*}
\ket{++}& =\frac{\ket{0}+\ket{1}}{\sqrt{2}}\otimes\frac{\ket{0}+\ket{1}}{\sqrt{2}} = \frac{\ket{00}+\ket{01}+\ket{10}+\ket{11}}{2}
\end{align*}
We can also check whether the state is entanglement or not by the quick test mentioned above.
\begin{align*}
\ket{++} & = \frac{1}{2}\begin{pmatrix}
1\\1\\1\\1
\end{pmatrix}\\
\end{align*}
Here, $\displaystyle{a=b=c=d=\frac{1}{2}\Rightarrow ad=bc=\frac{1}{4}}$. It means that the state is \emph{Separable}.
\end{example}
\begin{example}{}{label}
Show that set of vectors $\displaystyle{\{\ket{++},\ket{+-},\ket{-+},\ket{--}\}}$ form a basis.\\\\
\textbf{Solution: }We know that set forms a basis when they are normalized and are orthogonal to each other. First we have to show that all vectors are normalized and then that they form orthogonal basis.
\begin{multicols}{4}
\begin{align*}
\ket{++}&=\frac{1}{2}\begin{pmatrix}
1\\1\\1\\1
\end{pmatrix}
\end{align*}

\columnbreak

\begin{align*}
\ket{+-}&=\frac{1}{2}\begin{pmatrix}
1\\-1\\1\\-1
\end{pmatrix}
\end{align*}

\columnbreak

\begin{align*}
\ket{-+}&=\frac{1}{2}\begin{pmatrix}
1\\1\\-1\\-1
\end{pmatrix}
\end{align*}

\columnbreak

\begin{align*}
\ket{--}&=\frac{1}{2}\begin{pmatrix}
1\\-1\\-1\\1
\end{pmatrix}
\end{align*}
\end{multicols}
\begin{align*}
\Rightarrow\norm{\ket{++}}& = \sqrt{\braket{++|++}} \\
& = \sqrt{\left(\frac{1}{2}\right)^2+\left(\frac{1}{2}\right)^2+\left(\frac{1}{2}\right)^2+\left(\frac{1}{2}\right)^2}\\
& = 1
\end{align*}
Similarly, $\displaystyle{\norm{\ket{+-}}=\norm{\ket{-+}}=\norm{\ket{--}}=1}$, this means that all vectors are normalized. Now, it can be easily shown either by matrix multiplication or by knowing the fact that $\braket{+|-}=\braket{-|+}=0$, that $\braket{++|--}=\braket{+-|-+}=0$. Thus set of given vectors form basis in Hilbert space.
\end{example}
\subsection{Bell States}
Bell states are the specific quantum states of two qubits that are maximally entanglement. Bell states are an example of bipartite system. Let say first party is \emph{Alice} and second party is \emph{Bob}. The entanglement of Bell's state means: ``The qubit held by Alice can be 0 as well as 1. If Alice measured her qubit in the standard basis, the outcome would be perfectly random, either possibility 0 or 1 having probability 1/2. However, if Bob then measured his qubit, the outcome would be the same as for Alice. So, if Bob measured his qubit following Alice, he would also have a seemingly random outcome. But if Alice and Bob communicated they would find out that, although their outcomes seemed random, they were perfectly correlated." Bell's states are given by:
\begin{equation}\label{Bells}
\begin{split}
\ket{\psi}^{00} & = \frac{\ket{00}+\ket{11}}{\sqrt{2}} \\
\ket{\psi}^{01} & = \frac{\ket{01}+\ket{10}}{\sqrt{2}} \\
\ket{\psi}^{10} & = \frac{\ket{00}-\ket{11}}{\sqrt{2}} \\
\ket{\psi}^{11} & = \frac{\ket{01}-\ket{10}}{\sqrt{2}}
\end{split}
\end{equation}
Bell's states can be written compactly as: 
\begin{align*}
\ket{\psi}^{xy} & = \frac{\ket{0y}+(-1)^x\ket{1\bar{y}}}{\sqrt{2}}
\end{align*}
The set of the bell states is known as Bell's Basis. It can be easily shown that Bell's states also form basis in Hilbert space.
\section{Two qubit gates}
Two qubit gates are to perform joint operation on qubits. These are at the heart of quantum computation because many application are result of two qubit gates. These gates can be expressed in terms of single qubit gate like \emph{pauli gates, Hadamard gate, rotation gates etc}. In this section we will discuss some important 2-qubit gates.
\subsection{Controlled Not Gate}
\begin{figure}[H]
\centering
\begin{quantikz}
\lstick{$\ket{c}$} & \ctrl{1} & \rstick{$\ket{c}$} \qw\\
\lstick{$\ket{t}$} & \gate{X} & \rstick{$\ket{c\oplus t}$} \qw
\end{quantikz}
\caption{Controlled Not Gate}\label{ch1cnotfig}
\end{figure}
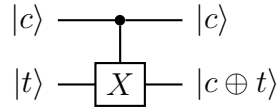

In this gate, we have controlled and target qubit. The controlled qubit remains invariant and the target qubit flips iff the controlled qubit is 1. Truth table for this gate is shown below: 
\begin{table}[H]
    \centering
    \begin{tabular}{|c|c|}
	\hline
	Input $\ket{ct}$ & Output $\ket{c\oplus t}$\\
	\hline
	$\ket{00}$ & $\ket{00}$\\
	\hline
	$\ket{01}$ & $\ket{01}$\\
	\hline
	$\ket{10}$ & $\ket{11}$\\
	\hline
	$\ket{11}$ & $\ket{10}$\\
	\hline
    \end{tabular}
    \caption{Truth table for Controlled Not gate}
    \label{cxgate}
\end{table}
Circuit representation for controlled not gate is illustrated in Fig.\ref{ch1cnotfig}.
Considering the truth table, $CX$ gate can be written in matrix form as: 
\begin{align*}
CX & = \begin{pmatrix}
1 & 0 & 0 & 0\\
0 & 1 & 0 & 0\\
0 & 0 & 0 & 1\\
0 & 0 & 1 & 0
\end{pmatrix}
\end{align*}
In the Dirac notation, $CX$ can be written as (referring to the truth table \ref{cxgate}:
\begin{align}\label{cnoteq}
CX & = \ketbra{00}{00}+\ketbra{01}{01} + \ketbra{11}{10} + \ketbra{10}{11}
\end{align}
Now we can counter verify our outputs in the truth table: 
\begin{align*}
CX\ket{00}& = \ket{00}\\
CX\ket{01}& = \ket{01}\\
CX\ket{10}& = \ket{11}\\
CX\ket{11}& = \ket{10}
\end{align*}
\begin{exercise}{}{label}
Apply Controlled Not gate on the state $\ket{++}$\\
\textbf{Hint: }Write $\ket{++}$ in computational basis i.e., $\{\ket{00},\ket{01},\ket{10},\ket{11}\}$
\end{exercise}
\subsection{Controlled $Z$ Gate}
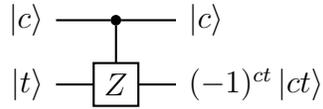
\begin{figure}[H]
\centering
\begin{quantikz}
\lstick{$\ket{c}$} & \ctrl{1} & \rstick{$\ket{c}$} \qw\\
\lstick{$\ket{t}$} & \gate{Z} & \rstick{$(-1)^{ct}\ket{ct}$} \qw
\end{quantikz}
\caption{Controlled $Z$ Gate}\label{ch1cz_fig}
\end{figure}

In this gate, we also have controlled and target qubit. The controlled qubit remains invariant and the target qubit gets phase iff the controlled qubit is 1. Truth table for this gate is shown below:
\begin{table}[H]
    \centering
    \begin{tabular}{|c|c|}
	\hline
	Input & Output\\
	\hline
	$\ket{00}$ & $\ket{00}$\\
	\hline
	$\ket{01}$ & $\ket{01}$\\
	\hline
	$\ket{10}$ & $\ket{10}$\\
	\hline
	$\ket{11}$ & $-\ket{11}$\\
	\hline
    \end{tabular}
    \caption{Truth table for Controlled $Z$ gate}
    \label{tableCZ}
\end{table}
Circuit representation for controlled $Z$ gate is illustrated in Fig.\ref{ch1cz_fig}.
The dirac notation for the controlled $Z$ gate is given by: 
\begin{align*}
CZ & = \ketbra{00}{00}+\ketbra{01}{01}+\ketbra{10}{10}-\ketbra{11}{11}
\end{align*}
\begin{example}{}{label}
Show that
\begin{figure}[H]
\centering
\begin{quantikz}
& \ctrl{1} & \midstick[2,brackets=none]{=}\qw&  & \gate{Z} & \qw \\
& \gate{Z} & \qw   & & \ctrl{-1} & \qw
\end{quantikz}
\end{figure}
\textbf{Solution: }
If we look at the Table. (\ref{tableCZ}), then the minus sign is added only when both control and target qubits are 1. So, the output will be the same for both $\ket{c}\ket{t}$ and $\ket{t}\ket{c}$. Hence both circuits are equivalent.
\end{example}
\subsection{Swap gate}
\begin{figure}[H]
\centering
\begin{quantikz}
\lstick{$\ket{t_1}$} & \swap{1} & \qw \rstick{$\ket{t_2}$} \\
\lstick{$\ket{t_2}$} & \targX{} & \qw \rstick{$\ket{t_1}$}
\end{quantikz}
\caption{Swap Gate}\label{ch1swap}
\end{figure}
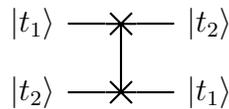

As clear from the name, this gate swaps the qubits. Circuit representation of swap gate is illustrated in Fig. \ref{ch1swap}.

\chapter{Quantum Logic}
\section{General controlled unitary operation}
`If $A$ is true, then do $B$'. This is sort of controlled operation which is one of the most useful in both classical and quantum computing.

Suppose $U$ is an arbitrary single qubit unitary operation. A controlled $U$ operation is a two qubit operation, with a control and target qubit. If the control qubit is set to 1, then $U$ is applied to the target qubit, otherwise target is left alone. In term of computational basis, the action of $C(U)$ is given by $\ket{c}\ket{t}\rightarrow \ket{c}U\ket{t}$. Circuit representation is illustrated in Fig. (\ref{cu}) below.
\begin{figure}[H]\label{cu}
\centering
\begin{quantikz}
& \ctrl{1} & \qw \\
& \gate{U} & \qw
\end{quantikz}
\caption{Controlled-Unitary Operation}
\end{figure}
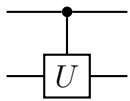
In general, hardware does not provide users an option to implement an arbitrary controlled-$U$ gate. It can be implemented using CNOTs and arbitrary single qubit operations. Since single qubit operations are rotations on the surface of Bloch sphere and much easier to implement rather than implementing arbitrary controlled-$U$ gate. An arbitrary unitary operation can be written as:
\begin{align}\label{unitaryequation}
U & = e^{i\alpha} AXBXC
\end{align}
The first step is to apply phase shift $e^{i\alpha}$ to target qubit, provided that control qubit is $1$.
\begin{align*}
e^{i\alpha}I & = \begin{pmatrix}
e^{i\alpha} & 0 \\
0 & e^{i\alpha}
\end{pmatrix}
\end{align*}
But this is essentially a rotation around $z-$axis on bloch sphere by an angle $2\alpha$. The circuit, that will implement phase shift $e^{i\alpha} $, is given below:
\begin{figure}[H]
\centering
\begin{quantikz}[ampersand replacement=\&]
\& \ctrl{1} \& \qw \midstick[2,brackets=none]{=} \& \ctrl{1} \& \qw \\
 \& \gate{\left(\begin{array}{cc} e^{i\alpha} & 0 \\ 0 & e^{i\alpha} \end{array}\right)} \& \qw \& \gate{R_z(2\alpha)} \& \qw
\end{quantikz}
\caption{Controlled phase shift gate}
\end{figure}
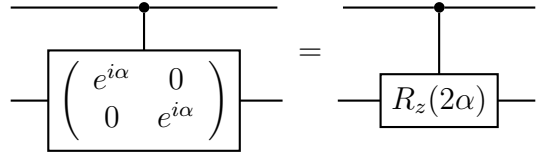
The effect of the circuit on the input qubits is as follow:
\begin{align*}
\ket{00} & \rightarrow \ket{00} \\
\ket{01} & \rightarrow \ket{01} \\
\ket{10} & \rightarrow e^{i\alpha}\ket{10} \\
\ket{11} & \rightarrow e^{i\alpha}\ket{11}
\end{align*}
We see that, phase is added only if the control qubit is 1. So we can replace the controlled phase shift gate by a single qubit gate: \emph{Phase gate} $P(\alpha)$, as illustrated below:
\begin{figure}[H]
\centering
\begin{quantikz}[ampersand replacement=\&]
\& \gate{\left(\begin{array}{cc}
1 & 0\\
0 & e^{i\alpha}
\end{array}\right)} \& \qw
\end{quantikz}
\caption{Equivalent circuit of Controlled phase shift gate}
\end{figure}
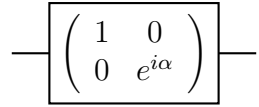
In Eq. (\ref{unitaryequation}) $A$, $B$ and $C$ are single qubit operations such that $ABC=I$. If control qubit is 1, then $e^{i\alpha}AXBXC$ is applied to target qubit, otherwise $ABC=I$ is applied, i.e., no change is made. The circuit that applies controlled-$U$ operation is given by: 
\begin{figure}[H]
\centering
\begin{quantikz}
& \ctrl{1}& \qw \midstick[2,brackets=none]{=}& \qw & \ctrl{1} & \qw & \ctrl{1} & \gate{P(\alpha)} & \qw \\
& \gate{U} & \qw & \gate{C} & \targ{} & \gate{B} & \targ{} & \gate{A} & \qw
\end{quantikz}
\caption{Circuit representation of controlled-$U$ operation using single qubit gates and CNOTs}
\end{figure}
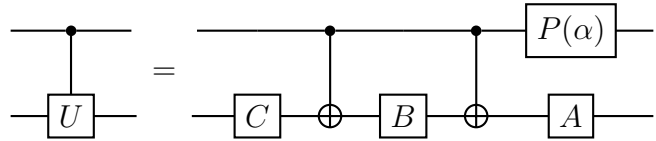

\subsection{Toffoli gate}
\begin{figure}[H]
\centering
\begin{quantikz}
& \ctrl{1} & \qw \\
& \ctrl{1} & \qw \\
& \targ{} & \qw
\end{quantikz}
\caption{Circuit representation of Toffoli gate}\label{ch2toffoli}
\end{figure}
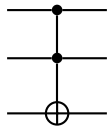

Toffoli gate is a unitary operation controlled by two qubits. It can be drawn as shown in Fig. (\ref{ch2toffoli}).
If both control qubits are set to 1, target will be flipped otherwise it will be unchanged. A general unitary operator controlled by two qubits ($C^2(U)$) is illustrated in figure below:
\begin{figure}[H]
\centering
\begin{quantikz}
& \ctrl{1} & \qw \\
& \ctrl{1} & \qw \\
& \gate{U} & \qw
\end{quantikz}
\caption{$C^2(U)$ gate}
\end{figure}
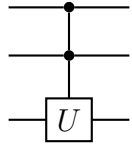
Suppose we have access to a unitary operator $V$ such that $U^2=V$, then $C^2(U)$ can be implemented using the circuit illustrated in Fig. (\ref{ccu}).
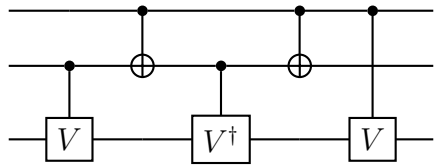
\begin{figure}[H]
\centering
\begin{quantikz}
& \qw      & \ctrl{1} & \qw & \ctrl{1} & \ctrl{2} & \qw \\
& \ctrl{1} & \targ{} & \ctrl{1} & \targ{} & \qw & \qw \\
& \gate{V} & \qw & \gate{V^{\dagger}} & \qw & \gate{V} & \qw
\end{quantikz}
\caption{Circuit for $C^2(U)$ gate}\label{ccu}
\end{figure}
For toffoli gate, $U=X$, thus $V=\sqrt{X}$, where, 
\begin{align*}
\sqrt{X} & = \frac{1+i}{2}\left(I-iX\right)
\end{align*}
\begin{example}{}{Proof}
Prove that if $U=X$, then
\begin{align*}
V & = \frac{1+i}{2}\left(I-iX\right)
\end{align*}
\textbf{Solution: }We will diagonalize $X$, for this, we have to find eigenvalues and eigenvectors. Eigenvalues are given by, 
\begin{align*}
\lambda_1 & = 1 \\
\lambda_2 & = -1
\end{align*}
This implies that the diagonal matrix $D$ is given by:
\begin{align*}
D & = \begin{pmatrix}
\lambda_1 & 0 \\
0 & \lambda_2
\end{pmatrix} = \begin{pmatrix}
1 & 0 \\ 0 & -1
\end{pmatrix}
\end{align*}
Eigenvectors corresponding to each eigenvalue is given by:
\begin{align*}
\mathbf{v}_1 & = \begin{pmatrix}
1\\1
\end{pmatrix}\\
\mathbf{v}_2 & = \begin{pmatrix}
1\\-1
\end{pmatrix}
\end{align*}
Thus, the matrix $P$ is given by
\begin{align*}
P & = \begin{pmatrix}
\mathbf{v}_1 & \mathbf{v}_2
\end{pmatrix} = \begin{pmatrix}
1 & 1 \\ 1 & -1
\end{pmatrix} \\
\Rightarrow P^{-1} & = \frac{1}{2}\begin{pmatrix}
1 & 1 \\ 1 & -1
\end{pmatrix}
\end{align*}
Thus, 
\begin{align*}
\sqrt{X} & = P \sqrt{D} P^{-1} \\
& = \frac{1}{2}\begin{pmatrix}
1+i & 1-i \\
1-i & 1+i
\end{pmatrix} = \frac{1+i}{2}\begin{pmatrix}
1 & -i \\ -i & 1
\end{pmatrix} \\
& = \frac{1+i}{2} \left(I - i X \right)
\end{align*}
It can also be shown that $VV^{\dagger}=V^{\dagger}V=I$
\end{example}
\subsection*{Controlled Unitary operation w.r.t $\ket{0}$}
The circuit that performs unitary operation to target qubit when control qubit is 0 and do nothing when control qubit is 1 is illustrated below:
\begin{figure}[H]
\centering
\begin{quantikz}
& \octrl{1} & \qw \midstick[2,brackets=none]{=} & \gate{X} & \ctrl{1} & \gate{X} & \qw \\
& \gate{U} & \qw & \qw & \gate{U} & \qw & \qw 
\end{quantikz}
\end{figure}
\begin{exercise}{}{label}
Perform a $C^2(X)$ gate with respect to $\ket{01}$.
\end{exercise}
\subsection{Multi-controlled Unitary Operation}
How can we implement an arbitrary single qubit unitary gate controlled by $n$ qubits? For this purpose, we need $n-1$ auxiliary qubits that start and end in the state $\ket{0}$. First,  Toffoli gates are applied in a such way that the final auxiliary qubit is in the state $\ket{c_1 \cdot c_2 \dotsc\cdot c_n}$ where $\ket{c_0,c_1,c_2,\dotsc}$ is the computational basis of control qubits. Next, a unitary operator is applied w.r.t last auxiliary qubit. Finally, Toffoli gates are applied in reverse order just to bring auxiliary qubits back to the state $\ket{0}$. The circuit that implements $C^4(U)$ is given by:
\begin{figure}[H]
\centering
\begin{quantikz}
\lstick{$\ket{c_0}$} & \ctrl{1} & \qw & \qw & \qw & \qw & \qw & \ctrl{1} & \qw \\
\lstick{$\ket{c_1}$} & \ctrl{3} & \qw & \qw & \qw & \qw & \qw & \ctrl{3} & \qw \\
\lstick{$\ket{c_2}$} & \qw & \ctrl{2} & \qw  & \qw & \qw & \ctrl{2} & \qw & \qw \\
\lstick{$\ket{c_3}$} & \qw & \qw & \ctrl{2} & \qw & \ctrl{2} & \qw & \qw & \qw \\
\lstick{$\ket{0}$} & \targ{} & \ctrl{1} & \qw & \qw & \qw & \ctrl{1} & \targ{} & \qw \\
\lstick{$\ket{0}$} & \qw & \targ{} & \ctrl{1} & \qw & \ctrl{1} & \targ{} & \qw & \qw \\
\lstick{$\ket{0}$} & \qw & \qw & \targ{} & \ctrl{1} & \targ{} & \qw & \qw & \qw \\
\lstick{$\ket{t}$} & \qw & \qw & \qw & \gate{U} & \qw & \qw & \qw & \qw
\end{quantikz}
\caption{Implementation of $C^n(U)$ for $n=4$}
\end{figure}
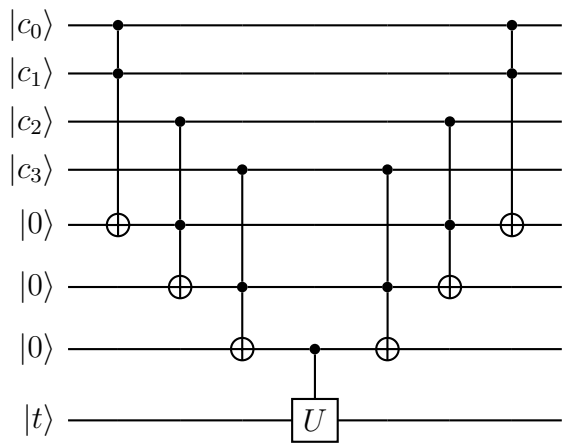

\section{Universal Quantum Gates}
Consider a unitary matrix $U$ that belong to $d-$dimensional hilbert space. Such unitary matrix can be decomposed into a product of \emph{two-level unitary matrices}, i.e., unitary matrices that acts non-trivially only on two or fewer state vector components.
\begin{align*}
U & = U_1U_2\dotsc U_k
\end{align*}
For the sake of simplicity, consider the case of $3\times 3$ unitary matrix given by:
\begin{align}\label{U}
U & = \begin{pmatrix}
a & d & g \\
b & e & h \\
c & f & j
\end{pmatrix}
\end{align}
We need to find two-level unitary matrices $U_1$, $U_2$ and $U_3$ such that $U_3U_2U_1U=I$. First of all, we have to come up with a matrix $U_1$  such that the matrix element: $(U_1U)_{2,1}$ of the matrix $U_1U$ is 0.
\begin{enumerate}
\item[1. ]Suppose $b=0$, then we can take $U_1$ simply an identity matrix.
\begin{align*}
U_1 & = \begin{pmatrix}
1 & 0 & 0 \\
0 & 1 & 0 \\
0 & 0 & 1
\end{pmatrix}
\end{align*}
\item[2. ] Now, if $b\neq 0$, then 
\begin{align*}
U_1 & = \begin{pmatrix}
a^*/m & b^*/m & 0 \\
b/m & -a/m & 0 \\
0 & 0 & 1
\end{pmatrix}
\end{align*}
where $m=\sqrt{\abs{a}^2+\abs{b}^2}$
\end{enumerate}
Thus the matrix $U_1U$ is given by:
\begin{align*}
U_1U & = \begin{pmatrix}
a' & d' & g' \\
0 & e' & h' \\
c' & f' & j'
\end{pmatrix}
\end{align*}
Now, we have to come up with a matrix $U_2$ such that the matrix element: $(U_2U_1U)_{3,1}$ is 0.
\begin{enumerate}
\item[1. ]Suppose $c=0$, then $U_2$ is simply identity matrix.
\begin{align*}
U_2 & = \begin{pmatrix}
1 & 0 & 0 \\
0 & 1 & 0 \\
0 & 0 & 1
\end{pmatrix}
\end{align*}
\item[2. ]If $c\neq 0$, then
\begin{align*}
U_2 & = \begin{pmatrix}
a'^*/n & 0 & c'^*/n \\
0 & 1 & 0 \\
c'^*/n & 0 & -a'^*/n
\end{pmatrix}
\end{align*}
where $n = \sqrt{\abs{a'}^2+\abs{c'}^2}$
\end{enumerate}
Thus the matrix $U_2U_1U$ is given by:
\begin{align*}
U_2U_1U & = \begin{pmatrix}
a'' & d'' & g'' \\
0 & e'' & h'' \\
0 & f'' & j''
\end{pmatrix}
\end{align*}
Take $U_2U_1U=V$. $V$ is unitary matrix because $U_1,\ U_2$ and $U$ are unitaries. Thus, we can write:
\begin{align*}
V^{\dagger}V & = I \\
\begin{pmatrix}
a''^* & 0 & 0 \\
d''^* & e''^* & f''^* \\
g''^* & h''^* & j''^*
\end{pmatrix}\begin{pmatrix}
a'' & d'' & g'' \\
0 & e'' & h'' \\
0 & f'' & j''
\end{pmatrix} & = \begin{pmatrix}
1 & 0 & 0\\
0 & 1 & 0\\
0 & 0 & 1
\end{pmatrix} \\
\begin{pmatrix}
\abs{a''}^2 & a''^*d'' & a''^*g'' \\
a''d''^* & \abs{d''}^2+\abs{e''}^2+\abs{f''}^2 & d''^*g''+e''^*h''+f''^*j''\\
a''g''^* & g''^*d''+h''^*e''+j''^*f'' & \abs{g''}^2+\abs{h''}^2+\abs{j''}^2
\end{pmatrix} & = \begin{pmatrix}
1 & 0 & 0 \\
0 & 1 & 0 \\
0 & 0 & 1
\end{pmatrix}
\end{align*}
The above equation shows that
\begin{align*}
a'' & = 1 \\
d'' & = g'' = 0 \\
\abs{e''}^2 + \abs{f''}^2 & = \abs{h''}^2+\abs{j''}^2 = 1
\end{align*}
Hence, 
\begin{align*}
U_2U_1U & = \begin{pmatrix}
1 & 0 & 0 \\
0 & e'' & h'' \\
0 & f'' & j''
\end{pmatrix}
\end{align*}
and finally, \begin{align*}
U_3 & = V^{\dagger} = \begin{pmatrix}
1 & 0 & 0 \\
0 & e''^* & f''^* \\
0 & h''^* & j''^*
\end{pmatrix}
\end{align*}
It can be easily shown that $U_3U_2U_1U=I$. It implies that
\begin{align*}
U & = U_1^{\dagger}U_2^{\dagger}U_3^{\dagger}
\end{align*}
or we can write like 
\begin{align*}
U & = V_1 V_2 V_3
\end{align*}
where $V_i=U_i^{\dagger}$ and are two-level unitary operators. More generally, suppose $U$ belongs to $d-$dimensional hilbert space, then it can be expressed as:
\begin{align*}
U_{d\times d} & = V_1^{(d-1)}\cdot V_2^{(d-1)}\cdots V_d^{(d-1)} 
\end{align*}
where each gate is \emph{$d-1$-level unitary} but we need to make them two level unitaries for which we need $k$ gates such that $U_{d\times d}=V_1\cdot V_2\cdots V_k$ where $V_i$ are two-level unitary operation and
\begin{align*}
k & = (d-1)+(d-2)+\cdots+1 = \frac{d(d-1)}{2}
\end{align*}
\subsection{Implementing general two-level unitary}
First of all, note that two-level unitary is not a single qubit gate because it is being applied jointly on all of the qubits. The difference, however, between the general $n-$bit unitary and two level $n-$bit unitary is that it only act non-trivially only on two computational basis states. So, if we have $n-$bit system, then computational basis states are $2^n$. For instance, $2-$bit system has four basis states, $\{\ket{00},\ket{01},\ket{10},\ket{11}\}$ then two-level unitary operator will act only on two of the basis states, whatever they are.
\paragraph*{}
In $3-$dimensional hilbert space with basis states$\displaystyle{\{\ket{000}, \ket{001}, \cdots, \ket{111}\}}$ suppose we wish to implement the two-level unitary transformation
\begin{align}\label{2levelU}
U & = \begin{pmatrix}
a & 0 & 0 & 0 & 0 & 0 & 0 & c \\
0 & 1 & 0 & 0 & 0 & 0 & 0 & 0 \\
0 & 0 & 1 & 0 & 0 & 0 & 0 & 0 \\
0 & 0 & 0 & 1 & 0 & 0 & 0 & 0 \\
0 & 0 & 0 & 0 & 1 & 0 & 0 & 0 \\
0 & 0 & 0 & 0 & 0 & 1 & 0 & 0 \\
0 & 0 & 0 & 0 & 0 & 0 & 1 & 0 \\
b & 0 & 0 & 0 & 0 & 0 & 0 & d
\end{pmatrix}
\end{align}
here $a$, $b$, $c$ and $d$ are complex numbers such that
\begin{align*}
\tilde{U} & = \begin{pmatrix}
a & c\\b & d
\end{pmatrix}
\end{align*}
Two-level unitary given in Eq. (\ref{2levelU}) can be implemented in terms of multi-controlled NOT gates and multi-controlled single qubit unitary gates, where $\tilde{U}$ is a single qubit unitary gate. Notice that $U$ acts non-trivially only on the states $\ket{000}$ and $\ket{111}$. We write a gray code connecting 111 and 000. So we start with $\ket{111}$ and keep on changing one qubit at a time until we reach $\ket{000}$.
\begin{align*}
\begin{matrix}
A & B & C \\
1 & 1 & 1 \\
0 & 1 & 1 \\
0 & 0 & 1 \\
0 & 0 & 0
\end{matrix}
\end{align*}
Let's build the circuit that performs $U$ given in Eq. (\ref{2levelU}).
\begin{enumerate}
\item[i. ]Going from $\ket{111}$ to $\ket{011}$, we see that $2^{nd}$ and $3^{rd}$ qubits are same in each state, so $C^2(X)$: NOT gate controlled by $2^{nd}$ and $3^{rd}$ qubit will be applied as illustrated below:
\begin{figure}[H]
\centering
\begin{quantikz}
& \targ{} & \qw \\
& \ctrl{-1} & \qw \\
& \ctrl{-1} & \qw 
\end{quantikz}
\caption{Step-1}
\end{figure}
\item[ii. ]Going from $\ket{011}$ to $\ket{001}$, we see that $1^{st}$ and $3^{rd}$ qubit are same in each state, so $C^2(X)$: NOT gate controlled by $1^{st}$ and $3^{rd}$ qubit will be applied as illustrated below:
\begin{figure}[H]
\centering
\begin{quantikz}
& \octrl{1} & \qw \\
& \targ{} & \qw \\
& \ctrl{-1} & \qw
\end{quantikz}
\caption{Step-2}
\end{figure}
\item[iii. ]Going from $\ket{001}$ to $\ket{000}$, $1^{st}$ and $2^{nd}$ qubits are same in each state, so $C^{2}(\tilde{U})$: $\tilde{U}$ gate controlled by $1^{st}$ and $2^{nd}$ qubit will be applied as illustrated below:
\begin{figure}[H]
\centering
\begin{quantikz}
& \octrl{1} & \qw \\
& \octrl{1} & \qw \\
& \gate{\tilde{U}} & \qw
\end{quantikz}
\caption{Step-3}
\end{figure}
\item[iv. ]Finally, the first two steps are reversed ensuring that $\ket{001}$ gets swapped back with the state $\ket{111}$. Therefore the circuit that implements the two-level unitary operation defined by Eq. (\ref{2levelU}) is illustrated in the figure below:
\begin{figure}[H]\label{2Ufig}
\centering
\begin{quantikz}
& \targ{} & \octrl{1} & \octrl{1} & \octrl{1} & \targ{} & \qw \\
& \ctrl{-1} & \targ{} & \octrl{1} & \targ{} & \ctrl{-1} & \qw \\
& \ctrl{-1} & \ctrl{-1} & \gate{\tilde{U}} & \ctrl{-1} & \ctrl{-1} & \qw
\end{quantikz}
\caption{Circuit implementation of two-level unitary operation defined in Eq. (\ref{2levelU})}
\end{figure}
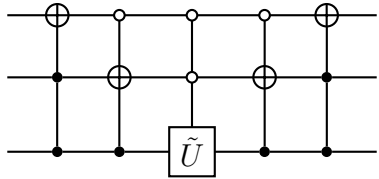
\end{enumerate}
\begin{exercise}{}{label}
Show that the circuit given in Fig. (\ref{2Ufig}) is a two-level.\\\\
\textbf{Hint: }Apply the circuit to each basis state of $3-$dimensional hilbert space. It must gives the same output as input except $\ket{000}$ and $\ket{111}$.
\end{exercise}
\paragraph*{}We see that circuit in Fig. (\ref{2Ufig}) implements two-level unitary in terms of single qubit gates and CNOTs. We know that $O(n)$ gates are required to perform $C^n(X)$. There are also $O(n)$ gates, if we count in series. For instance, the circuit in Fig. (\ref{2Ufig}) has $2n-1$ gates in series, where $n=3$ for this specific circuit. Hence, implementing $U$ will require $O(n\times n)=O(n^2)$ single qubit and $C(X)$ gates. Also, we see that an arbitrary unitary matrix operating on $d-$dimensional state space of $n-$qubits may be written as a product of $O(d^2)=O(2^{2n})=O(4^n)$ two-level unitary operations. Combining these result, we conclude that an arbitrary unitary operation on $n-$qubits can be implemented using a circuit containing at most $O(n^24^n)$ single qubit and $C(X)$ gates.

\paragraph*{} Any arbitrary unitary can be expressed as a product of two-level unitary operators.
\begin{align*}
U & = V_1 V_2 \cdots V_k
\end{align*}
where $U$ is $2^n\times 2^n$ matrix. Two-level unitary operator can be expressed as single qubit and $C(X)$ gates. To implement an arbitrary unitary operator, we require $O(n^24^n)$ single qubit and $C(X)$ gates, where single qubit gate $(\tilde{U})$ in general is: \begin{align*}
\tilde{U} & = \begin{pmatrix}
a & c \\
b & d
\end{pmatrix}
\end{align*}
We also know that, a single qubit unitary operator can be expressed in terms of rotation operators.
\begin{align*}
U & = e^{i\alpha} R_{\hat{n}}(\beta)R_{\hat{m}}(\gamma)R_{\hat{n}}(\delta)
\end{align*}
where $\hat{n}$ and $\hat{m}$ are orthogonal axes and the angles $\alpha$, $\beta$, $\gamma$ and $\delta$ are arbitrary. To implement such unitary, we must have hardware capability that perform rotation at any angle. For which, we will require a gate set that has infinite possibilities. Hence, a discrete set of gates cannot be used to perform an arbitrary unitary operation exactly. Rather, discrete set of gates can be used to approximate an arbitrary unitary operation.
\subsection{Approximation of Unitary Operation}
Our goal is to approximate $U$ with finite single qubit gates with the desired accuracy, $\epsilon$. Suppose $U$ and $V$ are two unitary operators that act on same state space. $U$ is a unitary operator that we wish to implement and $V$ is the unitary operator that is actually implemented in practice. Thus, the error when $V$ is operated instead of $U$ is given by:
\begin{align*}
E(U-V) & = _{max}\abs{\langle\psi|U|\psi\rangle - \langle\psi|V|\psi\rangle }
\end{align*}
where maximum is over all normalized states $\ket{\psi}$ in state space. We want unitary operator $U$ to approximated such that
\begin{align*}
E(U-V) & \leq \epsilon
\end{align*}
Let's start with a set of three gates: $\{ T, H, C(X) \}$ as universal gates. We begin the universality proof by showing that the Hadamard and $\pi/8$ gate can be used to approximate any single qubit unitary operation to an arbitrary accuracy. Consider the gate $HTH$
\begin{align*}
HTH & = \frac{1}{\sqrt{2}}\begin{pmatrix}
1 & 1\\
1 & -1
\end{pmatrix}\begin{pmatrix}
1 & 0\\
0 & e^{i\pi/4}
\end{pmatrix}\frac{1}{\sqrt{2}}\begin{pmatrix}
1 & 1\\
1 & -1
\end{pmatrix} \\
& = \frac{1}{2}\begin{pmatrix}
1+e^{i\pi/4} & 1-e^{i\pi/4} \\
1-e^{i\pi/4} & 1+e^{i\pi/4}
\end{pmatrix} \\
& = \frac{e^{i\pi/8}}{2}\begin{pmatrix}
e^{-i\pi/8}+e^{i\pi/8} & e^{-i\pi/8}-e^{i\pi/8} \\
e^{-i\pi/8}-e^{i\pi/8} & e^{-i\pi/8}-e^{i\pi/8}
\end{pmatrix}\\
& = e^{i\pi/8} \begin{pmatrix}
\cos \pi/8 & -i \sin \pi/8 \\
-i \sin \pi/8 & \cos \pi/8
\end{pmatrix} \\
& = e^{i\pi/8}R_x\left(\pi/4\right)
\end{align*}
Now, the rotation operator around any arbitrary axis: $\hat{n}$ is given by:
\begin{align*}
R_{\hat{n}}(\theta) & = e^{-i\frac{\theta}{2}\hat{n}\cdot \vec{\sigma}}
\end{align*}
where
\begin{align*}
\hat{n} & = \langle n_x, n_y, n_z \rangle \\
\vec{\sigma} & = \langle X,Y,Z \rangle
\end{align*}
Hence,
\begin{align}\label{eq1}
R_{\hat{n}}(\theta) & = \cos\frac{\theta}{2}I -i\sin \frac{\theta}{2} \hat{n}\cdot\vec{\sigma}
\end{align}
Now consider a gate $R_{\hat{m}}$
\begin{align*}
R_{\hat{m}} & = THTH \\
& = \begin{pmatrix}
1 & 0 \\
0 & e^{i\pi/4}
\end{pmatrix}e^{i\pi/8}\begin{pmatrix}
\cos \pi/8 & -i \sin \pi/8 \\
-i \sin \pi/8 & \cos \pi/8
\end{pmatrix} \\
& = \cos^2 \frac{\pi}{8}I -i\sin \frac{\pi}{8} \langle \cos \frac{\pi}{8} , \sin \frac{\pi}{8} , \cos \frac{\pi}{8} \rangle \cdot \vec{\sigma}
\end{align*}
take \begin{align*}
\vec{m} & = \langle \cos \frac{\pi}{8} , \sin \frac{\pi}{8} , \cos \frac{\pi}{8} \rangle \\
\Rightarrow \hat{m} & = \frac{1}{\sqrt{1+\cos^ \frac{\pi}{8}}}\langle \cos \frac{\pi}{8} , \sin \frac{\pi}{8} , \cos \frac{\pi}{8} \rangle \\
\Rightarrow \vec{m} & = \left(\sqrt{1+\cos^2 \frac{\pi}{8}}\right) \hat{m}
\end{align*}
Thus we can write, 
\begin{align}\label{eq2}
R_{\hat{m}} & = \cos^2 \frac{\pi}{8} I -i\sin \frac{\pi}{8\sqrt{1+\cos^ \frac{\pi}{8}}}\ \hat{m}\cdot\vec{\sigma}
\end{align}
Comparing Eq. (\ref{eq1}) and Eq. (\ref{eq2}) gives
\begin{align*}
\cos \frac{\theta}{2} & = \cos^2 \frac{\pi}{8} \\
\Rightarrow \theta & = 2 \cos^{-1}\left(\cos^2 \frac{\pi}{8} \right)
\end{align*}
$\theta$ comes out to be an irrational multiple of $2\pi$. Consider another gate $R_{\hat{n}}$
\begin{align*}
R_{\hat{n}} & = HR_{m}H \\
& = H \left( \cos^2 \frac{\pi}{8} I -i\sin \frac{\pi}{8} \vec{m}\cdot \langle X,Y,Z \rangle \right) H \\
& = \cos^2 \frac{\pi}{8} I -i\sin \frac{\pi}{8} \vec{m}\cdot \langle HXH,HYH,HZH \rangle \\
& = \cos^2 \frac{\pi}{8} I -i\sin \frac{\pi}{8} \langle \cos \frac{\pi}{8}, \sin \frac{\pi}{8} , \cos \frac{\pi}{8} \rangle \cdot \langle Z, -Y, X \rangle \\
& = \cos^2 \frac{\pi}{8} I -i\sin \frac{\pi}{8} \langle \cos \frac{\pi}{8}, -\sin\frac{\pi}{8}, \cos \frac{\pi}{8} \rangle \cdot \langle X,Y,Z \rangle
\end{align*}
Here, 
\begin{align*}
\vec{n} & = \langle \cos \frac{\pi}{8}, -\sin\frac{\pi}{8}, \cos \frac{\pi}{8} \rangle
\end{align*}
The repeated iteration of $R_{\hat{n}}(\theta)$ can be used to approximate any arbitrary rotation $R_{\hat{n}}(\alpha)$. If $\epsilon_1$ is desired accuracy, then after $k$ iterations, we will fall within desired range. 
Geometrically, 
\begin{figure}[H]
\centering
\includegraphics[scale=0.6]{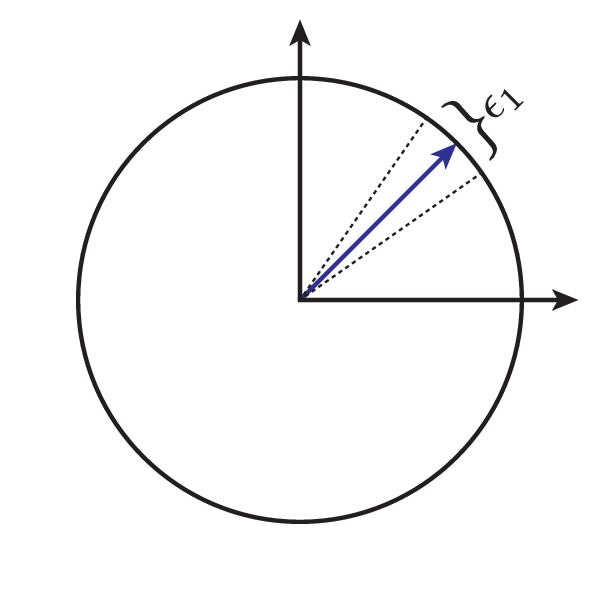}
\caption{After $k$ iterations the state vector will fall in desired range.}
\end{figure}
Now, divide the entire interval into small intervals of width $\epsilon_1$. Then the number of patches are $2\pi/\epsilon_1$. Now, if $k\geq 2\pi/\epsilon_1$, then we will fall in the desired interval. This implies that number of gates required to implement a single qubit arbitrary unitary operator in terms of $T$ and $H$ gate is or order of $1/\epsilon_1$.
\paragraph*{} Suppose, we have a quantum circuit with total of $m$ gates and each of these gate can be implemented using $1/\epsilon_1$ gates. If the whole circuit has error, $\epsilon$, then the error in individual gate is: $\epsilon/m=\epsilon_1$. To implement such circuit, we require $\displaystyle{k=m^2/\epsilon}$ iterations of single qubit unitary operator. We already know that implementing single qubit unitary operator required $O(n^24^n)$ gates. This means that to approximate an arbitrary unitary and implement it in terms of single qubit and $C(X)$ gates require $\displaystyle{O\left(\frac{n^416^n}{\epsilon}\right)}$ gates.
\subsection{Why is approximation of unitary operator so hard?}
Suppose we have a set of gates$\{g_1,g_2,\cdots g_g \}$ and each gate can take at most $f$ inputs. Suppose we have a quantum circuit containing $m$ gates with $n-$inputs, each initialized to $\ket{0}$. State,$\ket{\psi}$, at the output will be different each time for each arrangement of gates. Hence there are at most $\displaystyle{{\binom nf}^m}$ possible choices. In fact, these are the possible states that we can get from permutations of gate.
\paragraph*{} Suppose, we wish to approximate a particular state, $\ket{\psi}$, to within a distance, $\epsilon$, where $\epsilon$ is the radius of small patch on state space. Such patches cover the whole state space. The first observation is we need that the state space of $n$ qubits is $2^{n+1}-1$ dimensional\footnote{It makes sense, because, for 1 qubit, state space is bloch sphere, which is $2^{1+1}-1=3$ dimensional}. Hence the number of patches required to cover the state space is given by:
\begin{align*}
N \propto \frac{(1)^{2^{n+1}-1}}{\epsilon^{2^{n+1}}} = \frac{1}{\epsilon^{2^{n+1}}}
\end{align*}
In order to have states more than number of patches, we must have
\begin{align*}
{\binom nf}^m & > \frac{1}{\epsilon^{2^{n+1}}} \\
\left(\frac{n!}{f!(n-f)!}\right)^m & > \frac{1}{\epsilon^{2^{n+1}}} \\
\left(\frac{n(n-1)\cdots (n-(f+1))(n-f)!}{f!(n-f)!)} \right)^m & > \frac{1}{\epsilon^{2^{n+1}}}
\end{align*}
Here, on the right side, the leading term is $n^f$, so we can neglect other terms, hence we have
\begin{align*}
(n^f)^m & > \frac{1}{\epsilon^{2^{n+1}}} \\
\Rightarrow fm \log(n) & > 2^{n+1}\log\left(\frac{1}{\epsilon}\right) \\
\Rightarrow m & > \frac{2^{n+1}\log\left(\frac{1}{\epsilon}\right)}{f\log(n)}
\end{align*}
That is, there are states of $n$ qubits which take $\displaystyle{O\left(\frac{2^{n+1}\log\left(\frac{1}{\epsilon}\right)}{f\log(n)}\right)}$ operations to approximate to within a distance $\epsilon$. Since, this is exponential in $n$, thus it is difficult in sense of computational complexity.

\section{Quantum Oracles}
\subsection{Phase Kickback}
Phase kickback is very important and is used in almost every quantum algorithm. Kickback is where the eigenvalue added by a gate to a qubit is \emph{kicked back} into a different qubit via a controlled operation. For example, we saw that performing an $X$-gate on a $\ket{-}$ qubit gives it the phase $-1$.
\begin{align*}
X\ket{-}& = -\ket{-}
\end{align*}
When our control qubit is in either $\ket{0}$ or $\ket{1}$, this phase affects the whole state, however it is a global phase and has no observable effects:
\begin{align*}
C(X)\ket{-0} & = \ket{-0}\\
C(X)\ket{-1} & = X\ket{-}\otimes\ket{1}\\
& = -\ket{-}\otimes\ket{1}\\
& = -\ket{-1}
\end{align*}
The interesting effect is when our control qubit is in superposition, the component of the control qubit that lies in the direction of $\ket{1}$ applies this phase factor to the target qubit, which in turn kicks back a relative phase to our control qubit:
\begin{align*}
C(X)\ket{-+} & = \frac{1}{\sqrt{2}}\left(C(X)\ket{-0}+C(X)\ket{-1}\right)\\
& = \frac{1}{\sqrt{2}}\left(\ket{-0}+X\ket{-1}\right)\\
& = \frac{1}{\sqrt{2}}\left(\ket{-0}-\ket{-1}\right)\\
& = \ket{-}\frac{\ket{0}-\ket{1}}{\sqrt{2}}\\
& = \ket{--}
\end{align*}
Wrapping the $CNOT$ in $H$-gates transforms the qubits from the computational basis to the $\{+,-\}$ basis, where we see this effect. This identity is very useful in hardware, since some hardwares only allow for CNOTs in one direction between two specific qubits. We can use this identity to overcome this problem and allow CNOTs in both directions. Few example circuits with truth table are given below: 
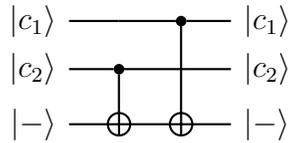
\begin{figure}[H]
\centering
\begin{quantikz}
\lstick{$\ket{c_1}$} & \qw & \rstick{$\ket{c_1}$}  \qw \\
\lstick{$\ket{c_2}$} & \ctrl{1} & \rstick{$\ket{c_2}$} \qw \\
\lstick{$\ket{-}$}  & \targ{} & \rstick{$\ket{-}$} \qw
\end{quantikz}
\caption{Circuit for Phase Kickback}
\label{cir 1}
\end{figure}
Truth table for the circuit in Fig.\ref{cir 1} is given by: 
\begin{table}[H]
    \centering
    \begin{tabular}{|c|c|}
	\hline
	Input & Output\\
	\hline
	$\ket{00}$ & $\ket{00}$\\
	\hline
	$\ket{01}$ & $-\ket{01}$\\
	\hline
	$\ket{10}$ & $\ket{10}$\\
	\hline
	$\ket{11}$ & $-\ket{11}$\\
	\hline
    \end{tabular}
	\label{pkbtable_1}
    \caption{Truth table for Circuit in Fig.\ref{cir 1}}

\end{table}
Here we see that, output is related to input in following manner: 
\begin{align*}
\ket{c_1c_2}\longrightarrow (-1)^{c_2}\ket{q_0q_1}
\end{align*}
Another circuit that also adds phase to output is illustrated below:
\begin{figure}[H]
\centering
\begin{quantikz}
\lstick{$\ket{c_1}$} & \qw & \ctrl{2} & \rstick{$\ket{c_1}$} \qw \\
\lstick{$\ket{c_2}$} & \ctrl{1} & \qw & \rstick{$\ket{c_2}$} \qw \\
\lstick{$\ket{-}$} & \targ{} & \targ{} & \rstick{$\ket{-}$} \qw
\end{quantikz}
\caption{Circuit for Phase Kickback}
\label{cir 2}
\end{figure}
Truth table for the circuit in Fig. (\ref{cir 2}) is given by: 
\begin{table}[H]
    \centering
    \begin{tabular}{|c|c|}
	\hline
	Input & Output\\
	\hline
	$\ket{00}$ & $\ket{00}$\\
	\hline
	$\ket{01}$ & $-\ket{01}$\\
	\hline
	$\ket{10}$ & $-\ket{10}$\\
	\hline
	$\ket{11}$ & $\ket{11}$\\
	\hline
    \end{tabular}
    \label{pkbt_2}
    \caption{Truth table for Circuit in Fig.\ref{cir 2}}
\end{table}
Here, we see that output is related to input in following manner: 
\begin{align*}
\ket{c_1c_2} \longrightarrow (-1)^{c_1+c_2}\ket{c_1c_2}
\end{align*}
Many quantum algorithms are based around the analysis of some function $f(x)$. Often these algorithms simply assume the existence of some `black box' implementation of this function, which we can give an input $x$ and receive the corresponding output $f(x)$. This is referred to as an oracle.\\
\indent The advantage of thinking of the oracle in this abstract way allows us to concentrate on the quantum techniques we use to analyze the function, rather than the function itself. Consider a black box illustrated in figure below:
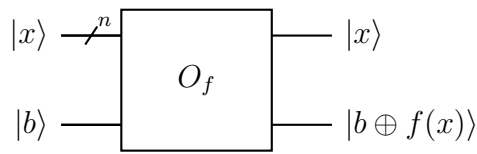
\begin{figure}[H]
\centering
\begin{quantikz}
\lstick{$\ket{x}$} &[3mm] \gate[wires=2][2cm]{O_f}\qwbundle{n} &[3mm] \rstick{$\ket{x}$} \qw\\
\lstick{$\ket{b}$} & \qw & \rstick{$\ket{b\oplus f(x)}$} \qw
\end{quantikz}
\caption{General Oracle}
\label{figure: 1}
\end{figure}
In general, $\ket{b}$ can be multiple qubits and is called \emph{ancillary qubit}, but usually, it is just one qubit input. We can set the input $\ket{b}$ in any way we want. Where $f(x)$ can be either 0 or 1.\\
Note that there can be some other \emph{garbage qubits} other than $\ket{x}$ and $\ket{b}$ to design an oracle.

\subsection{Amplitude Oracle} For instance, if we set $\ket{b}=\ket{0}$, then the output is always $\ket{b\oplus f(x)}=\ket{f(x)}$.

\subsection{Phase Oracle} 
Another form of the oracle is the phase oracle. So, if we set ancillary qubit $\ket{b}=\ket{-}$, then output will be:
\begin{equation*}
\ket{-\oplus f(x)} = \begin{cases}
\ \ \ \ket{-} & \text{if}\ f(x)=0 \\
-\ket{-}& \text{if}\ f(x)=1
\end{cases}
\Bigg\}= (-1)^{f(x)}\ket{-}
\end{equation*}
So, the total output of the phase oracle can be written as: 
\begin{align*}
\ket{x}\otimes\ket{-} & \longrightarrow \ket{x}\otimes(-1)^{f(x)}\ket{-} \\
\text{Or}\ \ket{x}\otimes\ket{-} & \longrightarrow (-1)^{f(x)}\ket{x}\otimes\ket{-} \\
\Rightarrow \ket{x} & \longrightarrow (-1)^{f(x)}\ket{x}
\end{align*}
So we can say that phase oracle has associated the phase with input $\ket{x}$ leaving the ancillary qubit invariant.
\subsubsection{One Qubit Phase Oracle}
\noindent\textbf{Case 1: }\newline
\noindent If the function has the form:
\begin{equation}\label{equation-1}
f(x) = \begin{cases}
0 & \forall\ \ x\in \{0,1\} \\
1 & \forall\ \ x\in \{0,1\}
\end{cases}
\end{equation}
\indent\textbf{Case 1(a): } For $f(x)=0$, the oracle looks like: 
\begin{figure}[H]
\centering
\begin{quantikz}
\lstick{$\ket{x}$} & \qw &  \qw \\
\lstick{$\ket{-}$} & \qw &  \qw 
\end{quantikz}
\caption{One qubit oracle for $f(x)=0$}
\label{One qubit oracle-case 1-a}
\end{figure}
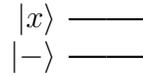
\textbf{Case 1(b): }For $f(x)=1$, the oracle looks like: 
\begin{figure}[H]
\centering
\begin{quantikz}
\lstick{$\ket{x}$} & \qw &  \qw \\
\lstick{$\ket{-}$} & \gate{X} &  \qw 
\end{quantikz}
\caption{One qubit oracle for $f(x)=1$}
\label{One qubit oracle-case 1-b}
\end{figure}
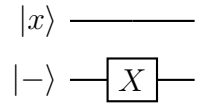
\noindent \textbf{Case 2: }\newline
\noindent Now, if the function has the form:
\begin{equation}\label{equation-2}
f(x) = \begin{cases}
0 & :\ x=0 \\
1 & :\ x=1
\end{cases}
\end{equation}
Then the oracle looks like: 
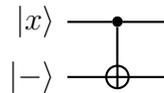
\begin{figure}[H]
\centering
\begin{quantikz}
\lstick{$\ket{x}$} & \ctrl{1} &  \qw \\
\lstick{$\ket{-}$} & \targ{} &  \qw 
\end{quantikz}
\caption{One qubit oracle for $f(x)$ in Eq.(\ref{equation-2})}
\label{One qubit oracle-case 2}
\end{figure}
\noindent\textbf{Case 3: }\newline
\noindent At last, if the function has the form: 
\begin{equation}\label{equation-3}
f(x) = \begin{cases}
1 & :\ x=0 \\
0 & :\ x=1
\end{cases}
\end{equation}
Then the oracle will have a form like:
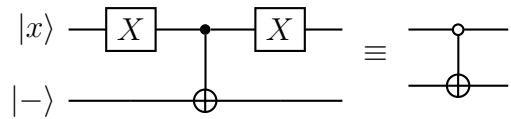
\begin{figure}[H]
\centering
\begin{quantikz}
\lstick{$\ket{x}$} &\gate{X}& \ctrl{1}&\gate{X} & \qw \\
\lstick{$\ket{-}$} &\qw & \targ{} &\qw & \qw 
\end{quantikz} $\ \equiv$ 
\begin{quantikz}
 & \octrl{1} &  \qw \\ 
 & \targ{} &  \qw\end{quantikz}
\caption{One qubit oracle for $f(x)$ in Eq.(\ref{equation-3})}
\label{One qubit oracle-case 3}
\end{figure}

\subsubsection{Two Qubits Phase Oracle}
\noindent\textbf{Case 1: }\newline
\noindent If the function has the form:
\begin{equation}\label{equation-4}
f(x) = \begin{cases}
0 & \forall\ \ x\in \{0,1\} \\
1 & \forall\ \ x\in \{0,1\}
\end{cases}
\end{equation}
\indent\textbf{Case 1(a): } For $f(x)=0$, the oracle looks like: 
\begin{figure}[H]
\centering
\begin{quantikz}
\lstick[wires=2]{$\ket{x}$} & \qw &  \qw \\
& \qw & \qw \\
\lstick{$\ket{-}$} & \qw &  \qw 
\end{quantikz}
\caption{Two qubit oracle for $f(x)=0$}
\label{Two qubit oracle-case 1-a}
\end{figure}
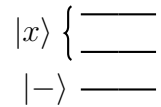
\textbf{Case 1(b): }For $f(x)=1$, the oracle looks like: 
\begin{figure}[H]
\centering
\begin{quantikz}
\lstick[wires=2]{$\ket{x}$} & \qw & \qw \\
& \qw & \qw \\
\lstick{$\ket{-}$} & \gate{X} & \qw 
\end{quantikz}
\caption{One qubit oracle for $f(x)=1$}
\label{Two qubit oracle-case 1-b}
\end{figure}
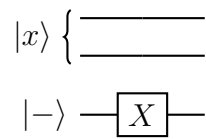
\noindent\textbf{Case 2:}\newline
If the function is balanced i.e., 
\begin{equation}\label{equation-5}
f(x) = \begin{cases}
0 & \text{for half of the inputs} \\
1 & \text{for otherhalf of the inputs} 
\end{cases}
\end{equation}
Then the oracle has a form: 
\begin{figure}[H]
\centering
\begin{quantikz}
\lstick[wires=2]{$\ket{x}$}& \qw & \octrl{2} & \qw\\
& \ctrl{1}& \qw & \qw \\
\lstick{$\ket{-}$} & \targ{} & \targ{} & \qw
\end{quantikz}
\caption{Two qubit oracle for balanced function}
\label{two-qubit oracle for balanced function}
\end{figure}
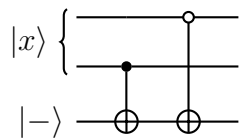
\noindent\textbf{Case 3:}\newline
Now, if we want to implement unbalanced function, for example, out of the states $\{\ket{00},\ket{01},\ket{10},\ket{11}\}$, we want to mark the state $\ket{00}$. Similarly, for three qubits, we want to mark even more than one state for which there are eight possible states. To accomplish this purpose, let's introduce \emph{Toffoli gate}.
\subsubsection*{Toffoli gate}
It is essentially controlled controlled not gate. It is represented by $CCX$. Circuit representation is given below: 
\begin{figure}[H]
\centering
\begin{quantikz}
\lstick{$c_1$} & \ctrl{1} & \qw \\
\lstick{$c_2$} & \ctrl{1} & \qw \\
\lstick{$t$} & \targ{}  & \qw
\end{quantikz}
\caption{Toffoli gate}
\label{Toffoli gate}
\end{figure}
\noindent Truth table of toffoli gate is given in the table below:
\begin{table}[H]
\centering
\begin{tabular}{|c|c|}
\hline
\textbf{Input} & \textbf{Output}\\
\hline
$\ket{000}$ & $\ket{000}$\\
$\ket{001}$ & $\ket{001}$\\
$\ket{010}$ & $\ket{010}$\\
$\ket{100}$ & $\ket{100}$\\
$\ket{011}$ & $\ket{011}$\\
$\ket{101}$ & $\ket{101}$\\
$\ket{110}$ & $\ket{110}$\\
$\ket{111}$ & $\ket{110}$\\
\hline
\end{tabular}
\caption{Truth table of Toffoli gate}
\end{table}
Now, if I initialize my target qubit $\ket{t}=\ket{-}$ and call it ancillary qubit, then input qubits will transform like: 
\begin{table}[H]
\centering
\begin{quantikz}
\lstick[wires=2]{$x$} & \ctrl{1} & \qw \\
& \ctrl{1} & \qw \\
\lstick{$\ket{-}$} & \targ{}  & \qw
\end{quantikz}\ \ \ 
\begin{tabular}{|c|c|}
\hline
\textbf{Input} & \textbf{Output}\\
\hline
$\ket{00}$ &\quad  $\ket{00}$\\
$\ket{01}$ &\quad  $\ket{01}$\\
$\ket{10}$ &\quad  $\ket{10}$\\
$\ket{11}$ &  $-\ket{11}$\\
\hline
\end{tabular}
\caption{Toffoli gate with target qubit set to $\ket{-}$}
\label{Toffoli Function}
\end{table}
Now, if we want to mark $\ket{10}$\footnote{Note that $\ket{10}$ means that $c_0=0$ and $c_1=1$ i.e., labeling of qubits is in reverse order}, I will build my circuit as: 
\begin{table}[H]
\centering
\begin{quantikz}
\lstick[wires=2]{$x$} & \ctrl{1} & \qw \\
& \octrl{1} & \qw \\
\lstick{$\ket{-}$} & \targ{}  & \qw
\end{quantikz}\ \ \ 
\begin{tabular}{|c|c|}
\hline
\textbf{Input} & \textbf{Output}\\
\hline
$\ket{00}$ &\quad  $\ket{00}$\\
$\ket{01}$ &  $-\ket{01}$\\
$\ket{10}$ &\quad  $\ket{10}$\\
$\ket{11}$ &\quad  $\ket{11}$\\
\hline
\end{tabular}
\caption{Toffoli gate to mark $\ket{10}$ with target qubit set to $\ket{-}$}
\end{table}
\subsubsection*{How can we mark two of the inputs using the Toffoli gate?}
Suppose that we want to mark $\ket{01}$ and $\ket{10}$, then the circuit that will do the job is illustrated below:
\begin{figure}[H]
\centering
\begin{quantikz}
\lstick[wires=2]{$x$} & \ctrl{1} & \qw \\
& \octrl{1} & \qw \\
\lstick{$\ket{-}$} & \targ{}  & \qw
\end{quantikz}\ \ \ $+$\ \ \ 
\begin{quantikz}
& \octrl{1} & \qw \\
& \ctrl{1} & \qw \\
& \targ{}  & \qw
\end{quantikz}\ \ \ $=$\ \ \
\begin{quantikz}
& \ctrl{1}  & \octrl{1} & \qw \\
& \octrl{1} & \ctrl{1}  & \qw \\
& \targ{}  & \targ{}  & \qw
\end{quantikz}\ \ \
\begin{tabular}{|c|c|}
\hline
\textbf{Input} & \textbf{Output}\\
\hline
$\ket{00}$ &\quad  $\ket{00}$\\
$\ket{01}$ &  $-\ket{01}$\\
$\ket{10}$ &  $-\ket{10}$\\
$\ket{11}$ &\quad  $\ket{11}$\\
\hline
\end{tabular}
\caption{Toffoli gate that mark two states along with its truth table}
\end{figure}
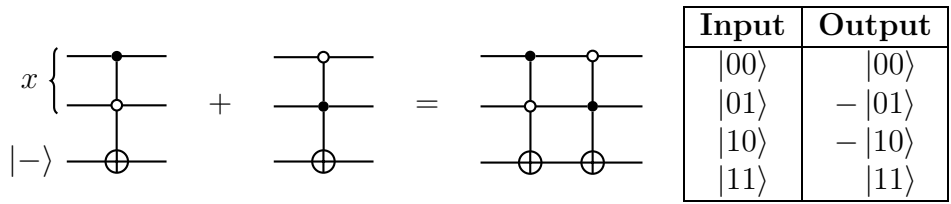
\subsubsection{Three Qubit Phase Oracle}
Similarly, for three qubits, circuit will have a form like: 
\begin{figure}[H]
\centering
\begin{quantikz}
\lstick[wires=3]{$\ket{x}$}
& \ctrl{1} & \qw \\
& \ctrl{1} & \qw \\
& \ctrl{1} & \qw \\
\lstick{$\ket{-}$}& \targ{}  & \qw
\end{quantikz}\ \ \ 
\begin{tabular}{|c|c|}
\hline
\textbf{Input} & \textbf{Output}\\
\hline
$\ket{000}$ &\quad $\ket{000}$\\
$\ket{001}$ &\quad $\ket{001}$\\
\vdots &\quad \vdots \\
$\ket{110}$ &\quad $\ket{110}$\\
$\ket{111}$ & $-\ket{111}$\\
\hline
\end{tabular}
\caption{3-qubits Toffoli gate that mark $\ket{111}$}
\end{figure}
But this gate is directly not available in \emph{qiskit}. So in order to implement this gate, we will simply add an garbage qubit initialized to 0 and perform the combination of controlled gates as shown in the figure below: 
\begin{figure}[H]
\centering
\begin{quantikz}
\lstick[wires=3]{$\ket{x}$}
& \ctrl{1} & \qw      & \ctrl{1} & \qw\\
& \ctrl{2} & \qw      & \ctrl{2} & \qw\\
& \qw      & \ctrl{1} & \qw      & \qw\\
\lstick{$\ket{0}$}
& \targ{}  & \ctrl{1} & \targ{}      & \qw\\
\lstick{$\ket{-}$}
& \qw      & \targ{}  & \qw  & \qw\\
\end{quantikz}\ \ \ $\equiv$\ \ \
\begin{quantikz}
\lstick[wires=3]{$\ket{x}$}
& \ctrl{1} & \qw \\
& \ctrl{1} & \qw \\
& \ctrl{1} & \qw \\
\lstick{$\ket{-}$}
& \targ{}  & \qw
\end{quantikz}
\caption{Toffoli gate implementation in qiskit}
\end{figure}
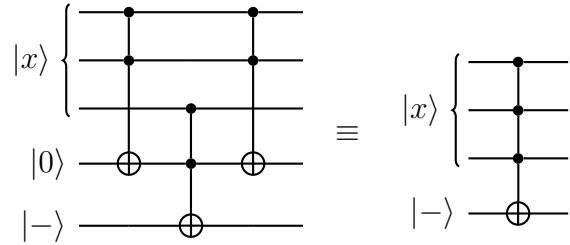
Similarly, we can create toffoli gates of higher orders by introducing garbage qubits.\\
So now, if we want to mark other than $\ket{111}$, let say, $\ket{101}$, then the circuit will have the form: 
\begin{figure}[H]
\centering
\begin{quantikz}
\lstick[wires=3]{$\ket{x}$}
& \ctrl{1} & \qw \\
& \octrl{1} & \qw \\
& \ctrl{1} & \qw \\
\lstick{$\ket{-}$}& \targ{}  & \qw
\end{quantikz}\ \ \
\begin{tabular}{|c|c|}
\hline
\textbf{Input} & \textbf{Output}\\
\hline
$\ket{000}$ & \quad$\ket{000}$\\
$\ket{001}$ & \quad$\ket{001}$\\
\vdots & \quad\vdots \\
$\ket{101}$ & $-\ket{101}$\\
\vdots & \quad\vdots \\
$\ket{111}$ &\quad $\ket{111}$\\
\hline
\end{tabular}
\caption{3-qubit Toffoli gate that mark $\ket{101}$}
\end{figure}
If we want to mark more than one states, we will combine the gates as we did in the two-qubit case. For example, if we want to mark, $\ket{011}$ and $\ket{001}$, the circuit will look like: 
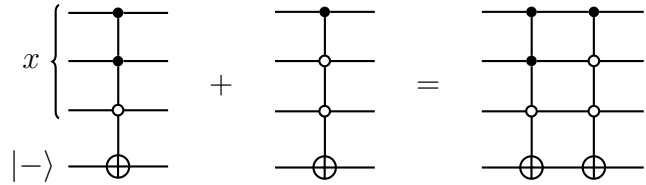
\begin{figure}[H]
\centering
\begin{quantikz}
\lstick[wires=3]{$x$} 
& \ctrl{1} & \qw \\
& \ctrl{1} & \qw \\
& \octrl{1} & \qw \\
\lstick{$\ket{-}$} 
& \targ{}  & \qw
\end{quantikz}\ \ \ $+$\ \ \ 
\begin{quantikz}
& \ctrl{1} & \qw \\
& \octrl{1} & \qw \\
& \octrl{1} & \qw \\
& \targ{}  & \qw
\end{quantikz}\ \ \ $=$\ \ \
\begin{quantikz}
& \ctrl{1}  & \ctrl{1} & \qw \\
& \ctrl{1} & \octrl{1}  & \qw \\
& \octrl{1} & \octrl{1}  & \qw \\
& \targ{}  & \targ{}  & \qw
\end{quantikz}\ \ \
\caption{3-qubit Toffoli gate that mark 011 and 001}
\end{figure}

\chapter{Simple Algorithms}

\section{Preparation of Bell States}
Our goal is to develop a circuit that creates Bell states given in Eq.\ref{Bells}. Following is the circuit that do the job.
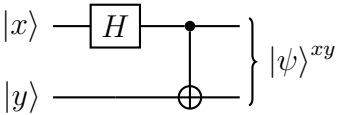
\begin{figure}[H]
\centering
\begin{quantikz}
\lstick{$\ket{x}$} & \gate{H} & \ctrl{1} & \rstick[wires=2]{$\ket{\psi}^{xy}$} \qw\\
\lstick{$\ket{y}$} & \qw & \targ{} & \qw
\end{quantikz}
\label{bellprep}
\caption{Circuit that create Bell's states}
\end{figure}
Let say, at input, we have a state $\ket{xy}=\ket{00}$. Here, first of all, \emph{hadamard gate} is applied on first qubit, resulting $\displaystyle{\frac{\ket{0}+\ket{1}}{\sqrt{2}}\ket{0}=\frac{\ket{00}+\ket{10}}{\sqrt{2}}}$. After that, \emph{controlled not gate} has been applied that gives us the ouput state: $\displaystyle{\ket{\psi}^{00}=\frac{\ket{00}+\ket{11}}{\sqrt{2}}}$ which is a bell state. Similarly, for other inputs, we can get other Bell states.
\subsection{Measurement in Bell's basis}
The circuit that helps us to do measurement in bell basis is illustrated in the figure below:
\begin{figure}[H]\label{bell measurement}
\centering
\begin{quantikz}
\lstick[wires=2]{$\ket{\psi}^{ij}$} & \ctrl{1} & \gate{H} & \rstick{$\ket{i}$}  \qw \\
& \targ{} & \qw & \rstick{$\ket{j}$}\qw
\end{quantikz}
\caption{Circuit for measurement in Bell Basis}
\end{figure}
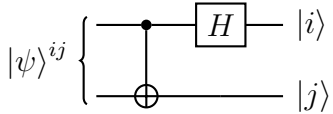
Let say, I have a state, $\ket{\psi}=a\ket{00}+b\ket{01}+c\ket{10}+d\ket{11}$. Also, every state can be written as the linear combination of the four bell states. So $\ket{\psi}$ can be written as: 
\begin{align}\label{state}
\ket{\psi} & = a'\ket{\psi}^{00}+b'\ket{\psi}^{01}+c'\ket{\psi}^{10}+d'\ket{\psi}^{11}
\end{align}
Now, I want to calculate the weights $\{a',b',c',d'\}$. But the problem is that I will each time get $\{a,b,c,d\}$ since measurement is always carried out in computational basis. So what I have to do is that apply the given circuit to the state $\ket{\psi}$ which will gives me: 
\begin{align}\label{newstate}
\ket{\psi} & = a'\ket{00}+b'\ket{01}+c'\ket{10}+d'\ket{11}
\end{align}
Now, if we perform measurement, we will get the desired results.
\subsection{Measurement in Hadamard basis}
The circuit that helps us to do measurement in hadamard basis is illustrated in the figure below:
\begin{figure}[H]\label{had}
\centering
\begin{quantikz}
\lstick[wires=2]{$\ket{\psi}^{\{+,-\}}$} & \gate{H} & \rstick[wires=2]{$\ket{\psi^{\{0,1\}}}$}  \qw \\
& \gate{H} & \qw 
\end{quantikz}
\caption{Circuit for measurement in Hadamard Basis}
\end{figure}
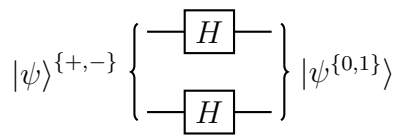
Same procedure will be followed. If the state is given in Hadamard basis, we first have to apply the circuit the given and then measure in computational basis. Because circuit transforms the state $$\displaystyle{\ket{\psi}=a\ket{++}+b\ket{+-}+b\ket{-+}+c\ket{--} \text{to}\ \ket{\psi}=a\ket{00}+b\ket{01}+b\ket{10}+c\ket{11}}$$ After measurement, we can easily get the values of $a,b,c,d$.

\section{Quantum Teleportation}
Quantum teleportation is a technique for transferring quantum information from a sender at one location to a receiver some distance away. While such a procedure is likely to remain in the realm of science fiction, quantum mechanics does allow us to do something almost as magical. To see how this works, let’s go through the basic formalism. The
the task at hand is that Alice wants to transmit an unknown quantum state $\ket{\phi}$ to Bob. The state is a qubit: 
\begin{align}\label{unknowstate}
\ket{\phi} & = \alpha \ket{0} + \beta \ket{1}
\end{align}
By saying the state is unknown, we are saying we don’t necessarily know what $\alpha$ and $\beta$ are. All we assume is that the state is normalized, so $\abs{a}^2+\abs{b}^2=1$.
\subsection{Protocol}
Teleportation takes place in a series of steps. We begin by creating an entangled EPR pair.
Alice and Bob create the entangled state through the following circuit.
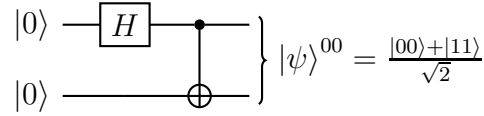
\begin{figure}[H]\label{bellfig}
\centering
\begin{quantikz}
\lstick{$\ket{0}$} & \gate{H} & \ctrl{1} & \rstick[wires=2]{$\ket{\psi}^{00}=\frac{\ket{00}+\ket{11}}{\sqrt{2}}$} \qw\\
\lstick{$\ket{0}$} & \qw & \targ{} & \qw
\end{quantikz}
\caption{Circuit that create Bell's states}
\end{figure}
\begin{align}\label{EPR}
\ket{\psi}^{00} & = \frac{\ket{0}_A\ket{0}_B+\ket{1}_A\ket{1}_B}{\sqrt{2}} = \frac{\ket{00}+\ket{11}}{\sqrt{2}}
\end{align}
Here we’ve decided that the first member of the pair belongs to Alice and the second member of the pair belongs to Bob. Now Alice and Bob physically separate. Alice decides that she wants to send the state $\ket{\phi}$ in Eq.(\ref{unknowstate}) to Bob. She can do it by letting it interact with her member of the EPR pair in Eq.(\ref{EPR}). Let’s begin by writing down the state of the entire system. It’s a product state of the unknown state Eq.(\ref{unknowstate}) and the EPR pair Eq.(\ref{EPR}).
\begin{align*}
\ket{\phi}\otimes \ket{\psi}^{00}&={}\left[\alpha\ket{0}+\beta\ket{1}\right]\otimes \frac{\ket{00}+\ket{11}}{\sqrt{2}}\\
& = \frac{\alpha\ket{000}+\alpha\ket{011}+\beta\ket{100}+\beta\ket{111}}{\sqrt{2}}\\
\end{align*}
Writing each term twice (with 2 in denominator to balance) and subtracting the appropriate term from each term so that we end up with Bob having the state $\ket{\phi}$.
\begin{align*}
\ket{\phi}\otimes \ket{\psi}^{00} & = \frac{1}{2\sqrt{2}}\bigg[\alpha\ket{000} \textcolor{red}{+\alpha\ket{110}}+\alpha\ket{011}\textcolor{violet}{+\alpha\ket{101}}\\
&{\quad\quad\;\;} +\alpha\ket{000}\textcolor{red}{-\alpha{\ket{110}}} + \alpha\ket{011}\textcolor{violet}{-\alpha\ket{101}}\\
&{\quad\quad\;\;} +\beta\ket{100} \textcolor{red}{+\beta\ket{010}}+\beta\ket{111}\textcolor{violet}{+\beta\ket{001}}\\
&{\quad\quad\;\;} +\beta\ket{100}\textcolor{red}{-\beta\ket{010}}+\beta\ket{111}\textcolor{violet}{-\beta\ket{001}}\bigg]\\
\end{align*}
Factoring out the last qubit and simplifying.
\begin{align*}
\ket{\phi}\otimes \ket{\psi}^{00}& = \frac{1}{2\sqrt{2}}\bigg[\left(\ket{00}+\ket{11}\right)\left(\alpha\ket{0}+\beta\ket{1}\right)\\
&{\quad\quad\quad } + \left(\ket{00}-\ket{11}\right)\left(\alpha\ket{0}-\beta\ket{1}\right)\\
&{\quad\quad\quad } + \left(\ket{01}+\ket{10}\right)\left(\alpha\ket{1}+\beta\ket{0}\right)\\
&{\quad\quad\quad} + \left(\ket{01}-\ket{10}\right)\left(\alpha\ket{1}-\beta\ket{0}\right)\bigg]\\
\end{align*}
We know that:
\begin{multicols}{4}
\begin{align*}
\ket{\psi}^{00}=\frac{\ket{00}+\ket{11}}{\sqrt{2}},
\end{align*}

\begin{align*}
\ket{\psi}^{01}=\frac{\ket{01}+\ket{10}}{\sqrt{2}},
\end{align*}

\begin{align*}
\ket{\psi}^{10}=\frac{\ket{00}-\ket{11}}{\sqrt{2}},
\end{align*}

\begin{align*}
\ket{\psi}^{11}=\frac{\ket{01}-\ket{10}}{\sqrt{2}},
\end{align*}
\end{multicols}
And we see that:
\begin{enumerate}[label=\roman*)]
\item \begin{align*}
\alpha\ket{0}+\beta\ket{1} & = \hat{I}\ket{\phi}
\end{align*}
\item \begin{align*}
\alpha\ket{0}-\beta\ket{1} & = \hat{Z}\ket{\phi}
\end{align*}
\item \begin{align*}
\alpha\ket{1}+\beta\ket{0} & = \hat{X}\ket{\phi}
\end{align*}
\item \begin{align*}
\alpha\ket{1}-\beta\ket{0} & = \hat{X}\hat{Z}\ket{\phi}
\end{align*}
\end{enumerate}
Therefore, the entire state of system can be written in the form of bell's state as: 
\begin{align*}
\ket{\phi}\otimes \ket{\psi}^{00}&=\frac{1}{2}\left[\ket{\psi}^{00}\otimes \hat{I}\ket{\phi} + \ket{\psi}^{10}\otimes\hat{Z}\ket{\phi} + \ket{\psi}^{01}\otimes\hat{X}\ket{\phi} + \ket{\psi}^{11}\otimes\hat{X}\hat{Z}\ket{\phi}\right]
\end{align*}
Thus if we measure qubits of Sender and Alice in Bell's Basis and get:
\begin{enumerate}[label=\roman*)]
\item $\ket{\psi}^{00} \rightarrow$ Bob will apply $\hat{I}$ to his state, to retrieve $\ket{\phi}$, or in other words, he will do nothing.
\item $\ket{\psi}^{01} \rightarrow$ Bob will apply $\hat{X}$ to his state, to retrieve $\ket{\phi}$.
\item $\ket{\psi}^{10} \rightarrow$ Bob will apply $\hat{Z}$ to his state, to retrieve $\ket{\phi}$
\item $\ket{\psi}^{11} \rightarrow$ Bob will apply $\hat{X}\hat{Z}$ to his state to, retrieve $\ket{\phi}$
\end{enumerate}
The circuit that measures in Bell's basis is given by: 
\begin{figure}[H]
\centering
\begin{quantikz}
\lstick[wires=2]{$\ket{\psi}^{ij}$} & \ctrl{1} & \gate{H} & \rstick{$\ket{i}$}  \qw \\
& \targ{} & \qw & \rstick{$\ket{j}$}\qw
\end{quantikz}
\caption{Circuit for measurement in Bell's Basis}
\label{measurement}
\end{figure}
Therefore the state of entire system after applying Circuit in Fig. \ref{measurement} to the first two qubits, is given as:
\begin{align*}
\ket{\phi}\otimes\ket{\psi}^{00} & = \frac{1}{2}\left[\ket{00}\otimes\ket{\phi}+\ket{10}\otimes\hat{Z}\ket{\phi}+\ket{01}\otimes\hat{X}\ket{\phi}+\ket{11}\otimes\hat{X}\hat{Z}\ket{\phi}\right]
\end{align*}
Now, Alice will measure both her qubits in a computational basis and tell the result of her measurement to Bob over the classical channel.
If:
\begin{enumerate}[label=\roman*)]
\item Alice gets $\ket{00}$, Bob has to do nothing.
\item Alice gets $\ket{10}$, Bob has to apply $Z$ gate.
\item Alice gets $\ket{01}$, Bob has to apply $X$ gate.
\item Alice gets $\ket{11}$, Bob has to apply $XZ$ gate.
\end{enumerate}
The lesson here is that quantum information-based communication can be
characterized by two key aspects—local operations and classical communications (LOCC). That is, each party has two tasks:
\begin{itemize}
\item Performs local quantum mechanical (local unitaries) operations on their respective states.
\item Uses classical communication to communicate measurement results.
\end{itemize}
If classical communications is not used, then the state will appear totally random
to Bob.
\subsection*{Circuit for Teleportation}
\begin{figure}[H]\label{teleportation}
\centering
\begin{quantikz}
\lstick{$\ket{\psi}$} & \ctrl{1} & \gate{H} & \meter{} & \cwbend{2} \\
\lstick[wires=2]{$\ket{\psi}^{00}$} & \targ{1} & \qw & \meter{} \\
\qw & \qw & \qw  & \gate{X}  \vcw{-1} & \gate{Z} & \qw \rstick{$\ket{\psi}$}
\end{quantikz}
\caption{Circuit for Quantum Teleportation}
\end{figure}
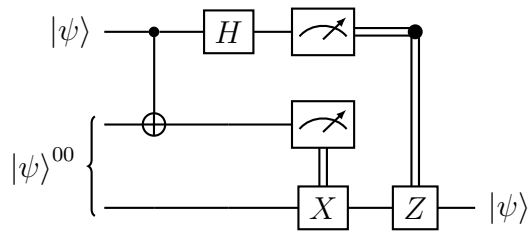
\begin{exercise}{}{label}
Work out for teleportation of a state $\ket{\psi}=\alpha\ket{0}+\ket{1}$ if both parties share
\begin{enumerate}[label=\roman*)]
\item $\ket{\psi}^{01}$
\item $\ket{\psi}^{10}$
\item $\ket{\psi}^{11}$
\end{enumerate}
\end{exercise}

\section{Deutsch-Jozsa Algorithm}
The Deutsch-Jozsa algorithm, first introduced in 1992 by David Deutsch and Richard Jozsa, was the first example of a quantum algorithm that performs better than the best classical algorithm. It showed that there could be advantages to using a quantum computer as a computational tool for a specific problem.
\subsection{Deutsch Problem}
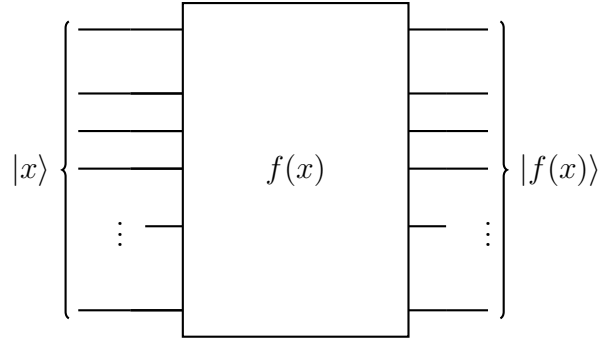
\begin{figure}[H]
\centering
\begin{quantikz}
\lstick[wires=6]{$\ket{x}$} & \qw & \gate[wires=6][3cm]{f(x)} & \qw & \rstick[wires=6]{$\ket{f(x)}$} \qw \\
& \qw & \qw & \qw & \qw\\
& \qw & \qw & \qw & \qw\\
& \qw & \qw & \qw & \qw\\
& \vdots{\ \ } & & \qw  & \vdots\\
& \qw & \qw & \qw & \qw\\
\end{quantikz}
\caption{Black Box: \emph{Oracle}}
\label{Black box}
\end{figure}

Suppose we have a \emph{Black box}, illustrated in Fig. \ref{Black box}, that implements function $f(x)$ to the input $\ket{x}$ and gives the output $\ket{f(x)}$.
This function $f(x)$ can be either balanced or constant function. If $f(x)$ is \emph{constant}, then the output is the same for all input values $x$. If function is \emph{balanced}, then output, $f(x)=0$ for half of the inputs and $f(x)=1$ for the other half of the inputs, and vice-versa. Mathematically, we can write it as,
\begin{equation*}
f(x) = \begin{cases}
\text{Constant}\begin{cases}
\forall\ x\ f(x)=0 \\
\forall\ x\ f(x)=1
\end{cases}\\\\
\text{Balanced}\begin{cases}
f(x)=0\ \text{for half of the inputs}\\
f(x)=1\ \text{for other half of the inputs}
\end{cases}
\end{cases}
\end{equation*}
Our task is to determine whether the given function is balanced or constant.
\subsection{Classical Solution}
Classically, in the best case, two queries to the oracle can determine if the hidden Boolean function $f(x)$: e.g. if we get both $f(0,0,0,\cdots)\rightarrow\ 0$ and $f(1,0,0,\cdots) \rightarrow\ 1$, then we know the function is balanced as we have obtained the two different outputs.\\
\indent In the worst case, if we continue to see the same output for each input we try, we will have to check exactly half of all possible inputs plus one to be certain that $f(x)$ is constant. Since the total number of possible inputs is $2^n$, this means that we need $2^{n-1}+1$ trial inputs to be certain that $f(x)$ is constant. For instance, for a 4-bit string, the maximum number of possible combinations is $2^4=16$. Now, if we check 8 out of 16 combinations, all getting $0'$s, then $9^{th}$ input will decide whether $f(x)$ is constant or balanced i.e., if $9^{th}$ input returns 0, it means function is constant, otherwise it is balanced. Thus the best deterministic classical algorithm to check whether function is constant or balanced requires $\displaystyle{\frac{2^n}{2}+1} \sim 2^{n-1}$ queries, where $n$ is the number of inputs.
\subsection{Quantum Solution}
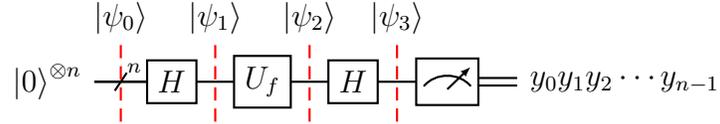
\begin{figure}[H]
\centering
\begin{quantikz}
\lstick{$\ket{0}^{\otimes n}$}\slice{$\ket{\psi_0}$} &[2mm] \gate{H}\slice{$\ket{\psi_1}$}\qwbundle{n} & \gate{U_f}\slice{$\ket{\psi_2}$} & \gate{H}\slice{$\ket{\psi_3}$} & \meter{} & \rstick{$y_0y_1y_2\cdots y_{n-1}$}\cw
\end{quantikz}
\caption{Deutsch-Jozsa Algorithm}
\label{DJ Circuit 01}
\end{figure}

The algorithm is to prepare a bunch of qubits in $\ket{0}$ states, and then we perform Hadamard gate to each qubit independently to create superposition states. After that, we apply phase oracle $U_f$ to all the superposition states. At last, Hadamard gates are applied to the oracle's output, followed by measuring each qubit. Based on the measurement results, we would be able to tell if the function is balanced or constant. The circuit for this algorithm is illustrated in Fig. \ref{DJ Circuit 01}.
Now, let's go through the steps of the algorithm:
\begin{enumerate}
\item Prepare a $n-$qubit register initialized to 0.
\begin{align}\label{equation-1}
\ket{\psi_0}& =\ket{0}\otimes\ket{0}\otimes\cdots\otimes\ket{0}
\end{align}
\item Apply Hadamard gate to each qubit.
\begin{equation}
\begin{split}
\ket{\psi_1} & = H\ket{0}\otimes H\ket{0}\otimes\cdots\otimes H\ket{0}\\
& = \frac{\ket{0}+\ket{1}}{\sqrt{2}} \otimes \frac{\ket{0}+\ket{1}}{\sqrt{2}} \otimes \cdots \otimes \frac{\ket{0}+\ket{1}}{\sqrt{2}}\\
& = \left(\frac{1}{\sqrt{2}}\right)^n\big[(\ket{0}+\ket{1})\otimes(\ket{0}+\ket{1})\otimes\cdots\otimes(\ket{0}+\ket{1})\big]\\
& = \frac{1}{\sqrt{2^n}}\Big(\ket{000\cdots 00}+\ket{000\cdots 01}+\ket{000\cdots 10}+\cdots+\ket{111\cdots 11}\Big)
\end{split}\label{ch3-equation-2}
\end{equation}
Take $N=2^n$ and since we know binary to decimal conversion\footnote{Suppose $\ket{j}=\ket{j_1,j_2,\cdots,j_n}$, where $j$ is a decimal number and $j_1,j_2,\cdots,j_n$ is a binary string then $j= j_12^{n-1}+j_22^{n-2}+\cdots+j_n2^{n-n}$}, we can write $\ket{000\cdots 00}$ as $\ket{0}$, $\ket{000 \cdots 01}$ as $\ket{1}$ and all the way upto $\ket{111\cdots 11}$ as $\ket{N-1}$. For example, if $n=3$ inputs, $N=8\ \Rightarrow \ket{111}=7$.\\
Thus, Eq.(\ref{ch3-equation-2}) can be written as: 
\begin{equation}
\begin{split}
\ket{\psi_1} & = \frac{1}{\sqrt{N}} \Big(\ket{0}+\ket{1}+\ket{2}+\cdots+\ket{N-1} \Big)\\
& = \frac{1}{\sqrt{N}} \sum_{x=0}^{N-1}\ket{x}
\end{split}\label{ch3-equation-3}
\end{equation}
Now we have a uniform superposition of all inputs which is going to be fed in the phase oracle, which is illustrated in figure below:
\begin{figure}[H]
\centering
\begin{quantikz}
\lstick{$\ket{x}$} & \gate[2,nwires={2},bundle={2}][2cm]{O_f}\qwbundle[alternate]{} & \qwbundle[alternate]{} \rstick{$\ket{x}$}\\
\lstick{$\ket{-}$} & \qw & \qw  \rstick{$\ket{-\oplus f(x)}$}
\end{quantikz}
\caption{Phase Oracle}
\label{Phase_Oracle}
\end{figure}
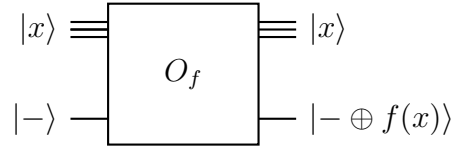
The phase oracle has introduced phase to input qubits whereas the ancilla qubit remained invariant.
\item Apply the phase oracle in Fig. \ref{Phase_Oracle} to the state $\ket{\psi_1}$ in Eq. \ref{ch3-equation-3}.
\begin{align*}
\ket{\psi_2} & = U_f\ket{\psi_1} = \frac{1}{\sqrt{N}}\sum_{x=0}^{N-1}U_f\ket{x}\\
& = \frac{1}{\sqrt{N}}\sum_{x=0}^{N-1} (-1)^{f(x)}\ket{x}
\end{align*}
\item Apply hadamard gate to the state $\ket{\psi_2}$.
\begin{align}\label{ch3-equation-6}
\ket{\psi_3} & = \frac{1}{\sqrt{N}}\sum_{x=0}^{N-1}(-1)^{f(x)}H^{\otimes n}\ket{x}\\
\end{align}
We know that, 
\begin{align*}
H\ket{y_0} & = \frac{\ket{0}+(-1)^{y_0}\ket{1}}{\sqrt{2}} \\
& = \frac{1}{\sqrt{2}}\sum_{k_0=0}^{1}(-1)^{k_0y_0}\ket{k_0}
\end{align*}
Thus for $n$-qubit,
\begin{align*}
H^{\otimes n}\ket{y} & = \frac{1}{\sqrt{2^n}}\sum_{k_0}^{1}\sum_{k_1}^1\cdots \sum_{k_{n-1}}^{1}(-1)^{k_{n-1}y_{n-1}\oplus\cdots\oplus k_0y_0}\ket{k_{n-1}k_{n-2}\cdots k_1k_0}\\
H^{\otimes n}\ket{y} & = \frac{1}{\sqrt{N}}\sum_{k=0}^{N-1}(-1)^{\vec{k}\cdot\vec{y}}\ket{k}
\end{align*}
Therefore, Eq. \ref{ch3-equation-6} can be written as: 
\begin{equation}
\begin{split}
\ket{\psi_3} & = \frac{1}{\sqrt{N}} \sum_{x=0}^{N-1}(-1)^{f(x)}\frac{1}{\sqrt{N}}\sum_{k=0}^{N-1}(-1)^{\vec{k}\cdot\vec{x}}\ket{k} \\
& = \frac{1}{N}\sum_{x=0}^{N-1}\sum_{k=0}^{N-1}(-1)^{f(x)+\vec{k}\cdot\vec{x}}\ket{k} \\
& = \sum_{k=0}^{N-1}\left(\sum_{x=0}^{N-1}\frac{1}{N}(-1)^{f(x)+\vec{k}\cdot\vec{x}}\right)\ket{k} \\
& = \sum_{k=0}^{N-1} C_k\ket{k}
\end{split}
\end{equation}
where
\begin{align*}
C_k & = \sum_{x=0}^{N-1}\frac{1}{N}(-1)^{f(x)+\vec{k}\cdot\vec{x}}
\end{align*}
We can write $\ket{\psi_3}$ as: 
\begin{align}\label{C}
\ket{\psi_3} & = C_0 \ket{000\cdots 00} + C_1 \ket{000\cdots 01} + \cdots + C_{N-1}\ket{111\cdots 11}
\end{align}
\end{enumerate} 
Let us compute the coefficient $C_0$ in Eq. (\ref{C}) that will quickly tell us about the nature of function $f(x)$.
\begin{align*}
C_0 = \frac{1}{N}\sum_{x=0}^{N-1}(-1)^{f(x)+\vec{x}\cdot0}=\frac{1}{N}(-1)^{f(x)}
\end{align*}
\textbf{Constant Function: }
\begin{enumerate}[label=\roman*)]
\item If, $f(x)=0\ \forall x$, \begin{align*}
C_0 & = \frac{1}{N}\sum_{x=0}^{N-1}(-1)^0=\frac{1}{N}\sum_{x=0}^{N-1}1=\frac{N}{N}=1\\
\Rightarrow  \abs{C_0}^2 & = 1
\end{align*}
\item If, $f(x)=1\ \forall x$, \begin{align*}
C_0 & = \frac{1}{N}\sum_{x=0}^{N-1}(-1)^1=\frac{1}{N}\sum_{x=0}^{N-1}(-1)=\frac{-N}{N}=-1\\
\Rightarrow  \abs{C_0}^2 & =1
\end{align*}
\end{enumerate}
\textbf{Balanced Function: }
\begin{enumerate}[label=\roman*)]
\item $f(x)=0$ for half inputs and $f(x)=1$ for another half of inputs.
\begin{align*}
\Rightarrow C_0 & = \frac{1}{N} \left(1-1+1-1+1-1+\cdots+1-1\right) = 0\\
\Rightarrow \abs{C_0}^2 & = 0
\end{align*}
\end{enumerate}
So we can identify the type of function $f(x)$ in only one run, which is because all possible inputs were processed once in phase oracle rather than one by one(classically).
\begin{exercise}{}{label}
Work out the same circuit but for input = $\ket{111\cdots 11}$  and develop the criteria for the function.
\end{exercise}

\section{Bernstein-Vazirani Algorithm}
The Bernstein-Vazirani algorithm, first introduced in 1997 by Ethan Bernstein and Umesh Vazirani, can be seen as an extension of the Deutsch-Jozsa algorithm.
\subsection{Bernstein-Vazirani Problem}
We are given a black-box function $f$ which takes as input a string of bits $x$, and return either 0 or 1, that is:
\begin{align*}
f(\{x_0,x_1,x_2,\cdots\}) & \longrightarrow\ 0\ \text{or}\ 1\ \text{where $x_n$ is 0 or 1}
\end{align*}
The Bernstein-Vazirani Oracle is illustrate in Fig. \ref{Oracle} below:
\begin{figure}[H]
\centering
\Large
\begin{quantikz}
\lstick{$\ket{x}$} &[2mm] \gate{f(x)} \qwbundle[alternate]{} & \qw \rstick{$s\cdot x$(mod 2)}
\end{quantikz}
\caption{Classical Bernstein-Vazirani Oracle}
\label{Oracle}
\end{figure}
As in Deutsch-Jozsa algorithm, the function was either balanced or constant  but here function is guaranteed to return the bit wise product of the input with some \emph{secret string\footnote{For a given black box, $s$ is fixed.}}, $s$. In other words, if input is $x$, the output is $s\cdot x(\text{mod 2})$. Let's have a look at an example.\\
Suppose secrete bit string is:\begin{align*}
s & = 101
\end{align*}
And we have a three bits input that has 8 possible combinations. So the input and output is given in the Table. \ref{table of black box}.
\begin{table}[H]
\centering
\begin{tabular}{|c|c|}
\hline
Input & Output \\
\hline
$x$ & $f(x)=s\cdot x$ (mod 2) \\
\hline
000 & 0 \\
\hline
001 & 1  \\
\hline
010 & 0 \\
\hline
100 & 1 \\
\hline
011 & 1 \\
\hline
101 & 0 \\
\hline
110 & 1 \\
\hline
111 & 0 \\
\hline
\end{tabular}
\caption{Input and Output of the Black Box}
\label{table of black box}
\end{table}
But our problem is to find the secret bit string, $s$.
\subsection{The Classical Solution}
Suppose, I have $n$-input state: $\ket{x_{n-1},x_{n-2},\dotsc,x_1,x_0}$, then $s$ will also be an $n$- bit i.e., $s=s_{n-1},s_{n-2},\dotsc,s_1,s_0$. We can reveal this secret bit string by querying the oracle with the sequence of inputs: 
\begin{table}[H]
\centering
\begin{tabular}{|c|}
\hline
Input ($x$) \\
\hline
$000\cdots 01$ \\
$000\cdots 10$ \\
$\vdots$ \\
$100\cdots 00$\\
\hline
\end{tabular}
\end{table}
\noindent Since,
\begin{align*}
f(x) & = x\cdot s \\
& = x_{n-1}s_{n-1} + x_{n-2}s_{n-2} + \cdots + x_1s_1 + x_0s_0
\end{align*}
So, applying oracle to the first qubit will give us $s_0$ at the output, the second qubit will give us $s_1$, and so on. This means we would need to call the function $f(x)$, $n$ times.
\subsection{The Quantum Solution}
Using a quantum computer, we can solve this problem with 100\% confidence after only one call to the function $f(x)$. The key steps to finding the hidden bit string in the quantum Bernstein-Vazirani algorithm are: 
\begin{enumerate}
\item Initialize each input qubit to $\ket{0}$.
\item Apply Hadamard gate to the input register.
\item Query the oracle.
\item Apply Hadamard gate to the output of oracle.
\item Measure
\end{enumerate}
Circuit for Quantum Bernstein-Vazirani algorithm looks like: 
\begin{figure}[H]
\centering
\Large
\begin{quantikz}
\lstick{$\ket{0}^{\otimes n}$}\slice{$\ket{\psi_0}$} &[4mm] \gate{H}\slice{$\ket{\psi_1}$}\qwbundle{n} & \qw & \gate{U_f}\slice{$\ket{\psi_2}$}& \qw & \gate{H}\slice{$\ket{\psi_3}$} & \qw & \meter{} & \cw \rstick{$s$}
\end{quantikz}
\caption{Quantum Bernstein-Vazirani Algorithm}
\label{QM Bernstein Algo}
\end{figure}
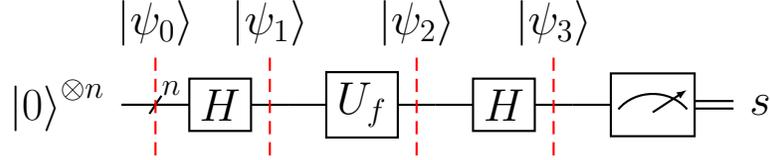
\noindent Input state is given by:
\begin{align}\label{input state}
\ket{\psi_0} & = \ket{0}\otimes\ket{0}\otimes\cdots\otimes\ket{0} = \ket{0}^{\otimes n}
\end{align}
After application of Hadamard gate, the state becomes: 
\begin{equation}\label{ch3-equation-2}
\begin{split}
\ket{\psi_1} & = H\ket{\psi_0} = H\ket{0}\otimes H\ket{0}\otimes\cdots\otimes H\ket{0} \\
& = \left(\frac{1}{\sqrt{2}}\right)^n\big[(\ket{0}+\ket{1})\otimes(\ket{0}+\ket{1})\otimes\cdots\otimes(\ket{0}+\ket{1})\big]\\
& = \frac{1}{\sqrt{2^n}}\Big(\ket{000\cdots 00}+\ket{000\cdots 01}+\ket{000\cdots 10}+\cdots+\ket{111\cdots 11}\Big)
\end{split}
\end{equation}
Take $N=2^n$ and since we know binary to decimal conversion\footnote{Suppose $\ket{j}=\ket{j_1,j_2,\cdots,j_n}$, where $j$ is a decimal number and $j_1,j_2,\cdots,j_n$ is a binary string then $j= j_12^{n-1}+j_22^{n-2}+\cdots+j_n2^{n-n}$}, we can write $\ket{000\cdots 00}$ as $\ket{0}$, $\ket{000 \cdots 01}$ as $\ket{1}$ and all the way upto $\ket{111\cdots 11}$ as $\ket{N-1}$. For example, if $n=3$ inputs, $N=8\ \Rightarrow \ket{111}=7$.\\
Thus, Eq.(\ref{ch3-equation-2}) can be written as: 
\begin{equation}\label{equation-3}
\begin{split}
\ket{\psi_1} & = \frac{1}{\sqrt{N}} \Big(\ket{0}+\ket{1}+\ket{2}+\cdots+\ket{N-1} \Big)\\
& = \frac{1}{\sqrt{N}} \sum_{x=0}^{N-1}\ket{x}
\end{split}
\end{equation}
Now we have a uniform superposition of all inputs which is going to be fed in the phase oracle. So, 
\begin{equation}\label{ch3-equation-4}
\begin{split}
\ket{\psi_2} & = U_f\ket{\psi_1} = \frac{1}{\sqrt{N}}\sum_{x=0}^{N-1}U_f\ket{x} \\
& = \frac{1}{\sqrt{N}}\sum_{x=0}^{N-1} (-1)^{f(x)}\ket{x}
\end{split}
\end{equation}
\begin{equation}\label{ch3-equation-5}
\because f(x)=\vec{s}\cdot\vec{x} \Rightarrow \ket{\psi_2}  = \frac{1}{\sqrt{N}}\sum_{x=0}^{N-1} (-1)^{\vec{s}\cdot\vec{x}}\ket{x}
\end{equation}
Now we will apply Hadamard gate to the state $\ket{\psi_2}$ in Eq.(\ref{ch3-equation-5}).
\begin{equation}\label{equation-6}
\begin{split}
\ket{\psi_3} & = H^{\otimes n}\ket{\psi_2} \\
& = \frac{1}{\sqrt{N}}\sum_{x=0}^{N-1}(-1)^{\vec{s}\cdot\vec{x}}H^{\otimes n}\ket{x} \\
& = \frac{1}{\sqrt{N}}\sum_{x=0}^{N-1}(-1)^{\vec{s}\cdot\vec{x}} \left(\frac{1}{\sqrt{N}}\sum_{k=0}^{N-1}(-1)^{\vec{x}\cdot \vec{k}}\ket{k}\right)\\
& = \frac{1}{N} \sum_{k=0}^{N-1} \left(\sum_{x=0}^{N-1} (-1)^{\vec{s}\cdot\vec{x}}(-1)^{\vec{x}\cdot \vec{k}}  \right)\ket{k} \\
& = \frac{1}{N} \sum_{k=0}^{N-1}  \left(\sum_{x=0}^{N-1} (-1)^{\vec{x}\cdot(\vec{s}\oplus\vec{k})} \right)\ket{k} \\
& = \frac{1}{N} \sum_{k=0}^{N-1} C_k \ket{k}
\end{split}
\end{equation}
where $C_k$ is given by: 
\begin{align}\label{equation-7}
C_k & = \sum_{x=0}^{N-1} (-1)^{\vec{x}\cdot(\vec{s}\oplus\vec{k})}
\end{align}
In Eq.(\ref{equation-7}), $s$ is fixed, $x$ is changing from $0$ to $N-1$ and $k$ can take any value. Consider two cases here:\newline\\
\textbf{Case 1: }
$$s \neq k$$
Let's take a look at an example here:\\
Consider a case of 3-input qubits and take $s=101$ and $k=001$.
\begin{align*}
\Rightarrow s\oplus k & = 101\oplus001 = 100
\end{align*}
Thus:
\begin{align*}
C_{001}&=\sum_{x=0}^7(-1)^{\vec{x}\cdot(\vec{s}\oplus\vec{k})} \\
& = (-1)^{000\cdot 100} + (-1)^{001\cdot 100} + (-1)^{010\cdot 100} + \cdots + (-1)^{101\cdot 100} + (-1)^{110\cdot 100} + (-1)^{111\cdot 100} \\
& = (-1)^0 + (-1)^0 + (-1)^0 + (-1)^0 + (-1)^1 + (-1)^1 + (-1)^1 + (-1)^1\\
& = 1 + 1 + 1 + 1 - 1 - 1 - 1 - 1\\
& = 0
\end{align*}
\noindent So, in general, whenever $s \neq k$, $C_k$ are 0.\newpage
\noindent\textbf{Case 2:} $$s = k$$
\noindent Considering the same example, take $k=101$
\begin{align*}
\Rightarrow s \oplus k & = 101 \oplus 101 = 000
\end{align*}
Thus: 
\begin{align*}
C_k & = \sum_{x=0}^{N-1} (-1)^{\vec{x}\cdot0} \\
& = \sum_{x=0}^{N-1} (-1)^0 \\
& = \sum_{x=0}^{N-1} 1 \\
& = N
\end{align*}
So the output state $\ket{\psi_3}$ can be written as: 
\begin{equation}\label{Final State}
\begin{split}
\ket{\psi_3} & = \frac{1}{N} \left(C_0\ket{0}+C_1\ket{1}+\cdots+C_s\ket{s}+\cdots+C_1\ket{1}+C_0\ket{0}\right)
\end{split}
\end{equation}
But \begin{equation*}
C_k =
\begin{cases}
0\ \text{if}\ s \neq k\\
N\ \text{if}\ s = k
\end{cases}
\end{equation*}
Therefore, final state is nothing but: 
\begin{equation}
\begin{split}
\ket{\psi_3} & =\frac{1}{N}\left(0+0+\cdots+N\ket{s}+\cdots+0+0\right) \\
& = \frac{1}{N}N\ket{s}\\
& = \ket{s}
\end{split}
\end{equation}
The measurement will reveal the hidden bit string.

\section{Simon's Algorithm}
Simon's algorithm was first introduced in 1997 by Daniel R. Simon. It was the first quantum algorithm to show an exponential speed-up versus the best classical algorithm in solving a specific problem. This inspired the quantum algorithms based on the quantum Fourier transform that was used in the most famous quantum algorithm: Shor's factoring algorithm.
\subsection{Simon's Problem}
We are given an unknown blackbox function $f$ such that $f:\{0,1\}^{\otimes n}\mapsto\{0,1\}^{\otimes n}$ where $f$ is guaranteed to be either \emph{one-to-one} or \emph{two-to-one}.
\begin{itemize}
\item \textbf{One-to-One function: }The function that maps exactly one unique output for every input is called one-to-one function. For example: 
$$f(1)\mapsto 1 ,\ f(2)\mapsto 2 ,\ f(3)\mapsto 3,\ f(4)\mapsto 4$$
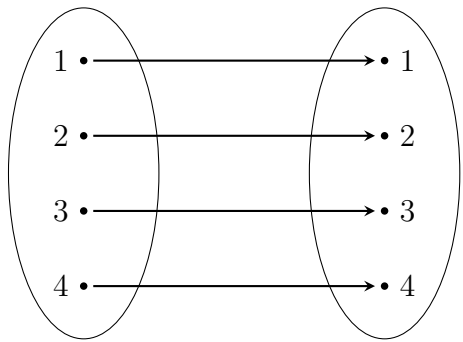
\begin{figure}[H]
\centering
  \begin{tikzpicture}[
    >=stealth,
    bullet/.style={
      fill=black,
      circle,
      minimum width=1pt,
      inner sep=1pt
    },
    projection/.style={
      ->,
      thick,
      shorten <=2pt,
      shorten >=2pt
    },
    every fit/.style={
      ellipse,
      draw,
      inner sep=0pt
    }
  ]
    \foreach \y/\l in {4/1,3/2/,2/3,1/4}
      \node[bullet,label=left:$\l$] (a\y) at (0,\y) {};

    \foreach \y/\l in {1/4,2/3/,3/2,4/1}
      \node[bullet,label=right:$\l$] (b\y) at (4,\y) {};

    \node[draw,fit=(a4) (a3) (a2) (a1),minimum width=2cm] {} ;
    \node[draw,fit=(b4) (b3) (b2) (b1),minimum width=2cm] {} ;

    \draw[projection] (a1) -- (b1);
    \draw[projection] (a2) -- (b2);
    \draw[projection] (a3) -- (b3);
    \draw[projection] (a4) -- (b4);
  \end{tikzpicture}
\caption{One-to-One function}
\end{figure}
There are $n!$ possible one-to-one relations for $n-$inputs.
\item \textbf{Two-to-One function: }The function that maps exactly two outputs for every input are called two-to-one function. For example:
$$f(1)\mapsto1,\ f(2)\mapsto2,\ f(3)\mapsto1,\ f(4)\mapsto2$$
\begin{figure}[H]
\centering
  \begin{tikzpicture}[
    >=stealth,
    bullet/.style={
      fill=black,
      circle,
      minimum width=1pt,
      inner sep=1pt
    },
    projection/.style={
      ->,
      thick,
      shorten <=2pt,
      shorten >=2pt
    },
    every fit/.style={
      ellipse,
      draw,
      inner sep=1pt
    }
  ]
    \foreach \y/\l in {4/1,3/2/,2/3,1/4}
      \node[bullet,label=left:$\l$] (a\y) at (0,\y) {};

    \foreach \y/\l in {4/1,3/2}
      \node[bullet,label=right:$\l$] (b\y) at (2,\y) {};

    \node[draw,fit=(a1) (a2) (a3) (a4),minimum width=2cm] {} ;
    \node[draw,fit= (b3) (b4),minimum width=1.5cm] {} ;

    \draw[projection] (a1) -- (b3);
    \draw[projection] (a2) -- (b3);
    \draw[projection] (a3) -- (b4);
    \draw[projection] (a4) -- (b4);
  \end{tikzpicture}
\caption{Two-to-One function}
\end{figure}
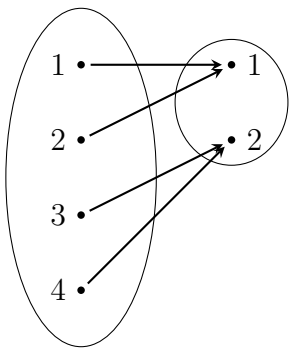
This two-to-one mapping is according to a hidden bit string, $p$, where:
$$\text{given}\ x_1,\ x_2:\quad f(x_1)=f(x_2)$$
$$\text{It is guaranteed:}\ \; x_1\oplus x_2=p$$
For example, if we have a bit string $p=010$, then the truth table is as follow: 
\begin{table}[H]
\centering
\begin{tabular}{|c|c|}
\hline
Input & Output \\
\hline
000 & $f_0$ \\
001 & $f_1$ \\
010 & $f_0$ \\
100 & $f_2$ \\
011 & $f_1$ \\
101 & $f_3$ \\
110 & $f_2$ \\
111 & $f_3$ \\
\hline
\end{tabular}
\end{table}
Here we can see that, Since
\begin{align*}
000\oplus010&=p \\
\Rightarrow f(000) &= f(010)=f_0
\end{align*}
And so on for other functions as well.
\end{itemize}
So given the blackbox $f$, the problem is to find whether the function is one-to-one or two-to-one. This can be done by finding the bit string $p$. If $p$ turns out to be $000$ (for 3-qubit case), then function is one-to-one otherwise two-to-one.

\subsection{Classical Solution}
Classically, if we want to know what secret bit string $p$ is for the given $f$, we have to check upto $\displaystyle{\frac{2^n}{2}+1}$ times, i.e., checking just over the half of the inputs. Here $n$ is number of bits in input. We can also solve our problem in first two queries, but in the worst case, we have to check just over the half of inputs to see whether function is either one-to-one or two-to-one.
\subsection{Quantum Solution}
First we prepare two sets of $\ket{0}^{\otimes n}$ and then we apply the circuit to these qubits as shown below:
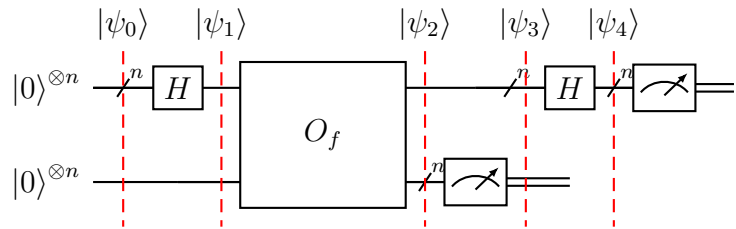
\begin{figure}[H]
\centering
\begin{quantikz}
\lstick{$\ket{0}^{\otimes n}$}\slice{$\ket{\psi_0}$} &[3mm] \gate{H}\qwbundle{n}\slice{$\ket{\psi_1}$} & \gate[wires=2][2.2cm]{O_f}\slice{$\ket{\psi_2}$} & \qw\slice{$\ket{\psi_3}$} & \gate{H}\qwbundle{n}\slice{$\ket{\psi_4}$} & \meter{}\qwbundle{n} & \cw \\
\lstick{$\ket{0}^{\otimes n}$} & \qw  & \qw & \meter{}\qwbundle{n}& \cw & 
\end{quantikz}
\caption{Quantum circuit that implements Simon's Algorithm}
\end{figure}
Where $O_f$ is an oracle such that: 
\begin{align*}
O_f\ket{x}\ket{a} & = \ket{x}\ket{a \oplus f(x)}
\end{align*}
Let's start working each step:
\begin{enumerate}
\item Input state is: 
\begin{equation}
\begin{split}
\ket{\psi_0}=\ket{0}^{\otimes n} \otimes \ket{0}^{\otimes n}
\end{split}
\end{equation}
\item Applying Hadamard gate to first input register:
\begin{equation}
\begin{split}
\ket{\psi_1} & = H^{\otimes n}\ket{0} \otimes \ket{0}^{\otimes n} \\
& = \frac{1}{\sqrt{N}}\sum_{x=0}^{N-1}\ket{x}\otimes \ket{0}^{\otimes n}
\end{split}
\end{equation}
where $N=2^n$.
\item Apply Oracle $O_f$ to the input register:
\begin{equation}
\begin{split}
\ket{\psi_2} & = O_f \ket{\psi_1} \\
& = \frac{1}{\sqrt{N}}\sum_{x=0}^{N-1} \ket{x}\otimes\ket{f(x)}
\end{split}
\end{equation}
$\ket{\psi_2}$ can be written as: 
\begin{equation}
\begin{split}
\ket{\psi_2}&=\frac{1}{\sqrt{N}} \Big(\ket{0}\ket{f(0)}+\ket{1}\ket{f(1)}+\cdots\\
&\cdots+\ket{m}\ket{f(m)+\cdots+\ket{m\oplus p}\ket{f(m)+\cdots+\ket{N-1}\ket{f(N-1)}}} \Big)
\end{split}
\end{equation}
Take $m\oplus p=q$ and writing the state in nicer way, we get: 
\begin{equation}
\begin{split}
\ket{\psi_2} & = \frac{1}{\sqrt{N}}\Bigg[ \left(\sum_{x\neq m}\ket{x}\ket{f(x)}\right)+\big(\ket{m}+\ket{q}\big)\ket{f(m)} \Bigg]
\end{split}
\end{equation}
\item Measuring the second register: After measuring the second register, the second register will collapse to $\ket{f(m)}$. Thus the state right after the measurement is: 
\begin{equation}
\begin{split}
\ket{\psi_3} & = \frac{\ket{m}+\ket{q}}{\sqrt{2}}\otimes \ket{f(m)}
\end{split}
\end{equation}
The factor $\displaystyle{\frac{1}{\sqrt{2}}}$ is just for the sake of normalization.
\item Apply Hadamard gate to the first register:
\begin{equation}
\begin{split}
\ket{\psi_4} & =\frac{H^{\otimes n}\ket{m}+H^{\otimes n}\ket{q}}{\sqrt{2}} \\
& = \frac{1}{\sqrt{2N}}\left(\sum_x(-1)^{\vec{m}\cdot \vec{x}}\ket{x}+\sum_y(-1)^{\vec{q}\cdot \vec{y}}\ket{y}\right) \\
& = \frac{1}{\sqrt{2N}} \left(\sum_x \left((-1)^{\vec{m}\cdot \vec{x}}\ket{x}+(-1)^{\vec{q}\cdot\vec{x}}\ket{x}\right) \right) \\
& = \frac{1}{\sqrt{2N}} \left[\sum_x\left( (-1)^{\vec{m}\cdot \vec{x}}+(-1)^{\vec{q}\cdot \vec{x}} \right)\ket{x} \right]
\end{split}
\end{equation}
\item Measuring the first register: It is clear that measuring the first register will give an output only if: 
\begin{align*}
(-1)^{\vec{m}\cdot \vec{x}} & = (-1)^{\vec{q}\cdot \vec{x}} \\
\Rightarrow \vec{m}\cdot \vec{x} & = \vec{q}\cdot \vec{x} \\
\Rightarrow \vec{m}\cdot \vec{x} & = (m \oplus p) \cdot \vec{x} \\
\Rightarrow \vec{m}\cdot \vec{x} & = \vec{m}\cdot \vec{x} \oplus \vec{p}\cdot \vec{x} \\
\Rightarrow \vec{p}\cdot\vec{x} & = 0\ (mod\ 2)
\end{align*}
So, the string $x$ will measured whose inner product with $p=0$. Thus repeating the algorithm $\approx n$ times, we will be able to find $n$ different values of $x$ and the following system of linear equation can be written: 
\begin{equation}
\begin{cases}
\vec{p}\cdot\vec{x_1} & = 0 \\
\vec{p}\cdot\vec{x_2} & = 0 \\
 \vdots \\
\vec{p}\cdot\vec{x_n} & = 0
\end{cases}
\label{equation-8}
\end{equation}
From Eq.(\ref{equation-8}), we can easily find $p$ by classical methods like Gaussian Elimination method. So if $p=000\cdots00$, function is one-to-one otherwise two-to-one.
\end{enumerate}

\section{Measurement Error Mitigation}
We know that the quantum computers available to us are very noisy. There are many sources of error that make computations deviate from ideal expectations. If the algorithm contains many qubits and gates, then most of the errors in the final result will be coming from the algorithm itself. So if we implement error mitigation techniques on algorithms having more extended depth, i.e., having many qubits, then the effect will be minimal but still visible. On the other hand, implementing this technique on shallow algorithms like Deutsch-Jozsa, Bernstein-Vazirani algorithm, etc., will significantly improve the final result.
\section{Sources of Error}
\begin{enumerate}[label=\roman*)]
\item \textbf{Qubits Initialization: } Qubits don't work exactly the way they should. They get poked by small external forces. So the initial configuration of the qubit may have small errors.
\item \textbf{Unitary Operations: } Each gate is slightly wrong, i.e., unitary operations are not perfect. Even if we do nothing to our qubit, i.e., errors are still present because of the noise present in the system. These types of errors can be characterized as decoherence, decay\footnote{Excited state can be decay to a ground state even if we are doing nothing} and bit flip.
\item \textbf{Measurement Errors: }We can have measurement error while doing computations. For example, if we have a state $\ket{\psi}=\sqrt{p_0}\ket{0}+e^{i\phi}\sqrt{p_1}\ket{1}$. Then ideally, probability of getting $\ket{0}$ is $p_0$ and that of getting $\ket{1}$ is $p_1$. But in real case, for instance, we can have bit flip error that can give the probabilities in reverse order i.e., Pr$(0)=p_1$ and Pr$(1)=p_0$. Similarly, phase flip and bit flip can give us error in the measurement results.
\end{enumerate}
\begin{tcolorbox}
\begin{center}
\large\textbf{Decoherence}
\end{center}
Quantum decoherence is the loss of quantum coherence. As long as there is a definite phase relation between the different states, the system is said to be coherent. A definite phase relationship is necessary for performing quantum computations. Coherence is preserved only if the quantum system is perfectly isolated. But generally, our quantum systems are not entirely isolated from the environment. They still interact with the environment. So, for example, during measurement, coherence is shared with the environment, and it appears to be lost with time. This process is called quantum decoherence.
\end{tcolorbox}
\section{Characterizing Errors}
Instead of characterizing individual error, we can simply characterize total measurement error. Suppose we create a state $\ket{00}$. So, the ideal measurements results are: 
\begin{equation*}
\text{Probabilities}= \begin{cases}
00\ \ :&1 \\
01\ \ :&0 \\
10\ \ :&0 \\
11\ \ :&0 \\
\end{cases}
\end{equation*}
However, in the real case, due to the errors like bit flip, decoherence etc. our measurement results may look like as: 
\begin{equation*}
\text{Probabilities}= \begin{cases}
00\ \ :& 0.90 \\
01\ \ :& 0.05\\
10\ \ :& 0.04\\
11\ \ :& 0.01
\end{cases}
\end{equation*}
The measurement results for each quantum computers will be different depending on the noise they have.\\
\subsection{Error Mitigation with Linear Algebra}
If we speak in term of matrices, then the ideal measurement of the state $\ket{00}$ can be written in the form of column vector as: 
\begin{equation*}
P_{ideal} = \begin{pmatrix}
1\\0\\0\\0
\end{pmatrix}
\end{equation*}
Similarly, non-ideal measurements can be represented as:
\begin{equation*}
P_{noisy}^{00}=\begin{pmatrix}
0.90\\0.05\\0.04\\0.01
\end{pmatrix}
\end{equation*}
Relation between actual and ideal measurements can be characterized as: 
\begin{equation}\label{eq-1}
P_{noisy} = MP_{ideal}
\end{equation}
Similarly, noisy probabilities for the states $\ket{01},\ket{10},\ket{11}$ are given by:
\begin{multicols}{3}
\begin{equation*}
P_{noisy}^{01}=\begin{pmatrix}
0.01 \\ 0.98 \\0.002 \\ 0.008
\end{pmatrix}
\end{equation*}

\begin{equation*}
P_{noisy}^{10}=\begin{pmatrix}
0.02\\0.04\\0.91\\0.03
\end{pmatrix}
\end{equation*}

\begin{equation*}
P_{noisy}^{11}=\begin{pmatrix}
0.01 \\ 0.03 \\ 0.04 \\ 0.92
\end{pmatrix}
\end{equation*}
\end{multicols}
\noindent So the matrix $M$ in Eq.(\ref{eq-1}) is given below: 
\begin{equation}\label{Matrix-M}
\begin{split}
M & = \begin{bmatrix}
P_{noisy}^{00} & P_{noisy}^{01} & P_{noisy}^{10} & P_{noisy}^{11}
\end{bmatrix} \\
M & = \begin{pmatrix}
0.90 & 0.01 & 0.02 & 0.01 \\
0.05 & 0.98 & 0.04 & 0.03 \\
0.04 & 0.002 & 0.91 & 0.04 \\
0.01 & 0.008 & 0.03 & 0.92
\end{pmatrix}
\end{split}
\end{equation}
The matrix $M$, transform my ideal measurement into noisy ones. So, before doing measurements on a quantum computer, if we can somehow know the matrix $M$ of that quantum computer, we can understand what that quantum computer will do with our final state. So we can therefore mitigate such errors in measurement.
\begin{example}{}{label}
Given the matrix $M$ of the quantum computer in Eq.(\ref{Matrix-M}). Calculate the noisy probabilities if we were to measure the state:
\begin{align*}
\ket{\psi}^{00} & = \frac{\ket{00}+\ket{11}}{\sqrt{2}}
\end{align*}
\textbf{Solution: }Upon measuring the state $\ket{\psi}^{00}$, the probability of getting $\ket{00}$ and $\ket{11}$ is $0.5$. Therefore $P_{ideal}$ is: 
\begin{align*}
P_{ideal} & = \begin{pmatrix}
0.5\\0\\0\\0.5
\end{pmatrix}
\end{align*}
Thus, the probabilities that we will actually get is: 
\begin{align*}
P_{noisy} & = \begin{pmatrix}
0.90 & 0.01 & 0.02 & 0.01 \\
0.05 & 0.98 & 0.04 & 0.03 \\
0.04 & 0.002 & 0.91 & 0.04 \\
0.01 & 0.008 & 0.03 & 0.92
\end{pmatrix}\begin{pmatrix}
0.5\\0\\0\\0.5 \end{pmatrix} = \begin{pmatrix}
0.455\\0.04\\0.04\\0.465
\end{pmatrix}
\end{align*}
It means that:
\begin{equation*}
\text{Probability of getting }\begin{cases}
\ket{00} & \text{is: }0.455 \\
\ket{01} & \text{is: }0.04 \\
\ket{10} & \text{is: }0.04 \\
\ket{11} & \text{is: }0.465
\end{cases}
\end{equation*}
\end{example}
\subsubsection{How to recover ideal measurement results?}
\noindent We can simply multiply inverse of $M$ to $P_{ideal}$ to recover our ideal measurements:
\begin{align*}
P_{ideal} & = M^{-1}P_{noisy}
\end{align*}
But remember that errors associated with algorithms cannot be mitigated by the method mentioned above.
\chapter{Algorithms for  Phase Estimation}
\section{Quantum Fourier Transform}\footnote{Copyrights: Muhammad Faryad, LUMS}
The Fourier transform occurs in many different versions throughout classical computing, ranging from signal processing to data compression to complexity theory. The Quantum Fourier Transform (QFT) is the quantum implementation of the discrete Fourier transform over the amplitudes of a wavefunction. It is part of many quantum algorithms, most notably Shor's factoring algorithm and quantum phase estimation.

The Quantum Fourier transform is the quantum implementation of the Discrete Fourier transform (DFT) over the amplitudes of a wave function. Suppose we have a N-dimensional vector of complex number: $(x_0,x_1,\dotsc,x_{N-1})$. Action of DFT on this vector will map this into another vector: $(y_0,y_1,\dotsc,y_{N-1})$ according to the formula:
\begin{equation}\label{ch4-equation-1}
y_k \equiv \frac{1}{\sqrt{N}}\sum_{j=0}^{N-1}x_je^{2\pi ijk/N}
\end{equation}
QFT is exactly the same transformation, although the conventional notation for the quantum Fourier transform is somewhat different. QFT is defined by equation:
\begin{equation}
QFT\ket{j}=\frac{1}{\sqrt{N}}\sum_{k=0}^{N-1}e^{2\pi ijk/N}\ket{k}
\label{ch4-equation-2}
\end{equation}
where
\begin{align*}
N & = 2^n \\
\ket{j} & = \ket{j_{n-1}j_{n-2}\cdots j_2j_1j_0}
\end{align*}
The label $j$ can be computed from binary representation like this:
\begin{align*}
j & = j_{n-1}2^{n-1} + j_{n-2}2^{n-2} + \cdots + j_22^2 + j_12^1 + j_02^0 
\end{align*}
For instance, if $\ket{j}=\ket{101}$ then:
\begin{align*}
j & = 1\cdot2^{3-1}+0\cdot2^{3-2}+1\cdot2^{3-3} \\
& = 4 + 0 + 1 \\
& = 5
\end{align*}
Same is the case for $\ket{k}$.

\subsection{Single Qubit Quantum Fourier Transform}
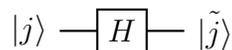
\begin{figure}[H]
\centering
\begin{quantikz}
\lstick{$\ket{j}$} & \gate{H} & \rstick{$\tilde{\ket{j}}$} \qw
\end{quantikz}
\caption{Circuit for Single Qubit QFT}\label{fig42}
\end{figure}

Suppose that $n=1$, then compute the single-qubit quantum Fourier transform of $\ket{j}$ can be found by just using the Hadamard gate. 
Since $n=1 \Rightarrow\ N=2^1=2$
\begin{align*}
\therefore QFT\ket{j} &= \frac{1}{\sqrt{2}}\sum_{k=0}^{1}e^{2\pi ijk/2}\ket{k}\\
\tilde{\ket{j}} & = \frac{1}{\sqrt{2}}\left( \ket{0} + e^{\pi i j}\ket{1} \right)
\end{align*}
For $\ket{j}=\ket{0}$,
\begin{align*}
QFT \ket{0} & = \frac{1}{\sqrt{2}}\left( \ket{0}+\ket{1}\right) \\
\tilde{\ket{0}} & = \ket{+}
\end{align*}
For $\ket{j}=\ket{1}$,
\begin{align*}
QFT \ket{1} & = \frac{1}{\sqrt{2}}\left( \ket{0}+e^{\pi i}\ket{1}\right) \\
& = \frac{1}{\sqrt{2}}\left( \ket{0}-\ket{1}\right) \\
\tilde{\ket{1}}& = \ket{-}
\end{align*}
$\Rightarrow$ For a single qubit, quantum Fourier transform is just a Hadamard gate. The circuit for single qubit QFT is illustrated in Fig. \ref{fig42}.
Bloch sphere visualization is illustrated in the figure below:
\begin{figure}[H]
\centering
\includegraphics[scale=0.4]{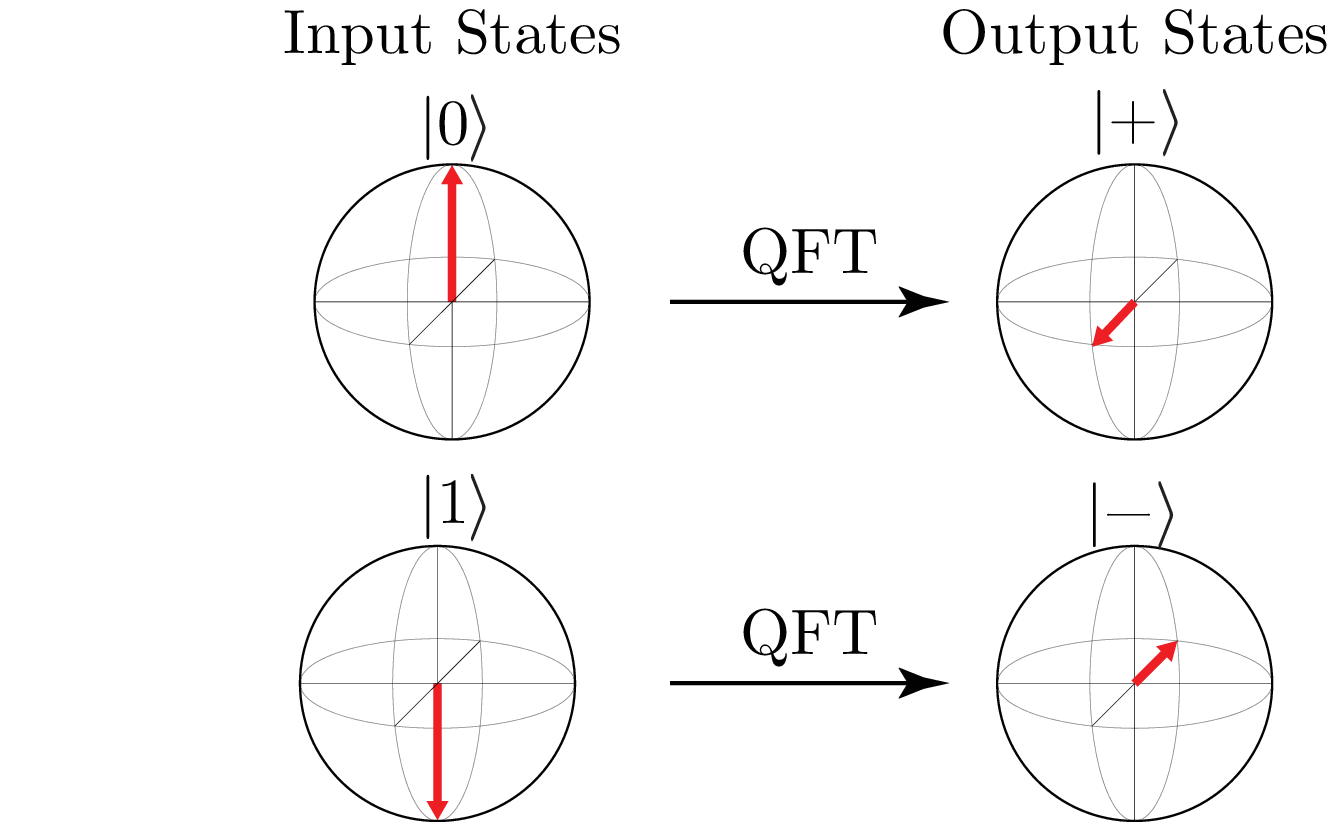}
\caption{Bloch Sphere representation of single qubit QFT}
\end{figure}

\subsection{Two Qubit Quantum Fourier Transform}
Generally, QFT of two qubit can be written as: 
\begin{align*}
\tilde{\ket{j}} & = QFT\ket{j_1j_0} \\
& = \frac{1}{\sqrt{4}}\sum_{k=0}^3e^{2\pi ijk/4}\ket{k} \\
& = \left(\frac{1}{\sqrt{2}}\right)^2\sum_{k_1=0}^1\sum_{k_0=1}^1e^{\frac{2\pi ij(2k_1+k_0)}{4}}\ket{k_1k_0}\\
& = \left(\frac{1}{\sqrt{2}}\right)^2\sum_{k_1=0}^1e^{2\pi ijk_1/2}\ket{k_1}\otimes \sum_{k_0=1}^1e^{2\pi ijk_0/4}\ket{k_0} \\
& = \frac{\ket{0}+e^{\pi ij}\ket{1}}{\sqrt{2}}\otimes \frac{\ket{0}+e^{\pi ij/2}\ket{1}}{\sqrt{2}}
\end{align*}
Substitute $j=2j_1+j_0$ in the above expression:
\begin{align*}
\Rightarrow \tilde{\ket{j}} & = \frac{\ket{0}+e^{2\pi ij_1}e^{i\pi j_0}\ket{1}}{\sqrt{2}} \otimes \frac{\ket{0}+e^{\pi ij_1}e^{i\pi j_0/2}\ket{1}}{\sqrt{2}} \\
& = \frac{\ket{0}+e^{i\pi j_0}\ket{1}}{\sqrt{2}} \otimes \frac{\ket{0}+e^{\pi ij_1}e^{i\pi j_0/2}\ket{1}}{\sqrt{2}}\ \ \because e^{2\pi ij}=1 \\
& = H\ket{j_0}\otimes P\left(\pi/2\right)H\ket{j_1} \\
& = \tilde{\ket{j_1}}\otimes \tilde{\ket{j_0}}
\end{align*}
Circuit for two-qubit quantum Fourier transform is given below: 
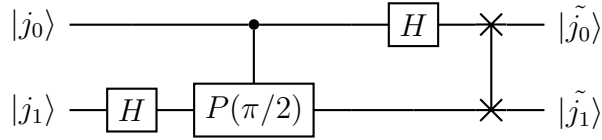
\begin{figure}[H]
\centering
\begin{quantikz}
\lstick{$\ket{j_0}$} 
& \qw & \ctrl{1} & \qw & \gate{H} & \swap{1} & \rstick{$\tilde{\ket{j_0}}$} \qw \\
\lstick{$\ket{j_1}$} & \gate{H} & \gate{P(\pi/2)} & \qw & \qw & \targX{} & \rstick{$\tilde{\ket{j_1}}$} \qw
\end{quantikz}
\caption{Circuit for 2-qubit QFT}
\end{figure}
\begin{example}{}{label}
Compute the quantum Fourier transform of $\ket{00}$ and $\ket{01}$.\\\\
\textbf{Solution:}
Here $n=2\Rightarrow N=2^2=4$
For $\ket{j}=\ket{00}$, $j_1=0,\ j_0=0 \Rightarrow\ j=0$.\\
Similarly, for $\ket{j}=\ket{01}$, $j_1=0,\ j_0=1 \Rightarrow\ j=1$.\\
\textbf{QFT of $\ket{00}$: }
\begin{align*}
\therefore \tilde{\ket{j}} & = \frac{1}{\sqrt{4}}\sum_{k=0}^{3}e^{2\pi ijk/4}\ket{k} \\
\tilde{\ket{00}}& = \frac{1}{\sqrt{4}}\sum_{k=0}^3\ket{k} \\
& = \frac{1}{\sqrt{4}}\left(\ket{0}+\ket{1}+\ket{2}+\ket{3}\right) \\
& = \frac{1}{\sqrt{4}}\left(\ket{00}+\ket{01}+\ket{10}+\ket{11}\right) \\
& = \frac{\ket{0}+\ket{1}}{\sqrt{2}}\otimes \frac{\ket{0}+\ket{1}}{\sqrt{2}}\\
& = \ket{++}
\end{align*}
\textbf{QFT of $\ket{01}$: }
\begin{align*}
\tilde{\ket{01}} & = \frac{1}{\sqrt{4}}\sum_{k=0}^3e^{\pi ik /2}\ket{k} \\
& = \frac{1}{\sqrt{4}}\left(\ket{0}+i\ket{1}-\ket{2}-i\ket{3}\right) \\
& = \frac{1}{\sqrt{4}}\left(\ket{00}+i\ket{01}-\ket{10}-i\ket{11}\right)\\
& = \frac{\ket{0}-\ket{1}}{\sqrt{2}}\otimes \frac{\ket{0}+i\ket{1}}{\sqrt{2}} \\
& = \ket{-}\otimes\ket{+i}
\end{align*}
\end{example}
\begin{exercise}{}{label}
Compute the QFT of $\ket{10},\ket{11}$ and explain the output on Bloch Sphere.
\end{exercise}
\subsection{General Formalism}
\noindent Suppose we have a state:
\begin{align*}
\ket{\phi}=\sum_{j=0}^{N-1}x_{j}\ket{j}
\end{align*}
Then,
\begin{align*}
\ket{\psi}=QFT\ket{\phi}=\sum_{j=0}^{N-1}x_j QFT\ket{j}
\end{align*}
Using Eq.(\ref{ch4-equation-2}) in the above expression will give:

$$\ket{\psi}=\sum_{j=0}^{N-1}x_{j}\frac{1}{\sqrt{N}}\sum_{k=0}^{N-1}e^{2\pi ijk/N}\ket{k}$$
\begin{equation}\label{ch4-equation-3}
\ket{\psi}=\sum_{k=0}^{N-1}\left(\frac{1}{\sqrt{N}}\sum_{j=0}^{N-1}x_je^{2\pi ijk/N}\right)\ket{k}
\end{equation}
Using Eq.(\ref{ch4-equation-1}), $\ket{\psi}$ can be written as:
\begin{equation}\label{ch4-equation-4}
\ket{\psi}=QFT\left(\sum_{j=0}^{N-1}x_j\ket{j}\right)=\sum_{k=0}^{N-1}y_k\ket{k}
\end{equation}
The amplitudes $y_k$ are the discrete Fourier transform of the amplitudes $x_j$ and $\ket{j}=\ket{j_{n-1}j_{n-2}\cdots j_1j_0}$. QFT of $\ket{j}$ in Eq.(\ref{ch4-equation-4}) is given by:
\begin{equation}
\displaystyle{QFT\ket{j}=\frac{1}{\sqrt{2^n}}\sum_{k_{n-1}=0}^1\cdots\sum_{k_1=0}^1\sum_{k_0}^1\exp\left(\frac{2\pi ij}{2^n}\left[k_{n-1}2^{n-1}+k_{n-2}2^{n-2}+\cdots+k_12^1+k_02^0\right]\right)\ket{k}}\label{ch4-equation-5}
\end{equation}
where
\begin{align*}
\tilde{\ket{j}} & = \ket{\tilde{j}_{n-1}\tilde{j}_{n-2}\cdots\tilde{j_1}\tilde{j_0}}
\end{align*}
\begin{equation}
\displaystyle{\tilde{\ket{j}}=\frac{1}{\sqrt{2^n}}\sum_{k_{n-1}=0}^1\cdots\sum_{k_1=0}^1\sum_{k_0}^1\exp\left(\pi ij\left[k_{n-1}+\frac{k_{n-2}}{2}+\frac{k_{n-3}}{4}+\cdots+\frac{k_1}{2^{n-2}}+\frac{k_0}{2^{n-1}}\right]\right)\ket{k_{n-1}k_{n-2}\dotsc k_1k_0}}\label{ch4-equation-6}
\end{equation}
The state $\tilde{\ket{j}}$ can be written in the form of tensor product as:
\begin{equation}\label{ch4-equation-8}
\tilde{\ket{j}}=\frac{1}{\sqrt{2}}\sum_{k_{n-1}=0}^1\exp(i\pi jk_{n-1})\ket{k_{n-1}}\otimes\frac{1}{\sqrt{2}}\sum_{k_{n-2}=0}^1\exp\left(\frac{i\pi jk_{n-2}}{2}\right)\ket{k_{n-2}}\otimes\cdots\otimes\frac{1}{\sqrt{2}}\sum_{k_{0}=0}^1\exp\left(\frac{i\pi jk_0}{2^{n-1}}\right)\ket{k_0}
\end{equation}
\begin{equation}\label{ch4-equation-10}
\displaystyle{\ket{\psi}=\frac{\ket{0}+e^{\pi ij}\ket{1}}{\sqrt{2}}\otimes\frac{\ket{0}+e^{\pi ij/2}\ket{1}}{\sqrt{2}}\otimes \cdots \otimes \frac{\ket{0}+e^{\pi ij/2^{n-1}}\ket{1}}{\sqrt{2}}}
\end{equation}
Now,
\begin{align*}
\exp\left(\pi ij\right) & = \exp\left[2\pi i\left(j_{n-1}2^{n-2}+j_{n-2}2^{n-3}+j_1+j_02^{-1}\right)\right] \\
& = \exp\left(j_0\pi i\right)\ \ \ \ \ \ \because \exp(2n\pi i)=1
\end{align*}
Similarly,
\begin{align*}
\exp\left(\pi ij/2\right) & = \exp\left[\frac{i\pi}{2}\left(j_{n-1}2^{n-1}+j_{n-2}2^{n-2}+\cdots+j_12+j_0\right)\right] \\
& = \exp\left(j_{1}\pi i\right)\ \exp\left(j_0\pi i/2\right)
\end{align*}

\begin{align*}
\exp\left(\pi ij/2^{n-1}\right) & = \exp\left[\frac{i\pi}{2^{n-1}}\left(j_{n-1}2^{n-1}+j_{n-2}2^{n-2}+\cdots+j_12+j_0\right)\right] \\
& = \exp\left(j_{n-1}\pi i\right)\ \exp\left(j_{n-2}\pi i/2\right)\dotsc \exp\left(j_0\pi i/2^{n-1}\right)
\end{align*}
And so on till the last product term in Eq.(\ref{ch4-equation-10}).
Therefore, Eq.(\ref{ch4-equation-10}) becomes:
\begin{equation}\label{ch4-equation-12}
\tilde{\ket{j}}=\frac{\ket{0}+e^{j_0\pi i}\ket{1}}{\sqrt{2}}\otimes\frac{\ket{0}+e^{j_{1}\pi i}e^{j_0\pi i/2}\ket{1}}{\sqrt{2}}\otimes\dotsc\otimes\frac{\ket{0}+e^{j_{n-1}\pi i}e^{j_{n-2}\pi i/2}\dotsc e^{j_0\pi i/2^{n-1}}\ket{1}}{\sqrt{2}}
\end{equation}
Here the output qubits are:
\begin{align*}
\ket{\tilde{j}_{n-1}} & = \frac{\ket{0}+e^{j_0\pi i}\ket{1}}{\sqrt{2}} \\
\ket{\tilde{j}_{n-2}} & =\frac{\ket{0}+e^{j_{1}\pi i}e^{j_0\pi i/2}\ket{1}}{\sqrt{2}}\\
\vdots\ \  & \ \ \ \ \ \ \ \vdots \\
\ket{\tilde{j}_{0}} & = \frac{\ket{0}+e^{j_{n-1}\pi i}e^{j_{n-2}\pi i/2}\dotsc e^{j_0\pi i/2^{n-1}}\ket{1}}{\sqrt{2}}
\end{align*}
Notice that we have two things going on here \textemdash superposition states and introduction to phases. First, we need to build a circuit with Hadamard Gates to introduce superposition states. As for the phases, we see that phase of output qubits depends on the input qubits. So we need to apply controlled phase gates $P(\theta)$. Matrix representation of discrete phase gate in computational basis is:
$$\displaystyle{R_k=\begin{pmatrix}\label{R_k}
1 & 0 \\
0 & e^{2\pi i/2^k}
\end{pmatrix}}$$\\
$R_1$ is just Hadamard Gate($H$), $R_2$ is $\displaystyle{P\left(\frac{\pi}{2}\right)}$, $R_3$ is $\displaystyle{P\left(\frac{\pi}{4}\right)}$ and so on. Thus we can write output states in terms of input states as follow:
\begin{align*}
\ket{\tilde{j}_{n-1}} & = \hat{H}\ket{j_0} \\
\ket{\tilde{j}_{n-2}} & = \hat{R}_2\ \hat{H}\ \ket{j_{1}} \\
\ket{\tilde{j}_{n-3}} & = \hat{R}_3\ \hat{R}_2\ \hat{H}\ \ket{j_{2}} \\
\vdots & \quad\quad \vdots\quad\vdots\quad\vdots \\
\ket{\tilde{j}_{0}}& = \hat{R}_n\ \hat{R}_{n-1}\ \hat{R}_{n-2}\dotsc\hat{R}_2\ \hat{H}\ket{j_{n-1}}
\end{align*}
\subsection*{Circuit implementation of QFT}
Following is the circuit for quantum Fourier transform:
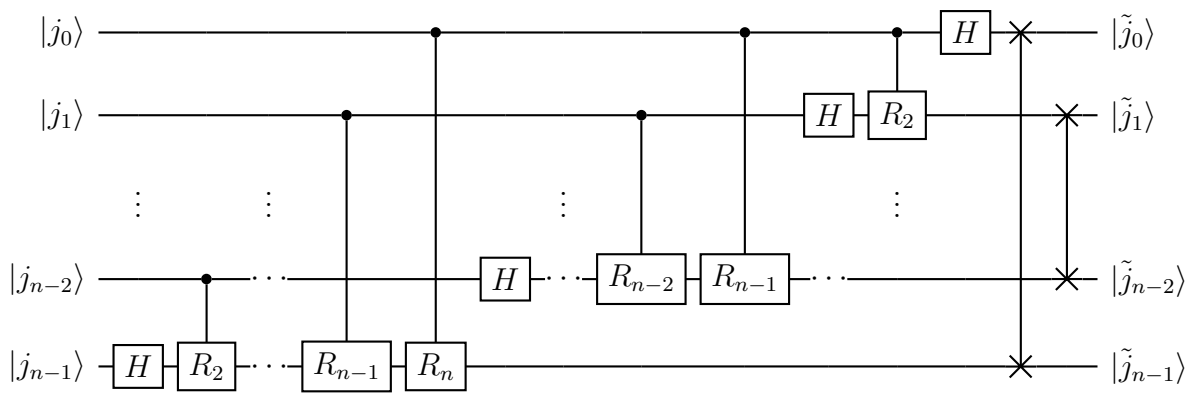
\begin{figure}[H]
\centering
\begin{quantikz}[column sep=0.2cm]
\lstick{$\ket{j_0}$}& \qw & \qw & \qw & \qw  & \ctrl{4} & \qw & \qw & \qw & \ctrl{3} & \qw & \ctrl{1} & \gate{H} & \swap{4} & \qw & \rstick{$\ket{\tilde{j}_{0}}$} \qw\\
\lstick{$\ket{j_{1}}$}& \qw & \qw & \qw & \ctrl{3} & \qw & \qw & \qw & \ctrl{2} & \qw & \gate{H} & \gate{R_{2}} & \qw & \qw & \swap{2} &\rstick{$\ket{\tilde{j}_{1}}$} \qw \\
& \vdots & & \vdots & & & & \vdots &  & & & \vdots & & & & \\
\lstick{$\ket{j_{n-2}}$}& \qw & \ctrl{1} & \qw\cdots & \qw & \qw & \gate{H} &\qw\cdots &\gate{R_{n-2}} & \gate{R_{n-1}} & \qw\cdots & \qw & \qw & \qw &\targX{} & \rstick{$\ket{\tilde{j}_{n-2}}$} \qw \\
\lstick{$\ket{j_{n-1}}$}& \gate{H} & \gate{R_2} & \qw\cdots & \gate{R_{n-1}} & \gate{R_n} & \qw & \qw & \qw & \qw & \qw & \qw & \qw & \targX{} & \qw  &\rstick{$\ket{\tilde{j}_{n-1}}$} \qw \\
\end{quantikz}
\label{circuit}
\caption{Circuit for Quantum Fourier Transform}
\end{figure}
\subsection*{How many gates does this circuit have ?}
We start with a Hadamard Gate and $n-1$ controlled phase gates on the first qubit, totaling $n$-gates. This is followed by a Hadamard gate and $n-2$ controlled phase gates on the second qubit, a total of $n-1$ gates. Continuing this way, we see that total number of gates are $\displaystyle{n+(n-1)+(n-2)+\dotsc+1=\frac{n(n+1)}{2}}$. In addition, we also need to swap the qubits to get output qubits in order with input qubits. Number of swap gate will be equal to $\displaystyle{\frac{n}{2}}$. Thus total number of gates are $\displaystyle{\frac{n^2}{2}+n}$. QFT circuit provides $O(n^2)$ algorithm for performing quantum Fourier transform.
\subsection{Graphical Interpretation}
We know that Hadamard Gate is used to force qubits in a superposition state. The phase gates rotate the qubit $\pi$ radians around the $x+z$ axis. It can be visualized in a Bloch sphere as $\pi/2$ radians rotation around $y-axis$ followed by $\pi$ radians rotation around $x-axis$. While the phase gates $\hat{R}_k$ rotates the qubits around $z-axis$.\\
In quantum Fourier transform, Hadamard gate is applied to input qubit. It is transformed in the superposition state, i.e., in another basis, followed by series of controlled phase gates that will rotate qubits around $z-axis$. Below is the Bloch sphere representation of 4-qubit quantum Fourier transform with input $\ket{1011}$.\\
\begin{figure}[H]
\centering
\includegraphics[scale=0.5]{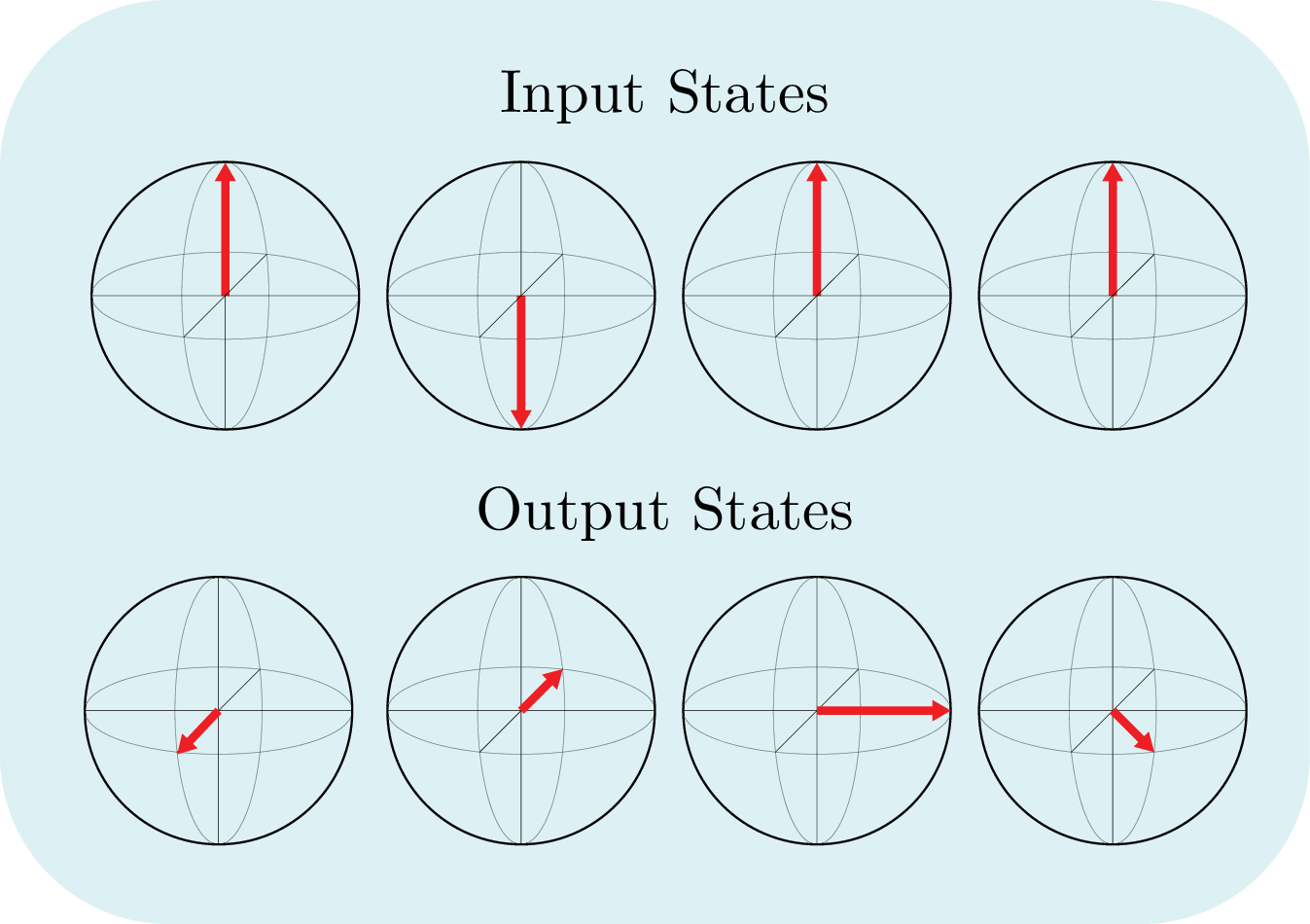}
\caption{4-Qubit QFT representation on Bloch Sphere}
\end{figure}
\subsection{Inverse Fourier Transform}
\noindent First, suppose that:
\begin{align*}
\ket{\psi '} & = ABCD\ket{\psi}
\end{align*}
where $A,B,C$ and $D$ are unitary operators. To get $\ket{\psi}$ back, we have to perform inverse:
\begin{align*}
\ket{\psi} & = (ABCD)^{-1}\ket{\psi '} \\
& = D^{\dagger}C^{\dagger}B^{\dagger}A^{\dagger}\ket{\psi '}
\end{align*}
$\because (XY)^{-1}=Y^{-1}X^{-1}=Y^{\dagger}X^{\dagger}$ in case of unitary operators. Thus the inverse phase gate can be written as: 
\begin{align*}
P^{\dagger} & = \begin{pmatrix}
1 & 0\\
0 & e^{i\theta}
\end{pmatrix}^{\dagger} \\
& = \begin{pmatrix}
1 & 0\\
0 & e^{-i\theta}
\end{pmatrix}
\end{align*}
\subsection*{Circuit for Inverse Fourier Transform}
For the inverse quantum Fourier transform, we have to run the circuit from the end and replace each gate by its hermitian conjugate. The circuit for IQFT is illustrated in the figure below:
\begin{figure}[H]
\centering
\begin{quantikz}[column sep=0.2cm]
\lstick{$\ket{\tilde{j}_{0}}$}& \qw & \swap{4} & \gate{H} & \ctrl{1} & \qw & \ctrl{3} & \qw & \qw & \qw & \ctrl{4} & \qw & \qw & \qw & \qw & \rstick{$\ket{j_0}$} \qw\\
\lstick{$\ket{\tilde{j}_{1}}$}&\swap{2} & \qw & \qw & \gate{R_2^{\dagger}} & \gate{H} & \qw & \ctrl{2} & \qw & \qw & \qw & \ctrl{3} & \qw & \qw & \qw & \rstick{$\ket{j_1}$} \qw \\
& \vdots & & & & \vdots & & & & \vdots & & & \vdots & & & \\
\lstick{$\ket{\tilde{j}_{n-2}}$}& \targX{} & \qw & \qw & \qw & \qw\cdots & \gate{R_{n-1}^{\dagger}} &\gate{R_{n-2}^{\dagger}} &\qw\cdots & \gate{H} & \qw & \qw & \qw\cdots & \ctrl{1} & \qw &\rstick{$\ket{j_{n-2}}$} \qw \\
\lstick{$\ket{\tilde{j}_{n-1}}$}& \qw & \targX{} & \qw & \qw & \qw & \qw & \qw & \qw & \qw & \gate{R_n^{\dagger}} & \gate{R_{n-1}^{\dagger}} &\qw\cdots & \gate{R_2^{\dagger}} & \gate{H} & \rstick{$\ket{j_{n-1}}$} \qw
\end{quantikz}
\caption{Circuit for IQFT}
\end{figure}
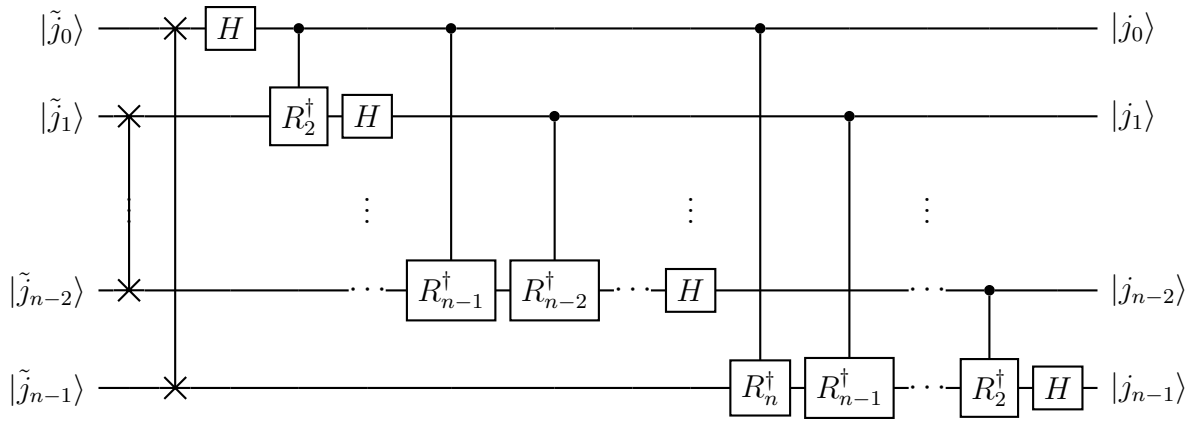
Note that $H^{\dagger}=H$.
\newpage
\section{Quantum Phase Estimation}
Quantum phase estimation is one of the most important subroutines in quantum computation. It serves as a central building block for many quantum algorithms. The objective of the algorithm is to determine the eigenvalue of unitary operator $U$ such that:
\begin{align*}
U\ket{\psi} & = e^{2\pi i\theta}\ket{\psi}
\end{align*}
\subsection*{Eigenvalues, Eigenvectors and Unitary Matrices}
Suppose a matrix $A$ of order $n\times n$ has eigenvalues, $\lambda_i$ and eigenvectors. $\ket{v_i}$, then the eigenvalue equation is: 
\begin{align}\label{equ-1}
A\ket{v_i} & = \lambda_i\ket{v_i}
\end{align}
Eigenvalues can be found by computing det$(A-\lambda \mathbb{I})=0$. Eq.(\ref{equ-1}) can be written as:
\begin{align*}
A\ket{v_i}-\lambda_i\ket{v_i} & = 0\\
(A-\lambda_i\mathbb{I})\ket{v_i} & = 0
\end{align*}
To find eigenvectors, we solve the above equation by Gaussian Elimination Method.
\begin{exercise}{}{label}
Find the eigenvalues and eigenvectors of the given matrix:
\begin{align*}
A & = \begin{pmatrix}
6 & -1 \\
2 & 3
\end{pmatrix}
\end{align*}
\end{exercise}
\subsection*{Unitary Matrices}
A matrix $U$ is called unitary if its inverse is equal to its hermitian conjugate and has eigenvalues with unit magnitude.
\begin{example}{}{label}
Prove that Unitary matrices have eigenvalues with unit magnitude.\\\\
\textbf{Solution:}\\\\
Suppose,
\begin{align}\label{equ-2}
U\ket{v} & = \lambda\ket{v} 
\end{align}
where $\lambda$ is eigenvalue of $U$ and $\ket{v}$ is its eigenvector.
Taking the hermitian conjugate of Eq.(\ref{equ-2}):
\begin{align}\label{equ-3}
\bra{v}U^{\dagger} & = \bra{v}\lambda^{*}
\end{align}
Taking the inner product of Eq.(\ref{equ-2}) and Eq.(\ref{equ-3}):
\begin{align*}
\bra{v}U^{\dagger}U\ket{v} & = \bra{v}\lambda^*\lambda\ket{v} \\
\bra{v}\mathbb{I}\ket{v} & = \abs{\lambda}^2\ip{v}{v}\ \ \ \because\ U\ \text{is unitary}\\
\ip{v}{v} & = \abs{\lambda}^2 \\
\Rightarrow \abs{\lambda}^2 & = 1 \\
\Rightarrow \abs{\lambda} & = 1
\end{align*}
\end{example}
\noindent Therefore, we can write the eigenvalues of unitary matrix in polar form:
\begin{align*}
\lambda & = e^{i\theta}\ \ :\ 0\leq\theta\leq2\pi
\end{align*}
We can also write it in the form:
\begin{align*}
\lambda & = e^{2\pi i j}\ \ :\ 0\leq j\leq1 \\
\Rightarrow U\ket{v} & = e^{2\pi i j}\ket{v}
\end{align*}
\subsection{Quantum Phase Estimation: Algorithm}
We have a unitary operator such that its $j$th eigenstate has phase $0\leq\theta_j\leq1$ so that
\begin{align*}
U\ket{x_j} & = e^{2\pi i\theta_j}\ket{x_j} \\
\Rightarrow U^2\ket{x_j} & = e^{2\pi i (2\theta_j)}\ket{x_j} \\
\Rightarrow U^3\ket{x_j} & = e^{2\pi i (3\theta_j)}\ket{x_j} \\
\vdots\ \ \ \ & \ \ \ \vdots \\
\Rightarrow U^x\ket{x_j} & = 2^{2\pi i (x\theta_j)}\ket{x_j} \\
\end{align*}
The general circuit for Quantum Phase Estimation is illustrated in a figure below:
\begin{figure}[H]
\centering
\begin{quantikz}
\lstick[wires=4]{$\ket{0}^{\otimes c}$} \slice{1}&[2mm] \gate{H}\slice{2} 
& \ctrl{4} & \qw & \qw & \qw\slice{3} & \gate[wires=4]{IQFT}\slice{4} & \meter{}\slice{5} & \rstick[wires=4]{$\ket{2^c\theta}$}\qw \\
& \gate{H} & \qw & \ctrl{3} & \qw & \qw & & \meter{} & \qw \\
& \vdots & & &\vdots & & & \vdots & \\
& \gate{H} & \qw & \qw & \qw & \ctrl{1} & & \meter{} & \qw  \\
\lstick{$\ket{x_j}^{\otimes n}$}
& \qwbundle[alternate]{}\qwbundle{n} & \gate{U}\qwbundle[alternate]{} & \gate{U^{2}}\qwbundle[alternate]{} & \qwbundle[alternate]{}\cdots &  \gate{U^{2^{c-1}}}\qwbundle[alternate]{}& \qwbundle[alternate]{} & \qwbundle[alternate]{} &\rstick{$\ket{x_j}$} \qwbundle[alternate]{} 
\end{quantikz}
\caption{Circuit of Quantum Phase Estimation}
\label{circuit}
\end{figure}
\subsection{Mathematical Formalism}
\begin{enumerate}
\item At step 1, composite state of the system is given by:
\begin{equation}
\begin{split}
\ket{\psi_1} & = \ket{x_j}^{\otimes n}\otimes \ket{0}^{\otimes c} \\
& = \ket{x_j}^{\otimes n}\otimes \ket{0}\otimes\ket{0}\otimes \cdots \otimes \ket{0}
\end{split}
\label{equ-4}
\end{equation}
\item At step 2, state after the action of Hadamard gate is given by:
\begin{equation}
\begin{split}
\ket{\psi_2} & = \ket{x_j}^{\otimes n}\otimes \frac{\ket{0}+\ket{1}}{\sqrt{2}} \otimes\frac{\ket{0}+\ket{1}}{\sqrt{2}} \otimes \cdots \otimes \frac{\ket{0}+\ket{1}}{\sqrt{2}}
\end{split}
\label{equ-5}
\end{equation}
\item Now, apply the controlled unitary operators:\\
First controlled unitary operator will result:
\begin{equation}
\begin{split}
CU\left( \ket{x_j}^{\otimes n}\frac{\ket{0}+\ket{1}}{\sqrt{2}}\right) & = CU\left( \frac{\ket{x_j}^{\otimes n}\ket{0}+\ket{x_j}^{\otimes n}\ket{1}}{\sqrt{2}}\right) \\
& = \frac{\ket{x_j}^{\otimes n}\ket{0}+e^{2\pi i j\theta}\ket{x_j}^{\otimes n}\ket{1}}{\sqrt{2}} \\
& = \ket{x_j}^{\otimes n}\otimes \frac{\ket{0}+e^{2\pi ij\theta}\ket{1}}{\sqrt{2}}
\end{split}
\end{equation}
Similarly, the rest of the unitary operator will result:
\begin{equation}
\begin{split}
\ket{\psi_3} & = \ket{x_j}^{\otimes n} \otimes \left(\frac{\ket{0}+e^{2\pi i2^{c-1}\theta_j}\ket{1}}{\sqrt{2}}\otimes\frac{\ket{0}+e^{2\pi i2^{c-2}\theta_j}\ket{1}}{\sqrt{2}}\otimes\cdots\otimes\frac{\ket{0}+e^{2\pi i\theta_j}\ket{1}}{\sqrt{2}}\right)
\end{split}
\end{equation}
After rearranging, we can write the state as:
\begin{equation}
\begin{split}
\ket{\psi_3} & = \ket{x_j}^{\otimes n} \otimes \left(\frac{\ket{0}+e^{\pi i2^c\theta_j}\ket{1}}{\sqrt{2}}\otimes\frac{\ket{0}+e^{\frac{\pi}{2}i2^c\theta_j}\ket{1}}{\sqrt{2}}\otimes\cdots\otimes \frac{\ket{0}+e^{\frac{\pi}{2^{c-1}}i2^c\theta_j}\ket{1}}{\sqrt{2}}\right)
\end{split}
\end{equation}
\item 
The state $\ket{\psi_3}$ is the quantum Fourier transform of the state $\ket{2^c\theta_j}$. To recover this state, we have to apply inverse quantum Fourier transform. So, at step 4, the state becomes:
\begin{equation}
\begin{split}
\ket{\psi_4} & = \ket{x_j}^{\otimes n}\otimes \ket{2^c\theta j}
\end{split}
\end{equation}
\item Now, measuring the state will give us the phase, $\theta$:
\begin{align*}
\theta & = \frac{\text{measured state}}{2^c}
\end{align*}
\end{enumerate}
\subsection*{What if we have superposition states of $\ket{x_j}$}
\noindent Suppose that we prepare the state $\ket{\psi}$ such that:
\begin{align*}
\ket{\psi} & = \sum_{j=1}^{\mu}c_j\ket{x_j}
\end{align*}
Then the final state will also be in superposition:
\begin{align*}
\ket{\psi_4} & = \ket{x_j}^{\otimes n}\otimes \sum_{j=1}^{\mu}c_j\ket{2^c \theta j}
\end{align*}
So without knowing the eigenvectors, we can find the eigenvalues of the unitary operator.
\begin{exercise}{}{label}
Instead of $\ket{0}^{\otimes n}$, if we prepare our input register in $\ket{1}^{\otimes n}$, are we still able to get the eigenvalues of the unitary operator? Prove or disprove by working out each step in the circuit given in Fig. \ref{circuit}
\end{exercise}
\paragraph*{} The above analysis applies to the ideal case, where $\theta$ can be written exactly with a $t$ bit binary expansion. What happens when this is not the case? The state $\ket{\psi_3}$ in Eq. (7) can be written as: 
\begin{align}
\ket{\psi_3} & = \ket{x_j}^{\otimes n}\otimes \frac{1}{\sqrt{2^c}}\sum_{k=0}^{2^c-1}e^{2\pi i\theta k}\ket{k}
\end{align}
Applying inverse Fourier transform to this state will give us:
\begin{align}
IQFT \ket{\psi_3} & = \ket{x_j}^{\otimes n}\otimes \frac{1}{\sqrt{N}}\sum_{k=0}^{N-1}e^{2\pi i\theta k}IQFT\ket{k} \\
& = \ket{x_j}^{\otimes n}\otimes \frac{1}{\sqrt{N}}\sum_{k=0}^{N-1}e^{2\pi i\theta k} \left( \frac{1}{\sqrt{N}}\sum_{l=0}^{N-1}e^{-2\pi ikl/N}\ket{l} \right) \\
& = \ket{x_j}^{\otimes n}\otimes \frac{1}{N} \sum_{l=0}^{N-1} \left( \sum_{k=0}^{N-1}e^{2\pi i (\theta - l/N)k}  \right)\ket{l}
\end{align}
Suppose, $\alpha_l$ be the coefficients of $\ket{l}$ and it can be written as:
\begin{align}\label{alphal}
\alpha_l & = \frac{1}{N} \sum_{k=0}^{N-1}\left[e^{2\pi i (\theta - l/N)}\right]^k
\end{align}
This is the sum of geometric series and we know that, 
\begin{align*}
\sum_{r=1}^{n}a^r & = a + a^2 + a^3 + \cdots \\
& = \frac{a(1-r^n}{1-r}
\end{align*}
Thus, Eq. (\ref{alphal}) can be written as: 
\begin{align*}
\alpha_l & = \frac{1}{N} \left( \frac{1-e^{2\pi i(\theta-l/N)N}}{1-e^{2\pi i (\theta -l/N)}} \right)
\end{align*}
Suppose the outcome of final measurement results $\tilde{l}$, from which we can get $\displaystyle{\tilde{\theta}=\frac{\tilde{l}}{N}}$. Thus, the probability amplitude of $\ket{\tilde{l}}$ is given by: 
\begin{align*}
\alpha_{\tilde{l}} & = \frac{1}{N} \left( \frac{1-e^{2\pi i(\theta-\tilde{\theta})N}}{1-e^{2\pi i (\theta -\tilde{\theta})}} \right)
\end{align*}
Let's call $\delta=\theta-\tilde{\theta}$. Thus we can write, 
\begin{align*}
\alpha_{\tilde{l}} & = \frac{1}{N} \left(\frac{1-e^{2\pi i\delta N}}{1-e^{2\pi i\delta}} \right) \\
& = \frac{1}{N} \frac{e^{\pi i\delta N}}{e^{\pi i\delta}} \left( \frac{e^{-\pi i\delta N}-e^{\pi i\delta N}}{e^{-\pi i\delta}-e^{\pi i\delta}} \right) \\
& = \frac{1}{N}e^{\pi i\delta (N-1)} \frac{\sin \left(\pi\delta N\right)}{\sin\left(\pi\delta\right)}
\end{align*}
The probability of getting $\ket{\tilde{l}}$ is given by: 
\begin{align*}
\abs{\alpha_{\tilde{l}}}^2 & = \frac{1}{N^2} \frac{\sin^2 \left(\pi\delta N\right)}{\sin^2\left(\pi\delta\right)}
\end{align*}
This function is periodic with a period equals 1. The plot of the function is shown in the figure below:
\begin{figure}[H]
\centering
\includegraphics[scale=1]{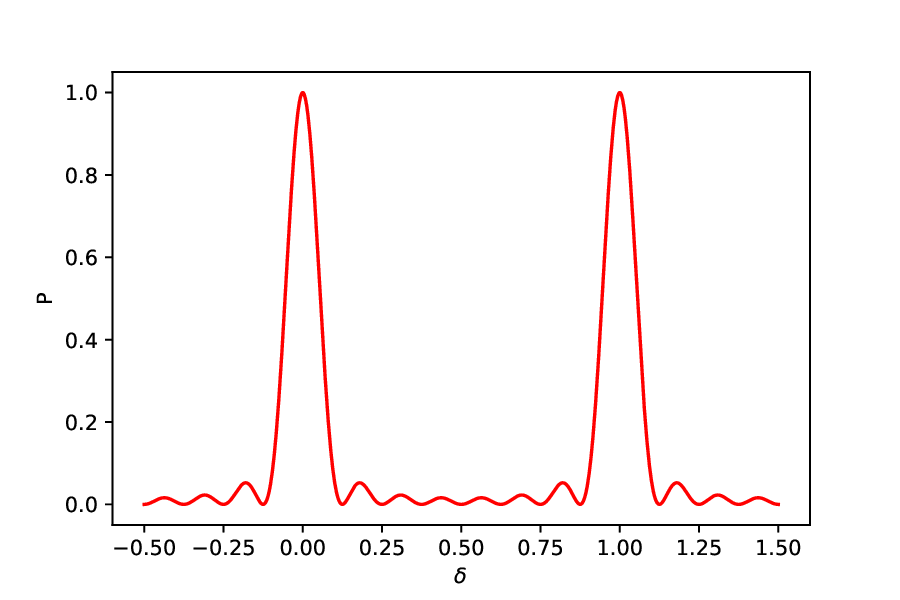}
\caption{Plot of probability: $|\alpha_{\tilde{l}}|^2$ vs $\delta$}
\end{figure}
We see that, the probability of getting $\ket{\tilde{l}}$ is maximum when $\theta-\tilde{\theta}$ is either 0 or 1.
For small $\delta$, $\abs{\alpha_{\tilde{l}}}^2$ is just a square of sinc function,
\begin{align*}
\abs{\alpha_{\tilde{l}}}^2 & = \left( \frac{\sin^2{\pi \delta N}}{\pi \delta N} \right)^2 = \text{sinc}^2(\pi\delta N)
\end{align*}
Now, the probability of getting $\ket{\tilde{l}}$ is maximum if $\delta \rightarrow 0$, i.e., probability is maximum if $\tilde{\theta}\rightarrow \theta$. So, we can say that, the circuit in Fig. (\ref{circuit}) gives us the best approximate of the phase.

\section{Iterative Phase Estimation}
The circuit of Quantum Phase Estimation is limited to the number of qubits necessary for algorithm precision. Every additional qubit add cost in terms of noise and hardware requirements. The IPE algorithm implements quantum phase estimation with only a single auxiliary qubit, and the accuracy of the algorithm is restricted by the number of iterations rather than the number of counting qubits. Therefore, IPE circuits are of practical interest and are of foremost importance for near-term quantum computing as QPE is an essential element in many quantum algorithms.
The circuit for Quantum Phase Estimation along with the components of inverse Fourier transform is illustrated in the figure below.
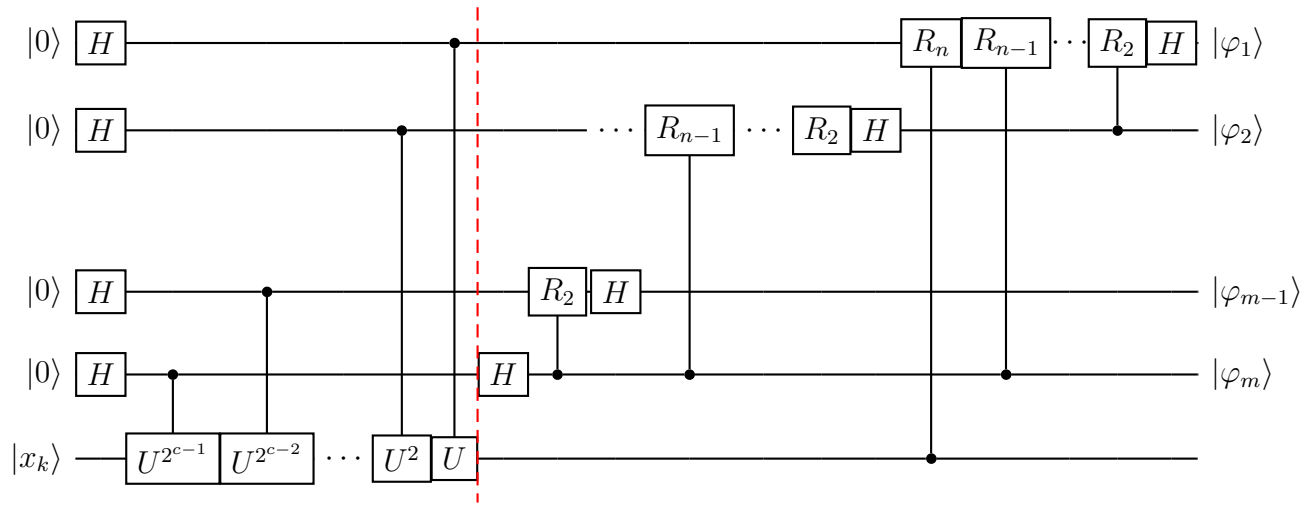
\begin{figure}[H]
\centering
\begin{quantikz}[column sep=0.01cm]
\lstick{$\ket{0}$} & \gate{H} & \qw & \qw & \qw & \qw & \ctrl{4} & \qw & \qw & \qw & \qw & \qw & \qw & \qw & \qw & \gate{R_n} & \gate{R_{n-1}} &\cdots & \gate{R_2} & \gate{H} & \qw & \qw& \rstick{$\ket{\varphi_1}$} \qw \\
\lstick{$\ket{0}$} & \gate{H} & \qw & \qw & \qw & \ctrl{3} & \qw & \qw & \qw & \qw & \qw \ \cdots\  & \gate{R_{n-1}} & \qw\ \cdots\ & \gate{R_2} & \gate{H} & \qw & \qw & \qw & \ctrl{-1} & \qw & \qw & \qw & \rstick{$\ket{\varphi_2}$} \qw \\[1 cm]
	\lstick{$\ket{0}$} & \gate{H} & \qw & \ctrl{2} & \qw & \qw & \qw & \qw & \qw & \gate{R_2} & \gate{H} & \qw & \qw & \qw & \qw & \qw & \qw & \qw & \qw & \qw & \qw & \qw  & \rstick{$\ket{\varphi_{m-1}}$} \qw \\
\lstick{$\ket{0}$} & \gate{H} & \ctrl{1} & \qw & \qw & \qw & \qw & \qw & \gate{H} & \ctrl{-1} & \qw & \ctrl{-2} & \qw  & \qw  & \qw & \qw & \ctrl{-3} & \qw  & \qw & \qw & \qw & \qw & \rstick{$\ket{\varphi_m}$} \qw \\
\lstick{$\ket{x_k}$} & \qw & \gate{U^{2^{c-1}}} & \gate{U^{2^{c-2}}} &\qw\ \cdots\ & \gate{U^2} & \gate{U} \slice{} & \qw  & \qw  & \qw  & \qw  & \qw  & \qw  & \qw  & \qw  & \ctrl{-4}  & \qw  & \qw  & \qw  & \qw  & \qw & \qw 
\end{quantikz}
\caption{Circuit for Quantum Phase Estimation}\label{iqpe}
\end{figure}
The problem of QPE is to find the phase, $\varphi$ of a unitary operator $U$, where $U\ket{u}=e^{2\pi i\varphi}$. Assume that $\varphi=0.\varphi_1\varphi_2\varphi_3...\varphi_m$ can be written as:
\begin{align*}
\varphi & = \varphi_1/2 + \varphi_2/4 + ... + \varphi_m/2^m
\end{align*}
where $\varphi_i$ are the bits of phase. In  Fig. (\ref{iqpe}), we see that the measurement result of the lease significant qubit is independent of the rest of the qubits. Circuit that will measure the least significant bit is given by: 
\begin{figure}[H]
\centering
\begin{quantikz}
\lstick{$\ket{0}$} & \gate{H} & \ctrl{1} & \gate{H} & \qw & \meter{} & \cw \rstick{$\varphi_m$} \\
\lstick{$\ket{x_k}$}\qw & \qw  & \gate{U^{N-1}} & \qw & \qw & \qw & \qw \rstick{$\ket{x_k}$}
\end{quantikz}
\caption{First Step of Iterative Phase Estimation}
\end{figure}
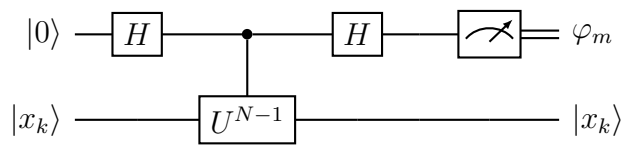
Similarly, the measurement result of second last qubit is independent of the rest of the qubits but the rotation operation depends on the phase bit, $\varphi_m$. Circuit that will measure the second last bit is given by: 
\begin{figure}[H]
\centering
\begin{quantikz}
\lstick{$\ket{0}$} & \gate{H} & \ctrl{1} & \gate{R_2(\phi_m)} & \gate{H} & \meter{} & \cw \rstick{$\varphi_{m-1}$} \\
\lstick{$\ket{x_k}$} & \qw & \gate{U^{N-2}} & \qw & \qw & \qw & \qw \rstick{$\ket{x_k}$} 
\end{quantikz}
\caption{Second Step of Iterative Phase Estimation}
\end{figure}
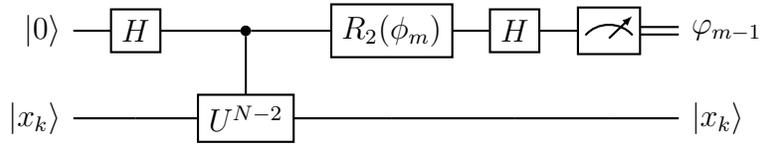
So on and so forth, circuit for measuring the phase bit $\varphi_1$, which is the measurement result of the first qubit is given by: 
\begin{figure}[H]
\centering
\begin{quantikz}
\lstick{$\ket{0}$} & \gate{H} & \ctrl{1} & \gate{R_n(\varphi_m)} & \gate{R_{n-1}(\varphi_{m-1})} & \qw\ \cdots\ & \gate{R_2(\varphi_2)} & \gate{H} & \meter{} & \cw \rstick{$\varphi_1$}\\
\lstick{$\ket{x_k}$} & \qw & \gate{U} & \qw & \qw & \qw & \qw & \qw & \qw & \qw \rstick{$\ket{x_k}$}
\end{quantikz}
\caption{$k-$th Step of Iterative Phase Estimation}
\end{figure}
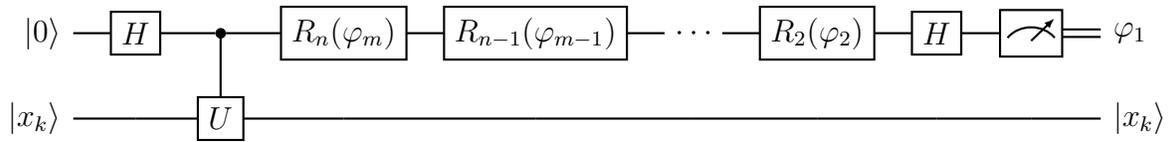
Note that, the rotation operators depends on the phase bits. For instance, in the second step, $\hat{R_2}$ depends on $\varphi_m$. Thus, if $\varphi_m=0$, $R_2$ will not act and if $\varphi_m=1$, the state will be rotated around $z-$axis, through angle $-\pi/2$.
\begin{exercise}{}{label}
Estimate the phase of $T$ gate using Iterative Phase Estimation Algorithm.
\end{exercise}

\chapter{Period Finding Algorithms}
\section{Introduction}
\footnote{Copyrights: Muhammad Faryad, LUMS}
\noindent Period finding algorithm is an application of quantum Fourier transform. Quantum period finding is a simple but fundamental problem in the realm of applications of quantum Fourier transform.
\subsection{Problem:}
\noindent Suppose there is a function $f(x)$ which is a transformation from a set of $t$ bits to single bit, where $x$ is general and output is taken to be 0 and 1 for simplicity and can be generalized later.
\begin{align*}
f(x):\ \{0,1\}^{\otimes t} \rightarrow \{0,1\}
\end{align*}
and suppose $f$ is periodic function such that $f(x+r)=f(x)$, for some unknown $0<r<2^L$, where $L\leq t$ and $x,r \in \{0,1,2,\cdots\}$. Consider an example of following periodic function:
\begin{table}[H]\label{table}
\centering
\begin{tabular}{|c|c|}
\hline
$\ket{x}$ & $\ket{f(x)}$ \\
\hline
$\ket{000}$ & $\ket{0}$ \\
$\ket{001}$ & $\ket{1}$ \\
$\ket{010}$ & $\ket{1}$ \\
$\ket{011}$ & $\ket{0}$ \\
$\ket{100}$ & $\ket{1}$ \\
$\ket{101}$ & $\ket{1}$ \\
$\ket{110}$ & $\ket{0}$ \\
$\ket{111}$ & $\ket{1}$ \\ \hline
\end{tabular}
\caption{Truth table of some periodic function $f(x)$}
\end{table}
The plot of the above function is illustrated below: 
\begin{figure}[H]
\centering
\includegraphics[scale=0.7]{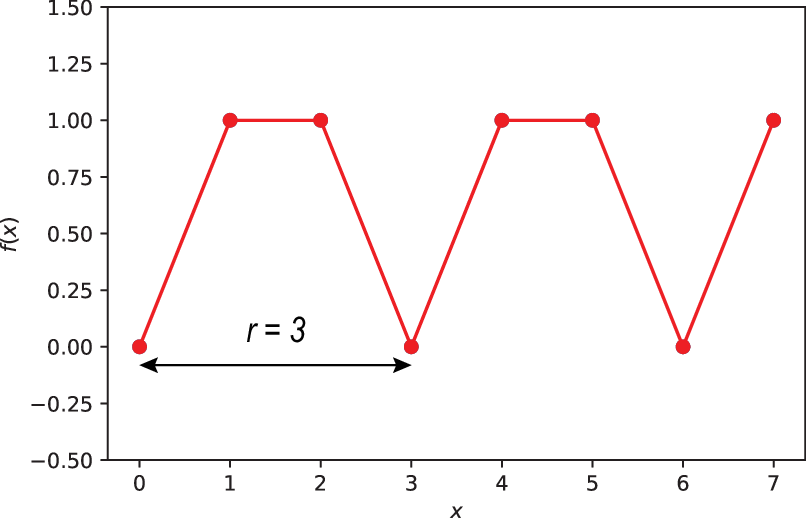}
\end{figure}
Given an oracle $U$ that is given $t$ bit input and an ancillia bit initially set to $\ket{y}$ , performs a unitary transformation $U\ket{x}\ket{y} \rightarrow \ket{x}\ket{y \oplus f(x)}$. For $\ket{y}=\ket{0}$, the output is $\ket{f(x)}$. We are not given any information about $f(x)$ except that it is periodic.
\subsection{Algorithm}
\noindent The circuit that will lead us to find the period of function is given below:
\begin{figure}[H]
\centering
\begin{quantikz}
\lstick{$\ket{0}^{\otimes t}$} & [2mm] \gate{H^{\otimes t}}\qwbundle{t} & \gate[wires=2]{U_f} & \gate{FT^{\dagger}} & \meter{} \\
\lstick{$\ket{0}$} & \qw & \qw & \qw & \qw \rstick{$\ket{f(x)}$}
\end{quantikz}
\end{figure}
\begin{enumerate}
\item We begin with the state,
\begin{align*}
\ket{\psi_{in}} & = \ket{0}^{\otimes t}\ket{0}
\end{align*}
\item Apply hadamard gate to the first register that will create the uniform superposition of all basis states.
\begin{align*}
\ket{\psi_2} & = H^{\otimes t}\ket{0}^{\otimes t} \otimes \ket{0} \\
& = \frac{1}{\sqrt{2^t}} \sum_{x=0}^{2^t-1}\ket{x}\ket{0}
\end{align*}
\item The state $\ket{\psi_2}$ is processed through oracle $U_f$.
\begin{align*}
\ket{\psi_3} & = \frac{1}{\sqrt{2^t}}\sum_{x=0}^{2^t-1} U_f \ket{x}\ket{0} \\
& = \frac{1}{\sqrt{2^t}}\sum_{x=0}^{2^t-1}\ket{x} \ket{f(x)}
\end{align*}
The state $\ket{\psi_3}$ is an entangled state, because we cannot write it in terms of product states. Recall the classical Fourier transform:
\begin{align*}
F(f) & \rightarrow \int f(t) e^{-2\pi i ft} dt \\
f(t) & \rightarrow \int F(f) e^{2\pi i ft} df
\end{align*}
where $F(f)$ is the Fourier transform of $f(t)$. Thus we can write, 
\begin{align}
\ket{\hat{f}(l)} & \equiv \frac{1}{\sqrt{r}} \sum_{x=0}^{r-1}e^{-2\pi ilx/r} \ket{f(x)} \\
\ket{f(x)} & = \frac{1}{\sqrt{r}} \sum_{l=0}^{r-1} e^{2\pi ilx/r} \ket{\hat{f}(l)}
\end{align}
Note that $\ket{\hat{f}(l)}$ is the Fourier transform of $\ket{f(x)}$. It is easy to verify that Eq. (1) and Eq. (2) holds true. Substitute $\ket{\hat{f}(l)}$ from Eq. (1) into Eq. (2)
\begin{align*}
\ket{f(x)} & = \frac{1}{r}\sum_{l=0}^{r-1}e^{2\pi ilx/r} \sum_{x'=0}^{r-1}e^{-2\pi ilx'/r} \ket{f(x')} \\
& = \frac{1}{r} \sum_{l=0}^{r-1}\sum_{x'=0}^{r-1} e^{2\pi i(x-x')l/r}\ket{f(x')}
\end{align*}
Since, $x$ and $x'$ are integers, thus $x-x'$ is also an integer. 
\begin{equation*}
\sum_{l=0}^{r-1} e^{2\pi i (x-x')l/r} = \begin{cases}
0 &\ \text{for }x\neq x' \\
r &\ \text{for }x=x'
\end{cases}
\end{equation*}
It follows that
\begin{equation*}
\ket{f(x)} = \begin{cases}
\ket{f(x)} &\ \text{for }x=x' \\
0 &\ \text{otherwise}
\end{cases}
\end{equation*}
The state $\ket{\psi_3}$ can be thus written as:
\begin{align*}
\ket{\psi_3} & = \frac{1}{\sqrt{r2^t}} \sum_{l=0}^{r-1}\sum_{x=0}^{2^t-1} e^{2\pi ilx/r}\ket{x}\ket{\hat{f}(l)} \\
& = \frac{1}{\sqrt{r2^t}} \sum_{l=0}^{r-1}\sum_{x=0}^{2^t-1} e^{\frac{2\pi ix}{2^t}(2^tl/r)}\ket{x}\ket{\hat{f}(l)}
\end{align*}
\item Now apply the inverse fourier transform.
\begin{align*}
\ket{\psi_4} & = \frac{1}{\sqrt{r}}\sum_{l=0}^{r-1}\left[ FT^{\dagger} \left(\frac{1}{\sqrt{2^t}}\sum_{x=0}^{2^t-1} e^{\frac{2\pi ix}{2^t}(2^tl/r)}\ket{x} \right)\right]\ket{\hat{f}(l)} \\
& = \frac{1}{\sqrt{r}}\sum_{l=0}^{r-1}\ket{2^tl/r}\ket{\hat{f}(l)}
\end{align*}
\item Measuring the first register will give us:
\begin{align}
m & = \frac{2^t l}{r} \\
\Rightarrow \frac{m}{2^t} & = \frac{l}{r}
\end{align}
We know that $\displaystyle{0 \leq \frac{m}{2^t} \leq 1}$ and is always a rational number. We can retrieve $r$ from $\displaystyle{\frac{l}{r}}$ by applying continued fractions algorithm.
\begin{example}{}{label}
What possible outputs can we get if $r=4$ and $t=4$?
\paragraph*{Solution:} If $r=4$, it means $l$ goes from 0 to 3. Using Eq. (3), the possible outputs are: 
\begin{enumerate}
\item For $l=0$,
\begin{align*}
m & = 0\ \text{or}\ 0000
\end{align*}
\item For $l=1$,
\begin{align*}
m & = 4\ \text{or}\ 0100
\end{align*}
\item For $l=2$,
\begin{align*}
m & = 8\ \text{or}\ 1000
\end{align*}
\item For $l=3$,
\begin{align*}
m & = 12\ \text{or}\ 1100
\end{align*}
\end{enumerate}
\end{example}
From the above example, we see that upon measurement we can get only one state, and each state corresponds to different $r$. Thus, we have to run the circuit few times so that we can sample other states as well and calculate $r$ for each. Once we have all the values of $r$, we can consider the least common multiple of them as the period of the function $f(x)$.
\end{enumerate}
\subsection{Implementation}
Consider the periodic function $f(x)$ whose truth table is given in Table. (\ref{table}). The oracle for such transformation is illustrated below: 
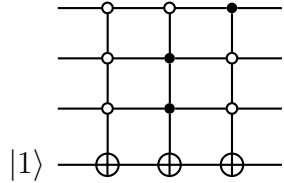
\begin{figure}[H]
\centering
\begin{quantikz}
& \octrl{1} & \octrl{1} & \ctrl{1} & \qw \\
& \octrl{1} & \ctrl{1} & \octrl{1} & \qw \\
& \octrl{1} & \ctrl{1} & \octrl{1} & \qw \\
\lstick{$\ket{1}$} & \targ{} & \targ{} & \targ{} & \qw
\end{quantikz}
\caption{Circuit that implements $U_f$ for the function $f(x)$ given in Table. (\ref{table})}
\end{figure}
We see that, this circuit gives us $\ket{f(x)}=\ket{0}$ when the states are $\{ \ket{000},\ket{011},\ket{100} \}$ and $\ket{1}$ otherwise. Similarly, we can implement oracle for other kind of functions as well.

\section{Number Theory for Factoring}
\paragraph{Prime Number:} An integer $p$ greater than 1 is called prime if it is divisible by only 1 and $p$.
\paragraph{Composite number:}A positive integer that is greater than 1 and is not prime is called composite.
\subsection*{Fundamental Theorem of Arithmetic}
\noindent Any composite number $N$ can be written in the form of prime numbers, 
\begin{align*}
N & = p_1^{\alpha_1}p_2^{\alpha_2}p_3^{\alpha_3}\cdots p_n^{\alpha_n}
\end{align*}
where $p_i$ are prime numbers and $\alpha_i$ are integers. This presentation of composite number is unique.
\paragraph{Proof:} If $N$ is prime, then 
\begin{align*}
N & = 1\times N
\end{align*}
If $N$ is not a prime, then
\begin{align*}
N & = m\times n
\end{align*}
If $m$ is a prime, then we are done here. Same is the case for $n$. If these are not prime, then we will continue to factorize them until we reach prime numbers,
\begin{align}\label{eq1}
N & = p_1 p_2 p_3 \cdots p_n
\end{align}
Now, suppose that there is another representation for $N$ as well, such that
\begin{align}\label{eq2}
N & = q_1 q_2 q_3 \cdots q_m
\end{align}
where $m\neq n$. Equating Eq. (\ref{eq1}) and Eq. (\ref{eq2}) will give us:
\begin{align*}
q_1 q_2 q_3 \cdots q_m & = p_1 p_2 p_3 \cdots p_n
\end{align*}
Suppose, $n<m$ then we are left with
\begin{align*}
q_k q_{k+1} q_{k+2} \cdots q_m & = 1
\end{align*}
and this hold true if $q_k=1$, $q_{k+1}=1$, $\cdots$, $q_m=1$. It shows that we can have only a unique representation of composites numbers in term of primes.
\subsubsection*{If $N$ is a composite integer, then $N$ has a prime divisor less than or equal to $\sqrt{N}$}
\paragraph*{Proof:}If $N$ is a composite number, we know that it has a factor $a$ such that $1<a<N$. Hence, by the definition of a factor of a positive integer, we have $N=ab$, where $b>1$. We want to show that $a\leq \sqrt{N}$ or $b\leq \sqrt{N}$. If $a>\sqrt{N}$ and $b>\sqrt{N}$, then $ab>\sqrt{N}\cdot\sqrt{N}=N$ which is a contradiction. Consequently, $a\leq \sqrt{N}$ or $b\leq\sqrt{N}$. Because both $a$ and $b$ are divisor of $N$, we see that $N$ has a positive divisor not exceeding $\sqrt{N}$. This divisor is either prime or, by the fundamental theorem of arithmetic,
has a prime divisor less than itself. In either case, $N$ has a prime divisor less than or equal to $\sqrt{N}$.

\subsection{Greatest common divisors}
The largest integer that divides both of two integers is called the \textbf{greatest common divisor} of these integers.
\paragraph*{Definition: }Suppose $a$ and $b$ are two non-zero integers. The largest integer $d$ such that $d|a$ and $d|b$ \footnote{$x|y$ means that $y$ is divisible by $x$.} is called the \textit{greatest common divisor} of $a$ and $b$. The greatest common divisor of $a$ and $b$ is denoted by $gcd(a,b)$.
\begin{example}{}{label}
What is the greatest common divisor of $18$ and $36$?
\paragraph*{Solution:}The positive common divisor of $18$ and $36$ are 1, 2, 3, 6 and 9. Hence, $gcd(18,36)=9$
\end{example}
\subsection*{The Euclidean Algorithm}
Computing the greatest common divisor of two integers directly from the prime factorizations of these integers is inefficient. The reason is that it is time-consuming to find prime factoriza- Links tions. We will give a more efficient method of finding the greatest common divisor, called the Euclidean algorithm. The Euclidean algorithm is based on the following result about greatest common divisors and the division algorithm.
\paragraph*{Lemma:} $gcd(a,b)=gcd(b,R)$ where $R$ is remainder when $a$ is divided by $b$.
\paragraph*{Proof:} Suppose that $d$ divides both $a$ and $b$. Then it follows that $d$ also divides $a-bQ=R$. Hence, any common divisor of $a$ and $b$ is also a common divisor of $b$ and $R$. Likewise, suppose that $d$ divides both $b$ and $r$. Then $d$ also divides $bQ+R=a$. Hence, any common divisor of $b$ and $r$ is also a common divisor of $a$ and $b$. Consequently, $gcd(a,b)=gcd(b,R)$
\paragraph*{Algorithm:}Suppose $a$ and $b$ are positive integers with $a\geq b$. Let $r_0=a$ and $r_1=b$. When we successively apply division algorithm, we obtain:
\begin{align*}
r_0 & = r_1q_1 + r_2 \\
r_1 & = r_2q_2 + r_3 \\
\vdots \\
r_{N-2} & = r_{N-1}q_{N-1} + r_N \\
r_{N-1} & = r_Nq_N 
\end{align*}
and $0\leq r_{i}<r_{i-1}$. Eventually a remainder of zero occurs in this sequence of successive divisions, because the sequence of remainders $a=r_0>r_1>r_2\cdots \geq 0$ cannot contain more than $a$ terms. Further more, it follows from Lemma that:
\begin{align*}
gcd(a,b) & = gcd(r_0,r_1) = gcd(r_1,r_2)= \cdots = gcd(r_i,r_{i+1})=\cdots=gcd(r_N,0)
\end{align*}
Hence, the greatest common divisor is the last nonzero remainder in the sequence of divisions.
\begin{example}{}{label}
Find the greatest common divisor of 6 and 21 using the Euclidean algorithm.
\paragraph*{Solution:} We know that $21=6\cdot 3 + 3$. Thus
\begin{align*}
gcd(21,6) & = gcd(6,3)
\end{align*}
Now, $6=3\cdot 2+0$. Thus
\begin{align*}
gcd(6,3) & = gcd(3,0)
\end{align*}
Hence, $gcd(21,6)=3$
\end{example}
\subsection*{Modular Arithmetic}
In some situations we care only about the remainder of an integer when it is divided by some specified positive integer. We generally use the notation $a$ \textbf{mod} $m$ to represent the remainder when an integer
$a$ is divided by the positive integer $m$. But there is also another notation that indicates that two integers have the same remainder when they are divided by the positive integer $m$. Any positive number can be written as:
\begin{align*}
x & = kN + R
\end{align*}
where $R$ is remainder, when $x$ is divided by $N$. Thus we can also write as:
\begin{align*}
x & = R\ \text{(mod N)}
\end{align*}
\subsubsection{GCD as Linear Combinations}
\paragraph*{Theorem:} The greatest common divisor of two integers $a$ and $b$ can be expressed in the form of linear combination:
\begin{align*}
gcd(a,b) & = ax + by
\end{align*}
where $x,y\in \mathbb{Z}$. For example, $gcd(6,14)=2$ and $2=6(-2)+14(1)$. This is also known as B\'{E}ZOUT'S THEOREM. $x$ and $y$ are called B\'{e}zout's coefficients.
\paragraph*{Proof:}
Suppose $s$ is a least positive integer such that, $s  = ax+by$. Now, $gcd(a,b)$ divides $a$ and $b$. It follows that $gcd(a,b)$ must divides $s$. Hence,
\begin{align}\label{1}
gcd(a,b)\leq s
\end{align}
Now, suppose that $s$ does not divide $a$. Then,
\begin{align*}
a & = ks + R \quad\quad ,\quad 0\leq R<s \\
\Rightarrow R & = a - ks \\
\Rightarrow R & = a - k(xa + yb) \\
\Rightarrow R & = a(1-kx) + b(-ky)
\end{align*}
Now, we get another linear combination of $a$ and $b$. But this time, $R<s$, which contradicts ``$s$ is least''. This concludes that $s$ divides $a$. With the same argument, we can easily prove that $s$ divides $b$. This means that,
\begin{align}\label{2}
s \leq gcd(a,b)
\end{align}
because $s$ divides both $a$ and $b$ and $gcd(a,b)$ is the greatest number that divides $a$ and $b$. Eq. (\ref{1}) and Eq. (\ref{2}) implies that 
\begin{align*}
gcd(a,b) & = xa + yb
\end{align*}
\subsubsection*{Corollary:}
\noindent Suppose $a>1$ and $b$, which is multiplicative inverse of $a$ (mod $N$). It means that
\begin{align*}
ab & = 1 \text{(mod N)}
\end{align*}
\begin{example}{}{label}
Find multiplicative inverse of $a=2$, if $N=5$. 
\paragraph*{Solution:} Clearly, if $b=3$, then $ab=6$ and $6$ mod $5$ is 1. Thus multiplicative inverse of $a=2$ is $b=3$.
\end{example}
\noindent In other words, if
\begin{align*}
gcd(a,N) & = 1\ \text({mod N)}
\end{align*}
then inverse of $a$ exists ans we can write
\begin{align*}
1 & = ax + Ny \\
\Rightarrow ax & = 1 - Ny \\
\Rightarrow ax & = 1\ \text{(mod N)}
\end{align*}
This means that $x$ is multiplicative inverse of $a$.
\paragraph*{Theorem:} Suppose $x^2=1$ (mod $N$), where $x\neq \pm 1$. Then we can write: 
\begin{align*}
x^2 - 1 & = 0\ \text{(mod N)} \\
(x-1)(x+1) & = 0\ \text{(mod N)} \\
(x-1)(x+1) & = kN
\end{align*}
This means that $(x-1)$ or $(x+1)$ or both are factors of $N$. It follows that $gcd(x-1,N)$ or $gcd(x+1,N)$ is a factor of $N$.
\paragraph*{What if} $x^r = 1$ (mod $N$)?\\
\noindent For this, we have to choose $x$ such that is it is co-prime with $N$, i.e., they must not share factors. In other words, $gcd(x,N)=1$. Using this we can easily find $r$. For example, if $N=5$, and we choose $x=2$, and a function such that $y(i)=x^i$ then 
\begin{align*}
x^1 & = 2 \\
x^2 & = 4 \\
x^3 & = 3 \\
x^4 & = 1 \\
x^5 & = 2 \\
x^6 & = 4
\vdots 
\end{align*}
We see that $y(i)$ is periodic with period, $r=4$. Generally, if $r$ is even then we can write:
\begin{align*}
\left(x^{\frac{r}{2}}\right)^2 & = 1\ \text{(mod N)} \\
\left(x^{\frac{r}{2}}\right)^2 - 1 & = 0\ \text{(mod N)} \\
(x^{r/2}-1)(x^{r/2}+1) & = 0\ \text{(mod N)}
\end{align*}
Thus, the factors of $N$ will be either $gcd(x^{r/2}-1,N)$ or $gcd(x^{r/2}+1,N)$ or both. If $r$ comes out to be odd, we have to choose $x$ co-prime with $N$ again and again until $r$ comes out to be even.

\section{Shor's Algorithm}
Shor's factoring is essentially a period finding algorithm in disguise. Shor's algorithm is famous for factoring integers in polynomial time. Suppose, we want to find the factors of a composite number $N$, then we choose $a$, such that $gcd(a,N)=1$, i.e., $a$ must be co-prime with $N$. Second step is to find the period of a function $f(x)=a^x\ \text{mod }N$. That essentially means, we have to solve $a^r=1\ (\text{mod }N)$. Remember that $a$ must be such that $r$ comes out to be even otherwise we have to choose another $a$. For an even $r$, we can write
\begin{align*}
\left( a^{r/2}\right)^2 & = 1\ \text{mod }N \\
\left(a^{r/2}-1 \right)\left(a^{r/2} + 1 \right) & = 0\ \text{mod }N
\end{align*}
This means, either $\left(a^{r/2}-1 \right)$ or $\left(a^{r/2}+1 \right)$ or both share factors with $N$. Therefore, factors of $N$ can be either $gcd\left(a^{r/2}-1,N \right)$ or $gcd\left(a^{r/2}+1,N \right)$ or both.
\section{Implementation}
We need an oracle $U_f$ that can implement a function $f(x)=a^x\ (\text{mod }N)$. Consider the circuit below: 
\begin{figure}[H]\label{fig}
\centering
\begin{quantikz}
\lstick{$\ket{x}$} & [4 mm] \gate[wires=2]{U_f}\qwbundle{t} & \qw  \rstick{$\ket{x}$} \\
\lstick{$\ket{y}$} & \qwbundle{n} & \qw \rstick{$\ket{a^x\ \text{mod }N}$}
\end{quantikz}
\caption{An oracle that implements $f(x)=a^x$ mod $N$}
\end{figure}
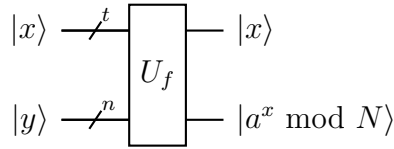
\noindent One way to think is that if $U$ has a form, 
\begin{align}\label{U}
U & = \begin{pmatrix}
I & 0 \\
0 & a^x
\end{pmatrix}
\end{align}
and second register in the Circuit. (\ref{fig}) is initialized to $\ket{1}=\ket{000\cdots 1}$, then we can get $\ket{a^x\ \text{mod }N}$ at output. But how to implement $U$ given in Eq. (\ref{U})? Note that we can write, 
\begin{align*}
a^x & = a^{x_0}\cdot a^{2x_1} \cdot a^{2^2x_2} \cdots\cdots a^{2^{t-1}x_{t-1}}
\end{align*}
Thus we can write $U$ as:
\begin{align*}
U & = \begin{pmatrix}
I & 0 \\
0 & a
\end{pmatrix}^x \\
& = \begin{pmatrix}
I & 0 \\
0 & a
\end{pmatrix}^{x_0}\begin{pmatrix}
I & 0 \\
0 & a
\end{pmatrix}^{2x_1}\cdots\cdots\begin{pmatrix}
I & 0 \\
0 & a
\end{pmatrix}^{2^{t-1}x_{t-1}}
\end{align*}
and take 
\begin{align}\label{V}
V & = \begin{pmatrix}
I & 0 \\
0 & a
\end{pmatrix}
\end{align}
where $V\ket{y}=\ket{ay\ \text{mod }N}$ and $\ket{y}=\ket{1}$ in our case. Thus, we can write
\begin{align*}
U & = V^{x_0} \left(V^{x_1}\right)^2\left(V^{x_2}\right)^{2^2}\cdots \left(V^{x_{t-1}}\right)^{2^{t-1}}
\end{align*}
We see that gates $V$ are controlled by the qubits of first register. The circuit that will find period of the function $f(x)=a^x$ mod $N$ is given by:
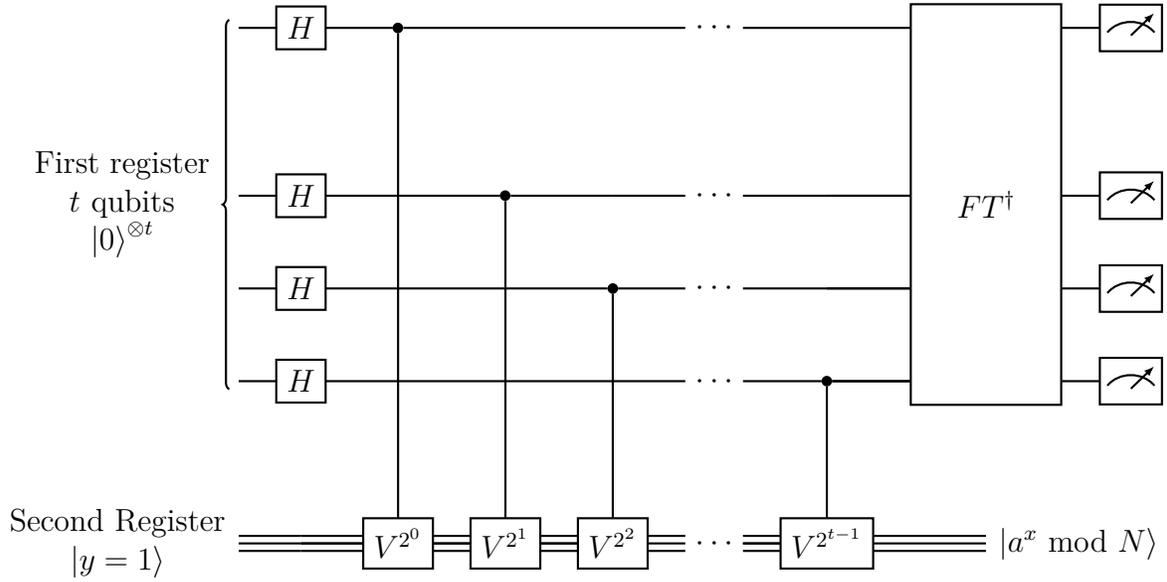
\begin{figure}[H]\label{circuit}
\centering
\begin{quantikz}
\lstick[wires=4]{First register\\ $t$ qubits\\ $\ket{0}^{\otimes t}$} & \gate{H} & \ctrl{4} & \qw & \qw & \qw \ \cdots\  & \qw  & \gate[wires=4][2 cm]{FT^{\dagger}} & \meter{} \\[1 cm]
& \gate{H} & \qw & \ctrl{3} & \qw & \qw \ \cdots\  & \qw  & \qw & \meter{}  \\
& \gate{H} & \qw & \qw & \ctrl{2} & \qw \ \cdots\  & \qw  & \qw & \meter{}  \\
& \gate{H} & \qw & \qw & \qw & \qw \ \cdots\  & \ctrl{1}  & \qw & \meter{} \\ [1 cm]
\lstick[wires=1]{Second Register\\ $\ket{y=1}$}  &  \qwbundle[alternate]{} & \gate{V^{2^0}}\qwbundle[alternate]{} &  \gate{V^{2^1}}\qwbundle[alternate]{} &  \gate{V^{2^2}}\qwbundle[alternate]{} &\qwbundle[alternate]{}\ \cdots\  & \gate{V^{2^{t-1}}}\qwbundle[alternate]{} & \qwbundle[alternate]{} \rstick{$\ket{a^x\ \text{mod }N}$}
\end{quantikz}
\caption{Circuit for period finding of the function $f(x)=a^x$ mod $N$}
\end{figure}
This is the same circuit as that of \emph{Quantum Phase Estimation}. So, at the output, we will obtain the eigenvalue of the operator $V$. It can be shown that eigenvector of unitary operator $V$ has the form,
\begin{align}\label{EV}
\ket{u_s} & = \frac{1}{\sqrt{r}} \sum_{k=0}^{r-1} e^{-2\pi isk/r} \ket{a^k}
\end{align}
with an eigenvalue,
\begin{align}
\label{eigval}
\lambda_s & = e^{2\pi is/r}
\end{align}
\begin{exercise}{}{label}
Verify that $\ket{u_s}$ and $\lambda_s$ are indeed eigenvector and eigenvalue of unitary operator $V$ given in Eq. (\ref{V}).
\paragraph*{Solution:}
Let's apply $V$ to the state $\ket{u_s}$,
\begin{align*}
V\ket{u_s} & = \frac{1}{\sqrt{r}} \sum_{k=0}^{r-1} e^{-2\pi isk/r} V\ket{a^k} \\
& = \frac{1}{\sqrt{r}} \sum_{k=0}^{r-1} e^{-2\pi isk/r} \ket{a^{k+1}}
\end{align*}
Take $m=k+1$, it means, $k=m-1$, thus
\begin{align*}
V\ket{u_s} & = \frac{1}{\sqrt{r}}  \sum_{m=1}^{r} e^{-2\pi is(m-1)/r} \ket{a^{m}} \\
& = e^{2\pi is/r} \left( \frac{1}{\sqrt{r}}  \sum_{m=0}^{r-1} e^{-2\pi ism/r} \ket{a^{m}} \right) \\
& = e^{2\pi is/r} \ket{u_s}
\end{align*}
\end{exercise}
We know that, in quantum phase estimation, we have to initialize second register to a state which is a linear superposition of eigenfunctions of operator $V$. Therefore, the state $\ket{y=1}$ must be a linear super position of $\ket{u_s}$. 
\begin{exercise}{}{label}
Prove that $\ket{y=1}$ is a linear superposition of $\ket{u_s}$.
\paragraph*{Solution:} We know that
\begin{align*}
\ket{u_s} & = \frac{1}{\sqrt{r}} \sum_{k=0}^{r-1} e^{-2\pi isk/r} \ket{a^k}\\
\Rightarrow \sum_{s=0}^{r-1} \ket{u_s}& = \frac{1}{\sqrt{r}}\sum_{s=0}^{r-1} \sum_{k=0}^{r-1} e^{-2\pi isk/r} \ket{a^k} \\
& = \frac{1}{\sqrt{r}}\sum_{k=0}^{r-1} \left( \sum_{s=0}^{r-1} e^{-2\pi isk/r} \right) \ket{a^k}
\end{align*}
where
\begin{equation*}
\sum_{k=0}^{r-1} \left( \sum_{s=0}^{r-1} e^{-2\pi isk/r} \right) = \begin{cases}
r\ket{1} &\ \text{for }k=0\\
0 &\ \text{otherwise}
\end{cases}
\end{equation*}
Thus, 
\begin{align*}
\sum_{s=0}^{r-1} \ket{u_s} & = \frac{1}{\sqrt{r}} \left(r\right) \ket{1} \\
& = \sqrt{r} \ket{1} \\
\Rightarrow \ket{1} & = \frac{1}{\sqrt{r}} \sum_{s=0}^{r-1} \ket{u_s}
\end{align*}
\end{exercise}
\section{Implementation of $V\ket{y}=\ket{ay\ \text{mod }N}$}
Suppose we want to find factors of $N=15$. So we have to take $a$ such that it is co-prime with $N$. Let's take $a=2$. Then,
\begin{align*}
V\ket{1} & = \ket{2\ \text{mod }15}\text{ } = \ket{0010} \\
V^2\ket{1} & = \ket{4\ \text{mod }15}\text{ } = \ket{0100} \\
V^3\ket{1} & = \ket{8\ \text{mod }15} \text{ }= \ket{1000} \\
V^4\ket{1} & = \ket{16\ \text{mod }15}= \ket{0001} \\
V^5\ket{1} & = \ket{32\ \text{mod }15}= \ket{0010} 
\end{align*}
The circuit that will do this kind of transformation is given by: 
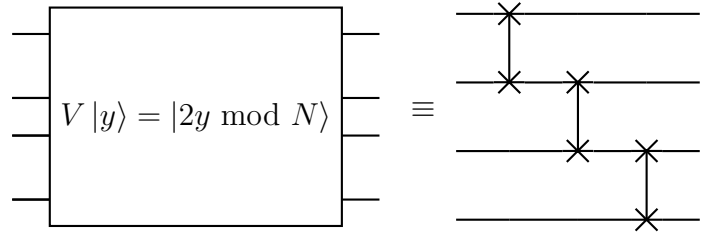
\begin{figure}[H]
\centering
\begin{quantikz}
& \gate[4][3 cm]{V\ket{y}=\ket{2y\ \text{mod }N}} & \qw \\
& \qw & \qw \\
& \qw & \qw \\
& \qw & \qw
\end{quantikz}\ \  $\equiv$
\begin{quantikz}
& \swap{1} & \qw & \qw & \qw \\
& \swap{} & \swap{1} & \qw & \qw \\
& \qw & \swap{} & \swap{1} & \qw \\
& \qw & \qw & \swap{} & \qw
\end{quantikz}
\caption{Circuit implementation of $V$}
\end{figure}
where the swap gate can be implemented in following way:
\begin{figure}[H]
\centering
\begin{quantikz}
& \swap{1} & \qw \\
& \swap{} & \qw
\end{quantikz}\ $\equiv$\ 
\begin{quantikz}
& \ctrl{1} & \targ{} & \ctrl{1} & \qw \\
& \targ{} & \ctrl{-1} & \targ{} & \qw
\end{quantikz}
\caption{Swap Gate}
\end{figure}
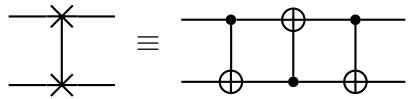
\begin{exercise}{}{label}
Draw a circuit that implements $V\ket{y}=\ket{ay\ \text{mod }N}$ for $N=15$, where $a=7$ and $y=1$.
\end{exercise}
\section{Breaking RSA Encryption with Shor's Factoring}
RSA (Rivest–Shamir–Adleman) is an algorithm used by modern computers to encrypt and decrypt messages. The basic idea behind RSA is to use two keys,
one that is public and one that is private. To decrypt a message we must have the private key. Generally, the security of the system is based on generating quite large prime numbers that are difficult to factorize classically. Thus, we need to factor such numbers in order to crack the system.
\begin{enumerate}
\item We begin with two large prime numbers $p$ and $q$.
\item We make a composite number $n$ such that
\begin{align*}
n & = pq
\end{align*}
\item We compute another product that some imaginative number theorists have denoted the totient:
\begin{align*}
\phi & = (p-1)(q-1)
\end{align*}
\item We choose another number $e$ such that $gcd(e,\phi)=1$, i.e., $e$ is co-prime with $\phi$.
\item Public key is being generated using $n$ and $e$.
\begin{align*}
\text{P-key} & = [n , e]
\end{align*}
\item Secret key is created by computing the multiplicative inverse of $e$, i.e., 
\begin{align*}
de & = 1\ \text{mod }\phi \\
\Rightarrow d & = \frac{1\ \text{mod }\phi}{e}
\end{align*}
Thus the secret key is: 
\begin{align*}
\text{S-key} & = [d,n]
\end{align*}
\end{enumerate}
Now, if Bob wants to send some secure message $M$ to Alice then she must send him the public key so that Bob can encrypt his message $M$. Encryption can be done in the following way:
\begin{align*}
C & = M^e\ \text{mod }n
\end{align*}
Now, Alice will have to decrypt this message in the following way:
\begin{align*}
C^d & = (M^e)^d\ \text{mod }n
\end{align*}
Since $n=pq$, it follows that
\begin{align*}
M^{ed} & = M\ \text{mod }p \\
M^{ed} & = M\ \text{mod }q \\
\Rightarrow C^d & = M\ \text{mod }n
\end{align*}
\paragraph*{}The RSA encryption scheme works for many applications in the present day. However, Shor’s algorithm demonstrates that a workable
quantum computer can easily crack an RSA encryption scheme, since $n =pq$ can
be readily factored.
\begin{exercise}{}{label}
How can we encrypt and decrypt a message $M=6$ using RSA Scheme, where $p=7$ and $q=13$.
\paragraph{Solution:} First we create a product, 
\begin{align*}
n & = pq \\
& = 7\times 13 = 91
\end{align*}
Then we create a totient, 
\begin{align*}
\phi & = (p-1)(q-1) \\
& = 6\times 12 =72
\end{align*}
Let's take $e$ such that $gcd(e,\phi)=1$. The smallest number that satisfy this criterion is $e=5$. To find $d$, we must use the relation
\begin{align*}
de & = 1\ \text{mod }\phi
\end{align*}
It means, we must find a smallest number $k$ such that
\begin{align*}
de & = 1+k\phi \\
\Rightarrow d & = \frac{1+k(72)}{5}
\end{align*}
The smallest value of $k=2$, giving 
\begin{align*}
d & = \frac{145}{5} = 29
\end{align*}
\paragraph{Encryption:} Encrypted message is given by: 
\begin{align*}
C & = M^e\ \text{mod n} \\
& = 6^5\ \text{mod 91} \\
& = 41
\end{align*}
\paragraph{Decryption:} Decrypted message is given by: 
\begin{align*}
M & = C^d\ \text{mod }n \\
& = 41^{29}\ \text{mod }91 \\
& = 6
\end{align*}
\end{exercise}

\section{Discrete Log Algorithm}
The problem is to find $s$, given the two integers $a$ and $b$ such that
\begin{align}\label{eq1}
\log_ab & = s
\end{align}
Three input registers are initialized to the states $\ket{0}^{\otimes t}$, $\ket{0}^{\otimes t}$ and $\ket{m}^{\otimes n}$ respectively where, $2^n=N$. Hadamard gate is applied to first two registers to obtain linear superposition of the basis states. Oracle $U_f$ is then performed that essentially implements a function $\displaystyle{f(x,y)=a^{sx+y}}$ where all variables are integers. Inverse fourier transform is applied to first two registers followed by measurement. From the results of measurement, we can easily obtain $s$. The circuit for this job is illustrated in the figure below: 
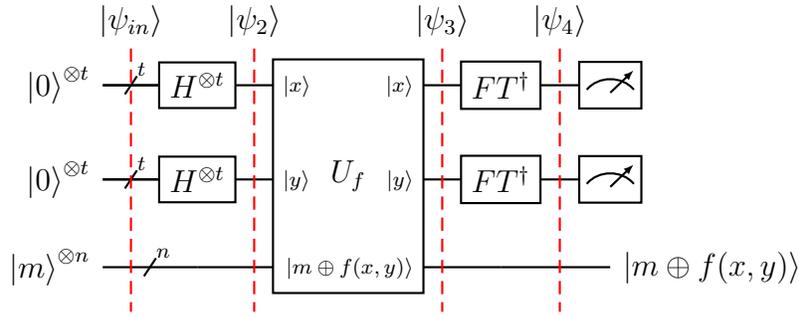
\begin{figure}[H]
\label{circuit}
\centering
\begin{quantikz}
\lstick{$\ket{0}^{\otimes  t}$} \slice{$\ket{\psi_{in}}$} & [2.5 mm]\gate{H^{\otimes t}}\qwbundle{t}\slice{$\ket{\psi_2}$} & \gate[wires=3][2 cm]{U_f}\gateinput{$\ket{x}$}\gateoutput{$\ket{x}$}\slice{$\ket{\psi_3}$} & \gate{FT^{\dagger}}\slice{$\ket{\psi_4}$} & \meter{} \\
\lstick{$\ket{0}^{\otimes t}$} & \gate{H^{\otimes t}}\qwbundle{t} &\gateinput{$\ket{y}$}\gateoutput{$\ket{y}$} & \gate{FT^{\dagger}} & \meter{} \\
\lstick{$\ket{m}^{\otimes n}$} & \qwbundle{n} &\gateoutput{$\ket{m\oplus f(x,y)}$} & \qw & \qw \rstick{$\ket{m\oplus f(x,y)}$}
\end{quantikz}
\caption{Circuit for Discrete Logarithm Algorithm}
\end{figure}
\noindent Now consider the function $f(x,y)$,
\begin{align}\label{eq2}
f(x,y) & = a^{sx+y}\ \text{mod }N
\end{align}
This function is periodic, since
\begin{align*}
f(x+l,y-ls) & = a^{s(x+l)+(y-ls)}\\
& = a^{sx+y}
\end{align*}
But now the period is 2-tuple, $\left(l,-ls\right)$ for any integer $l$. For any $a$, we can find the period, $r$, of the function $f(x,y)$, such that:
\begin{align*}
a^r & = 1\ \text{mod }N
\end{align*}
Also, from Eq. (\ref{eq1}), we can write
\begin{align*}
b & = a^s \\
\Rightarrow f(x,y) & = (a^s)^x\cdot a^y \\
& = b^x\cdot a^y
\end{align*}
Thus, to implement such function, we only need $a$ and $b$.
\section{Implementation of Oracle $U_f$}
Implementation of a function $f(i)=a^i$ is exactly in the same way as that of $f(x)=a^x\ \text{mod }N$ that is explained quite rigorously in Shor's Factoring Algorithm. The circuit that will implement a function $f(x,y)$ is illustrated in the figure below: 
\begin{figure}[H]
\label{oracle_implementation}
\centering
\begin{quantikz}
\lstick{$\ket{x}$} & \gate[wires=3][2 cm]{U_f} & \qw \rstick{$\ket{x}$} \\
\lstick{$\ket{y}$} & & \qw \rstick{$\ket{y}$} \\
\lstick{$\ket{1}$} & & \qw \rstick{$\ket{f(x,y)}$}
\end{quantikz}
\ $\equiv$\ \ 
\begin{quantikz}
\lstick{$\ket{x}$} & \qw & \ctrl{2} & \qw \rstick{$\ket{x}$} \\
\lstick{$\ket{y}$} & \ctrl{1} & \qw & \qw \rstick{$\ket{y}$} \\
\lstick{$\ket{1}$} & \gate{a^y} & \gate{b^x} & \qw \rstick{$\ket{f(x,y)}$}
\end{quantikz}
\caption{Circuit implementation of $f(x,y)$}
\end{figure}
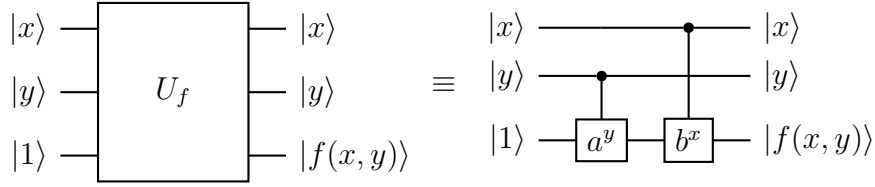
\section{Algorithm: Discrete Logarithm}
\begin{enumerate}
\item Input state is given by: 
\begin{align*}
\ket{\psi_{in}} & = \ket{0}^{\otimes t}\ket{0}^{\otimes t}\ket{0}^{\otimes n}
\end{align*}
\item Hadamard gate is applied to first two registers.
\begin{align*}
\ket{\psi_2} & = H^{\otimes t}\ket{0}^{\otimes t} H^{\otimes t}\ket{0}^{\otimes t}\ket{0}^{\otimes n} \\
& = \frac{1}{\sqrt{2^t}} \sum_{x=0}^{2^t-1}\ket{x} \frac{1}{\sqrt{2^t}} \sum_{y=0}^{2^t-1}\ket{y} \ket{0}^{\otimes n} \\
& = \frac{1}{2^t} \sum_{x=0}^{2^t-1}\sum_{y=0}^{2^t-1} \ket{x}\ket{y}\ket{0}^{\otimes n}
\end{align*}
\item The state $\ket{\psi_2}$ is processed by an oracle that implements a function $f(x,y)$ given in Eq. (\ref{eq2}). State after application of oracle is given by: 
\begin{align*}
\ket{\psi_3} & = \frac{1}{2^t}\sum_{x=0}^{2^t-1}\sum_{y=0}^{2^t-1} \ket{x}\ket{y}\ket{f(x,y)}
\end{align*}
Note that, we can also write $f(x,y)$ as: 
\begin{align*}
f(x,y) & = a^{sx+y} \\
& = f(0,sx+y)
\end{align*}
Take $z=sx+y$ and write function $f(0,z)$ in terms of fourier basis.
\begin{align*}
f(0,z) & = \frac{1}{\sqrt{r}} \sum_{k=0}^{r-1} e^{2\pi izk/r} \ket{\hat{f}(0,k)}
\end{align*}
where $\ket{\hat{f}(0,k)}$ is the the fourier transform of $f(0,z)$ and is given by: 
\begin{align*}
\ket{\hat{f}(0,k)} & = \frac{1}{\sqrt{r}}\sum_{j=0}^{r-1}e^{-2\pi ijk/r}\ket{f(0,j)}
\end{align*}
Thus, the state $\ket{\psi_3}$  can be written as: 
\begin{align*}
\ket{\psi_3} & = \frac{1}{2^t} \sum_{x=0}^{2^t-1}\sum_{y=0}^{2^t-1}\ket{x}\ket{y}\frac{1}{\sqrt{r}}\sum_{k=0}^{r-1}e^{2\pi i(sx+y)k/r}\ket{\hat{f}(0,k)} \\
& = \frac{1}{\sqrt{r}} \sum_{k=0}^{r-1} \left(\frac{1}{\sqrt{2^t}}\sum_{x=0}^{2^t-1}e^{2\pi isxk/r}\ket{x} \right)\left(\frac{1}{\sqrt{2^t}}\sum_{y=0}^{2^t-1}e^{2\pi iyk/r}\ket{x} \right) \ket{\hat{f}(0,k)} \\
& = \frac{1}{\sqrt{r}} \sum_{k=0}^{r-1} \left(QFT  \bigg|\frac{2^tsk}{r}\bigg\rangle \right) \left(QFT  \bigg|\frac{2^tk}{r}\bigg\rangle \right) \ket{\hat{f}(0,k)}
\end{align*}
\item Applying the Inverse Fourier transform on the first two registers will give us: 
\begin{align*}
\ket{\psi_4} & = \frac{1}{\sqrt{r}} \sum_{k=0}^{r-1} \bigg|\frac{2^tsk}{r}\bigg\rangle\bigg|\frac{2^tk}{r}\bigg\rangle\ket{\hat{f}(0,k)}
\end{align*}
\item Last step is to measure the first two registers which will give us (upon dividing by $2^t$): 
\begin{align*}
m_1 & = \frac{sk}{r} \\
m_2 & = \frac{k}{r}
\end{align*}
The generalized continued fraction algorithm is used to find $k$ and $r$ that will determine the value of $s$\footnote{We have to run algorithm several times so that we can obtain correct value of $r$ and $k$}.
\end{enumerate}

\chapter{Amplitude Amplification and Estimation Algorithms}
\section{Grover's Algorithm}
One of the many advantages a quantum computer has over a classical computer is its superior speed of searching databases. Grover's algorithm demonstrates this capability. This algorithm can speed up an unstructured search problem quadratically, but its uses extend beyond that; it can serve as a general trick or subroutine to obtain quadratic run time improvements for various other algorithms.
\subsection{Search Problem: Classical Approach}
Grover’s algorithm is an unstructured search quantum procedure to find an entry of $n$ bits on a digital haystack of $N$ elements. Grover’s quantum algorithm provides significant speedup at $\mathcal{O}(\sqrt{N})$ steps. It may not seem much compared to the classical solution, but when we talk about millions of entries, then $\sqrt{10^6}$ is much faster than $10^6$.\\
\indent Suppose we are given a large list of $N$ items. We assign each item an index. Among these items, there is one item with index \textbf{$w$}, that we want to locate. So, our \emph{database} is comprised of all the possible computational basis states our qubits can be in. For example, if we have 3 qubits\footnote{Number of items: $N$ and number of qubits: $n$ are related as: $2^n=N$, We can also add some garbage items if equality doesn't hold true}, our list is the states $\{\ket{000},\ket{001},\cdots,\ket{111}\}$ (i.e., the states from $\ket{0}\rightarrow \ket{7}$).\\
So we apply some function $f(x)$ that converts $n-$bit string to a single bit i.e., $f(x)\{0,1\}^{\otimes n}\rightarrow {0,1}$ such that: 
\begin{equation*}
f(x) =
\begin{cases}
1 & \text{if $x$ is a solution} \\
0 & \text{if $x$ is not a solution}
\end{cases}
\end{equation*}
Classically, we have to implement $f(x)$ to each index one by one until we reach a solution. But the worst-case requires us to make $N$ queries.\newpage
\subsection{Quantum Solution}
First, let's consider the circuit for Grover's Search algorithm.
\begin{figure}[H]
\centering
\begin{quantikz}
\lstick{$\ket{0}^{\otimes n}$} \slice{$\ket{\psi_0}$} & \gate{H^{\otimes n}}\qwbundle{n}\slice{$\ket{s}$} & \gate{U_f}\slice{$\ket{s'}$}\gategroup[1,steps=4,style={dashed,
rounded corners,fill=blue!20, inner xsep=2pt},
background,label style={label position=below,anchor=
north,yshift=-0.2cm}]{{Repeat $r$ times}} & \gate{H^{\otimes n}} & \gate{U_{f0}} & \gate{H^{\otimes n}}\slice{$\ket{\psi}$} & \meter{} & \cw
\end{quantikz}
\caption{Circuit for Grover Search Algorithm}
\label{Main Circuit}
\end{figure}
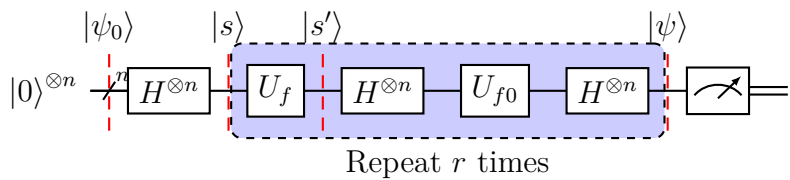
Highlighted gates can be considered as a single gate that is usually called Grover Operator, $G$. It has basically two parts:
\begin{enumerate}[label=\roman*)]
\item Phase Oracle: $U_f$
\item Diffuser: $V=H^{\otimes n}U_{f0}H^{\otimes n}$
\end{enumerate}
\noindent Diffuser is illustrated in the circuit below: 
\begin{figure}[H]
\centering
\begin{quantikz}
& \gate{H^{\otimes n}}\qwbundle{n} & \gate{U_{f0}} & \gate{H^{\otimes n}} & \qw
\end{quantikz}
\caption{Diffuser}
\label{Diffuser}
\end{figure}
Let's workout each step for the circuit given in Fig. \ref{Main Circuit}.
\begin{enumerate}
\item Write the input state: 
\begin{align}
\ket{\psi_0} & = \ket{0}^{\otimes n} = \ket{0}\otimes \cdots \otimes \ket{0}
\end{align}
\item After applying Hadamard gate to the input state, we get $\ket{s}$:
\begin{align}\label{equation-2}
\ket{s} & =H^{\otimes n}\ket{0}^{\otimes n} =  \frac{1}{\sqrt{N}} \sum_{x=0}^{N-1}\ket{x}
\end{align}
where $\displaystyle{\ket{x}\in \{\ket{0},\ket{1},\cdots,\ket{N-1}\}}$.
\subsection*{Phase Oracle: $U_f$}
Suppose $\ket{w}$ is the solution we are looking for, then phase oracle implements the function $f(x)$ to the input states such that: 
\begin{equation*}
f(x) =
\begin{cases}
1 & \text{if $\ket{x}=\ket{w}$} \\
0 & \text{if $\ket{x}\neq\ket{w}$}
\end{cases}
\end{equation*}
That is: \begin{equation*}
\ket{x} \longrightarrow
\begin{cases}
-\ket{x} & \text{if}\ \ x=w \\
\ \ \ \ket{x} &  \text{if}\ \ x\neq w
\end{cases}
\end{equation*}
Consider an example. Suppose that a particular item that we are looking for has index $101$. Then the truth table for the Phase Oracle is given by:
\begin{table}[H]
\centering
\begin{tabular}{|c|c|}
\hline
\textbf{$\ket{x}$} & \textbf{$(-1)^{f(x)}\ket{x}$} \\
\hline
$\ket{000}$ & \ \; $\ket{000}$ \\
$\ket{001}$ & \ \; $\ket{001}$ \\
$\ket{010}$ & \ \; $\ket{010}$ \\
$\ket{100}$ & \ \; $\ket{100}$ \\
$\ket{011}$ & \ \; $\ket{011}$ \\
$\ket{101}$ & $-\ket{101}$ \\
$\ket{110}$ & \ \; $\ket{110}$ \\
$\ket{111}$ & \ \; $\ket{111}$ \\
\hline
\end{tabular}
\caption{Truth table of the Oracle for 3-input qubits}
\label{Table: 1}
\end{table}
Table tells us that, oracle is the diagonal matrix, where the entry that correspond to the marked item will have a negative phase. For example, if we have three qubits and $w=101$, our oracle will have the matrix:
\begin{equation}\label{Oracle Matrix Form}
U_f = \begin{pmatrix}
1 & 0 & 0 & 0 & 0 & 0 & 0 & 0 \\
0 & 1 & 0 & 0 & 0 & 0 & 0 & 0 \\
0 & 0 & 1 & 0 & 0 & 0 & 0 & 0 \\
0 & 0 & 0 & 1 & 0 & 0 & 0 & 0 \\
0 & 0 & 0 & 0 & 1 & 0 & 0 & 0 \\
0 & 0 & 0 & 0 & 0 & -1 & 0 & 0 \\
0 & 0 & 0 & 0 & 0 & 0 & 1 & 0 \\
0 & 0 & 0 & 0 & 0 & 0 & 0 & 1 \\
\end{pmatrix}
\end{equation}
Operating the oracle on each input will result the state itself except for the solution i.e., 
\begin{align*}
U_f \ket{000} & = \ket{000}\\
U_f \ket{001} & = \ket{001}\\
\vdots\ \ \ \ \  & \ \ \ \ \ \ \vdots\\
U_f \ket{101} & = -\ket{101}\\
\vdots\ \ \ \ \  & \ \ \ \ \ \ \vdots\\
U_f \ket{111} & = \ket{111}
\end{align*}
In Dirac notation, $U_f$ can be written as: 
\begin{align}\label{Oracle Dirac Notation Form}
U_f = \op{0}{0} + \op{1}{1} + \cdots - \op{w}{w} + \cdots + \op{N-1}{N-1}
\end{align}
where $\ket{w}$ is the solution. Adding and subtracting $\op{w}{w}$ in Eq.(\ref{Oracle Dirac Notation Form}) gives us: 
\begin{align*}
U_f & = \op{0}{0} + \op{1}{1} + \cdots - \op{w}{w} + \cdots + \op{N-1}{N-1} +  \op{w}{w} - \op{w}{w} \\
& =\Big( \op{0}{0} + \op{1}{1} + \cdots + \op{w}{w} + \cdots + \op{N-1}{N-1} \Big) - \op{w}{w} - \op{w}{w} \\
& = \mathbb{I} - 2\op{w}{w}
\end{align*}
Let's write the state in Eq.(\ref{equation-2}) in a special way: 
\begin{align*}
\ket{s} & = \frac{1}{\sqrt{N}} \sum_{x=0}^{N-1}\ket{x}\\
& = \frac{1}{\sqrt{N}} \Big(\ket{0}+\ket{1}+\cdots+\ket{w}+\cdots+\ket{N-1} \Big) \\
& = \frac{1}{\sqrt{N}} \Big(\ket{0}+\ket{1}+\cdots\ket{w-1}+\ket{w+1}+\cdots+\ket{N-1}\Big) + \frac{1}{\sqrt{N}}\ket{w} \\
& = \frac{1}{\sqrt{N}} \sum_{x\neq w}^{N-1}\ket{x} + \frac{1}{\sqrt{N}}\ket{w}
\end{align*}
Multiply and divide $\displaystyle{\sqrt{N-1}}$ to the first term in above equation to normalize the superposition state.
\begin{align*}
\ket{s} & = \sqrt{\frac{N-1}{N}}\left(\frac{1}{\sqrt{N-1}}\sum_{x\neq w}^{N-1}\ket{x}\right) + \frac{1}{\sqrt{N}}\ket{w}
\end{align*}
Since, $\ket{w}$ is orthogonal to every other state, therefore we can write:
\begin{align*}
\ket{w^{\perp}} & = \frac{1}{\sqrt{N-1}}\sum_{x\neq w}^{N-1}\ket{x}
\end{align*}
And thus state $\ket{s}$ can be written as: 
\begin{align*}
\ket{s} & = \sqrt{\frac{N-1}{N}}\ket{w^{\perp}} + \frac{1}{\sqrt{N}}\ket{w}
\end{align*}
The state is normalized\footnote{$\because \left(\sqrt{\frac{N-1}{N}}\right)^2+\left(\frac{1}{\sqrt{N}}\right)^2=1=\cos^2(\alpha)+\sin^2(\alpha)$}, we can also write it like: 
\begin{align}\label{s-state}
\ket{s} & = \cos(\alpha)\ket{w^{\perp}} + \sin(\alpha)\ket{w}
\end{align}
where $\displaystyle{\cos(\alpha) = \sqrt{\frac{N-1}{N}}}$ and $\displaystyle{\sin(\alpha)=\frac{1}{\sqrt{N}}}$. The state $\ket{s}$ can be represented geometrically as in Fig. \ref{figure-2}.
\begin{figure}[H]
\centering
\includegraphics[scale=0.3]{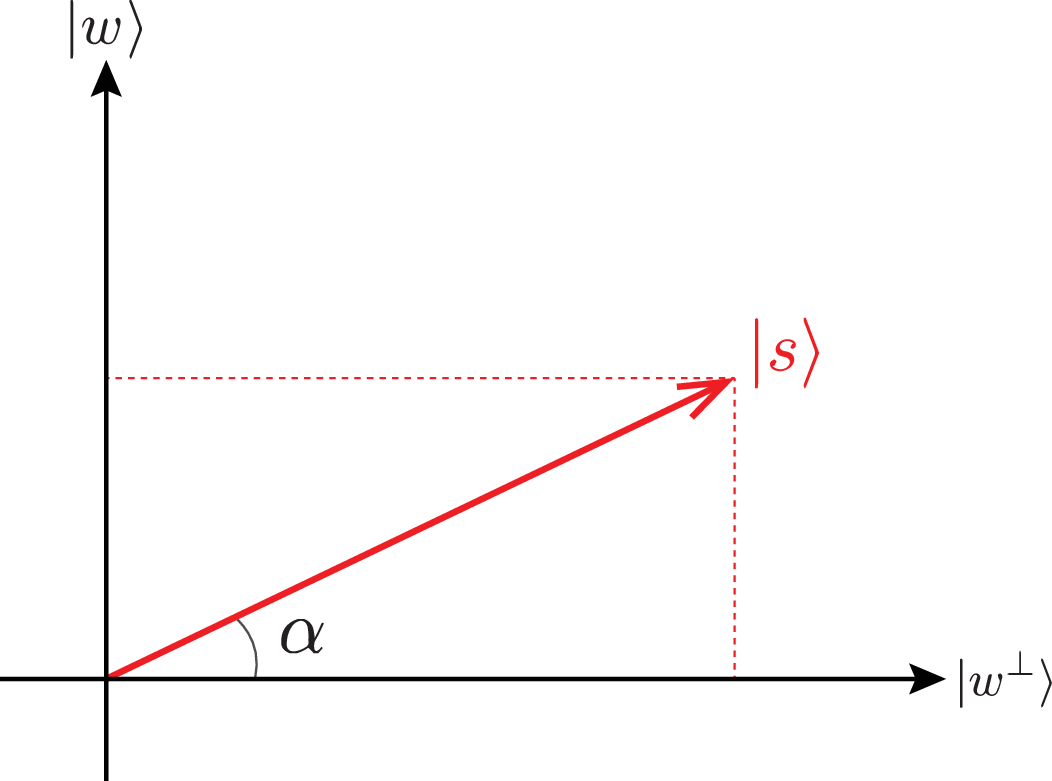}
\caption{Geometrical representation of $\ket{s}$}
\label{figure-2}
\end{figure}
\item Let's apply phase oracle, $U_f$ on the state $\ket{s}$:
\begin{equation}\label{equation-6}
\begin{split}
\ket{s'} & = U_f\ket{s} \\
& = ( \mathbb{I} -2 \op{w}{w} )(\cos(\alpha)\ket{w^{\perp}} + \sin(\alpha)\ket{w} \\
& = \cos(\alpha)\ket{w^{\perp}} + \sin(\alpha)\ket{w} + 0 - 2 \sin(\alpha)\ket{w} \\
& = \cos(\alpha)\ket{w^{\perp}} - \sin(\alpha)\ket{w}
\end{split}
\end{equation}
The state $\ket{s'}$ can be geometrically represented as: 
\begin{figure}[H]
\centering
\includegraphics[scale=0.3]{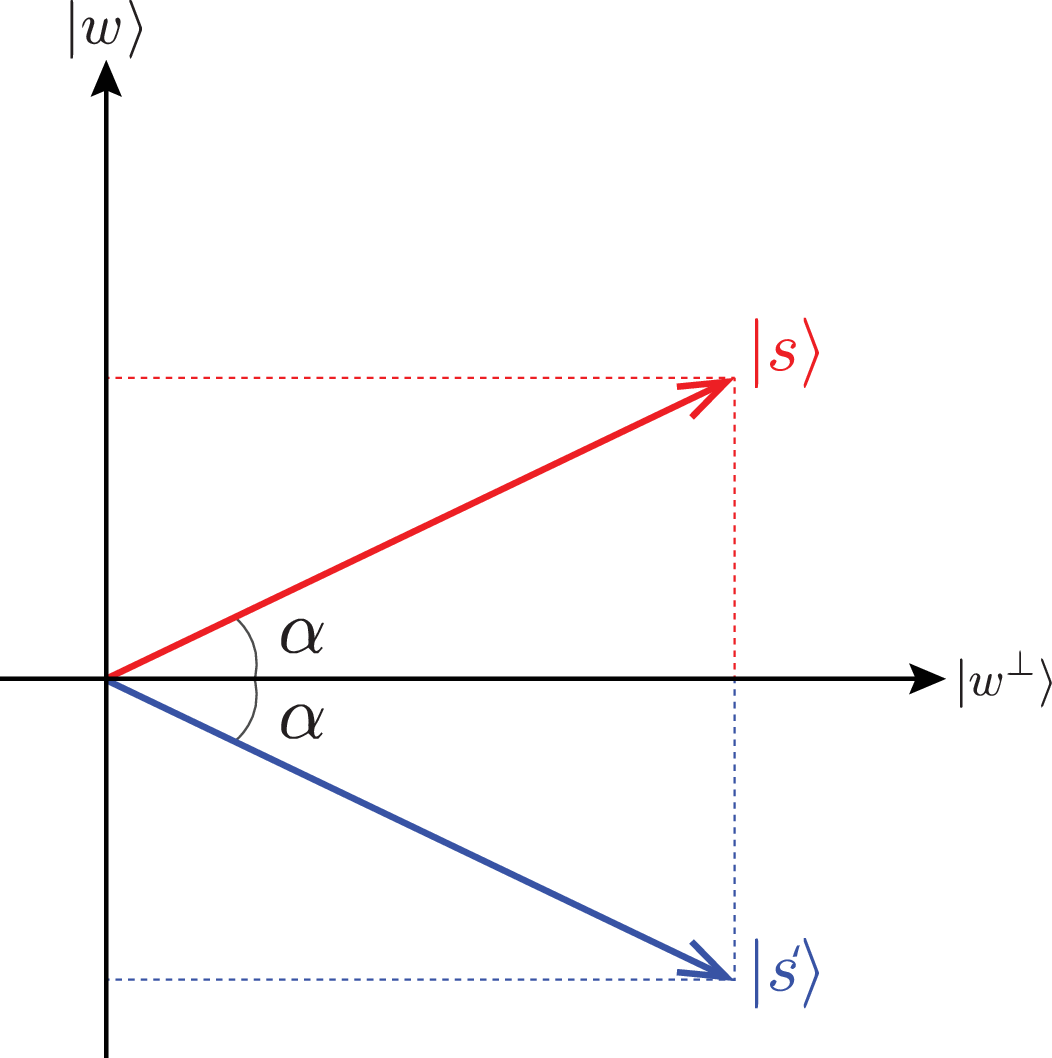}
\caption{Geometrical representation of $\ket{s'}$}
\label{figure-3}
\end{figure}
So we can say that, Oracle rotated the state by an angle of $2\alpha$ in clockwise direction. There is also another way of representing the action of oracle on the state $\ket{s}$.
\begin{figure}[H]
\centering
\includegraphics[scale=0.3]{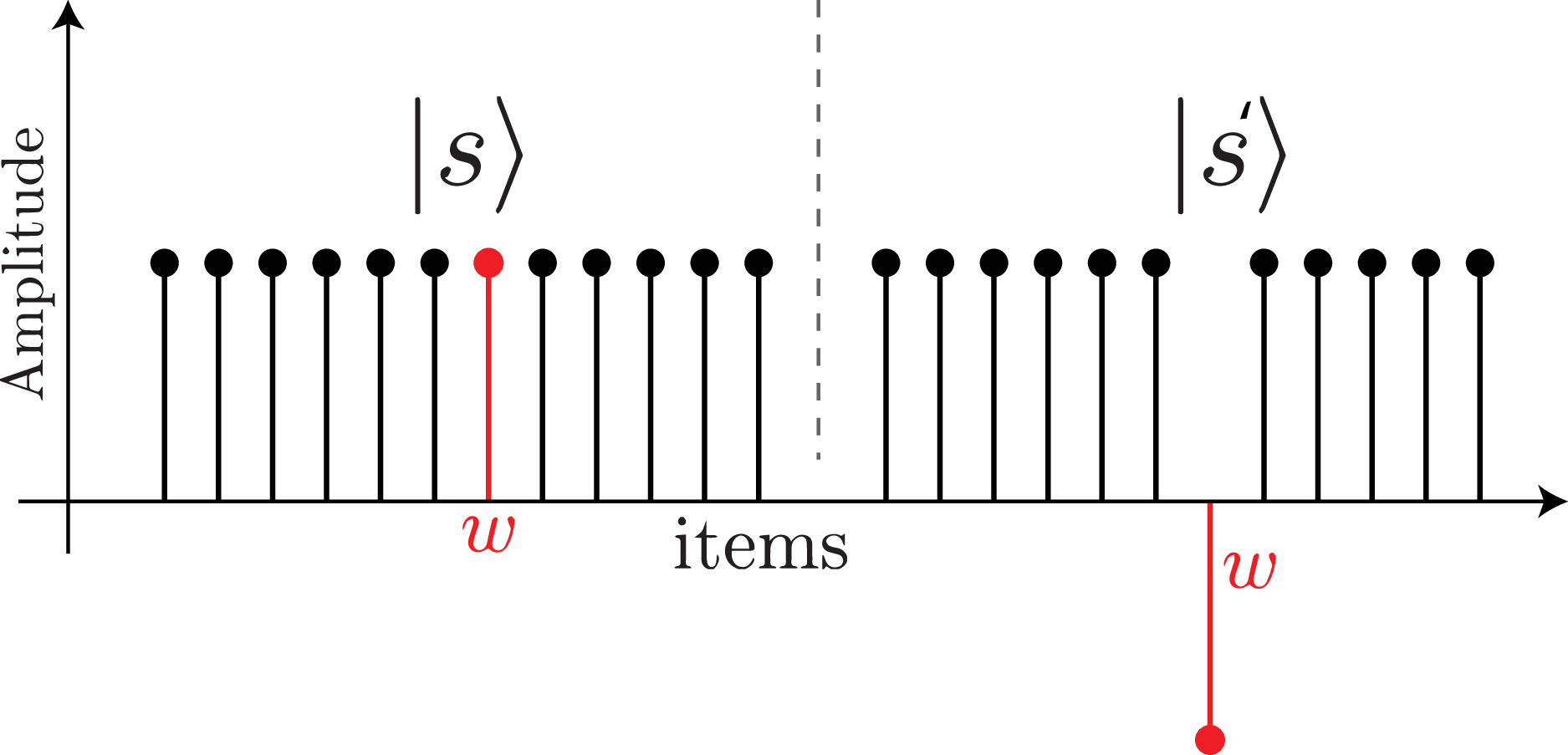}
\caption{Representation of state $\ket{s}$ and $\ket{s'}$}
\label{histogram}
\end{figure}
So, from Fig. \ref{histogram}, we can see that the amplitude of the solution state,$\ket{w}$ has been changed from $+$ to $-$.
\subsection*{Oracle: $U_{f0}$}
The oracle, $U_{f0}$ implements a function $f(x)$ such that: 
\begin{equation}\label{Function_old}
f(x) = \begin{cases}
1 & x=0 \\
0 & x\neq0
\end{cases}
\end{equation}
In other words, 
\begin{align*}
U_{f0}\ket{x} & = \ \ \ket{x}\ :\ x\neq0 \\
U_{f0}\ket{0} & = -\ket{0}
\end{align*}
A function equivalent to $f(x)$ can be thought like:
\begin{equation}\label{Function_new}
f_0(x) = \begin{cases}
0 & x=0 \\
1 & x\neq0
\end{cases}
\end{equation}
In other words, \begin{align*}
U_{f0}\ket{x} & = -\ket{x}\ :\ x\neq0 \\
U_{f0}\ket{0} & = \ \ \ \ket{0}
\end{align*}
We will pick $f_0(x)$, because it is easy to handle it analytically. Since it is a matter of choice. Like $U_f$ in Eq.(\ref{Oracle Matrix Form}), $U_{f0}$ will also be diagonal matrix like: 
\begin{equation}\label{New Oracle Matrix Form}
U_{f0} = 
\begin{pmatrix}
1 & 0  & \cdots & 0 \\
0 & -1 & \cdots & 0 \\
\vdots & \vdots & \ddots & 0\\
0 & 0 & \cdots & -1
\end{pmatrix}
\end{equation}
Clearly, applying oracle will leave the state $\ket{0}$ invariant and will multiply -1 to the rest of the states. In operator form, $U_{f0}$ can be written as: 
\begin{align*}
U_{f0} & = \op{0}{0} - \op{1}{1} - \op{2}{2} - \cdots - \op{N-1}{N-1}
\end{align*}
Add and subtract $\op{0}{0}$, and we can easily get $U_{f0}$ in a very nice form: 
\begin{align}\label{New Oracle Equation}
U_{f0} & = 2\op{0}{0} - \mathbb{I}
\end{align}
\subsection*{Diffuser}
Operator form of diffuser is given by: 
\begin{align*}
V & = H^{\otimes n} U_{f_0} H^{\otimes n} \\
& = H^{\otimes n} \big(2\op{0}{0} - \mathbb{I} \big) H^{\otimes n} \\
& = 2H^{\otimes n}\op{0}{0}H^{\otimes n}-H^{\otimes n}H^{\otimes n} \\
\end{align*}
But $H^{\otimes n}\ket{0}=\ket{s}$ and $H^{\otimes n}H^{\otimes n}=\mathbb{I}$. Thus, diffuser operator can be written as: 
\begin{align}\label{equation-11}
V & = 2\op{s}{s}-\mathbb{I}
\end{align}
\item Let's apply $V$ to $\ket{s'}$:
\begin{equation}\label{equation-12}
\begin{split}
V\ket{s'} & = \big(2\op{s}{s}-\mathbb{I}\big)\ket{s'}\\
& = 2\ket{s}\ip{s}{s'}-\ket{s'}
\end{split}
\end{equation}
where
\begin{align*}
\ip{s}{s'} & = \big(\cos\alpha\bra{w^{\perp}} + \sin\alpha\bra{w} \big)\big(\cos\alpha\ket{w^{\perp}} - \sin\alpha\ket{w}\big) \\
& = \cos^2\alpha- \sin^2\alpha
\end{align*}
Thus Eq. (\ref{equation-12}) becomes:
\begin{equation}\label{equation-13}
\begin{split}
V\ket{s'}& = 2 \big(\cos^2\alpha - \sin^2\alpha\big)\big(\cos\alpha\ket{w^{\perp}} + \sin\alpha\ket{w}\big)-\big(\cos\alpha\ket{w^{\perp}} - \sin\alpha\ket{w}\big)
\end{split}
\end{equation}
Simplifying this equation will give us: 
\begin{align*}
V\ket{s'}& = \big(2\cos^3\alpha-2\cos\alpha\sin^2\alpha-\cos\alpha\big)\ket{w^{\perp}}\\
&\  + \big(2\sin\alpha\cos^2\alpha-2\sin^3\alpha+\sin\alpha\big)\ket{w}
\end{align*}
It can be easily shown that: 
\begin{align}\label{equation-14}
\Rightarrow V\ket{s'} & = \cos(3\alpha)\ket{w^{\perp}}+\sin(3\alpha)\ket{w}
\end{align}
Geometrically, the state in Eq.(\ref{equation-14}) is illustrated in figure below: 
\begin{figure}[H]
\centering
\includegraphics[scale=0.3]{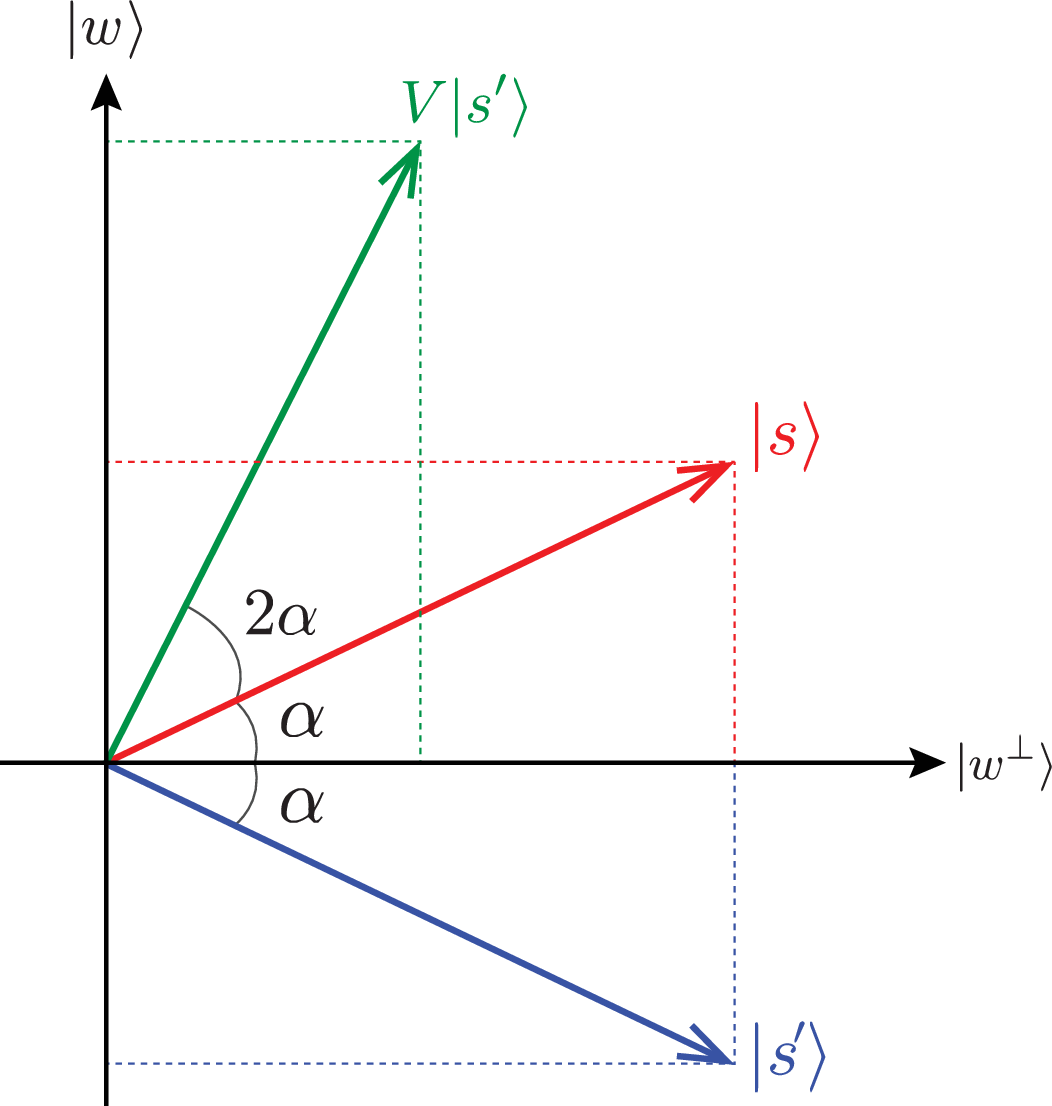}
\caption{Geometrical Representation of the state $V\ket{s'}$}
\label{VS}
\end{figure}
So diffuser actually increases the amplitude\footnote{As angle $\theta$ increases, $\sin(\theta)$ increases but $\cos(\theta)$ decreases.} of the solution state, $\ket{w}$ and decrease the amplitude of others. Geometrically, it is illustrated in figure below: 
\begin{figure}[H]
\centering
\includegraphics[scale=0.3]{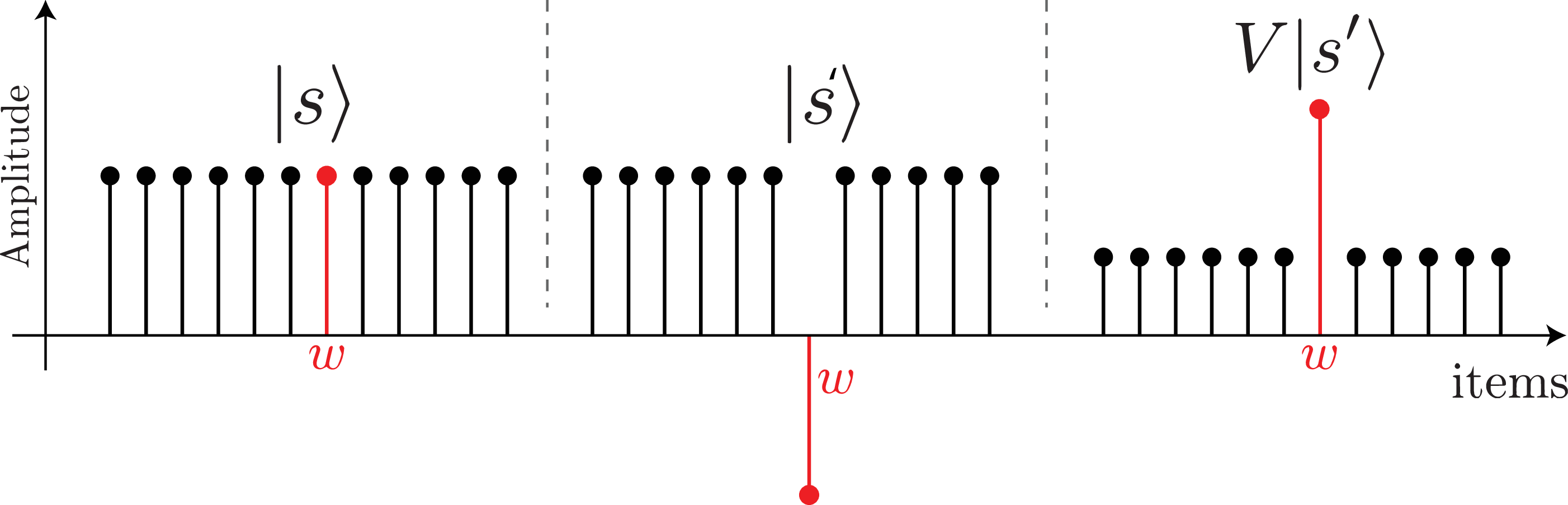}
\caption{Geometrical Representation of $V\ket{s'}$}
\label{VSHisto}
\end{figure}
\item But we want to apply the Grover Operator $r-$times. We know that: 
\begin{align*}
VU_{f}\ket{s} & = \cos(3\alpha)\ket{w^{\perp}}+\sin(3\alpha)\ket{w}\\
& = \cos(2\alpha+\alpha)\ket{w^{\perp}}+\sin(2\alpha+\alpha)\ket{w}
\end{align*}
Thus we can write: 
\begin{align*}
\ket{\psi}&=(VU_f)^r\ket{s} = \cos(2r\alpha+\alpha)\ket{w^{\perp}}+\sin(2r\alpha+\alpha)\ket{w}
\end{align*}
The final state $\ket{\psi}$ is geometrically shown in figure below: 
\begin{figure}[H]
\centering
\includegraphics[scale=0.3]{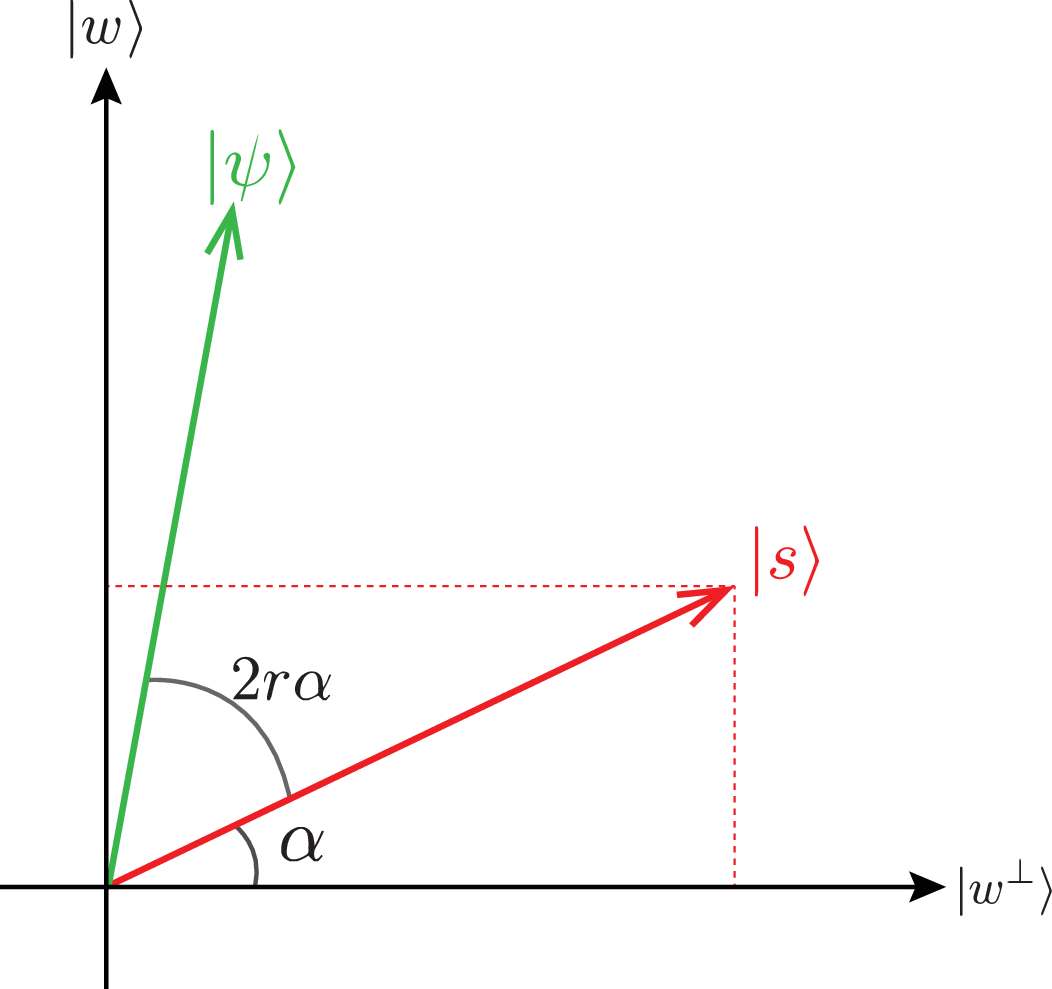}
\caption{Geometrical Representation of $\ket{\psi}$}
\label{psi_final}
\end{figure}
As shown in Fig. \ref{psi_final}, if the angle, $2r\alpha+\alpha$ becomes equal to $\frac{\pi}{2}$, then the state $\ket{\psi}$ exactly aligns with $\ket{w}$ and upon measuring, we get the desired state. Mathematically, 
\begin{align*}
2r\alpha+\alpha & = \frac{\pi}{2} \\
(2r+1)\alpha & = \frac{\pi}{2} \\
2r+1 & = \frac{\pi}{2\alpha} \\
2r & = \frac{\pi}{2\alpha}-1 \\
r & = \frac{\pi}{4\alpha} - \frac{1}{2}
\end{align*}
From Eq. (\ref{s-state}), we know that $\displaystyle{\alpha=\arcsin\left(\frac{1}{\sqrt{N}}\right)\sim \frac{1}{\sqrt{N}}}$. Therefore, 
\begin{align*}
r & = \frac{\pi\sqrt{N}}{4}-\frac{1}{2} \\
& \sim \sqrt{N}
\end{align*}
\end{enumerate}\newpage
\subsection*{Order of error}
In general, our state $\ket{\psi}$ will be closer to the vertical axis by an angle of the order $\alpha$ as shown in figure below: 
\begin{figure}[H]
\centering
\includegraphics[scale=0.3]{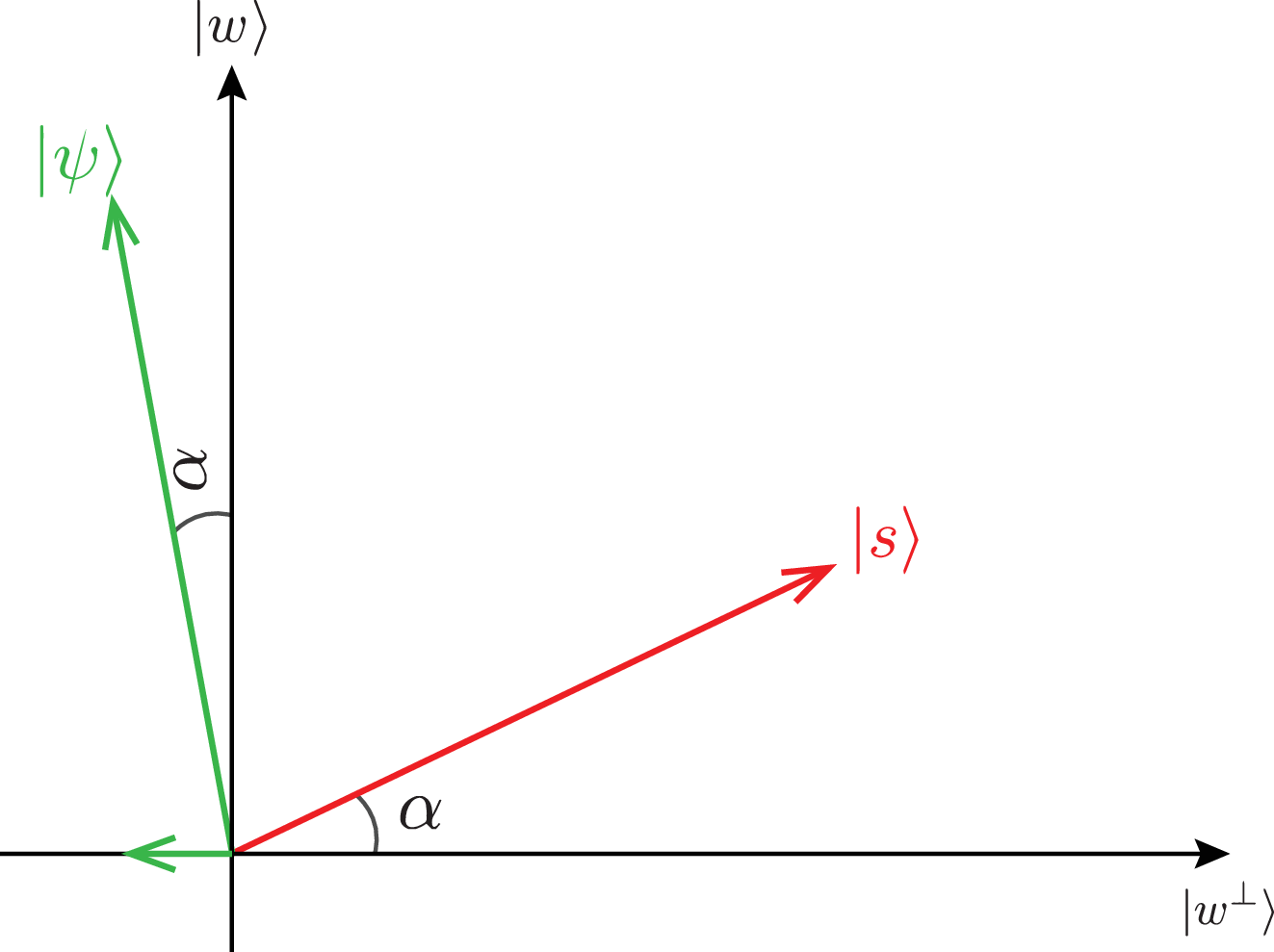}
\caption{Getting $\ket{\psi}$ in the real case}
\label{error}
\end{figure}
The error in this case will be actually equal to probability of measuring $\ket{w^{\perp}}$ which is given by: 
\begin{align*}
\text{Pr}(w^{\perp}) & = \sin^2{\alpha}\\
& = \frac{1}{N}
\end{align*}
This means that for large number of items, the error is extremely small.
\subsubsection{What if we have more than 1 solutions ?}
Suppose we have $\mu$ number of solutions. In this case, the function $f(x)$ that oracle will implement will have a form like:
\begin{equation}
f(x) = \begin{cases}
1 & x=w_i,\ i=1,2,\cdots,\mu \\
0 & x\neq w_i
\end{cases}
\end{equation}
Similarly we can write $\ket{s}$ state as: 
\begin{equation*}
\begin{split}
\ket{s} & = \frac{1}{\sqrt{N}}\sum_{x\neq w_i}^{N-1}\ket{x} + \frac{1}{\sqrt{N}}\sum_{i=1}^{\mu}\ket{w_i} \\
& = \sqrt{\frac{N-\mu}{N}}\left(\frac{1}{\sqrt{N-\mu}}\sum_{x\neq w_i}^{N-1}\ket{x}\right) + \sqrt{\frac{\mu}{N}}\left(\frac{1}{\sqrt{\mu}}\sum_{i=1}^{\mu}\ket{w_i}\right) \\
& = \sqrt{\frac{N-\mu}{N}} \ket{w^{\perp}} + \sqrt{\frac{\mu}{N}}\ket{w}
\end{split}
\end{equation*}
Now, I can define:
\begin{equation*}
\begin{split}
\cos(\alpha) & = \sqrt{\frac{N-\mu}{N}} \\
\sin(\alpha) & = \sqrt{\frac{\mu}{N}}
\end{split}
\end{equation*}
Since, 
\begin{align*}
r & = \frac{\pi}{4\alpha} - \frac{1}{2} \\
\Rightarrow r & = \frac{\pi}{4}\sqrt{\frac{N}{\mu}}-\frac{1}{2} \\
\Rightarrow r & \sim \sqrt{\frac{N}{\mu}}
\end{align*}

\section{Quantum Counting}
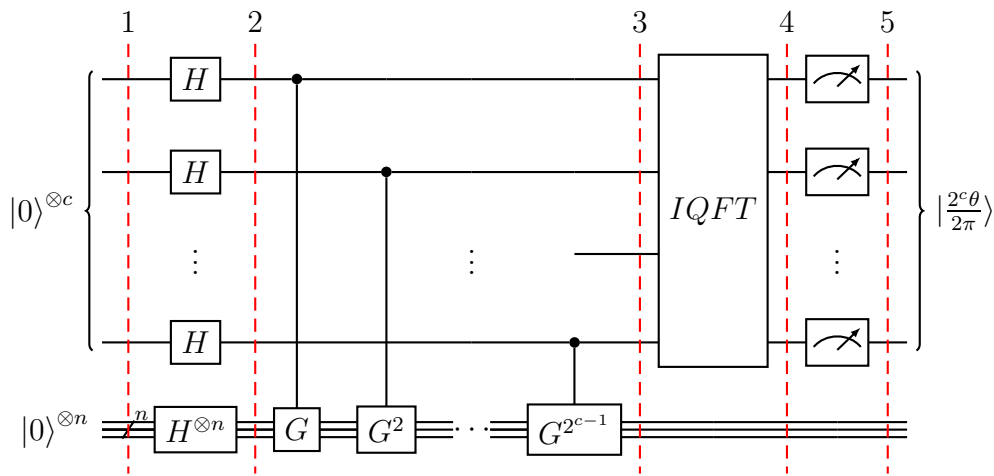
\begin{figure}[h]
\centering
\begin{quantikz}
\lstick[wires=4]{$\ket{0}^{\otimes c}$} \slice{1}&[2mm] \gate{H}\slice{2} 
& \ctrl{4} & \qw & \qw & \qw\slice{3} & \gate[wires=4]{IQFT}\slice{4} & \meter{}\slice{5} & \rstick[wires=4]{$\ket{\frac{2^c\theta}{2\pi}}$}\qw \\
& \gate{H} & \qw & \ctrl{3} & \qw & \qw & & \meter{} & \qw \\
& \vdots & & &\vdots & & & \vdots & \\
& \gate{H} & \qw & \qw & \qw & \ctrl{1} & & \meter{} & \qw  \\
\lstick{$\ket{0}^{\otimes n}$}
& \gate{H^{\otimes n}}\qwbundle[alternate]{}\qwbundle{n} & \gate{G}\qwbundle[alternate]{} & \gate{G^{2}}\qwbundle[alternate]{} & \qwbundle[alternate]{}\cdots &  \gate{G^{2^{c-1}}}\qwbundle[alternate]{}& \qwbundle[alternate]{} & \qwbundle[alternate]{} & \qwbundle[alternate]{} 
\end{quantikz}
\caption{Circuit for Quantum Counting}
\label{Circuit}
\end{figure}

In quantum counting, we simply use quantum phase estimation algorithm to find the eigenvalues of Grover operator, $G$ instead of unitary operator, $U$. Recall that:
\begin{align*}
G & = VU_f
\end{align*}
where $U_f$ is phase oracle and $V$ is the diffuser.
\begin{align*}
U_f & = \mathbb{I}-2\op{w}{w} \\
V & = H^{\otimes n}\otimes\big(2\op{0}{0}-\mathbb{I}\big)\otimes H^{\otimes n}
\end{align*}
Recall from Grover's algorithm that state $\ket{s}$ was:
\begin{align*}
\ket{s} & = \cos(\alpha)\ket{w^{\perp}} + \sin(\alpha)\ket{w}
\end{align*}
Let's take $\displaystyle{\alpha=\frac{\theta}{2}}$. So the state becomes:
\begin{align*}
\ket{s} & = \cos\left(\frac{\theta}{2}\right)\ket{w^{\perp}} + \sin\left(\frac{\theta}{2}\right)\ket{w}
\end{align*}
This state can also be written as:
\begin{align*}
\ket{s} & = \begin{pmatrix}
\cos\frac{\theta}{2}\\[6pt]
\sin\frac{\theta}{2}
\end{pmatrix}
\end{align*}
Action of Grover operator on the state is: 
\begin{align*}
G\ket{s} & = \cos\left(\theta + \frac{\theta}{2}\right)\ket{w^{\perp}} + \sin\left(\theta+\frac{\theta}{2}\right)\ket{w}
\end{align*}
$G$ is rotation operator whose matrix form is:
\begin{align*}
G & = \begin{pmatrix}
\cos\theta & -\sin\theta \\
\sin\theta & \cos\theta
\end{pmatrix}
\end{align*}
The matrix $G$ has eigenvectors:
\begin{align*}
\begin{pmatrix}
i\\1
\end{pmatrix},\begin{pmatrix}
-i\\1
\end{pmatrix}
\end{align*}
with eigenvalues $\displaystyle{e^{i\theta},e^{-i\theta}}$ respectively.
\begin{exercise}{}{label}
\begin{enumerate}
\item Prove that:
\begin{align*}
\begin{pmatrix}
\cos\left(\theta+\frac{\theta}{2}\right) \\[7pt]
\sin\left(\theta+\frac{\theta}{2}\right)
\end{pmatrix} & = \begin{pmatrix}
\cos\theta & -\sin\theta \\
\sin\theta & \cos\theta
\end{pmatrix}\begin{pmatrix}
\cos\frac{\theta}{2}\\[6pt]
\sin\frac{\theta}{2}
\end{pmatrix}
\end{align*}
\item Show that the state $\ket{s}$ can be written as:
\begin{align*}
\ket{s} & = \frac{1}{2i}e^{i\theta/2}\begin{pmatrix}
i\\1
\end{pmatrix} - \frac{1}{2i}e^{-i\theta/2}\begin{pmatrix}
-i\\1
\end{pmatrix}
\end{align*}
\end{enumerate}
\end{exercise}
We can prepare the state, $\ket{s}$ by applying the Hadamard gate to $\ket{0}^{\otimes n}$ and then applying Grover's operator. The circuit for quantum counting algorithm is illustrated in the Fig. \ref{Circuit}.
Since one eigenvalue of $G$ is: 
\begin{align*}
e^{i\theta} & = e^{2\pi i \frac{\theta}{2\pi}}
\end{align*}
The second eigenvalue of $G$ is: 
\begin{align*}
e^{-i\theta} & = e^{i(2\pi-\theta)}\\
& = e^{2\pi i(1-\frac{\theta}{2\pi})}
\end{align*}
As a result, the output of the Quantum Phase Estimation algorithm will be a superposition of the two states:
\begin{align*}
\left|2^c\frac{\theta}{2\pi}\right\rangle\ \ ,\ \left|2^c\left(1-\frac{\theta}{2\pi}\right)\right\rangle
\end{align*}
After measuring, we will obtain one of these states, from which we can easily find $\theta$. Once we have $\theta$, we can find the number of solution in search space. Recall that, in general,
\begin{align*}
\sin\left(\frac{\theta}{2}\right) & = \sqrt{\frac{\mu}{N}} \\
\end{align*}
where $\mu$ is total number of solution. Therefore, 
\begin{align*}
\mu & = \sin^2\left(\frac{\theta}{2}\right)N
\end{align*}

\section{Quantum search as quantum simulation}
Instead of probabilistic result of Grover's Search algorithm, we can obtain a \emph{deterministic} quantum search algorithm. For this, we make a guess as to a Hamiltonian $H$ which solves the search problem. Hamiltonian is written in such a way that it both depends on the solution state $\ket{x}$ as well as initial state $\ket{\psi}$, and evolves the quantum system from $\ket{\psi}$ to $\ket{x}$ after some time. Finally, we have to simulate the action of Hamiltonian using quantum circuit.

Suppose that algorithm starts with a quantum computer in a state $\ket{\psi}$ where $\ket{\psi}$ is the linear superposition of the all the basis states belonging to $n-$dimensional Hilber space.
\begin{align*}
\ket{\psi} & = \frac{1}{\sqrt{N}} \sum_{j=0}^{N-1}\ket{j} \\
\Rightarrow \ket{\psi} & = \frac{1}{\sqrt{N}}\ket{x} + \frac{1}{\sqrt{N}}\sum_{j\neq x}^{N-1}\ket{j} \\
& = \frac{1}{\sqrt{N}} \ket{x} + \sqrt{\frac{N-1}{N}} \left( \frac{1}{\sqrt{N-1}}\sum_{j\neq x}^{N-1}\ket{j}\right)
\end{align*}
Take, \begin{align*}
\alpha & = \frac{1}{\sqrt{N}} \\
\beta & = \sqrt{\frac{N-1}{N}} \\
\ket{y} & = \frac{1}{\sqrt{N-1}}\sum_{j\neq x}^{N-1} \ket{j}
\end{align*}
Thus, 
\begin{align}\label{psi}
\ket{\psi} & = \alpha \ket{x} + \beta \ket{y}
\end{align}
Now, the goal of quantum searching is to transform $\ket{\psi}$ into $\ket{x}$. Simplicity suggests that Hamiltonian must be of the form,
\begin{align}\label{hamiltonian}
H & = \op{x}{x} + \op{\psi}{\psi}
\end{align}
Substitute $\ket{\psi}$ from Eq. (\ref{psi}) into Eq. (\ref{hamiltonian}),
\begin{align*}
H & =\op{x}{x}+ \left(\alpha\ket{x}+\beta\ket{y}\right)\left(\alpha\bra{x}+\beta\bra{y}\right) \\
& = \left(1+\alpha^2\right)\op{x}{x} + \alpha\beta\op{y}{x} + \alpha\beta\op{x}{y} + \beta^2\op{y}{y}
\end{align*}
The matrix form of Hamiltonian in $\{\ket{x},\ket{y}\}$ basis is given by: 
\begin{align*}
H & = \begin{pmatrix}
1+\alpha^2 & \alpha\beta \\
\alpha\beta & 1-\alpha^2
\end{pmatrix}
\end{align*}
But since, $\alpha^2+\beta^2=1$, thus
\begin{align*}
H & = \begin{pmatrix}
1+\alpha^2 & \alpha\beta \\
\alpha\beta & 1-\alpha^2
\end{pmatrix} \\
& = \begin{pmatrix}
1&0\\ 0&1
\end{pmatrix} + \alpha^2 \begin{pmatrix}
1&0\\ 0&-1
\end{pmatrix} + \alpha\beta\begin{pmatrix}
0&1\\ 1&0
\end{pmatrix} \\
& = I + \alpha^2Z+\alpha\beta X \\
H & = I + \alpha\left( \alpha Z + \beta X \right)
\end{align*}
After a time $t$, the state of a quantum system evolving according to the Hamiltonian $H$ and initially in the state $\ket{\psi}$ is given by:
\begin{align*}
\ket{\psi(t)} & = e^{-iHt}\ket{\psi} \\
& = e^{-iIt} \left( e^{-i\alpha(\alpha Z+\beta X)t}\right)\ket{\psi} \\
& =e^{-it} \left[\left( I - i\alpha t(\alpha Z + \beta X\right) + \frac{\left( -i\alpha t(\alpha Z+ \beta X)\right)^2}{2!} + \frac{\left( -i\alpha t(\alpha Z+ \beta X)\right)^3}{3!} + \cdots \right] \ket{\psi}
\end{align*}
Now, 
\begin{align*}
\left(\alpha Z+\beta X\right)^2 & = \alpha^2Z^2 + \beta^2X^2 + \alpha\beta(ZX+XZ) \\
& = \alpha^2 + \beta^2 + \alpha\beta (ZX - ZX)\ \ \ \ \because ZX = -XZ \\
& = 1+ \alpha\beta(0) \\
& = 1
\end{align*}
Similarly, 
\begin{align*}
\left(\alpha Z+\beta X\right)^3 & = \alpha Z + \beta X
\end{align*}
Therfore,
\begin{align*}
\ket{\psi(t)} & = \left[ I - i\alpha t(\alpha Z+\beta X) + \frac{(i\alpha t)^2}{2!} + \frac{(-i\alpha t)^3(\alpha Z + \beta X)}{3!} + \cdots \right]\ket{\psi} \\
& =e^{-it}\left( I\left[ 1 + \frac{(i\alpha t)^2}{2!} + \cdots \right] -i (\alpha Z+\beta X) \left[ \alpha t - \frac{(\alpha t)^3}{3!}+ \cdots \right)\ket{\psi} \right] \\
& = e^{-it} \left[ \cos (\alpha t) I - i\sin (\alpha t) (\alpha Z + \beta X) \right]\ket{\psi} \\
& = e^{-it} \left[ \cos(\alpha t)\ket{\psi} - i\sin(\alpha t) (\alpha Z +\beta X)\ket{\psi} \right]
\end{align*}
The global phase factor can be ignored, since it does not have effect on measurement. Since, 
\begin{align*}
I + \alpha(\alpha Z +\beta X) & = H \\
\Rightarrow \alpha Z + \beta X & = \frac{H-I}{\alpha} \\
\Rightarrow (\alpha Z + \beta X)\ket{\psi} & = \frac{H\ket{\psi}-\ket{\psi}}{\alpha} \\
& = \frac{\ket{x}\ip{x}{\psi}+\ket{\psi}\ip{\psi}{\psi}-\ket{\psi}}{\alpha} \\
& = \frac{\alpha \ket{x}}{\alpha} \\
& = \ket{x}
\end{align*}
Thus, 
\begin{align*}
\ket{\psi(t)} & = \cos(\alpha t)\ket{\psi} - i\sin(\alpha t)\ket{x}
\end{align*}
This suggests that if we simulate our quantum system according to the Hamiltonian $H$ for time $t=\pi/2\alpha$, where $\alpha = 1/\sqrt{N}$, we will get the state $\ket{x}$ upon measurement.
\section{Quantum Circuit to simulate the Hamiltonian}
Consider an operator $e^{-iH\Delta t}$ is applied $r-$times, such that $r\Delta t=t$. The Hamiltonian given in Eq. (\ref{hamiltonian}) is sum of two terms. Take $H_1=\op{x}{x}$ and $H_2=\op{\psi}{\psi}$. Also, $H_1$ and $H_2$ do not commute. Thus we can write, 
\begin{align*}
e^{-iH\Delta t} & = e^{-iH_1\Delta t}e^{-iH_2\Delta t} \\
& = e^{-iH_1\Delta t} e^{-i\op{\psi}{\psi}\Delta t}
\end{align*}
Now,
\begin{align*}
e^{-i\op{x}{x}\Delta t} & = I - i\Delta t\op{x}{x} + (-i\Delta t)^2\op{x}{x} + \cdots \\
& = I + \op{x}{x} \left( e^{-i\Delta t}-1 \right)
\end{align*}
Similarly, 
\begin{align*}
e^{-iH_2\Delta t} & = I + \op{\psi}{\psi}\left( e^{-i\Delta t}-1\right)
\end{align*}
\begin{exercise}{}{label}
Show that, $H_1$ and $H_2$ do not commute, i.e., $[H_1,H_2]\neq 0$
\end{exercise}
\subsection{Circuit Implementation of $e^{-i\op{x}{x}\Delta t}$}
\noindent Application of this operator on the state $\ket{x}$ is given by: 
\begin{align*}
e^{-i\op{x}{x}\Delta t} \ket{x} & =  \left[I + \op{x}{x} \left( e^{-i\Delta t}-1 \right) \right]\ket{x} \\
& = e^{-i\Delta t} \ket{x}
\end{align*}
Similarly, 
\begin{align*}
e^{-i\op{x}{x}\Delta t} \ket{y} & = \ket{y}
\end{align*}
Circuit implementation of the given operator is given by: 
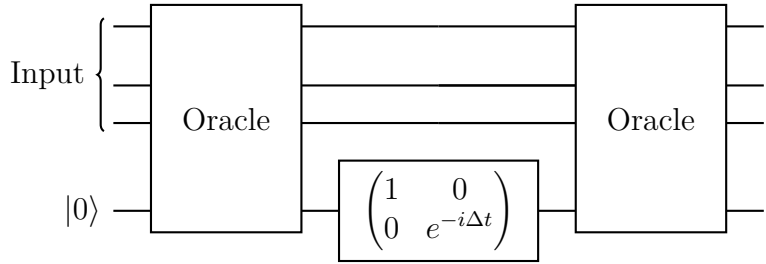
\begin{figure}[H]\label{circuit1}
\centering
\begin{quantikz}
\lstick[wires=3]{Input} & \gate[wires=4][2 cm]{\text{Oracle}} & \qw & \gate[wires=4][2 cm]{\text{Oracle}} & \qw \\
& &\qw & \qw &  \qw \\
& &\qw & \qw &  \qw \\
\lstick{$\ket{0}$} & & \gate{\begin{pmatrix}
1&0\\0&e^{-i\Delta t}\end{pmatrix}}  & & \qw 
\end{quantikz}
\caption{Circuit implementing the operator $e^{-i\Delta t \op{x}{x}}$}
\end{figure}
The \emph{Oracle} in the Fig. (\ref{circuit1}) is essentially a multi-qubit controlled NOT gate, that activates only if, the input is solution state, $\ket{x}$, otherwise it acts as an identity. When the input is $\ket{x}$, auxiliary qubit is transformed to $\ket{1}$, followed by phase gate applied to it, and at the end, $\ket{1}$ is again transformed to $\ket{0}$.
\subsection{Circuit Implementation of $e^{-i\op{\psi}{\psi}\Delta t}$}
Circuit implementation of this operator is illustrated in a figure below:
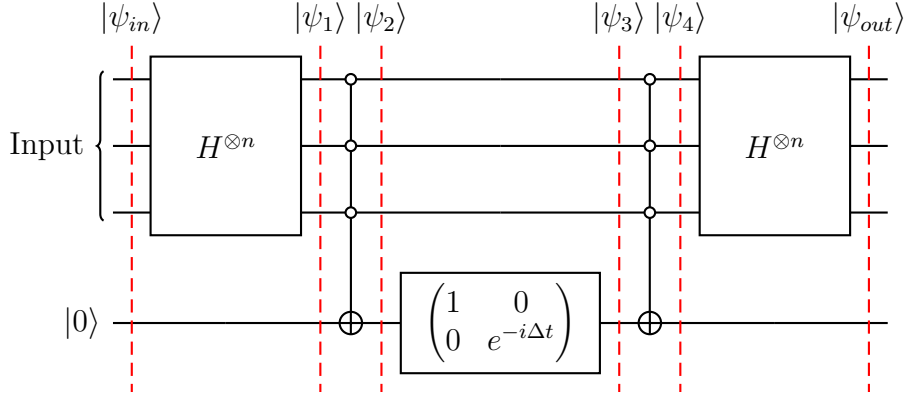
\begin{figure}[H]\label{c2}
\centering
\begin{quantikz}
\lstick[wires=3]{Input} \slice{$\ket{\psi_{in}}$}& \gate[wires=3][2 cm]{H^{\otimes n}}\slice{$\ket{\psi_{1}}$} & \octrl{1}\slice{$\ket{\psi_{2}}$} & \qw \slice{$\ket{\psi_{3}}$}& \octrl{1}\slice{$\ket{\psi_{4}}$} & \gate[wires=3][2 cm]{H^{\otimes n}}\slice{$\ket{\psi_{out}}$} & \qw \\
& & \octrl{1} & \qw & \octrl{1} & & \qw \\
& & \octrl{1} & \qw & \octrl{1} & & \qw \\
\lstick{$\ket{0}$}& \qw & \targ{} & \gate{\begin{pmatrix} 1&0\\0&e^{-i\Delta t}\end{pmatrix}} & \targ{} & \qw & \qw 
\end{quantikz}
\caption{Circuit implementing the operator $e^{-i\Delta \op{\psi}{\psi}}$}
\end{figure}
\subsection*{Analysis}
\noindent Input can be either $\ket{x}$ or $\ket{y}$. So we have two cases:
\paragraph*{Case: 1}
\begin{enumerate}
\item Input state is given by:
\begin{align*}
\psi_{in} & = \ket{x}\ket{0}
\end{align*}
\item State after the Hadamard gate is given by:
\begin{align*}
\ket{\psi_1} & = \frac{1}{\sqrt{N}} \sum_{j=0}^{N-1}\left(-1\right)^{\overrightarrow{x}\cdot\overrightarrow{j}}\ket{j}\ket{0}\\
& = \frac{1}{\sqrt{N}}\ket{000\cdots 0}\ket{0} + \frac{1}{\sqrt{N}}\sum_{j=1}^{N-1}\left(-1\right)^{\overrightarrow{x}\cdot\overrightarrow{j}}\ket{j}\ket{0}
\end{align*}
\item State after the action of multi-qubit controlled NOT gate with respect to the state $\ket{000\cdots 0}$ is given by: 
\begin{align*}
\ket{\psi_3} & = \frac{1}{\sqrt{N}}\ket{000\cdots 0}\ket{1} + \frac{1}{\sqrt{N}}\sum_{j=1}^{N-1}\left(-1\right)^{\overrightarrow{x}\cdot\overrightarrow{j}}\ket{j}\ket{0}
\end{align*}
\item State after phase gate is given by: 
\begin{align*}
\ket{\psi_4} & = \frac{e^{-i\Delta t}}{\sqrt{N}}\ket{000\cdots 0}\ket{1} + \frac{1}{\sqrt{N}}\sum_{j=1}^{N-1}\left(-1\right)^{\overrightarrow{x}\cdot\overrightarrow{j}}\ket{j}\ket{0}
\end{align*}
\item Multi-qubit controlled NOT gate is applied again to keep auxiliary qubit in the state $\ket{0}$. The state after apply this gate is given by: 
\begin{align*}
\ket{\psi_5} & = \frac{e^{-i\Delta t}}{\sqrt{N}}\ket{000\cdots 0}\ket{0} + \frac{1}{\sqrt{N}}\sum_{j=1}^{N-1}\left(-1\right)^{\overrightarrow{x}\cdot\overrightarrow{j}}\ket{j}\ket{0} \\
& =\left(\frac{e^{-i\Delta t}}{\sqrt{N}}\ket{000\cdots 0}- \frac{1}{\sqrt{N}}\ket{000\cdots 0}+\frac{1}{\sqrt{N}}\ket{000\cdots 0} + \frac{1}{\sqrt{N}}\sum_{j=1}^{N-1}\left(-1\right)^{\overrightarrow{x}\cdot\overrightarrow{j}}\ket{j} \right)\ket{0} \\
& = \frac{e^{-i\Delta t}}{\sqrt{N}}\ket{000\cdots 0}- \frac{1}{\sqrt{N}}\ket{000\cdots 0} + \frac{1}{\sqrt{N}}\sum_{j=0}^{N-1}\left(-1\right)^{\overrightarrow{x}\cdot\overrightarrow{j}}\ket{j} \\
& = \left( \frac{e^{-i\Delta t}-1}{\sqrt{N}}\ket{000\cdots 0} + H\ket{x} \right)\ket{0}
\end{align*}
\item State after the last Hadamard gate is given by: 
\begin{align*}
\ket{\psi_{out}} & =\left( \frac{e^{-i\Delta t}-1}{\sqrt{N}}H\ket{000\cdots 0}+HH\ket{x} \right)\ket{0} \\
& =\left[  \frac{e^{-i\Delta t}-1}{\sqrt{N}} \ket{\psi} + \ket{x}\right]\ket{0}\\
& = \alpha \left(e^{-i\Delta t}-1\right)\ket{\psi} + \ket{x}
\end{align*}
\end{enumerate}
\paragraph*{Case: 2} For the input, $\ket{y}$. Sovle the following exercise.
\begin{exercise}{}{label}
Prove that, the ouput of circuit in Fig. (\ref{c2}) for the input $\ket{y}$ is given by: 
\begin{align*}
\ket{\psi_{out}} & = \ket{y} + \beta \left(e^{i\Delta t}-1\right)\ket{\psi}
\end{align*}
\end{exercise}
\subsection*{Unitary time evolution Operator corresponding to Hamiltonian}
\noindent Unitary time evolution operator that transform $\ket{\psi}$ to $\ket{x}$ is given by: 
\begin{align}\label{unitary}
U(\Delta t) & \equiv e^{-i\op{\psi}{\psi}\Delta t}e^{-i\op{x}{x}\Delta t}
\end{align}
In the basis $\{\ket{x},\ket{y}\}$, 
$
\op{x}{x}  = \frac{I+Z}{2} = \frac{I + \hat{z}\cdot \vec{\sigma}}{2}
$
and
$
\op{\psi}{\psi}= \frac{I + \vec{\psi}\cdot\vec{\sigma}}{2}
$
where $\hat{z}\equiv (0,0,1)$ and $\vec{\psi}\equiv(2\alpha\beta,0,\alpha^2-\beta^2)$.
\begin{example}{}{label}
Express $\op{x}{x}$ and $\op{\psi}{\psi}$ in Bloch vector representation.
\paragraph*{Solution:} In $\{\ket{x},\ket{y}\}$ basis, 
\begin{align*}
\ket{\psi} & = \begin{pmatrix}
\alpha \\ \beta
\end{pmatrix}\\
\Rightarrow \ket{x} & = \begin{pmatrix}
1\\0
\end{pmatrix}
\end{align*}
Also note that, $\rho=(I + n_xX+n_yY+n_zZ)/2$, where $\rho$ is a density matrix and $\vec{n}=(n_x,n_y,n_z)$ is a bloch vector. and $n_i = Tr(\rho\sigma_i)$.
\begin{enumerate}
\item For $\op{x}{x}$, 
\begin{align*}
n_x & = Tr\left( \op{x}{x}X\right) = 0\\
n_y & = Tr\left( \op{x}{x}Y\right) = 0\\
n_z & = Tr\left( \op{x}{x}Z\right) = 1
\end{align*}
Thus, 
\begin{align*}
\op{x}{x} & = \frac{I + \hat{z}\cdot\vec{\sigma}}{2}
\end{align*}
\item For $\op{\psi}{\psi}$, 
\begin{align*}
n_x & = Tr(\op{\psi}{\psi}X)=2\alpha\beta \\
n_y & = Tr(\op{\psi}{\psi}Y)=0 \\
n_z & = Tr(\op{\psi}{\psi}Z)=\alpha^2-\beta^2
\end{align*}
Thus, 
\begin{align*}
\op{\psi}{\psi} & = \frac{I + \vec{\psi}\cdot\vec{\sigma}}{2}
\end{align*}
\end{enumerate}
Equivalently, 
\begin{align*}
\op{x}{x} & = (0,0,1) \\
\op{\psi}{\psi} & = (2\alpha\beta,0,\alpha^2-\beta^2)
\end{align*}
\end{example}
\noindent The unitary operator in Eq. (\ref{unitary}) can be written as: 
\begin{align*}
U(\Delta t) & = \text{exp}\left(-i \frac{I + \vec{\psi}\cdot\vec{\sigma}}{2}\Delta t\right)\text{exp}\left(-i \frac{I + \hat{z}\cdot\vec{\sigma}}{2}\Delta t\right) \\
& = \text{exp}\left(-i\frac{\Delta t}{2}\right) \text{exp}\left(-i\frac{\vec{\psi}\cdot\vec{\sigma}}{2}\Delta t \right)\text{exp}\left(-i\frac{\Delta t}{2}\right)\text{exp}\left(-i\frac{\hat{z}\cdot\vec{\sigma}}{2}\Delta t \right) \\
& = \text{exp}\left(-i\Delta t\right) \text{exp}\left(-i\frac{\vec{\psi}\cdot\vec{\sigma}}{2}\Delta t \right)\text{exp}\left(-i\frac{\hat{z}\cdot\vec{\sigma}}{2}\Delta t \right)
\end{align*}
The term $\text{exp}\left(-i\Delta t\right)$ is global phase factor. Therefore, the unitary operator can be written as: 
\begin{align*}
U(\Delta t) & = \left[ \cos\left(\frac{\Delta t}{2}\right) - i\sin\left(\frac{\Delta t}{2}\right) \vec{\psi}\cdot\vec{\sigma} \right]\times \left[ \cos\left(\frac{\Delta t}{2}\right) - i\sin\left(\frac{\Delta t}{2}\right) \hat{z}\cdot\vec{\sigma} \right] \\
& = \cos^2\left(\frac{\Delta t}{2}\right) -i\cos\left(\frac{\Delta t}{2}\right)\sin\left(\frac{\Delta t}{2}\right)\hat{z}\cdot\vec{\sigma} -i\cos\left(\frac{\Delta t}{2}\right)\sin\left(\frac{\Delta t}{2}\right)\vec{\psi}\cdot\vec{\sigma}-\sin^2\left(\frac{\Delta t}{2}\right)(\vec{\psi}\cdot\vec{\sigma})(\hat{z}\cdot\vec{\sigma}) \\
& = \cos^2\left(\frac{\Delta t}{2}\right) -i\cos\left(\frac{\Delta t}{2}\right)\sin\left(\frac{\Delta t}{2}\right)\left[\vec{\psi}+\hat{z} \right]\cdot\vec{\sigma}-\sin^2\left(\frac{\Delta t}{2}\right)(\vec{\psi}\cdot\vec{\sigma})(\hat{z}\cdot\vec{\sigma})
\end{align*}
Dot product terms are calculated following:
\begin{align*}
\vec{\psi}\cdot\vec{\sigma} & = (2\alpha\beta,0,\alpha^2-\beta^2)\cdot(X,Y,Z) = 2\alpha\beta X + \alpha^2Z-\beta^2Z \\
\hat{z}\cdot\vec{\sigma} & = (0,0,1)\cdot(X,Y,Z) = Z \\
\Rightarrow (\vec{\psi}\cdot\vec{\sigma})(\hat{z}\cdot\vec{\sigma}) & = \left(2\alpha\beta X + \alpha^2Z-\beta^2Z\right)Z \\
& = 2\alpha\beta XZ + \alpha^2-\beta^2 \\
& = -i2\alpha\beta Y + \alpha^2 -\beta^2\ \ \ \ \because\ XZ=-iY
\end{align*}
Now, 
\begin{align*}
\vec{\psi}\times\hat{z} & = \begin{vmatrix}
\hat{i} & \hat{j} & \hat{k} \\
2\alpha\beta & 0 & \alpha^2-\beta^2 \\
0 & 0 & 1
\end{vmatrix} = (0,-2\alpha\beta,0) \\
\Rightarrow (\vec{\psi}\times\hat{z})\cdot \vec{\sigma} & = (0,-2\alpha\beta,0)\cdot(X,Y,Z) = -2\alpha\beta Y
\end{align*}
And, 
\begin{align*}
\vec{\psi}\cdot\hat{z} & = (2\alpha\beta , 0 , \alpha^2-\beta^2 )\cdot(0,0,1)= \alpha^2-\beta^2
\end{align*}
Thus, 
\begin{align*}
(\vec{\psi}\cdot\vec{\sigma})(\hat{z}\cdot\vec{\sigma}) & = i (\vec{\psi}\times\hat{z})\cdot \vec{\sigma} + \vec{\psi}\cdot \hat{z}
\end{align*}
Therefore, the unitary operator becomes: 
\begin{align*}
U(\Delta t) &= \cos^2\left(\frac{\Delta t}{2}\right) -i\cos\left(\frac{\Delta t}{2}\right)\sin\left(\frac{\Delta t}{2}\right)\left[\vec{\psi}+\hat{z} \right]\cdot\vec{\sigma}-\sin^2\left(\frac{\Delta t}{2}\right)(\vec{\psi}\cdot\vec{\sigma})(\hat{z}\cdot\vec{\sigma}) \\
& = \cos^2\left(\frac{\Delta t}{2}\right) -i\cos\left(\frac{\Delta t}{2}\right)\sin\left(\frac{\Delta t}{2}\right)\left[\vec{\psi}+\hat{z} \right]\cdot\vec{\sigma}-\sin^2\left(\frac{\Delta t}{2}\right)\left[i (\vec{\psi}\times\hat{z})\cdot \vec{\sigma} + \vec{\psi}\cdot \hat{z} \right] \\
& = \left(\cos^2\left(\frac{\Delta t}{2}\right) - \sin^2\left(\frac{\Delta t}{2}\right)\vec{\psi}\cdot \hat{z} \right)I -2i\sin \left(\frac{\Delta t}{2}\right) \left(\cos\left(\frac{\Delta t}{2}\right)\frac{\vec{\psi}+\hat{z}}{2} + \sin\left(\frac{\Delta t}{2}\right)\frac{\vec{\psi}\times\hat{z}}{2}\right)\cdot\vec{\sigma}
\end{align*}
The above equation show that $U(\Delta t)$ is a rotation on the Bloch Sphere about an axis of rotation $\vec{r}$ defined by: 
\begin{align*}
\vec{r} & = \cos\left(\frac{\Delta t}{2}\right)\frac{\vec{\psi}+\hat{z}}{2} + \sin\left(\frac{\Delta t}{2}\right)\frac{\vec{\psi}\times\hat{z}}{2}
\end{align*}
and through the angle defined by: 
\begin{align*}
\cos\left(\frac{\theta}{2}\right) & = \cos^2\left(\frac{\Delta t}{2}\right) - \sin^2\left(\frac{\Delta t}{2}\right)\vec{\psi}\cdot \hat{z} \\
& = \cos^2\left(\frac{\Delta t}{2}\right) - \sin^2\left(\frac{\Delta t}{2}\right) (\alpha^2-\beta^2) \\
& = \cos^2\left(\frac{\Delta t}{2}\right) - \sin^2\left(\frac{\Delta t}{2}\right)\left(\frac{2}{N}-1\right) \\
& = \cos^2\left(\frac{\Delta t}{2}\right) + \sin^2\left(\frac{\Delta t}{2}\right) - \frac{2}{N}\sin^2\left(\frac{\Delta t}{2}\right) \\
& = 1 - \frac{2}{N}\sin^2\left(\frac{\Delta t}{2}\right)
\end{align*}
Summarizing, $U(\Delta t)$ rotates $\op{\psi}{\psi}$ about an axis $\vec{r}$, through an angle $\theta$. We perform rotations until $\ket{\psi}$ is transformed to $\ket{x}$. The above equation shows that, if $\Delta t=\pi$, rotation angle can be maximized. Thus, for $\Delta t=\pi$, we obtain
\begin{align*}
\cos\left( \frac{\theta}{2}\right) & = 1-\frac{2}{N}
\end{align*}
For large $N$, $\theta\approx 4/\sqrt{N}$, and the number of oracle calls required to reach the solution $\ket{x}$ is $O(\sqrt{N})$, just like the original Grover's Algorithm.

\section{Amplitude Estimation and Amplification}
\noindent Given a unitary operator $\mathcal{A}$ that act as: 
\begin{align}\label{s_state}
\mathcal{A}\ket{0} & = \sqrt{1-g}\ket{w^{\perp}} + \sqrt{g}\ket{w}
\end{align}
where $\ket{w}$ is the solution state from a search problem. Quantum Amplitude Estimation is the task of finding the estimate for the amplitude $g$ of the state $\ket{w}$. The generalized Grover's operator is: 
\begin{align}\label{groveroperator}
\mathcal{Q} & = \mathcal{A}\mathcal{S}_0\mathcal{A}^{\dagger}\mathcal{S}_g
\end{align}
where $\mathcal{S}_g$ marks the good states and $\mathcal{S}_0$ marks the $\ket{0}$ state.
\begin{align*}
\mathcal{S}_g & = \mathbb{I} - 2\op{w}{w} \\
\mathcal{S}_0 & = 2\op{0}{0} - \mathbb{I}
\end{align*}
The operator $\mathcal{Q}$ is same as the Grover's operator $G=HU_{f_0}HU_f$, in the Grover's algorithm, where $U_{f_0}$ marked the $\ket{0}$ state and $U_f$ marked the solution state. Eq. (\ref{groveroperator}) can be written as: 
\begin{align*}
\mathcal{Q} & = \mathcal{A}\left(2\op{0}{0} - \mathbb{I} \right)\mathcal{A}^{\dagger}\mathcal{S}_g \\
& = \left(2\mathcal{A}\ket{0}\bra{0}\mathcal{A}^{\dagger} - \mathcal{A}\mathbb{I}\mathcal{A}^{\dagger} \right) \mathcal{S}_g \\
& = \left(2\mathcal{A}\ket{0}\bra{0}\mathcal{A}^{\dagger} - \mathbb{I} \right) \mathcal{S}_g
\end{align*}
Take $\displaystyle{\sqrt{1-g}=\cos(\theta)}$ and $\displaystyle{\sqrt{g}=\sin(\theta)}$, and using Eq. (\ref{s_state}) in the above expression, it can be easily proved that:
\begin{align*}
\mathcal{Q} \ket{s} & = \cos(2\theta+\theta)\ket{w^{\perp}} + \sin(2\theta+\theta)\ket{w}
\end{align*}
where $\displaystyle{\ket{s}=\cos(\theta)\ket{w^{\perp}}+\sin(\theta)\ket{w}}$. If we apply $\mathcal{Q}$ $r-$times, we get
\begin{align*}
\mathcal{Q}^r \ket{s} & = \cos(2r\theta+\theta)\ket{w^{\perp}} + \sin(2r\theta+\theta)\ket{w}
\end{align*}
\begin{exercise}{}{label}
Show that:
\begin{align*}
\mathcal{Q}\ket{s} & = \cos(2\theta+\theta)\ket{w^{\perp}} + \sin(2\theta+\theta)\ket{w}
\end{align*}
\end{exercise}
We can obtain the state $\ket{w}$, if $2r\theta+\theta=\pi/2$, which means that $r=\pi/4\theta-1/2$. But in general, $r$ is not an integer. It means that upon measurement, there is a probability that we get the state $\ket{w^{\perp}}$ and is proportional to $\theta$. Smaller the $\theta$, smaller is the probability we collapse on $\ket{w^{\perp}}$. So, applying the $\mathcal{Q}$ operator $r-$times, such that we end up on $\ket{w}$, we can calculate $\theta$, from which we can easily estimate the amplitude $\sqrt{g}=\sin(\theta)$ of the state $\ket{w}$.
\subsection*{Derandomizing the Grover's Algorithm}
\begin{figure}[H]\label{newcircuit}
\centering
\begin{quantikz}
\lstick{$\ket{0}$} & \gate{H} & \gate[2][1.5 cm]{U_f} & \gate{H} & \gate[2][1.5 cm]{U_{f_0}} & \gate{H} & \qw \\
\lstick{$\ket{0}$} & \gate{R} & & \gate{R} & & \gate{R} & \qw 
\end{quantikz}
\caption{Modified Grover's Circuit}
\end{figure}
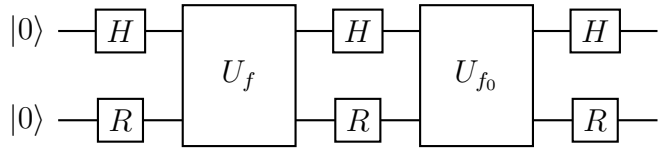

\noindent The Grover's Algorithm starts with a state $\ket{s}$, given by:
\begin{align*}
\ket{s} & = \sqrt{\frac{N-\mu}{N}}\ket{w^{\perp}} + \sqrt{\frac{N}{\mu}}\ket{w}
\end{align*}
Let's take $\displaystyle{\sqrt{\frac{N-\mu}{N}}=\sqrt{1-g_0}}$ and $\displaystyle{\sqrt{\frac{\mu}{N}}=\sqrt{g_0}}$, then the state becomes
\begin{align*}
\ket{s} & = \sqrt{1-g_0} \ket{w^{\perp}} + \sqrt{g_0}\ket{w}
\end{align*}
We know that, after applying Grover's operator $r-$times, we end up with the state $\ket{w}$and in general, $r$ is not an integer. But we can modify the state $\ket{s}$ in such a way that $r$ comes out to be an integer. For this, we have to add rotation operator $R$ in the circuit of Grover's algorithm, as illustrated in Fig. (\ref{newcircuit}), such that
\begin{align*}
R\ket{0} & = \sqrt{1-\frac{g_0'}{g_0}}\ket{0} + \sqrt{\frac{g_0'}{g_0}}\ket{1}
\end{align*}
where $g_0'$ is such that $r$ is an integer.
\noindent In this modified circuit, $\mathcal{A}=H\otimes R$. Let's apply this to the state $\ket{0}\otimes\ket{0}$.
\begin{align*}
\mathcal{A}\left(\ket{0}\otimes\ket{0}\right) & = H\ket{0}\otimes R\ket{0} \\
\ket{s_{new}}& = \ket{s} \otimes  \left(\sqrt{1-\frac{g_0'}{g_0}}\ket{0} + \sqrt{\frac{g_0'}{g_0}}\ket{1}\right) \\
& = \left(\sqrt{1-g_0} \ket{w^{\perp}} + \sqrt{g_0}\ket{w} \right) \otimes \left(\sqrt{1-\frac{g_0'}{g_0}}\ket{0} + \sqrt{\frac{g_0'}{g_0}}\ket{1}\right) \\
& = \sqrt{1-g_0}\sqrt{1-\frac{g_0'}{g_0}}\ket{w^{\perp}}\ket{0} + \sqrt{1-g_0}\sqrt{\frac{g_0'}{g_0}}\ket{w^{\perp}}\ket{1} + \sqrt{g_0}\sqrt{1-\frac{g_0'}{g_0}}\ket{w}\ket{0} + \sqrt{g_0'}\ket{w}\ket{1}
\end{align*}
Let's call $\ket{w}\ket{1}=\ket{u}$ the new good state. Thus, $\ket{s_{new}}$ becomes
\begin{align*}
\ket{s_{new}} & = \sqrt{1-g_0'}\ket{u^{\perp}} + \sqrt{g_0'}\ket{u}
\end{align*}
So, the oracle $U_f$ needs to be modified in such a way that it marks the state $\ket{u}=\ket{w\otimes 1}$. Using $\mathcal{A}=H\otimes R$, we can obtain the good state after applying the Modified Grover's Operator $r-$times such that we get $\ket{u}$ with 100\% probability on measurement.
\chapter{Methods for Hamiltonian Simulation}
\footnote{Copyrights: Muhammad Faryad, LUMS}
Simulation of quantum systems is the key motivation behind quantum computers. Simulation of quantum systems by classical computers, in general, cannot be done efficiently. The dynamical behavior of a quantum system is governed by Schr\"{o}dinger's equation,
\begin{align}\label{eq1}
i\hbar \frac{d}{dt}\ket{\psi} & = H\ket{\psi}
\end{align}
where the Hamiltonian\footnote{In general, $H$ may change with time, but we assumed that $H$ is time-independent.} $H$ is an observable corresponding to the total energy of a quantum system. This total energy is typically the sum of kinetic energy, potential energy, and many other local terms that act only on a few particles. Hamiltonian describes the physical characteristics of a system. It determines the evolution of the state in time governed by Schr\"{o}dinger's equation as in Eq. (\ref{eq1}). Thus, the state $\ket{\psi(t)}$ at time $t$ given the initial state $\ket{\psi(0)}$ is given by:
\begin{align}\label{eq2}
\ket{\psi(t)} & = e^{-iHt}\ket{\psi(0)}
\end{align}
where $e^{-iHt}$ is $2^n\times 2^n$ matrix and $t$ does not need to be an integer, i.e., evolution is continuous in time. The key challenge in simulating quantum systems is the exponential number of equations that must be solved. For instance, one qubit evolving according to Schr\"{o}dinger's equation requires us to solve two differential equations; for two qubits, we need to solve $2^2=4$ equations and so on. Generally, for a system of $n$ qubits, we need to solve $2^n$ differential equations. However, all the important physical Hamiltonian of quantum systems in which physicists are interested in having a form that can be efficiently implemented.
\section{Hamiltonian Simulations}
Since, in Eq. (\ref{eq2}), $H$ is extremely difficult to exponentiate. So a good beginning will be a first order solution.
\begin{align*}
\ket{\psi(\Delta t)} & = e^{-iH\Delta t}\ket{\psi(0)} \\
& = \left( I - iH\Delta t + \mathcal{O}(\Delta t^2) \right)\ket{\psi(0)} \\
& = U_{\Delta t}\ket{\psi(0)}
\end{align*}
However, such approach is generally not very satisfactory. An efficient approximation of the solution is possible to a higher order. For example, the Hamiltonian can be written as the sum over many local interactions in most systems. Specifically, for a system of $n$ particles, 
\begin{align*}
H = \sum_{k=1}^L H_k
\end{align*}
where $H_k$ acts on at most $f$ number of systems, such that $f<<n$ and $L$ is a polynomial in $n$. $H_k$ is called $f$-local. For fixed $t$ $\displaystyle{e^{-iH_k t}}$ is an $f$-qubit gates that acts like identity on the other $n-f$ qubits \footnote{$H$ can thus be described by $L$ $2^f\times 2^f$ matrices}. Such gates can be constructed from CNOTs and single-qubit gates. In other words, $H_k$ acts on a much smaller subsystem and thus is straightforward to approximate. But since $[H_j,H_k]\neq 0$. So, in general:
\begin{align}
e^{-iHt} & \neq \prod_k e^{-iH_k t}
\end{align}
\subsection{Lie-Suzuki-Trotter Method}
Consider a unitary operator $U=e^{i(A+B)t}$ where $A$ and $B$ are two hermitian operators, then for any real $t$, in general, 
\begin{align}
e^{i(A+B)t} & \neq e^{iAt}e^{iBt}
\end{align}
unless $A$ and $B$ commutes. So, if $A$ and $B$ commutes, then
\begin{align*}
e^{iAt} & = I + iAt + \frac{i^2A^2t^2}{2} + \cdots \\
e^{iBt} & = I + iAt + \frac{i^2B^2t^2}{2} + \cdots \\
\Rightarrow e^{iAt}e^{iBt} & = I + i(A+B)t + O(t^2) \\
& = e^{i(A+B)t}  + \mathcal{O}(t^2)
\end{align*}
More precisely and generally, for every integer $r\geq 1$, we have: 
\begin{align*}
U & = e^{i(A_1+A_2+\cdots+A_N)t} = (e^{i(A_1+A_2+\cdots+A_N)t/r})^r = (e^{iA_1t/r}e^{iA_2t/r}\cdots e^{iA_Nt/r} + \mathcal{E} )^r = (\tilde{U}+\mathcal{E})^r
\end{align*}
where $\displaystyle{\mathcal{E}=\mathcal{O}\left( \frac{N^2t^2}{r^2} \right)}$. Since we are going to operate $\tilde{U}$ $r$-times, thus the approximation error will be: 
\begin{align*}
\mathcal{E}_{tot} & = \mathcal{O}\left(\frac{N^2t^2}{r} \right) \\
\Rightarrow r & = \mathcal{O}\left(\frac{N^2t^2}{\mathcal{E}} \right)
\end{align*}
Choosing $r$ wisely, we can make approximation error less than $\mathcal{E}$. Modification of Trotter's method allows us to derive higher order approximations as well for performing quantum simulation. For example, it can be shown that:
\begin{align}\label{Trotter}
e^{i(A+B)t} & = e^{iAt/2}e^{iBt}e^{iAt/2} + \mathcal{O}(t^3)
\end{align}
\begin{exercise}{}{label}
Provide the proof of Eq. (\ref{Trotter}). \\

\textbf{Hint: }Expand the right hand side and combine the relevant terms to get $e^{i(A+B)t}$.
\end{exercise}
\subsection{Simulation of Sparse Hamiltonian}
In physics, we are interested with those problems that fortunately corresponds to a situation where the Hamiltonian is sparse. A $d$-sparse matrix is such a matrix that contains $d$ non-zero entries in each row. We will be dealing with a restricted class of Hamiltonian which are not only $d$-sparse, but are also row computable. By row computable, it means that we have access to an oracle $O_f$ that determines the position of non-zero entry and an oracle $O_H$ that determines the value of that entry in a sparse-matrix. Mathematically, 
\begin{align}
O_f(j)\ket{x}\ket{0} & = \ket{x}\ket{y=f(x_j)}
\end{align}
where $x$ is row number and $f(x_j)$ is the column number of a non-zero entry. Similarly, 
\begin{align}
O_H\ket{x}\ket{f(x_j)}\ket{0} & = \ket{x}\ket{y=f(x_j)}\ket{H_{xy}}
\end{align}
where $H_{xy}$ is the matrix element where the non-zero entry is located. Consider an example of a matrix $A$ given below:
\begin{align}
A & = \begin{pmatrix}
0 & 1 & 0 & a \\
1 & b & 0 & 2 \\
0 & 0 & c & 0 \\
a & 2 & 0 & 0
\end{pmatrix}
\end{align}
The position of non-zero entries in the first row can be determined by an oracle $O_f$ as: 
\begin{align*}
O_f(0)\ket{0}\ket{0} & = \ket{0}\ket{1} \\
O_f(1)\ket{0}\ket{0} & = \ket{0}\ket{3}
\end{align*}
This shows that the position of the first and second non-zero entry in the first row is: 01 and 03, where the indexing starts from 0. Now, the value of these entries can be determined by an oracle $O_H$ as:
\begin{align*}
O_H\ket{0}\ket{1}\ket{0} & = \ket{0}\ket{1}\ket{1} \\
O_H\ket{0}\ket{3}\ket{0} & = \ket{0}\ket{3}\ket{a}
\end{align*}
This shows that the $A_{01}=1$ and $A_{03}=a$.
\subsection{Simulating 1-sparse Hamiltonian}
If we somehow come up with a method to simulate 1-sparse Hamiltonian, then we can also simulate $d$-sparse Hamiltonian, since a $d$-sparse Hamiltonian can be expressed as the sum of $1$-sparse Hamiltonian, $\displaystyle{H=\sum_{j=1}^dH_j}$, such that $H_j$ is still a hermitian matrix.
\subsubsection{Intuition}
Consider a 1-sparse Hamiltonian, 
\begin{align}\label{hamiltonian}
H & = \begin{pmatrix}
0 & 0 & 0 & a \\
0 & b & 0 & 0 \\
0 & 0 & c & 0 \\
a & 0 & 0 & 0
\end{pmatrix}
\end{align}
Let's apply oracle $O_f$ that determines the position of non-zero entries in each row:
\begin{equation}\label{OF}
\begin{split}
O_f \ket{0}\ket{0} = \ket{0}\ket{3} \\
O_f \ket{1}\ket{0} = \ket{1}\ket{1} \\
O_f \ket{2}\ket{0} = \ket{2}\ket{2} \\
O_f \ket{3}\ket{0} = \ket{3}\ket{0}
\end{split}
\end{equation}
From Eq. (\ref{OF}), it is evident that for 1-sparse Hamiltonian, $f[f(x)]=x$. Now consider the states in computational basis\footnote{the 0,1,2 and 3 are the decimal representation of binary state in computational basis} $\ket{0},\ket{1},\ket{2}\ \text{\&}\ \ket{3}$. Operation of Hamiltonian from Eq. (\ref{hamiltonian}) on these state is as follow:
\begin{equation}\label{HamiltonianEquations}
\begin{split}
H\ket{0} = a\ket{3} \\
H\ket{1} = b\ket{1} \\
H\ket{2} = c\ket{2} \\
H\ket{3} = a\ket{0} 
\end{split}
\end{equation}
We see that some of the states are the eigenstates of Hamiltonian, and some of the states in pairs are being interchanged after the application of Hamiltonian. The Hamiltonian in Eq. (\ref{hamiltonian}) can also be written in the block-diagonal form after basis transformation as: 
\begin{equation}
H = \begin{pmatrix}
0 & a & 0 & 0 \\
a & 0 & 0 & 0 \\
0 & 0 & b & 0 \\
0 & 0 & 0 & c
\end{pmatrix} = \begin{pmatrix}
\begin{array}{c c}
a\hat{X} & O \\
O & \begin{array}{c c}
b & 0 \\
0 & c
\end{array}
\end{array}
\end{pmatrix}
\end{equation}
Thus the unitary time evolution of this Hamiltonian is given by: 
\begin{equation}
U = e^{-iHt} = \begin{pmatrix}
\begin{array}{c c}
e^{-iaXt} & O \\
O & \begin{array}{c c}
e^{-ibt} & 0 \\
0 & e^{-ict}
\end{array}
\end{array}
\end{pmatrix}
\end{equation}
where $e^{-iaXt}$ is the $X$-rotation gate $R_x(2a)$ and $e^{-ibt}$ and $e^{-ict}$ are the $Z$-rotation gate $R_z(b)$ and $R_z(c)$ respectively. Therefore, in order to simulate 1-sparse Hamiltonian, we apply $R_z$ to those computational basis states that are the eigen states of $H$ and apply $R_x$ to the rest of the states.
\subsubsection*{Algorithm}
We see that $R_z$ is applied to those states that correspond to diagonal elements of $H$, i.e., $x=f(x)$ and to the rest of states, $R_x$ is applied, for which $x\neq f(x)$.\\
\begin{center}
\textbf{Case I}
\end{center}
Take $\ket{\psi}=\alpha\ket{x}+\beta\ket{f(x)}$.
\paragraph{Step 1:}Apply $O_f$
\begin{align}
\ket{\psi_2} & = O_f\ket{\psi}\ket{0}  = \alpha O_f\ket{x}\ket{0} + \beta O_f\ket{f(x)}\ket{0} 
\end{align}
\begin{align}
\ket{\psi_2} & = \alpha \ket{x}\ket{f(x)} + \beta\ket{f(x)}\ket{x}
\end{align}
\paragraph{Step 2:}The state $\ket{\psi_2}$ is passed through comparison circuit given below:
\begin{figure}[H]
\centering
\begin{quantikz}
&\lstick{$\ket{x}$} & \gate[wires=3][2cm]{Compare} & \qw & \rstick{$\ket{x}$}\qw \\
&\lstick{$\ket{y}$} & & \qw & \rstick{$\ket{y}$} \\
&\lstick{$\ket{0}$} & & \qw & \rstick{$\ket{q}$}
\end{quantikz}
\caption{Circuit representation of \emph{Comparison Circuit}}
\end{figure}
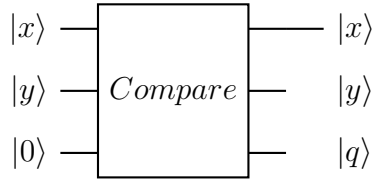
Mathematically, 
\begin{equation}
\text{COMP}\ket{x}\ket{y}\ket{0} = \begin{cases}
\ket{x}\ket{y}\ket{0} &\ \text{if $x<y$} \\
\ket{x}\ket{y}\ket{1} &\ \text{if $x>y$}
\end{cases}
\end{equation}
Assume that $f(x)>x$, then the state we get after passing $\ket{\psi_2}$ through comparison circuit will be: 
\begin{align}
\ket{\psi_3} & = \text{COMP}\ket{\psi_2}\ket{0}  = \alpha\text{COMP}\ket{x}\ket{f(x)}\ket{0} + \beta\text{COMP}\ket{f(x)}\ket{x}\ket{0} 
\end{align}
\begin{align}
\ket{\psi_3} & = \alpha \ket{x}\ket{f(x)}\ket{0} + \beta \ket{f(x)}\ket{x}\ket{1}
\end{align}
\paragraph{Step 3:} We swap the first two qubits controlled by the third qubit, by doing \emph{Controlled Swap Operation}. The state thus becomes:
\begin{align}
\ket{\psi_4} & = C_{SWAP}\ket{\psi_3} = \alpha C_{SWAP}\ket{x}\ket{f(x)}\ket{0} + \beta C_{SWAP}\ket{f(x)}\ket{x}\ket{1}
\end{align}
\begin{align}
\ket{\psi_4} & = \ket{x}\ket{f(x)} \left[ \alpha\ket{0}+\beta\ket{1}\right]
\end{align}
\paragraph{Step 4:}Apply $O_H$ to determine the value of matrix element corresponding to the state $\alpha\ket{0}+\beta\ket{1}$.
\begin{align}
\ket{\psi_5} & = O_H\ket{\psi_4}  \mapsto \ket{x}\ket{f(x)}H_{x,f(x)}\left[ \alpha\ket{0}+\beta\ket{1} \right]
\end{align}
\paragraph{Step 5:} Apply $X$-Rotation controlled by $H_{x,f(x)}$. The final state thus becomes:
\begin{align}
\ket{\psi_f} & \mapsto \ket{x}\ket{f(x)}e^{-iH_{x,f(x)}Xt}\left[ \alpha\ket{0}+\beta\ket{1}\right]
\end{align}
\paragraph{Step 6:} Uncompute except $X$-Rotation. By doing so, we get
\begin{align}
\ket{\Psi} & \rightarrow e^{-iH_{x,f(x)}Xt} \left[ \alpha\ket{0}+\beta\ket{1}\right]
\end{align}
\begin{center}
\textbf{Case II}
\end{center}
Consider the case when $x=f(x)$. This means that initial state is $\ket{\psi}=\ket{x}$.
\paragraph{Step 1:}Apply $O_f$
\begin{align}
O_f\ket{x} \ket{f(x)} & \rightarrow \ket{x}\ket{f(x)} = \ket{x}\ket{x}
\end{align}
\paragraph{Step 2:}State is passed through comparison circuit:
\begin{align}
\text{COMP}\ket{x}\ket{x} & \rightarrow \ket{x}\ket{x}\ket{0}
\end{align}
\paragraph{Step 3:}Controlled Swap operation is carries out. But since the last qubit is 0, thus the state will be unchanged, $\ket{x}\ket{x}\ket{0}$
\paragraph{Step 4:}Apply $O_H$.
\begin{align}
O_H\ket{x}\ket{x}\ket{0} & \rightarrow \ket{x}\ket{x}\ket{H_{x,x}}
\end{align}
\paragraph{Step 5:}Apply $Z$-Rotation controlled by $H_{x,x}$.
\begin{align*}
CR_z\ket{x}\ket{x}\ket{H_{x,x}} & \rightarrow \ket{x}\ket{x}e^{-iZH_{x,x}t}\ket{x}
\end{align*}
\paragraph{Step 6:}Uncompute except $Z$-Rotation. By doing so, we get
\begin{align}
e^{-iZH_{x,x}t}\ket{x}
\end{align}
\begin{center}
\textbf{Circuit Model}
\end{center}
\begin{figure}[H]
\centering
\begin{quantikz}
&\lstick{$\ket{\psi}$} & \gate[wires=2][1.5cm]{O_f} & \gate[wires=3][2cm]{Compare} & \swap{1} & \gate[4, nwires={3}][2cm]{O_H} & \ctrl{1}\gategroup[4,steps=5,style={dashed,
rounded corners,fill=blue!20, inner xsep=2pt},
background]{{\sc Controlled Rotation}} & \gate{H} & \gate{R_x} & \gate{H} & \ctrl{1} & \qw & \gate[wires=4]{\text{Un-Compute}} & \qw \\
&\lstick{$\ket{0}$} & & & \targX{} & & \targ{} & \octrl{-1} & \qw & \octrl{-1} & \targ{} & \qw & & \qw \\
&\lstick{$\ket{0}$} & \qw & &  \ctrl{-1} & \qw & \qw & \qw & \qw & \qw & \qw & \qw & \qw & \qw \\
&\lstick{$\ket{0}$} & \qw & \qw & \qw & \qw & \qw & \qw & \ctrl{-3} &\qw & \qw & \qw & & \qw
\end{quantikz}
\caption{Circuit that simulate 1-sparse Hamiltonian}
\end{figure}
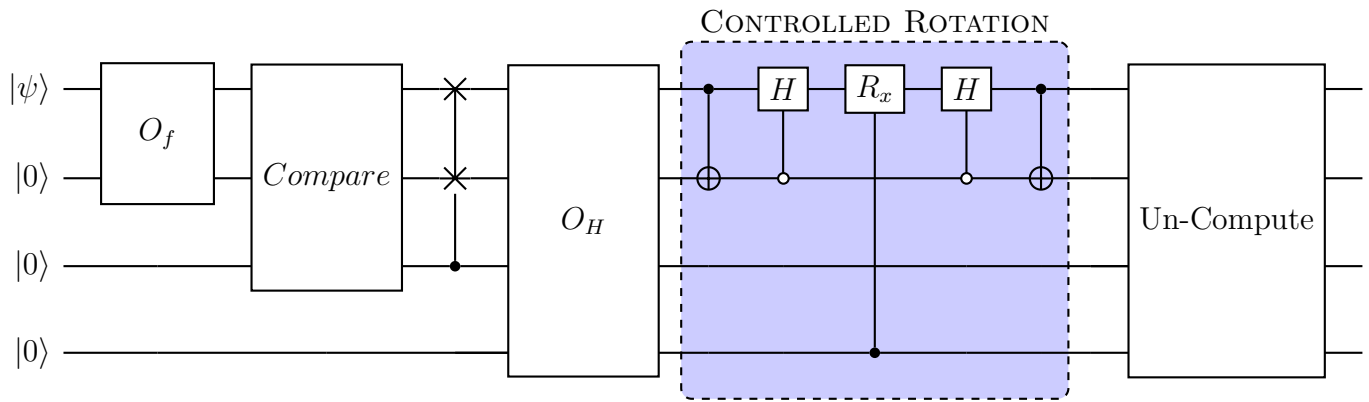
It shows that with only four queries, we can simulate 1-sparse Hamiltonian.
\begin{exercise}{}{label}
Draw the circuit for $O_f$ for the case of Eq. (\ref{OF}).
\end{exercise}
\subsection{Simulating Linear Combination of Unitaries}
Suppose we like to prepare $A\ket{\psi}$ where $A$ is $2^n\times 2^n$ matrix and $\ket{\psi}$ is an $n$-qubit state. Here, $A$ is not unitary in general but a linear combination of unitaries:
\begin{align}
A & = \sum_{i=0}^{m-1}a_i U_i
\end{align}
where $a_i$ are the positive real numbers.\footnote{Even if $a_i$ is complex, the imaginary part (phase) can be absorbed into $U_i$}. Suppose $O$ is a unitary operator, whose matrix elements are the function of $a_i$, that acts on $\log(m)$ qubits such that 
\begin{align}\label{phi}
O\ket{0}^{\otimes\log(m)} & = \frac{1}{\sqrt{\norm{a_i}_1}}\sum_i \sqrt{a_i}\ket{i} = \ket{\phi}
\end{align}
where $\norm{a_i}_1=\sum_ia_i$. We can implement the unitaries $U_i$ in a controlled way, having access to two register unitary $V = \sum_{i=0}^{m-1}\op{i}{i}\otimes U_i$ where $V$ maps $\ket{i}\ket{\psi_{inp}}$ $\mapsto$ $\ket{0}U_i\ket{\psi_{inp}}$. The first register can be think of ``selection register" since it decides which unitary $U_i$ to apply to $\ket{\psi_{inp}}$.
\subsubsection*{Algorithm}
\begin{enumerate}
\item Prepare the input state:
\begin{align}
\ket{\Psi_1} & = \ket{0}^{\otimes\log(m)}\ket{\psi_{inp}}
\end{align}
\item Apply $O$ to the first register to prepare $\ket{\phi}$, as shown in Eq. (\ref{phi}).
\begin{align}
\ket{\Psi_2} & = O\ket{0}^{\otimes\log(m)} \otimes \ket{\psi_{inp}}
\end{align}
\begin{align}
\ket{\Psi_2} & = \frac{1}{\sqrt{\norm{a_i}_1}}\sum_{i=0}^{m-1}\sqrt{a_i}\ket{i} \otimes \ket{\psi_{inp}}
\end{align}
\item Apply two register unitary $V$.
\begin{align}
\ket{\Psi_3} & = \frac{1}{\sqrt{\norm{a_i}_1}}\sum_{i=0}^{m-1}\sqrt{a_i}\ket{i} \ip{i}{i} \otimes U_i\ket{\psi_{inp}}
\end{align}
\begin{align}
\ket{\Psi_3} & = \frac{1}{\sqrt{\norm{a_i}_1}}\sum_{i=0}^{m-1}\sqrt{a_i}\ket{i}  \otimes U_i\ket{\psi_{inp}}
\end{align}
\item Apply $O^{\dagger}$ to the first register.
\begin{align}
\ket{\Psi_4} & = \frac{1}{\sqrt{\norm{a_i}_1}}\sum_{i=0}^{m-1} \ket{0}^{\otimes \log(m)} \otimes  a_iU_i\ket{\psi_{inp}} + \text{other terms}
\end{align}
\begin{align}\label{final}
\ket{\Psi_4} & = \frac{1}{\norm{a_i}_1} \ket{0}^{\otimes \log(m)} \otimes A\ket{\psi_{inp}} + \text{other terms}
\end{align}
\end{enumerate}
``Other terms'' in Eq. (\ref{final}) are the terms that has no support of the basis state $\ket{0}$. Measuring the first register is the final step and if the measurement result is outcome `0'', then the second register will collapse to $\displaystyle{\frac{A\ket{\psi_{inp}}}{\norm{A\ket{\psi_{inp}}}}}$ as wished. The success probability: probability of getting the outcome ``0'' is $\displaystyle{p=\frac{\norm{A\ket{\psi_{inp}}}^2}{\norm{a_i}_1^2}}$ which may be small, but we can use \emph{oblivious amplitude amplification} to amplify the part that contains $\ket{0}$.
\paragraph{Oblivious Amplitude Amplification:} The state in Eq. (\ref{final}) can be thought of prepared from $\log(m)+n$-qubit unitary $M$ applied to $\ket{0}^{\otimes\log(m)}$ and can be written as:
\begin{equation}
W\ket{0}^{\otimes\log(m)}\otimes \ket{\psi_{inp}} = \sqrt{p}\ket{\text{``Good''}} + \sqrt{1-p}\ket{\text{``Bad''}}
\end{equation}
The circuit for amplitude amplification is given by: 
\begin{figure}[H]
\centering 
\begin{quantikz}
&\lstick{$\ket{0}^{\log(m)}\ket{\psi_{inp}}$} & \gate{W} & \gate{U_f}\gategroup[1,steps=4,style={dashed,
rounded corners,fill=blue!20, inner xsep=2pt},
background]{{ Repeat $r$-times}} & \gate{W^{\dagger}} & \gate{U_{f_0}} & \gate{W} & \qw
\end{quantikz}
\caption{Circuit for Oblivious Amplitude Amplification, that amplifies the Good state.}
\label{circuitoblivious}
\end{figure}
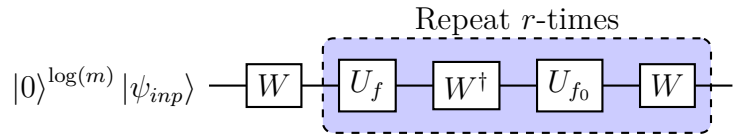
Since we are interested in amplifying the state with $\ket{0}$, thus $U_f=U_{f_0}=\left( I - 2\op{0}{0} \right) \otimes I$ and the highlighted part in the circuit given in Fig. (\ref{circuitoblivious}) is repeated $r\sim \mathcal{O}\left(1/\sqrt{p}\right)$ times.
\begin{example}{}{label}
Work out and draw the circuit for the implementation of $A = a_0U_0 + a_1U_1$. Also provide the matrix representation of $O$ and $O^{\dagger}$.\\
\paragraph{Solution:}\begin{enumerate}
\item The input state is:
\begin{align*}
\ket{\Psi_1} & = \ket{0}\ket{\psi_{inp}}
\end{align*}
\item Prepare $\ket{\phi}=O\ket{0}$.
\begin{align*}
\ket{\Psi_2} & = O\ket{0} \otimes \ket{\psi_{inp}} = \frac{1}{\sqrt{a_0+a_1}}\left( \sqrt{a_0}\ket{0}+\sqrt{a_1}\ket{1} \right)\otimes \ket{\psi_{inp}}
\end{align*}
\item Apply two register unitary: $V = \op{0}{0}\otimes U_0 + \op{1}{1}\otimes U_1$.
\begin{align*}
\ket{\Psi_3} & = \sqrt{\frac{a_0}{a_0+a_1}} \ket{0}U_0\ket{\psi_{inp}} + \sqrt{\frac{a_1}{a_0+a_1}} \ket{1}U_1\ket{\psi_{inp}}
\end{align*}
\item Apply $O^{\dagger}$.
\begin{align*}
\ket{\Psi_4} & = \frac{1}{a_0+a_1}\left( \ket{0}\otimes \left( a_0U_0 + a_1U_1 \right)\ket{\psi_{inp}} \right) + \text{other terms} \\
& = \frac{1}{a_0+a_1} \left(\ket{0}\otimes A\ket{\psi_{inp}} \right) + \text{other terms}
\end{align*}
\item Measure the first register. If the outcome is ``0'', then second register collapses to $A\ket{\psi_{inp}}$ as desired.
\end{enumerate}
\paragraph{Circuit:}The circuit is given by:
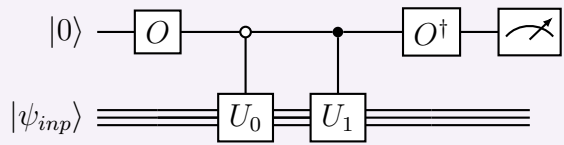
\begin{figure}[H]
\centering
\begin{quantikz}
&\lstick{$\ket{0}$} & \gate{O} & \octrl{1} & \ctrl{1} & \gate{O^{\dagger}} & \meter{} \\
&\lstick{$\ket{\psi_{inp}}$} & \qwbundle[alternate]{} & \gate{U_0}\qwbundle[alternate]{} & \gate{U_1}\qwbundle[alternate]{} &  \qwbundle[alternate]{} & \qwbundle[alternate]{}
\end{quantikz}
\caption{Circuit representation for the implementation of $A=a_0U_0+a_1U_1$}
\end{figure}
\paragraph{Matrix representation of $O$ and $O^{\dagger}$:}
Since the matrix elements of $O$ must be function of $a_0$ and $a_1$, and it must be unitary, thus:
\begin{align*}
O & = \frac{1}{\sqrt{a_0+a_1}} \begin{pmatrix}
\sqrt{a_0} & -\sqrt{a_1} \\
\sqrt{a_1} & \sqrt{a_0}
\end{pmatrix}
\end{align*}
Thus $O^{\dagger}$ will be:
\begin{align*}
O^{\dagger} & = \frac{1}{\sqrt{a_0+a_1}} \begin{pmatrix}
\sqrt{a_0} & \sqrt{a_1} \\
-\sqrt{a_1} & \sqrt{a_0}
\end{pmatrix}
\end{align*}
\end{example}
\begin{exercise}{}{label}
Verify that $O$ in Example 1.1 is unitary.
\end{exercise}
\begin{exercise}{}{label}
Work out and draw the circuit for the implementation of $\displaystyle{H = X+Z}$. Also provide the matrix representation of $O$ and $O^{\dagger}$.
\end{exercise}
\subsubsection{Hamiltonian simulation via LCU}
Since our goal is to efficiently implement the unitary $\displaystyle{U=e^{-iHt}}$ where $H$ is a Hamiltonian under an assumption that it is a linear combination of unitaries: $H=\sum_j a_jH_j$. Using Taylor series, we can write unitary as: 
\begin{align}
e^{-iHt} & = \sum_{k=0}^{\infty}\frac{(-iHt)^k}{k!} = \sum_{k=0}^{\infty} \frac{(-it)^k}{k!} \left(\sum_j a_jH_j \right)^k = \sum_{k=0}^{\infty} \frac{(-it)^k}{k!} \left(\sum_{j_1j_2\cdots j_{k}}a_{j_1}a_{j_2}\cdots a_{j_k} H_1H_2\cdots H_k \right)
\end{align}
We can define $\displaystyle{V_k = (-i)^k\sum_{j_1 j_2\cdots j_k}H_{j_1}H_{j_2}\cdots H_{j_k}}$ and $\displaystyle{a_k=a_{j_1}a_{j_2}\cdots a_{j_k}}$, thus $e^{-iHt}$ becomes:
\begin{align}
e^{-iHt} & = \sum_{k=0}^{\infty} \frac{t^k}{k!}a_k V_k 
\end{align}
It was assumed that $H_j$ were unitaries; this means that $V_k$ is also a unitary since the product of the unitary operators is always a unitary. Thus, $e^{-iHt}$ is now a linear combination of unitaries. It can be implemented using the same method as discussed earlier.

\section{Examples of Hamiltonian Simulation}
Any hermitian operator $H$ can be expressed as a linear combination of pauli matrices, 
\begin{align*}
H & = \sum_j a_jP_j
\end{align*}
where,
\begin{align*}
a_j & = \frac{1}{d} Tr[HP_j] \\
P_j & = \bigotimes_{i=1}^n \sigma_i^{(k)}
\end{align*}
where $d$ is the dimension of the operator $H$ and $\sigma^{(k)}\in \{I,X,Y,Z\}$. 
\subsection{Simulation of $\boldmath{H=\hat{Z}}$}
According to the Eq. (\ref{eq2}), the unitary time evolution operator corresponding to the Hamiltonian $H=\hat{Z}$ is given by:
\begin{align*}
U & = e^{-i\hat{Z} t} \\
& = R_z(2t)
\end{align*}
The circuit that will simulate a state $\ket{\psi}$ for time, $t$ under the Hamiltonian is given by: 
\begin{figure}[H]
\centering
\begin{quantikz}
\lstick{$\ket{\psi(0)}$} & \gate{R_z(2t)}& \qw \rstick{$\ket{\psi(t)}$}
\end{quantikz}
\caption{Simulation of $H=\hat{Z}$}
\end{figure}
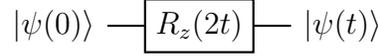
The circuit given below simulate exactly the same Hamiltonian, but by using the ancilla bit.
\begin{figure}[H]
\centering
\begin{quantikz}
&\lstick{$\ket{q}$}& \ctrl{1} & \qw & \ctrl{1} & \qw \\
&\lstick{$\ket{0}$}& \targ{} & \gate{R_z(2t)} & \targ{} & \qw 
\end{quantikz}
\end{figure}
\subsection{Simulation of $\boldmath{H=\hat{Z}\otimes \hat{Z}}$}
According to the Eq. (\ref{eq2}), the unitary time evolution operator corresponding to the Hamiltonian $H=\hat{Z}\otimes\hat{Z}$ is given by:
\begin{align*}
U & = e^{-i\hat{Z}\otimes\hat{Z}t} \neq e^{-i\hat{Z} t}\otimes e^{-i\hat{Z} t} \\
\Rightarrow U & = \hat{I} - i(\hat{Z}\otimes\hat{Z})t + \frac{[-i(\hat{Z}\otimes\hat{Z})^2t]^2}{2!} + \frac{[-i(\hat{Z}\otimes\hat{Z})^3t]^3}{3!} + \cdots
\end{align*}
Now, 
\begin{align*}
(\hat{Z}\otimes\hat{Z})^2 & = \hat{Z}\hat{Z}\otimes \hat{Z}\hat{Z} \\
& = \hat{I} \otimes \hat{I}
\end{align*}
Therefore, 
\begin{align*}
U & = \hat{I} - i(\hat{Z}\otimes\hat{Z})t - \frac{t^2}{2!}\hat{I} + i \frac{\hat{Z}\otimes\hat{Z}}{3!} - \cdots  \\
& = \hat{I}\cos(t) - i\sin(t) \hat{Z}\otimes\hat{Z} \\
\Rightarrow U\ket{q_0q_1} & = \cos(t)\ket{q_0q_1}-i(-1)^{q_0+q_1}\sin(t)\ket{q_0q_1} \\
& =\left[ \cos(t) + i(-1)^{q_0+q_1+1}\sin(t) \right]\ket{q_0q_1} \\
& = \text{exp}\left[ i(-1)^{q_0+q_1+1}t\right] \ket{q_0q_1}
\end{align*}
Truth table for such unitary operator is given below: 
\begin{table}[H]
\centering
\begin{tabular}{|c|c|}
\hline
Input & Output \\
\hline
$\ket{00}$ & $e^{-it}\ket{00}$ \\
\hline
$\ket{01}$ & $e^{it}\ket{01}$ \\
\hline
$\ket{10}$ & $e^{it}\ket{10}$ \\
\hline
$\ket{11}$ & $e^{-it}\ket{11}$ \\
\hline
\end{tabular}
\caption{Truth table for $U=e^{-it \hat{Z}\otimes\hat{Z}}$}
\end{table}
The circuit implementation of this unitary operator is illustrated in the figure below:
\begin{figure}[H]
\centering
\begin{quantikz}
&\qw & \ctrl{1} & \qw & \ctrl{1}  & \midstick[2,brackets=none]{=} \qw & \ctrl{1} & \qw & \ctrl{1} & \qw \\
&\qw & \targ{} & \gate{\begin{pmatrix}
e^{-it} & 0 \\ 0 & e^{it}
\end{pmatrix}} & \targ{} & \qw  & \targ{} & \gate{R_z(2t)} & \targ{} & \qw 
\end{quantikz}
\caption{Circuit implementation of $e^{-it\hat{Z}\otimes\hat{Z}}$}
\end{figure}
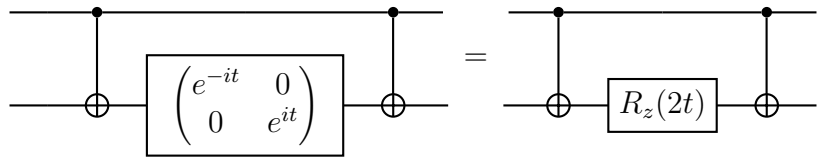
Same Hamiltonian can be simulated by the circuit given below: 
\begin{figure}[H]
\centering
\begin{quantikz}
&\lstick{$\ket{q_0}$} & \ctrl{2} & \qw & \qw & \qw & \ctrl{2} & \qw \\
&\lstick{$\ket{q_1}$} & \qw & \ctrl{1} & \qw & \ctrl{1} & \qw & \qw \\
& \lstick{$\ket{0}$} & \targ{} & \targ{} & \gate{R_z(2t)} & \targ{} & \targ{} & \qw 
\end{quantikz}
\end{figure}
\subsection{Simulation of $H=\hat{I}\otimes\hat{X}\otimes\hat{Z}$}
According to the Eq. (\ref{eq2}), the unitary time evolution operator corresponding to the Hamiltonian $H=\hat{I}\otimes\hat{X}\otimes\hat{Z}$ is given by:
\begin{align*}
U & = e^{-i\hat{I}\otimes\hat{X}\otimes\hat{Z}t}
\end{align*}
Since, $\hat{X}=\hat{H}\hat{Z}\hat{H}$, thus we can write $U$ as:
\begin{align*}
U & = \hat{I}\otimes\hat{H}\otimes\hat{I} e^{-i\hat{I}\otimes\hat{Z}\otimes\hat{Z}t}\hat{I}\otimes\hat{H}\otimes\hat{I}
\end{align*}
We already know the method of implementing $e^{-i\hat{Z}\otimes\hat{Z}t}$. Thus, the circuit implementation of this unitary operator is given by:
\begin{figure}[H]
\centering
\begin{quantikz}
& \qw & \qw & \qw & \qw & \qw & \qw  \\
& \gate{H} & \ctrl{1} & \qw & \ctrl{1} & \gate{H} & \qw \\
& \qw & \targ{} & \gate{R_z(2t)} & \targ{} & \qw & \qw 
\end{quantikz}=\begin{quantikz}[column sep=0.3cm]
&\lstick{$\ket{q_0}$} & \qw & \qw & \qw & \qw & \qw & \qw & \qw & \qw & \\
&\lstick{$\ket{q_1}$} & \gate{H} & \ctrl{2} & \qw & \qw & \qw & \ctrl{2} & \gate{H}  & \qw \\
&\lstick{$\ket{q_1}$} & \qw & \qw & \ctrl{1} & \qw & \ctrl{1} & \qw & \qw & \qw  \\
&\lstick{$\ket{0}$} & \qw & \targ{} & \targ{} & \gate{R_z(2t)} & \targ{} & \targ{} & \qw & \qw 
\end{quantikz}
\caption{Circuit implementation for simulation of $H=\hat{I}\otimes\hat{X}\otimes\hat{Z}$}
\end{figure}
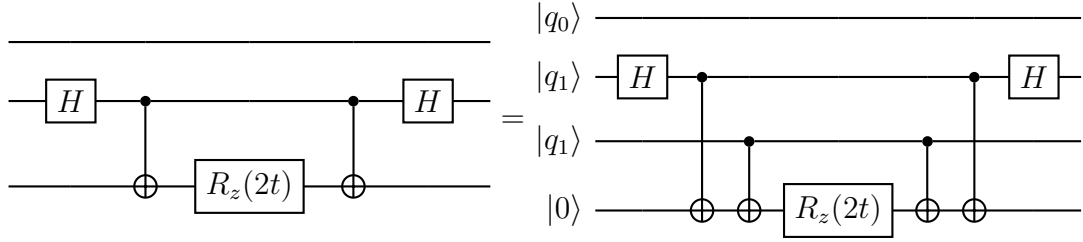
\begin{example}{}{label}
Given the set of universal gates $\{H, R_z, CX\}$, draw the circuit that simulates the Hamiltonian given by $
H  = \hat{X}\otimes\hat{Y}\otimes\hat{Z}
$.
\paragraph*{Solution:}
We know that, $\hat{X}=\hat{H}\hat{Z}\hat{H}$ and $\hat{Y}=\hat{R}_z\left(\frac{\pi}{2}\right)X\hat{R}_z\left(-\frac{\pi}{2}\right)$. Thus the circuit that simulate this hamiltonian is given by:
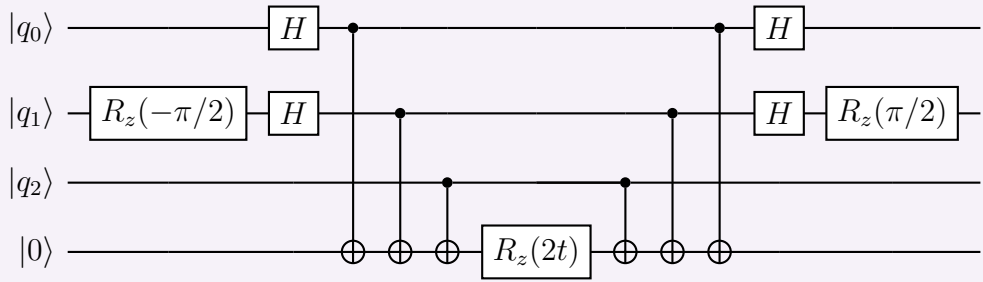
\begin{figure}[H]
\centering
\begin{quantikz}[column sep=0.3cm]
&\lstick{$\ket{q_0}$} & \qw & \gate{H} & \ctrl{3} & \qw & \qw & \qw & \qw & \qw & \ctrl{3} & \gate{H} & \qw & \qw  \\
&\lstick{$\ket{q_1}$} & \gate{R_z(-\pi/2)} & \gate{H} & \qw & \ctrl{2} & \qw & \qw & \qw & \ctrl{2} & \qw & \gate{H} & \gate{R_z(\pi/2)} & \qw  \\
&\lstick{$\ket{q_2}$} & \qw & \qw & \qw & \qw & \ctrl{1} & \qw & \ctrl{1} \qw & \qw & \qw & \qw & \qw & \qw  \\
&\lstick{$\ket{0}$} & \qw & \qw & \targ{} & \targ{} & \targ{} & \gate{R_z(2t)} & \targ{} & \targ{} & \targ{} & \qw & \qw & \qw 
\end{quantikz}
\caption{Circuit for simulating the Hamiltonian $H=\hat{X}\otimes\hat{Y}\otimes\hat{Z}$}
\end{figure}
\end{example}
\subsection{Simulation of the linear combination of Pauli Operators}
Suppose we want to simulate the Hamiltonian, $H = \hat{X}\otimes\hat{I}+\hat{Z}\otimes\hat{Z}$, then the unitary time evolution operator corresponding this Hamiltonian is given by: 
\begin{align*}
U & = e^{-i(\hat{X}\otimes\hat{I})t-i(\hat{Z}\otimes\hat{Z})t} \neq e^{-i(\hat{X}\otimes\hat{I})t}e^{-i(\hat{Z}\otimes\hat{Z})t}
\end{align*}
Recall \emph{Trotter Formula} that, if $A$ and $B$ are Hermitian oprators, then for any real $t$, 
\begin{align*}
\lim_{r\to\infty}\left( e^{iAt/r}e^{iB/r} \right)^r & = e^{i(A+B)}
\end{align*}
It means that, we have to simulate the terms of Hamiltonian $r-$times for very short interval of time. Thus, 
\begin{align*}
e^{-i(\hat{X}\otimes\hat{I})t-i(\hat{Z}\otimes\hat{Z})t} & = \left( e^{-i(\hat{X}\otimes\hat{I})\Delta t}e^{-i(\hat{Z}\otimes\hat{Z})\Delta t} \right)^r
\end{align*}
where $\Delta t = t/r$. This approximation will have error of $\Delta t^2$.
\chapter{Variational Quantum Algorithms}
\footnote{Copyrights: Muhammad Faryad, LUMS}
In many problems, it is important to solve for the minimum eigenvalue of a matrix. For example, in quantum chemistry, the minimum eigenvalue of Hamiltonian of a system corresponds to the ground state of that system. Variational quantum eigensolver provides an estimate $E(\theta)$ bounding $E_0$: actual ground state of Hamiltonian. 
\begin{align}
\langle \psi(\theta) | H | \psi(\theta) \rangle & = E(\theta) \geq E_0
\end{align}
where $\ket{\psi(\theta)}$ is an eigenstate associated with $E(\theta)$ which we can get by operating a parameterized unitary operator $U(\theta)$ to an arbitrary state $\ket{\psi}$, $U(\theta)\ket{\psi}=\ket{\psi(\theta)}$. Hence, $E(\theta)$ is calculated which is send to classical optimizer to get an updated parameter $\theta_{i+1}$. This process is repeated again and again until $E(\theta)$ is minimized. $[E(\theta)]_{min}$ is an approximate of the lowest energy of a Hamiltonian $H$. This variational method is used in many optimization algorithms, for instance MaxCut, MaxSat etc. 

Quantum Approximate Optimization Algorithm (QAOA) is a variational algorithm that uses a unitary $U(\beta,\gamma)$ to prepare a quantum state $\ket{\beta,\gamma}$, where $\beta$ and $\gamma$ are the parameters. This unitary comes from $H_C$: Problem Hamiltonian and $H_M$: Mixing Hamiltonian. Goal is to find parameters $\beta_{max}$ and $\gamma_{max}$ such that $\ket{\beta_{max},\gamma_{max}}$ encodes the solution. In other word, $\langle \beta_{new},\gamma_{new} | H_C | \beta_{new},\gamma_{new} \rangle$ is the best approximate of the solution.

\section{QUBO: Quantum Unconstraint Binary Optimization}
QUBO is optimization problem with a quadratic objective function. The problem is to optimize a binary cost function $C(x)$ given by:
\begin{align}\label{eq1}
C(x) & = x^TQx + c^Tx
\end{align}
where $x\in \{0,1\}$, $Q\in \mathbb{R}^{n\times n}$ and $c\in \mathbb{R}^n$. Considering this problem as variational problem, $C(x)$ is an eigenvalue corresponding to some cost Hamiltonian $H_c$ which we want to optimize.
\subsection{MaxCut Problem}
\begin{figure}[H]
\centering
\includegraphics[scale=1.3]{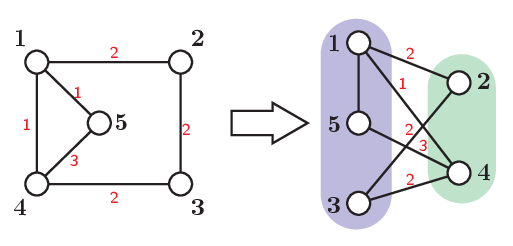}
\caption{Illustration of weighted MaxCut problem, MaxCut=5, Weight=10}
\label{maxcut}
\end{figure}

Let's consider a concrete example of QUBO, a \emph{weighted MaxCut problem}. As an input, the MaxCut problem takes a graph $G=G(N,E,W)$ characterized by $N$ nodes and $E$ undirected edges having weight $W$. The goal is to partition nodes into two groups such that the sum of weights of edges connecting these two groups is maximum. Fig. (\ref{maxcut}) provides an illustrative example with 5 nodes and 6 edges that has weights, as shown in the figure.
Let's construct a weight matrix $W$ for the Graph given in Fig. (\ref{maxcut}). Since there are 5 nodes, thus the size of the matrix would be $5\times 5$. The matrix element $W_{ij}$ is:
\begin{align}
W_{ij} & = \begin{cases}
w(ij) &\ \text{if $i^{th}$ and $j^{th}$ nodes are connected with weigth $w(ij)$} \\
0 &\ \text{otherwise}
\end{cases}
\end{align}
Thus, 
\begin{align}
W & = \begin{pmatrix}
0 & 2 & 0 & 1 & 1 \\
2 & 0 & 2 & 0 & 0 \\
0 & 2 & 0 & 2 & 0 \\
1 & 0 & 2 & 0 & 3 \\
1 & 0 & 0 & 3 & 0 
\end{pmatrix}
\end{align}
Suppose ${\textbf{x}}$ and $\mathbf{1}-{\textbf{x}}$ are the two groups, then the sum of weights of the edges connecting these two groups is:
\begin{align}\label{Cost}
C(x) & = \sum_{i,j} W_{ij} x_i(1-x_j)
\end{align}
For example, in Fig. (\ref{maxcut}), one group contains the nodes $N_1$, $N_3$ and $N_5$, whereas, the second group contains $N_2$ and $N_4$. Thus
\begin{align*}
{\textbf{x}} & = \begin{bmatrix}
1 & 0 & 1 & 0 & 1
\end{bmatrix} \\
\Rightarrow \mathbf{1}-{\textbf{x}} & = \begin{bmatrix}
0 & 1 & 0 & 1 & 0
\end{bmatrix}
\end{align*}
\subsubsection*{Converting to quantum mechanical problem}
We want to maximize $C(x)$, given in Eq. (\ref{Cost}), for which we have to convert it into the form given in Eq. (\ref{eq1}). We can write Eq. (\ref{Cost}) as: 
\begin{align}
C(x) & = -\sum_{i,j}x_j W_{ij} x_j + \sum_i W_{ij} x_i \\
& \rightarrow x^T(-W)x + \left(\sum_j W_{ij}\right)x \\
\Rightarrow Q & = -W\ \ ;\ \ c^T = \sum_j W_{ij}
\end{align}
Suppose $C(x)$ corresponds to the eigenvalue of a cost Hamiltonian $H_c$ with an eigenstate $\ket{x}$, i.e., 
\begin{align}\label{eq3}
H_c\ket{x} & = C(x) \ket{x}
\end{align}
Consider an operator $Z_i=I\otimes \cdots \otimes Z_i\otimes\cdots\otimes I$. Then
\begin{align*}
Z_i\ket{x} & = (-1)^{x_i}\ket{x} \\
& = (1-2x_i)\ket{x} \\
& = I\ket{x} - 2x_iI\ket{x}
\end{align*}
Therefore, 
\begin{align}\label{newform}
\frac{1}{2}\left( I - Z_i\right) \ket{x} & = x_i\ket{x}
\end{align}
We can re-write the cost function as: 
\begin{align*}
C(x) & = \sum_i\sum_j x_i Q_{ij} x_j + \sum_i c_ix_i
\end{align*}
The Hamiltonian $H_c$ can therefore be written as:
\begin{align*}
H_c & = \sum_{i,j} Q_{ij} \left(\frac{I-Z_i}{2} \right)\left(\frac{I-Z_j}{2} \right) + \sum_i \left(\frac{I-Z_i}{2} \right)c_i
\end{align*}
Rearranging the terms will lead us to write:
\begin{align}
H_c & = \sum_{i,j}\frac{1}{4}Q_{ij}Z_iZ_j - \sum_i \frac{1}{2}\left( c_i + \sum_j Q_{ij} \right)Z_i + \left( \sum_{ij} \frac{Q_{ij}}{4} + \sum_i \frac{1}{2}c_i \right)
\end{align}
Unitary operator corresponding to $H_c$ is given by:
\begin{align}
U_C(\gamma) & = \exp\left(-iH_c\gamma\right) \\
& \approx \prod_{i,j}^n R_{Z_iZ_j}\left(Q_{ij}/4\right) \prod_i^n R_{Z_i}\left(\frac{1}{2}\left( c_i + \sum_j Q_{ij} \right) \right)
\end{align}
Consider a mixer Hamiltonian $H_M$:
\begin{align}
H_M & = \sum_i X_i
\end{align}
Unitary operator corresponding to $H_M$ is given by:
\begin{align}
U_M(\beta) & = \exp\left(-iH_M\beta\right) \\
& = \prod_i^n R_x\left(2\beta\right)
\end{align}
\subsubsection*{Algorithm}
\paragraph{Input:}To implement the algorithm, we have to convert cost function into Hamiltonian and then select number of rounds $p\geq 1$ and two angles per round, $\beta_i\in [0,\pi]$ and $\gamma_i \in [0,2\pi]$ for $i$-th round.
\paragraph{Procedure:}\begin{enumerate}
\item Construct uniform superposition state of $n$-qubit by  applying Hadamard gate.
\item For $0\leq i \leq p$, Apply
\begin{itemize}
\item $U_C(\gamma_i)$
\item $U_M(\beta_i)$ 
\end{itemize}
\end{enumerate}
\paragraph{Output:}Let's call the state at the output $\ket{\beta,\gamma}$. The expectation value $\langle \beta,\gamma | H_C | \beta,\gamma \rangle$ is the approximate solution to the problem.\\

Choosing appropriate $p,\beta,\gamma$ will give the best approximate of the solution. The circuit represenation is given in the figure below:
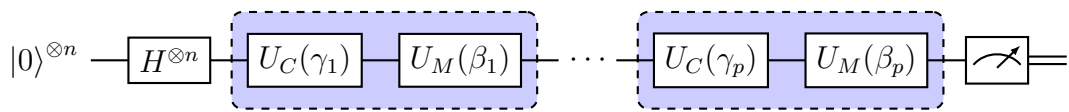
\begin{figure}[H]
\centering
\begin{quantikz}
&\lstick{$\ket{0}^{\otimes n}$} & \gate{H^{\otimes n}} & \gate{U_C(\gamma_1)}\gategroup[1,steps=2,style={dashed,
rounded corners,fill=blue!20, inner xsep=2pt},
background]{} & \gate{U_M(\beta_1)} & \qw\ \cdots\ & \gate{U_C(\gamma_p)}\gategroup[1,steps=2,style={dashed,
rounded corners,fill=blue!20, inner xsep=2pt},
background]{} & \gate{U_M(\beta_p)} & \meter{} & \cw 
\end{quantikz}
\caption{Illustration of Circuit of QAOA}
\end{figure}
\begin{exercise}{}{label}
Based on previous chapter, draw a circuit that implements the following operators:
\begin{enumerate}
\item $\displaystyle{\prod_{i=1}^3R_{Z_i}(\alpha)}$ \\
\item $\displaystyle{\prod_{i=1}^3R_{x}(k)}$ \\
\item $\displaystyle{R_{Z_1Z_2}(\beta)}$
\end{enumerate}
\end{exercise}

\section{Adiabatic Quantum Computing}
Adiabatic quantum computation is an alternative approach to quantum computation based on \emph{adiabatic theorem}: adiabatic(slow) evolution of a quantum system. According to the adiabatic theorem, a quantum system initially in the ground state of a time-dependent Hamiltonian will remain in the ground state if and only if Hamiltonian changes adiabatically(slowly).
\section{Description}
First, a Hamiltonian is found whose ground state encodes the problem's solution. Next, a system is prepared corresponding to a simple Hamiltonian and initialized to the ground state. This Hamiltonian has evolved adiabatically to the Hamiltonian of interest, such that the state remains in the ground state. The time of evolution depends on the minimum energy gap between the ground state and the first excited state of Hamiltonian $H(t)$, $t\sim \mathcal{O}\left(1/\Delta_{min}^2\right)$. 
\section{Adiabatic quantum computation in computational problems}
Adiabatic quantum computation can solve most of the problems such as optimization, search problem, etc. The Adiabatic process starts with preparing a ground state corresponding to an easy known Hamiltonian. The most typical example of such Hamiltonian is the combination of Pauli's operator:
\begin{align}
H_{ini} & = -\sum_i X_i
\end{align}
where $\ket{+}$ is the ground state of $H_{i}$ which we can easily prepare from from Hadamard gate applied to $\ket{0}$. Next step is to select a \emph{problem Hamiltonian} $H_{f}$, where our goal is to find a ground state of $H_f$. Hamiltonian that connects both $H_{i}$ and $H_f$ is given by:
\begin{align}\label{eq1}
H(t) & = \left[ 1- f(t) \right] H_f + f(t) H_i
\end{align}
where
\begin{align}
f(t) & = \begin{cases}
1 & t=0 \\
0 & t=T
\end{cases}
\end{align}
Thus $f(t) = 1 - t/T$ and Eq. (\ref{eq1}) can be therefore written as:
\begin{align}\label{hamil}
H(t) & = \frac{t}{T}H_f + \left(1-\frac{t}{T}\right)H_i
\end{align}
Unitary time evolution operator that we want to simulate is given by:
\begin{align}
U(t) & = \exp \left[ -i\int H(t) dt \right] \\
& = \exp \left[-i \left(H(t_1)\Delta t + H(t_2)\Delta t + \cdots \right) \right]
\end{align}
Using Trotter decomposition, we can approximate $U(t)$ as:
\begin{align}
U(t) & \approx  e^{-iH(t_1)\Delta t} e^{-iH(t_2)\Delta t} \cdots e^{-iH(t_p)\Delta t} + \mathcal{O}(\Delta t^2)
\end{align}
Expand $\displaystyle{e^{-iH(t_j)\Delta t}}$ as:
\begin{align}
e^{-iH(t_j)\Delta t} & = \exp \left[\frac{t_j}{T}H_f \Delta t + \left(1-\frac{t_j}{T}\right)H_i\Delta t \right] \\
\end{align}
Take $\beta_j = (t_j/T)\Delta t$ and $\gamma_j=(1-t_j/T)\Delta t$, where the index $j$ runs from 1 to $p$ such that $t_p=T$. Hence
\begin{align}\label{oper}
e^{-iH(t_i)\Delta t} & = e^{-i\beta_jH_f} e^{-i\gamma_jH_i}
\end{align}
The unitary operators in Eq. (\ref{oper}) can be simulated using the techniques described previously. The circuit representation for operating $U(t)$ is illustrated below:
\begin{figure}[H]
\centering
\begin{quantikz}
&\lstick{$\ket{0}$} & \qwbundle{n} & \gate{H} & \gate{e^{-i\beta_1H_f}}\gategroup[1,steps=2,style={dashed,
rounded corners,fill=blue!20, inner xsep=2pt},
background]{\sc {$t=0$}} & \gate{e^{-i\gamma_2H_i}} & \qw\ \cdots\ & \gate{e^{-i\beta_pH_f}}\gategroup[1,steps=2,style={dashed,
rounded corners,fill=blue!20, inner xsep=2pt},
background]{\sc {$t=T$}} & \gate{e^{-i\gamma_pH_i}} & \qw
\end{quantikz}
\caption{Circuit representation for simulating the Hamiltonian in Eq. (\ref{hamil}) }
\label{figure}
\end{figure}
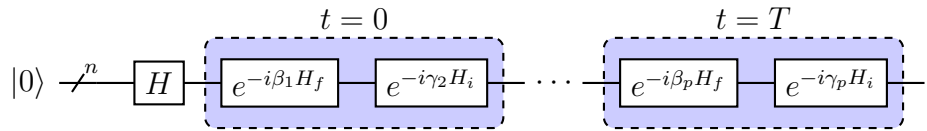
\subsection{Variational approach to AQC}
Instead of having $p$ layers of gates as in Fig. (\ref{figure}), we can solve this problem using variational approach. First, we select initial parameters 
\begin{align*}
\vec{\gamma} & = \{\gamma_1, \gamma_2, \cdots, \gamma_p \} \\
\vec{\beta} & = \{\beta_1, \beta_2, \cdots, \beta_p \}
\end{align*}
We run the algorithm and compute the cost which is function of these parameters: $C(\vec{\gamma}, \vec{\beta})$. Based on the result, we select $\{\vec{\gamma}, \vec{\beta}\}$ again and again, such that $C(\vec{\gamma}, \vec{\beta})$ tends to minimize.

\section{HHL Algorithm}
Linear system of equations appear in wide range of problems from different disciplines of science. For example, we come across such systems while solving differential equations. The problem of solving a system of linear
equations is the following: Given a system $A\text{\textbf{x}}=\text{\textbf{b}}$, find $\text{\textbf{x}}$ for given a matrix $A$ and $\text{\textbf{b}}$. In classical domain, there is no such limit on $A$\footnote{If $A$ is singular, the solution is unique.}. Where as, in the domain of quantum computing, we assume that $A$ is Hermitian\footnote{$A$ is Hermitian if $A^{\dagger}=A$} and the vectors $\text{\textbf{x}}$ and $\text{\textbf{b}}$ are represented by the quantum states $\ket{x}$ and $\ket{b}$ respectively, such that $\norm{\ket{x}}=\norm{\ket{b}}=1$. Thus the classical problem can be pose as finding $\ket{x}$ such that
\begin{align*}
A\ket{x} & = \ket{b}
\end{align*}
with the solution $\ket{x}$ being
\begin{align*}
\ket{x} & = \frac{A^{-1}\ket{b}}{\norm{A^{-1}\ket{b}}}
\end{align*}

Harrow, Hassidim, and Lloyd (HHL) proposed the quantum algorithm for the linear system. The main idea of the algorithm is as follows. Suppose $\{\lambda_j\}$ and $\{\ket{u_j}\}$ are the eigenvalues and eigenstates of the matrix $A$ respectively such that $0<\lambda_j<1$. Then the state $\ket{b}$ can be represented as the linear combination of eigenstates of the matrix $A$, i.e., $\ket{b}=\sum_jb_j\ket{u_j}$. We know that the matrix $A$ can be written as: 
\begin{align*}
A & = P \Lambda P^{\dagger}
\end{align*}
where $\Lambda$ is the diagonal matrix containing the eigenvalues $\lambda_j$ of $A$ and $P=[\text{\textbf{u}}_1,\text{\textbf{u}}_2,\cdots,\text{\textbf{u}}_n]$, i.e., every column of $P$ is an eigenstate of $A$. Hence, 
\begin{align*}
A^{-1} & = P \Lambda^{-1} P^{\dagger}
\end{align*}
The spectral decomposition of $A^{-1}$ allows us to write
\begin{align*}
A^{-1} & = \sum_j \frac{1}{\lambda_j}\op{u_j}{u_j}
\end{align*}
thus the solution $\ket{x}$ can be written as: 
\begin{equation}\label{x}
\begin{split}
\ket{x} & = A^{-1}\ket{b} \\
& = \left( \sum_j \frac{1}{\lambda_j}\op{u_j}{u_j} \right) \sum_i b_i \ket{u_i} \\
& = \sum_j \frac{1}{\lambda_j} b_j \ket{u_j}\ \ \ \ \ \ \ \because \ip{u_j}{u_i}=\delta_{ij}
\end{split}
\end{equation}
The goal of HHL Algorithm is to obtain $\ket{x}$ in the form given in Eq. (\ref{x}). The HHL Algorithm in a nutshell involves performing a set of operations that essentially performs the three steps:
$$P\Lambda P^{\dagger}\ket{x}=\ket{b} \xrightarrow{\text{Step-}01} \Lambda P^{\dagger}\ket{x}=P^{\dagger}\ket{b}\xrightarrow{\text{Step-}02} P^{\dagger}\ket{x}=\Lambda^{-1}P^{\dagger}\ket{b}\xrightarrow{\text{Step-}03}\ket{x}=P\Lambda^{-1}P^{\dagger}\ket{b}$$
This method requires us to find the eigenvalues of $A$. This can be done by exponentiating the matrix $A$ as $U=e^{iA}$ and using a quantum subroutine, \emph{quantum phase estimation}\footnote{Remember that eigenstates of $U$ and $A$ are same}.
\subsection{Algorithm}
The HHL Algorithm requires four sets of qubits: a single ancilla qubit, a register of $n_1$ qubits to store the $\displaystyle{\lambda_j^{-1}}$  in binary format, a register of $n_2$ qubits to store $\lambda_j$ in binary format and a memory of $\mathcal{O}(\log 2^n)$ that initially stores $\ket{b}$ and eventually store $\ket{x}$. The circuit diagram of the HHL Algorithm is illustrated in Fig. (\ref{HHL_Figure}). 
\begin{figure}[H]\label{HHL_Figure}
\centering
\begin{quantikz}
\lstick{$\ket{0}_a$} & \qw \gategroup[3,steps=3,style={dashed,
rounded corners,fill=blue!20, inner xsep=2pt}, background]{{\sc QPE}} & \qw & \qw & \qw & \qw & \gate{R}\gategroup[wires=3,steps=1,style={dashed,
rounded corners,fill=green!20, inner xsep=2pt}, background]{{$R(\lambda^{-1})$ Rotations}} & \qw & \qw &  \qw \gategroup[3,steps=3,style={dashed, rounded corners,fill=blue!20, inner xsep=2pt}, background]{{QPE$^{\dagger}$}} & \qw & \qw & \meter{}   \\
\lstick{$\ket{0}_r$} & \gate{H^{\otimes n}} & \ctrl{1} & \gate{FT^{\dagger}} & \qw & \qw & \ctrl{-1} & \qw & \qw &  \gate{FT} & \ctrl{1} & \gate{H^{\otimes n}}& \qw \rstick{$\ket{0}_r$} \\
\lstick{$\ket{b}_m$} & \qw & \gate{U} & \qw & \qw & \qw & \qw & \qw & \qw & \qw & \gate{U} & \qw  & \qw \rstick{$\ket{x}$}
\end{quantikz}
\caption{Schematic of circuit for HHL Algorithm. The first step involves phase estimation which maps eigenvalues $\lambda_j$ of $A$ into the register in binary form. The second step involves the controlled rotations of the ancilla qubit so that $\lambda_j^{-1}$ shows up in the state. The third step is the inverse of phase estimation to disentangle the system and restore the registers to $\ket{0}$ states. The memory register now stores $\ket{x}$ which is the solution of the linear system. It can be then post processed to get the expectation value of a quantum observable.} 
\end{figure}
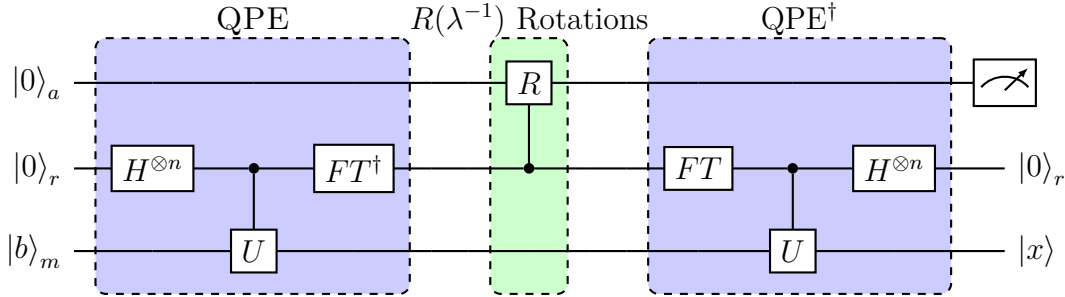
\begin{enumerate}
\item[\textbf{Step 1:}] Perform quantum phase estimation(QPE) to find the eigenvalues $\lambda_j$ of a unitary operator $U=e^{iA}$ in the binary form, 
\begin{align*}
\ket{0}_a\ket{0}_{r_1}\ket{0}_{r_2}\ket{b}_m & \longrightarrow \sum_j b_j \ket{0}_a\ket{0}_{r_1} \ket{\tilde{\lambda_j}}_{r_2}\ket{u_j}_m
\end{align*}
where the subscripts $a$, $r$ ,$m$, denote the sets of ancilla, register and memory qubits, respectively and $\tilde{\lambda_j}$ is an estimate for $\lambda_j$.
\item[\textbf{Step 2:}] Use controlled rotations on $r_1-$register, to get $\displaystyle{\ket{\lambda_j^{-1}}}$. Hence, the system will evolve to: 
\begin{align*}
\sum_j b_j \ket{0}_a \ket{\tilde{\lambda}_j^{-1}}_{r_1}\ket{\tilde{\lambda}_j}_{r_2}\ket{u_j}_{m}
\end{align*}
\item[\textbf{Step 3:}] Rotate the single ancilla qubit $\ket{0}_a$ around $y-$axis on bloch sphere by an angle, $\displaystyle{\theta=\cos^{-1}\left(\frac{C}{\lambda_j}\right)}$ for each $\lambda_j$. This is performed via controlled $R_y$ rotations on $\ket{0}_a$. Mathematically, 
\begin{align*}
R_y\left(\theta\right)\ket{0} & = \begin{pmatrix}
\cos(\theta) & -\sin(\theta) \\
\sin(\theta) & \cos(\theta)
\end{pmatrix}\begin{pmatrix}
1\\0
\end{pmatrix} \\
& = \cos(\theta)\ket{0} + \sin(\theta)\ket{1}
\end{align*}
For given $\theta$, $\displaystyle{\cos(\theta) = \frac{C}{\lambda_j}}$ and $\displaystyle{\sin(\theta)=\sqrt{1-\cos^2(\theta)} = \sqrt{1-\frac{C^2}{\lambda_j^2}}}$. Hence the system will transform to: 
\begin{align*}
\sum_j b_j \left( \frac{C}{\lambda_j}\ket{0}_a+ \sqrt{1-\frac{C^2} {\lambda_j^2}}\ket{1}_a \right)\ket{\tilde{\lambda}_j^{-1}}_{r_1} \ket{\tilde{\lambda_j}}_{r_2} \ket{u_j}_m
\end{align*}
\item[\textbf{Step 3:}] Perform the inverse of Step-2. This will give us the following system:
\begin{align*}
\sum_j b_j \ket{0}_a \ket{0}_{r_1}\ket{\tilde{\lambda}_j}_{r_2}\ket{u_j}_{m}
\end{align*}
\item[\textbf{Step 4:}] Perform inverse QPE. This will evolve the system to:
\begin{align*}
\sum_j b_j \left( \frac{C}{\lambda_j}\ket{0}_a+ \sqrt{1-\frac{C^2} {\lambda_j^2}}\ket{1}_a \right)\ket{0}_{r_1} \ket{0}_{r_2} \ket{u_j}_m
\end{align*}
\item[\textbf{Step 4:}] Measuring the single ancilla qubit will give us: 
\begin{align*}
\ket{x} & \approx \sum_j b_j \frac{C}{\lambda_j}\ket{u_j}_m
\end{align*}
if and only if the outcome is $\ket{0}$.
\end{enumerate}
\subsection{Implementation of Controlled Rotations}
In the second step, we have to do the controlled rotations so that we have $\lambda_j^{-1}$ in the state. After the QPE, the state of the register is $\ket{\lambda_0\lambda_1\cdots\lambda_{n-1}}$. Then the circuit that will inverse the eigenvalues is given by: 
\begin{figure}[H]
\centering
\begin{quantikz}
\lstick{$\ket{0}$} & \gate{R_y\left(\frac{2\pi}{2}\right)} & \gate{R_y\left(\frac{2\pi}{2^2}\right)} & \gate{R_y\left(\frac{2\pi}{2^3}\right)} &\qw & \cdots  \ \ & \gate{R_y\left(\frac{2\pi}{2^{n-1}}\right)} & \qw  \\
\lstick{$\ket{\lambda_0}$} & \ctrl{-1} & \qw  & \qw & \qw & \cdots  \ \ & \qw & \qw   \\
\lstick{$\ket{\lambda_1}$} & \qw  & \ctrl{-2} & \qw & \qw & \cdots  \ \ & \qw & \qw  \\
\lstick{$\ket{\lambda_2}$} & \qw & \qw & \ctrl{-3} & \qw & \cdots  \ \ & \qw & \qw    \\[1cm]
\lstick{$\ket{\lambda_{n}}$} & \qw & \qw & \qw & \qw & \qw  & \ctrl{-4} & \qw 
\end{quantikz}
\caption{Controlled rotations to get $\lambda_j^{-1}$}
\end{figure}
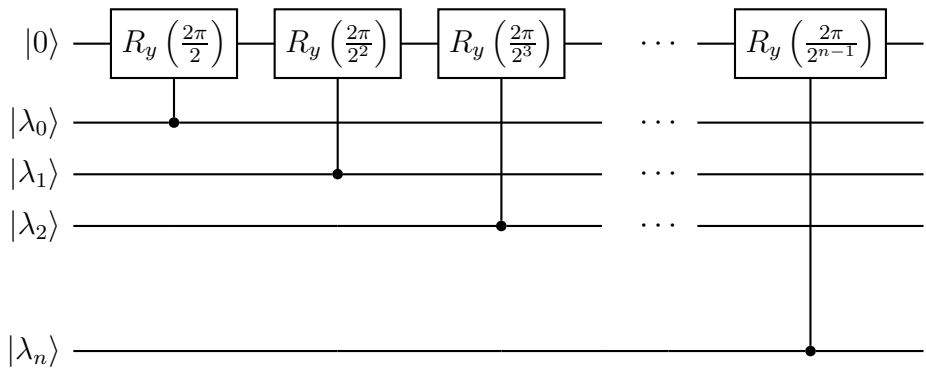
\noindent Mathematically, $\lambda$ can be written as: 
\begin{align*}
\lambda & = \lambda_02^0 + \lambda_12^1 + \lambda_22^2 + \cdots + \lambda_{n-1}2^{n-1} \\
& = 2^{n-1} \left(\frac{\lambda_0}{2^{n-1}} + \frac{\lambda_1}{2^{n-2}} + \frac{\lambda_2}{2^{n-3}} + \cdots + \lambda_{n-1} \right)
\end{align*}
The angle of controlled rotation as a whole can be written as: 
\begin{align*}
\theta & = \frac{2\pi}{2}\lambda_0 + \frac{2\pi}{2^2}\lambda_1 + \frac{2\pi}{2^3}\lambda_2 + \frac{2\pi}{2^n}\lambda_{n-1} \\
& = \pi \left( \lambda_0 + \frac{\lambda_1}{2} + \frac{\lambda_2}{2^2} + \cdots + \frac{\lambda_{n-1}}{2^n-1} \right)
\end{align*}
\noindent Here we see that smaller the $\lambda$, greater the $\theta$ and vice-versa. Thus, this circuit will map approximately $\lambda$ to $\lambda^{-1}$.
\section{HHL Algorithm vs Classical Algorithm}
A system of linear equations is called $d-$sparse if $A$ has $d$ non-zero entries per row. Solving $d-$sparse system of size $2^n$ classically would require $\mathcal{ O }(2^ns\kappa\log(1/\epsilon))$ running time using the conjugate gradient method. Here $\kappa$ denotes the condition number of the system, and $\epsilon$ represents the accuracy of the approximation. On the other hand, HHL Algorithm estimates the solution with a running time of order $\mathcal{ O }(\log(2^n)s^{2}\kappa^{2}/\epsilon)$ where $A$ is a Hermitian matrix. This is an exponential speed-up in the size of the system.
\section{Variational approach to solve Linear Equations}
Our goal is to solve $A\ket{x}=\ket{b}$ for $\ket{x}$. For this, we somehow create a state $\ket{\psi(\alpha)}$, where $\alpha$ is a variational parameter. We want to solve for $\alpha=\alpha_0$ such that $\ket{\psi(\alpha_0}\approx\ket{x}$. Method is to find the cost function $C(\alpha)$ and minimize it classically, such that $C(\alpha_0)$ is minimum.
\subsection*{Cost function}
Cost function for $A\ket{x}=\ket{b}$ is given by: 
\begin{align}\label{CF}
C(\alpha) & = \ip{\Phi}{\Phi} - \ip{\Phi}{b}\ip{b}{\Phi}
\end{align}
where $\ket{\Phi}=A\ket{\psi(\alpha)}$. Eq. (\ref{CF}) can be re-written in an operator form as:
\begin{align}
C(\alpha) & = \langle \Phi | I - \op{b}{b} | \Phi \rangle
\end{align}
Take $H_p=I - \op{b}{b}$, so $C(\alpha)=\langle \Phi | H_p | \Phi \rangle$. We minimize this Cost function \footnote{An alternate cost function for this problem can be: $C(\alpha) = \norm{A\ket{\psi(\alpha)}-\ket{b}}$} classically to find $\ket{\psi(\alpha_0)}\approx \ket{x}$.
\begin{example}{}{label}
At the end, the cost function has the form $C(\alpha)=\langle \Phi | WUV | \Phi \rangle$. Work out and draw the quantum circuit that can calculate $C(\alpha)$.\\
\paragraph*{Solution:} We compute $C(\alpha)$ by Hadamard test. Consider the circuit:
\begin{figure}[H]
\centering
\begin{quantikz}
&\lstick{$\ket{0}$} & \gate{H} & \ctrl{1} & \ctrl{1} & \ctrl{1} & \gate{H} & \meter{} \\
&\lstick{$\ket{\Phi}$} & \qw & \gate{V} & \gate{U} & \gate{W} & \qw & \qw
\end{quantikz}
\caption{Circuit to compute $C(\alpha)=\langle \Phi | WUV | \Phi \rangle$}
\label{circuitEXP}
\end{figure}
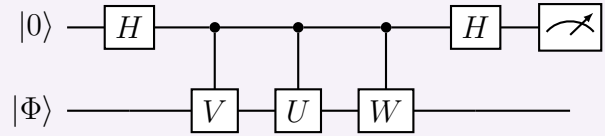
Let's work out the state at each step of the circuit in Fig. (\ref{circuitEXP}).
\begin{enumerate}
\item Input state is: 
\begin{align*}
\ket{0}\ket{\Phi}
\end{align*}
\item After the Hadamard gate, state becomes:
\begin{align*}
\frac{\ket{0}+\ket{1}}{\sqrt{2}} \ket{\Phi} \\
\frac{\ket{0}\ket{\Phi}+\ket{1}\ket{\Phi}}{\sqrt{2}}
\end{align*}
\item State after Controlled operations becomes:
\begin{align*}
\frac{\ket{0}\ket{\Phi}+\ket{1}WUV\ket{\Phi}}{\sqrt{2}}
\end{align*}
\item State after the last Hadamard is: 
\begin{align*}
\frac{1}{2}\left[ (\ket{0}+\ket{1})\ket{\Phi} + (\ket{0}-\ket{1})WUV\ket{\Phi} \right] \\
\frac{1}{2}\left[\ket{0}(I+WUV)\ket{\Phi} + \ket{1}(I-WUV)\ket{\Phi} \right]
\end{align*}
\item Measure the probabilities of getting the outcome ``0'' and ``1''.
\begin{align*}
p(0) & = \frac{\langle \Phi | I+WUV |\Phi \rangle }{4} \\
p(1) & = \frac{\langle \Phi | I-WUV |\Phi \rangle }{4} \\
\Rightarrow p(0)-p(1) & = \frac{\langle \Phi | WUV | \Phi \rangle}{2} \\
\Rightarrow C(\alpha) & = 2 \left[ p(0) - p(1) \right]
\end{align*}

\end{enumerate}
\end{example}

\chapter{Quantum Chemistry}

\section{Many-Particle System}
Most physical systems like molecules, which will be our primary focus, involve many particles. They are known as \emph{many-particle system}.
\subsection{Schr\"{o}dinger Equation}
The state\footnote{At the moment, spin is ignored} of a system of $N$ particles is described by a wavefunction:
\begin{equation}\label{State}
\Psi\left(\text{\textbf{r}}_1,\text{\textbf{r}}_2,\dotsc,\text{\textbf{r}}_N,t\right)
\end{equation}
Evolution of the state given in Eq. \ref{State} is according to the time-dependent Schr\"odinger equation:
\begin{equation}\label{TDSE}
i\hbar \frac{\partial}{\partial t}\Psi\left(\text{\textbf{r}}_1,\text{\textbf{r}}_2,\dotsc,\text{\textbf{r}}_N,t\right) = \hat{H} \Psi\left(\text{\textbf{r}}_1,\text{\textbf{r}}_2,\dotsc,\text{\textbf{r}}_N,t\right)
\end{equation}
where $\hat{H}$ is the $N$-particle Hamiltonian that is obtained by generalizing the one-particle Hamiltonian to $N$ particles.
\begin{equation}\label{hamiltonian}
\hat{H} = -\sum_{i=1}^{N}\frac{\hbar^2}{2m_i}\nabla_i^2+\hat{V}\left(\text{\textbf{r}}_1,\text{\textbf{r}}_2,\dotsc,\text{\textbf{r}}_N,t\right)
\end{equation}
Here, $m_i$ is the mass of $i$th particle, and $\hat{V}$ is the operator corresponding to the total potential energy that accounts for mutual interactions between the particles of the system. Here we also assume that the system is isolated.
\subsubsection{Multielectron atoms}
Suppose an atom with $Z$ electrons at positions $\text{\textbf{r}}_1,\text{\textbf{r}}_2,\dotsc,\text{\textbf{r}}_Z$ and a nucleus at position \textbf{R}. State of such system is represented by:
\begin{equation}\label{multielectron_system}
\Psi\left(\text{\textbf{r}}_1,\text{\textbf{r}}_2,\dotsc,\text{\textbf{r}}_Z,\text{\textbf{R}},t\right)
\end{equation}
Hamiltonian for such system is given by:
\begin{equation}\label{hamiltonian_of_multielectron_atom}
\hat{H} = -\frac{\hbar^2}{2m_e}\sum_{i=1}^Z\nabla_i^2 - \frac{\hbar^2}{2M}\nabla_R^2-\sum_{i=1}^Z\frac{Ze^2}{4\pi\epsilon_0 r_{iR}}+ \sum_{i>j}\frac{e^2}{4\pi\epsilon_0r_{ij}}
\end{equation}
First term is the kinetic energy of electrons. Second term corresponds to kinetic energy of nucleus that has mass $M$. Third term represents the attractive Coulomb interaction of each electron with nucleus and the last term represents the repulsive Coulomb interaction between the $i$th and $j$th electron; $r_{ij}$ is the distance separating them.
\subsubsection{Molecular Hamiltonian}\label{Mol_Ham}
Suppose that molecule contains $N$ electrons and $M$ nuclei. Molecular Hamiltonian can be obtained by generalizing the Hamiltonian of Multielectron atom to multi-atoms.
\begin{equation}
\hat{H} = -\sum_{i=1}^N\frac{\hbar^2}{2m_e}\nabla_i^2-\sum_{A=1}^M\frac{\hbar^2}{2M_A}\nabla_A^2-\sum_{i=1}^N\sum_{A=1}^M\frac{(Z_Ae)e}{4\pi\epsilon_0r_{iA}}+\sum_{j>i}\frac{e^2}{4\pi\epsilon_0 r_{ij}}+\sum_{B>A}\frac{(Z_Ae)(Z_Be)}{4\pi\epsilon_0 R_{AB}}
\end{equation}
where,
\begin{itemize}
\item $m_e$ is the mass of Electron.
\item $M_A$ is the mass of $A$th nucleus.
\item $Z_A$ is the no. of protons in $A$th nucleus.
\item $r_{ij}$ is the distance between $i$th and $j$th electron.
\item $r_{iA}$ is the distance between $i$th electron and $A$th nucleus.
\item $R_{AB}$ is the distance between $A$th and $B$th nucleus.
\end{itemize}
In the Hamiltonian, first and second terms correspond to the kinetic energy of electrons and nuclei, respectively. The third term represents attractive Coulomb interaction between electrons and protons. The last two terms represent repulsive Coulomb interaction between electrons and protons, respectively.
\subsubsection*{Born-Oppenheimer Approximation}
In Quantum Chemistry, Born-Oppenheimer Approximation is the best known mathematical approximation in molecular dynamics. It is the assumption that the wave functions of atomic nuclei and electrons can be treated separately. This assumption is based on the fact that nuclei are much heavier than electrons. We will use this approximation to simplify molecular Hamiltonian.\\
So, if nuclei are fixed, it means that their kinetic energies are zero and they will not repel each other as well. Therefore, we will be left out with Electronic Hamiltonian which is given below:
\begin{equation}\label{Electronic_H}
\hat{H}_e = -\sum_{i=1}^N\frac{\hbar^2}{2m_e}\nabla_i^2-\sum_{i=1}^N\sum_{A=1}^M\frac{(Z_Ae)e}{4\pi\epsilon_0r_{iA}}+\sum_{j>i}\frac{e^2}{4\pi\epsilon_0 r_{ij}}
\end{equation}

\section{Review of identical particles}
In classical mechanics, when a system is made of identical particles, it is possible to distinguish each particle from the other. But, in quantum mechanics, identical particles are truly indistinguishable, i.e., there exists no particular mechanism to distinguish particles as in classical mechanics.
\subsection{Exchange and Symmetry}
Consider $N$ identical particles, at positions $\text{\textbf{r}}_1,\text{\textbf{r}}_2,\dotsc,\text{\textbf{r}}_N$, the wave function that describes this is $\Psi\left(\text{\textbf{r}}_1,\text{\textbf{r}}_2,\dotsc,\text{\textbf{r}}_N,t\right)$. We now define an exchange operator $\hat{P}_{ij}$  that exchange particles $i$ and $j$. Thus,
\begin{equation}\label{exchange_operator}
\hat{P}_{ij}\Psi\left(\text{\textbf{r}}_1,\text{\textbf{r}}_2,\dotsc,\text{\textbf{r}}_i,\dotsc,\text{\textbf{r}}_j,\dotsc,\text{\textbf{r}}_N,t\right) = \Psi\left(\text{\textbf{r}}_1,\text{\textbf{r}}_2,\dotsc,\text{\textbf{r}}_j,\dotsc,\text{\textbf{r}}_i,\dotsc,\text{\textbf{r}}_N,t\right)
\end{equation}
If we apply exchange operator twice, we get back the same state, i.e., 
\begin{equation}
\hat{P}_{ij}^2\Psi\left(\text{\textbf{r}}_1,\text{\textbf{r}}_2,\dotsc,\text{\textbf{r}}_i,\dotsc,\text{\textbf{r}}_j,\dotsc,\text{\textbf{r}}_N,t\right) = \Psi\left(\text{\textbf{r}}_1,\text{\textbf{r}}_2,\dotsc,\text{\textbf{r}}_i,\dotsc,\text{\textbf{r}}_j,\dotsc,\text{\textbf{r}}_N,t\right)
\end{equation}
This means that $\hat{P}_{ij}^2=\hat{I}$\\
Since the particles are identical, so Hamiltonian must be invariant of particle exchange, in other words, Hamiltonian must commute with exchange operator, i.e,
\begin{equation}\label{hamiltonian_commutes}
\left[\hat{H},\hat{P}_{ij}\right]=0
\end{equation}
Now the particles are identical, so the probability must remain unchanged by interchange of the particles, i.e., 
\begin{equation}
\abs{\Psi_{ij}}^2 = \abs{\Psi_{ji}}^2
\end{equation}
hence we have
\begin{equation}
\Psi\left(\text{\textbf{r}}_1,\text{\textbf{r}}_2,\dotsc,\text{\textbf{r}}_i,\dotsc,\text{\textbf{r}}_j,\dotsc,\text{\textbf{r}}_N,t\right) = \pm \Psi\left(\text{\textbf{r}}_1,\text{\textbf{r}}_2,\dotsc,\text{\textbf{r}}_j,\dotsc,\text{\textbf{r}}_i,\dotsc,\text{\textbf{r}}_N,t\right)
\end{equation}
Also, if we expect that:
\begin{align}
\hat{P}_{ij}\Psi_{ij} & = \lambda \Psi_{ij} \\
\hat{P}_{ij}^2\Psi_{ij} & = \lambda^2 \Psi_{ij}
\end{align}
But $\hat{P}^2_{ij}=\hat{I}$, therefore,
\begin{align}
\hat{P}^2_{ij}\Psi_{ij} & = \hat{I}\Psi_{ij} = \Psi_{ij}
\end{align}
From Eq. 14 and Eq. 15, 
\begin{align*}
\lambda^2 & = 1 \\
\Rightarrow \lambda & = \pm 1 \\
\Rightarrow \hat{P}_{ij}\Psi_{ij} & = \pm \Psi_{ij}
\end{align*}
This means that the wave function of a system of N identical particles must have two types of \textbf{exchange symmetries}.
\begin{itemize}
\item The wave function is symmetric under exchange of particles, i.e., 
\begin{align*}
\Psi_{ij} & = \Psi_{ji}
\end{align*}
Such particles are called \textbf{Bosons} and they have \emph{integral spin} such as photons, pions, alpha particles.
\item The wave function is anti-symmetric under exchange of particles, i.e., 
\begin{align*}
\Psi_{ij} & = -\Psi_{ji}
\end{align*} 
Such particles are called \textbf{Fermions} and they have \emph{half-odd-integer spin} such as electrons, protons, neutrons etc.
\end{itemize}
\subsection{Wave function of identical particles}
Since the wave functions of systems of identical particles are either symmetric or anti-symmetric, it is appropriate to study the formalism of how to construct wave functions for such particles. For simplicity, consider a system of two identical particles. \\
Suppose particle 1 is in the state $\psi_{k_1}(\text{\textbf{r}}_1)$ and particle 2 in in the state $\psi_{k_2}(\text{\textbf{r}}_2)$. For the case of distinguishable particles, total wave function will be a simple product:
\begin{equation}
\Psi(\text{\textbf{r}}_1,\text{\textbf{r}}_2) = \psi_{k_1}(\text{\textbf{r}}_1)\psi_{k_2}(\text{\textbf{r}}_2)
\end{equation}
This means that we can tell which particle is in which state. But if particles 1 and 2 are identical, we cannot tell which one is in which state. There is a possibility that particle 1 is in the state $\psi_{k_2}$ and particle 2 is in state $\psi_{k_1}$, since both are identical. Thus, the total wave function will be merely a superposition of both possibilities.
\begin{equation}
\Psi(\text{\textbf{r}}_1,\text{\textbf{r}}_2) = \frac{1}{\sqrt{2}}\left[\psi_{k_1}(\text{\textbf{r}}_1)\psi_{k_2}(\text{\textbf{r}}_2) \pm \psi_{k_2}(\text{\textbf{r}}_1)\psi_{k_1}(\text{\textbf{r}}_2)\right]
\end{equation}
where plus sign is used for bosons and minus sign is used for fermions. We are interested in fermions, so the wave function for two identical fermions is:
\begin{equation}\label{Fermion_Equation}
\Psi_f(\text{\textbf{r}}_1,\text{\textbf{r}}_2) = \frac{1}{\sqrt{2}}\left[\psi_{k_1}(\text{\textbf{r}}_1)\psi_{k_2}(\text{\textbf{r}}_2) - \psi_{k_2}(\text{\textbf{r}}_1)\psi_{k_1}(\text{\textbf{r}}_2)\right]
\end{equation}
It can be easily proved that $\Psi_f(\text{\textbf{r}}_1,\text{\textbf{r}}_2) = -\Psi_f(\text{\textbf{r}}_2,\text{\textbf{r}}_1)$
\subsection*{Pauli exclusion principle}
\noindent If we put both the fermions in the same state, say $\psi_{k_1}$
\begin{equation*}
\Psi_f(\text{\textbf{r}}_1,\text{\textbf{r}}_2) = \frac{1}{\sqrt{2}}\left[\psi_{k_1}(\text{\textbf{r}}_1)\psi_{k_1}(\text{\textbf{r}}_2) - \psi_{k_1}(\text{\textbf{r}}_1)\psi_{k_1}(\text{\textbf{r}}_2)\right]=0
\end{equation*}
then we are left with no wave function at all. It means that two identical fermions cannot occupy the same state. This is the famous \textbf{Pauli exclusion principle}.\\
We rewrite fermionic wave function in Eq. \ref{Fermion_Equation} as:
\begin{equation}
\Psi_f(\text{\textbf{r}}_1,\text{\textbf{r}}_2) = \frac{1}{\sqrt{2!}}\sum_P(-1)^P\hat{P}\psi_{k_1}(\text{\textbf{r}}_1)\psi_{k_2}(\text{\textbf{r}}_2)
\end{equation}
where $\hat{P}$ is permutation operator and the sum is over all possible permutations. $(-1)^P$ is equal to $+1$ for even permutation (i.e., when we interchange both $k_1$ and $k_2$ and also $\text{\textbf{r}}_1$ and $\text{\textbf{r}}_2$) and $-1$ for odd permutations (i.e., when we interchange either $k_1$ with $k_2$ or $r_1$ with $r_2$). Also note that we can write the fermionic wave function given in Eq. \ref{Fermion_Equation} in the form of determinant as well.
\begin{equation}
\Psi_f(\text{\textbf{r}}_1,\text{\textbf{r}}_2) = \frac{1}{\sqrt{2!}} \begin{vmatrix}
\psi_{k_1}(\text{\textbf{r}}_1) & \psi_{k_1}(\text{\textbf{r}}_2) \\
\psi_{k_2}(\text{\textbf{r}}_1) & \psi_{k_2}(\text{\textbf{r}}_2)
\end{vmatrix}
\end{equation}
Similarly, wave function of $N$ non-interacting identical fermions is:
\begin{equation}
\Psi_f(\text{\textbf{r}}_1,\text{\textbf{r}}_2,\dotsc,\text{\textbf{r}}_N) = \frac{1}{\sqrt{N!}}\sum_P(-1)^P\hat{P}\psi_{k_1}(\text{\textbf{r}}_1)\psi_{k_2}(\text{\textbf{r}}_2)\dotsc\psi_{k_N}(\text{\textbf{r}}_N)
\end{equation}
or
\begin{equation}\label{slater}
\Psi_f(\text{\textbf{r}}_1,\text{\textbf{r}}_2,\dotsc,\text{\textbf{r}}_N) = \frac{1}{\sqrt{N!}} \begin{vmatrix}
\psi_{k_1}(\text{\textbf{r}}_1) & \psi_{k_1}(\text{\textbf{r}}_2) & \cdots & \psi_{k_1}(\text{\textbf{r}}_N)\\
\psi_{k_2}(\text{\textbf{r}}_1) & \psi_{k_2}(\text{\textbf{r}}_2) & \cdots & \psi_{k_2}(\text{\textbf{r}}_N) \\
\vdots & \vdots & \ddots & \vdots \\
\psi_{k_N}(\text{\textbf{r}}_1) & \psi_{k_N}(\text{\textbf{r}}_2) & \cdots & \psi_{k_N}(\text{\textbf{r}}_N)
\end{vmatrix}
\end{equation}
This is $N\times N$ determinant that involves single-particle states only and is called \textbf{Slater Determinant}.

\section{Second Quantization}
In the standard formulation of quantum mechanics, we represent observables by operators and the state of a particle by function. In the language of second quantization, wave functions are also defined by operators - the creation and annihilation operators working on vacuum state. We will also be able to get Hamiltonian in the form of creation and annihilation operators using the formalism of second quantization.
\subsection{Fock Space: Occupation Number Representation of Fermionic Wave Function}
Suppose ${\psi_{k}(\text{\textbf{x}})}$ be the basis of $M$ orthonormal spin orbitals, where the coordinates \textbf{x} collectively represents the spatial \textbf{r} and spin coordintates $\sigma$ of an electron. Then the normalized wave function of $N$ electrons can be written in the form of Slater Determinant as:
\begin{equation}
\Psi_f(\text{\textbf{x}}_1,\text{\textbf{x}}_2,\dotsc,\text{\textbf{x}}_N) = \frac{1}{\sqrt{N!}} \begin{vmatrix}
\psi_{k_1}(\text{\textbf{x}}_1) & \psi_{k_1}(\text{\textbf{x}}_2) & \cdots & \psi_{k_1}(\text{\textbf{x}}_N)\\
\psi_{k_2}(\text{\textbf{x}}_1) & \psi_{k_2}(\text{\textbf{x}}_2) & \cdots & \psi_{k_2}(\text{\textbf{x}}_N) \\
\vdots & \vdots & \ddots & \vdots \\
\psi_{k_N}(\text{\textbf{x}}_1) & \psi_{k_N}(\text{\textbf{x}}_2) & \cdots & \psi_{k_N}(\text{\textbf{x}}_N)
\end{vmatrix}
\end{equation}
We now introduce an abstract linear vector space - \emph{Fock Space}\footnote{Fock space of $N$-orbitals is the collection of all the sub-spaces that has fixed no. of electrons, $M\leq N$} - where each determinant is represented by occupation number vector $\ket{\text{\textbf{k}}}$\footnote{Rather than asking, which particle is in which state, we instead rephrased problem to a question asking how many particles are there in the state $\psi_{k}$}. Suppose that spin orbital, $\psi_{k_i}(\text{\textbf{x}})$ contains $n_i$ electrons. Then the state of entire system of $N$ particles can be written as:
\begin{equation}
\Psi_f(\text{\textbf{x}}_1,\text{\textbf{x}}_2,\dotsc,\text{\textbf{x}}_N) \longrightarrow \ket{k_1, k_2, \dotsc, k_N}
\end{equation}
where the occupation number, 
\begin{equation}
n_i = \begin{cases}
1\ \text{if}\ \psi_{k_i}\ \text{is occupied} \\
0\ \text{if}\ \psi_{k_i}\ \text{is empty}
\end{cases}
\end{equation}
Also,
\begin{align*}
\ket{k_1,k_2,\cdots,k_i,\cdots,k_j,\cdots,k_N} \longrightarrow -\ket{k_1,k_2,\cdots,k_j,\cdots,k_i,\cdots,k_N}
\end{align*}
The state in which there are no particles is called \textbf{vacuum state}, represented by
\begin{equation*}
\ket{0} = \ket{0_{k_1},0_{k_2},\cdots,0_{k_N}}
\end{equation*}
\subsubsection{Inner Product}
\noindent Consider two states:
\begin{align*}
\ket{\text{\textbf{k}}} & = \ket{k_1,k_2,\cdots,k_N} \\
\ket{\text{\textbf{k'}}} & = \ket{k'_1,k'_2,\cdots,k'_N}
\end{align*}
then the inner product is defined as:
\begin{align*}
\ip{\text{\textbf{k}}}{\text{\textbf{k'}}} & = \ip{k_1,k_2,\cdots,k_N}{k'_1,k'_2,\cdots,k'_N} \\
& = \delta_{k_1k'_1}\delta_{k_2k'_2}\cdots\delta_{k_Nk'_N} \\
& = \prod_{i=1}^N\delta_{k_ik'_i}
\end{align*}
\subsubsection{Completeness}
Fock States form a complete set of basis. We have a completeness relation:
\begin{align*}
\sum_{k_1=0}^1\sum_{k_2=0}^1\cdots\sum_{k_N=0}^1\op{k_1,k_2,\cdots,k_N}{k_1,k_2,\cdots,k_N} & = \mathbb{I}
\end{align*}
\subsection{Creation and Annihilation operators}
In the second quantization, all the operators and states can be constructed from the set of creation and annihilation operators. We know that
\begin{equation}
\begin{split}
\hat{a}\ket{0} & = 0 \\
\hat{a}^{\dagger}\ket{0} & = \ket{1} \\
\hat{a}\ket{1} & = \ket{0} \\
\hat{a}^{\dagger}\ket{1} & = 0
\end{split}
\end{equation}
\subsubsection*{Why is $\hat{a}\ket{0}=0$?}
The annihilation operator destroys the electron state. Applying it to the vacuum state will give us 0 because we cannot destroy the electron state if there is already no electron state.
\subsubsection*{Why is $\hat{a^{\dagger}}\ket{1}=0$?}
Here we attempt to create another electron of the same state in the same orbital, which is impossible according to the Pauli exclusion principle, i.e., no two electrons can occupy the same state. Thus $\hat{a^{\dagger}}\ket{1}=0$.\\

We can create an arbitrary occupation state $\ket{\text{\textbf{k}}}=\ket{k_1,k_2,\cdots,k_N}$ by applying creation operations on vacuum state $\ket{0}$. Furthermore, the order in which they are applied must play an essential role.
\begin{equation*}
\ket{k_1,k_2,\cdots,k_N} = (a_1^{\dagger})^{k_1}(a_2^{\dagger})^{k_2}\cdots(a_N^{\dagger})^{k_N}\ket{0}
\end{equation*}
\begin{example}{}{label}
Create $\ket{01011}$ from vacuum state.\\
\begin{align*}
\ket{01011} & = (a_1^{\dagger})^{0}(a_2^{\dagger})^{1}(a_3^{\dagger})^{0}(a_4^{\dagger})^{1}(a_5^{\dagger})^{1}\ket{00000} \\
& = \mathbb{I}\ket{0}\otimes a_2^{\dagger}\ket{0}\otimes \mathbb{I}\ket{0}\otimes a_4^{\dagger}\ket{0}\otimes a_5^{\dagger}\ket{0} \\
& = \ket{01011}
\end{align*}
\end{example}
\subsubsection{Anti-commutation Relations}
Suppose we have a state $\ket{k_i,k_j}$ which is prepared from the vacuum state.
\begin{align*}
\ket{k_i,k_j} & = a_i^{\dagger}a_j^{\dagger}\ket{0}
\end{align*}
Similarly,
\begin{align*}
\ket{k_j,k_i} & = a_j^{\dagger}a_i^{\dagger}\ket{0} \\
\Rightarrow -\ket{k_i,k_j} & = a_j^{\dagger}a_i^{\dagger}\ket{0} \\
\Rightarrow \ket{k_i,k_j} & = -a_j^{\dagger}a_i^{\dagger}\ket{0} \\
\Rightarrow a_i^{\dagger}a_j^{\dagger} & = -a_j^{\dagger}a_i^{\dagger}
\end{align*}
which is nothing but anti-commutator.
\begin{equation}\label{anticommutator}
\begin{split}
a_i^{\dagger}a_j^{\dagger} + a_j^{\dagger}a_i^{\dagger} & = 0 \\
\Rightarrow \{a_i^{\dagger},a_j^{\dagger}\} & = 0
\end{split}
\end{equation}
Likewise, we can prove that $\{a_i,a_j\}=0$
\begin{example}{}{label}
\textbf{Annihilate $0_2$ in the state $\ket{1011}$.}\\

First write the state $\ket{1011}$ in term of creation and annihilation operators.
\begin{align*}
\ket{1011} & = (a_1^{\dagger})^1(a_2^{\dagger})^0(a_3^{\dagger})^1(a_4^{\dagger})^1\ket{0}
\end{align*}
Now we want to create an electron in position 2.
\begin{align*}
a_2^{\dagger}\ket{1011} & = a_2^{\dagger}(a_1^{\dagger})^1(a_2^{\dagger})^0(a_3^{\dagger})^1(a_4^{\dagger})^1\ket{0}\\
& = a_2^{\dagger}a_1^{\dagger}a_3^{\dagger}a_4^{\dagger})\ket{0}
\end{align*}
Since order plays an important role, so we must move $a_2^{\dagger}$ to position 2. We know that, $a_2^{\dagger}a_1^{\dagger}=-a_1^{\dagger}a_2^{\dagger}$. Thus,
\begin{align*}
a_2^{\dagger}\ket{1011} & = -a_1^{\dagger}a_2^{\dagger}a_3^{\dagger}a_4^{\dagger})\ket{0} \\
& = (-1)^{k_1}a_1^{\dagger}a_2^{\dagger}a_3^{\dagger}a_4^{\dagger})\ket{0}\\
& = -\ket{1111}
\end{align*}
\end{example}
\subsubsection*{Creation Operator}
\noindent Generally, action of creation operator $a^{\dagger}_i$ on $k_i$ where $k_i=0,1$ in the state $\ket{k_1,k_2,\cdots,k_i,\cdots,k_N}$ is:
\begin{align}
a^{\dagger}_i\ket{\cdots,k_i,\cdots} & = (-1)^{k_1+k_2+\cdots+k_{i-1}}(1-k_i)\ket{\cdots,k_{i}+1,\cdots} \\
& = (-1)^{\sum_{j<i}k_j}(1-k_i)\ket{\cdots,k_{i}+1,\cdots}
\end{align}
The factor $(-1)^{\sum_{j<i}k_j}$ accounts for number of anti-commutations necessary to bring $a^{\dagger}_i$ to $i$th position. And the factor $(1-k_i)$ preserves Pauli exclusion principle, i.e., it yields zero, if the state $k_i$ is already occupied.
\subsubsection*{Annihilation Operator}
\noindent Now, take the adjoint of Eq. 29
\begin{equation}
\bra{\cdots,k_i,\cdots}a_i  = (-1)^{\sum_{j<i}k_j}(1-k_i)\bra{\cdots,k_{i}+1,\cdots}
\end{equation}
We can write the matrix element of annihilation operator $a_i$ as:
\begin{equation*}
a_{ii'} = \bra{\cdots,k_i,\cdots}a_i\ket{\cdots,k_i',\cdots} = (-1)^{\sum_{j<i}k_j}(1-k_i)\delta_{k_i',k_i+1}
\end{equation*}
We can write operator in the form of its basis as:
\begin{align*}
a_i & = \sum_{k_i}a_{ii'}\op{k_i}{k_i'} \\
& = \sum_{k_i} \bra{k_i}a_{i}\ket{k_i'}\op{k_i}{k_i'}
\end{align*}
If we apply this operator on $k_i'$, we will get:
\begin{align*}
a_i\ket{k_i'} & = \left(\sum_{k_i} \bra{k_i}a_{i}\ket{k_i'}\op{k_i}{k_i'}\right)\ket{k_i'} \\
& = \sum_{k_i} \bra{k_i}a_{i}\ket{k_i'} \ket{k_i} \\
& = (-1)^{\sum_{j<i}k_j}(1-k_i)\delta_{k_i',k_i+1} \ket{k_i}
\end{align*}
Clearly, if $k_i'=0$, the Kronecker delta $\delta_{k_i',k_i+1}=0$ always gives zero. So, instead of Kronecker delta, we can also use $k_i'$. Furthermore, coefficient will be non-zero if, $k_i'=k_i+1$, i.e., $k_i=k_i'-1 \Rightarrow (1-k_i)=(2-k_i')$. Thus above equation can be written as:
\begin{equation*}
a_i\ket{k_i'} = (-1)^{\sum_{j<i}k_j}(2-k_i')k_i'\ket{k_i'-1}
\end{equation*}
We can omit $(2-k_i')$ because, if $k_i'=0$, whole expression vanishes, and if $k_i'=1$, $(2-k_i')=1$. So,
\begin{equation*}
a_i\ket{k_i'} = (-1)^{\sum_{j<i}k_j}k_i'\ket{k_i'-1}
\end{equation*}
To summarize, the effect of creation and annihilation operators are:
\begin{equation}\label{a_adagger}
\begin{split}
a_i^{\dagger}\ket{\cdots,k_i,\cdots} & = (-1)^{\sum_{j<i}k_j}(1-k_i)\ket{\cdots,k_{i}+1,\cdots} \\
a_i\ket{\cdots,k_i,\cdots} & = (-1)^{\sum_{j<i}k_j}k_i\ket{\cdots,k_i-1,\cdots}
\end{split}
\end{equation}
It follows from this that:
\begin{align*}
a_ia_i^{\dagger}\ket{\cdots,k_i,\cdots} & = a_i\left[(-1)^{\sum_{j<i}k_j}(1-k_i)\ket{\cdots,k_{i}+1,\cdots}\right] \\
& = (-1)^{\sum_{j<i}k_j}(1-k_i)\left[a_i\ket{\cdots,k_{i}+1,\cdots}\right] \\
& = \left[(-1)^{\sum_{j<i}k_j}(1-k_i)\right]\left[(-1)^{\sum_{j<i}k_j}(k_i+1)\right]\ket{\cdots,k_i,\cdots} \\
& = (1-k_i)(k_i+1)\ket{\cdots,k_i,\cdots}
\end{align*}
We can omit $(k_i+1)$ because if $k_i=1$, the expression vanishes already, and if $k_i=0$, then it is 1. Thus, 
\begin{equation}\label{1-numberoperator}
a_ia_i^{\dagger}\ket{\cdots,k_i,\cdots} = (1-k_i)\ket{\cdots,k_i,\cdots}
\end{equation}
Similarly, 
\begin{align*}
a_i^{\dagger}a_i\ket{\cdots,k_i,\cdots} & = a_i^{\dagger}\left[(-1)^{\sum_{j<i}k_j}k_i\ket{\cdots,k_i-1,\cdots}\right]\\
& = (-1)^{\sum_{j<i}k_j}k_i \left[a_i^{\dagger}\ket{\cdots,k_i-1,\cdots}\right]\\
& = \left[(-1)^{\sum_{j<i}k_j}k_i\right]\left[(-1)^{\sum_{j<i}k_j}(1-(k_i-1))\right]\ket{\cdots,k_i,\cdots} \\
& = k_i(2-k_i)\ket{\cdots,k_i,\cdots}
\end{align*}
Here, we can omit $2-k_i$, because if $k_i=0$, the expression vanishes already and if $k_i=1$, then it is equal to 1. Thus, 
\begin{equation}\label{numoperator}
a_i^{\dagger}a_i\ket{\cdots,k_i,\cdots} = k_i\ket{\cdots,k_i,\cdots}
\end{equation}
Comparing Eq. \ref{1-numberoperator} and Eq. \ref{numoperator}, 
\begin{align*}
(\mathbb{I}-a_ia_i^{\dagger})\ket{\cdots,k_i,\cdots} & = k_i\ket{\cdots,k_i,\cdots} \\
& = a_i^{\dagger}a_i\ket{\cdots,k_i,\cdots}
\end{align*}
This means that,
\begin{align*}
1-a_ia_i^{\dagger} & = a_i^{\dagger}a_i\\
a_ia_i^{\dagger} + a_i^{\dagger}a_i = 1
\end{align*}
which is nothing but another anti-commutation relation.
\begin{equation}\label{anti2}
\{a_i , a_i^{\dagger}\} = 1
\end{equation}
In the anti-commutator, $\{a_i,a_j^{\dagger}\}$, with $i\neq j$, the phase factor of two terms is different:
\begin{align*}
\{a_i,a_j^{\dagger}\} & = a_ia_j^{\dagger}+a_j^{\dagger}a_i \\
& = (-1)^{S'_i}k_i(-1)^{S'_j}(1-k_j) + (-1)^{S_j}(1-k_j)(-1)^{S_i}k_i \\
& = k_i(1-k_j)(1-1)\\
& = 0
\end{align*}
To summarize, the Anti-Commutation relations for fermions are:
\begin{equation}\label{AC_Relations}
\begin{split}
\{a_i,a_j\} & = 0 \\
\{a_i^{\dagger},a_j^{\dagger}\} & = 0 \\
\{a_i , a_j^{\dagger}\} & = \delta_{ij}
\end{split}
\end{equation}

\subsection{Single and Many-body Operators}
For fermions, single and many-body operators can be written in terms of creation and annihilation operators.
\subsubsection{Single-Body Operator}\label{SBO}
An operator $\hat{T}$ is called single-body operator if the action of $\hat{T}$ on the state $\ket{k_1,k_2,\cdots,k_N}$ of $N$ particles is the sum of action of $\hat{T}$ on each particle, i.e., 
\begin{equation}\label{onebodyoperator}
\hat{T}\ket{k_1,k_2,\cdots,k_N} = \sum_{i=1}^N\hat{t}_i\ket{k_1,k_2,\cdots,k_N}
\end{equation}
Single operator in terms of creation and annihilation operator is given by \footnote{The proof has been skipped. One may find the proof in advanced quantum mechanics books.}
\begin{align}
\hat{T} & = \sum_{i,j}t_{ij}a_i^\dagger a_j
\end{align}

\subsubsection{Two-Body Operator}
An operator $\hat{F}$ is called two-body operator if the action of $\hat{F}$ on the state $\ket{k_1,k_2,\cdots,k_N}$ of $N$-particles  is the sum of the action of $\hat{F}$ on all distinct pairs of particles:
\begin{equation}
\hat{O}\ket{k_1,k_2,\cdots,k_N} = \frac{1}{2}\sum_{i,j}\hat{f}_{ij}\ket{k_1,k_2,\cdots,k_N}
\end{equation}
The factor $\displaystyle{\frac{1}{2}}$ in the above equation is to ensure that each interaction is included only once. Since particles are identical and symmetry implies that $\hat{F}(\text{\textbf{x}}_i,\text{\textbf{x}}_j)=\hat{F}(\text{\textbf{x}}_j,\text{\textbf{x}}_i)$. It can be prove that:
\begin{equation}\label{twobodyoperator}
\hat{F} = \frac{1}{2}\sum_{i,j,k,m}\hat{f}_{ijkm}a_i^{\dagger}a_j^{\dagger}a_ka_m
\end{equation}

\subsection{Molecular Hamiltonian in Second Quantized Form}
In Section. \ref{Mol_Ham}, we derived the expression for Molecular Hamiltonian which we can rewrite as:
\begin{equation}
\hat{H}_e = -\sum_{i=1}^N\frac{\hbar^2}{2m_e}\nabla_i^2-\sum_{i=1}^N\sum_{A=1}^M\frac{(Z_Ae)e}{4\pi\epsilon_0r_{iA}}+\frac{1}{2}\sum_{i\neq j}\frac{e^2}{4\pi\epsilon_0 r_{ij}}
\end{equation}
Here, the first two terms are One-Body Operators since, they involve only one coordinate and act on a single particle whereas the last term is Two-Body Operator because it involves two coordinates and act on all distinct pair of particles. We can get more simplified form of Hamiltonian by writing it in Atomic Units aka \emph{Hartree Units} as:
\begin{equation}
\hat{H}_e = -\sum_{i=1}^N\frac{\nabla_i^2}{2}-\sum_{i=1}^N\sum_{A=1}^M\frac{Z_A}{r_{iA}}+\frac{1}{2}\sum_{i\neq j}\frac{1}{r_{ij}}
\end{equation}
The One-Body Operator from the Hamiltonian is given by:
\begin{align*}
\hat{H}_1 & = -\sum_{i=1}^N\frac{\nabla_i^2}{2}-\sum_{i=1}^N\sum_{A=1}^M\frac{Z_A}{r_{iA}} \\
& = \sum_{i,j} \left\langle i \left|-\frac{\nabla_i^2}{2}-\sum_{A=1}^M\frac{Z_A}{r_{iA}}\right|j\right\rangle  a_i^{\dagger}a_j
\end{align*}
Two-Body Operator from the Hamiltonian is given by:
\begin{align*}
\hat{H}_2 & = \frac{1}{2}\sum_{i\neq j}\frac{1}{r_{ij}} \\
& = \frac{1}{2}\left\langle i,j \left| \frac{1}{r_{ij}}\right| k,m \right\rangle a_i^{\dagger}a_j^{\dagger}a_ka_m 
\end{align*}
So, Hamiltonian in the form of creation and annihilation operator is:
\begin{equation}\label{H2Q}
\hat{H} = \sum_{pq}h_{pq}a_p^{\dagger}a_q + \frac{1}{2}\sum_{pqrs}h_{pqrs}a_p^{\dagger}a_{q}^{\dagger}a_ra_s
\end{equation}
where $h_{pq}$ and $h_{pqrs}$ is one-body and two-body integrals respectively:
\begin{equation}
\begin{split}
h_{pq} & \equiv \int d\text{\textbf{x}}\  \chi_p^*(\text{\textbf{x}}) \left[-\frac{\nabla^2}{2}-\sum_{\alpha}\frac{Z_{\alpha}}{r_{\alpha, \text{\textbf{x}}}}\right] \chi_q(\text{\textbf{x}}) \\
h_{pqrs} & \equiv \int d\text{\textbf{x}}_1 \int d\text{\textbf{x}}_2\ \chi_p^*(\text{\textbf{x}}_1)\chi_q^*(\text{\textbf{x}}_2) \left[\frac{1}{r_{ij}}\right]\chi_r(\text{\textbf{x}}_1)\chi_s(\text{\textbf{x}}_2)
\end{split}
\end{equation}

\section{Jordan-Wigner Transformation}
Our goal is to map occupation number basis vector: $\ket{k_1 \cdots k_n}$ to the qubit state $\ket{q_1\cdots q_n}$ and the fermionic creation and annihilation operators onto the operations on qubits. We know that
\begin{align}
a_i^\dagger \ket{\cdots k_i \cdots} & = \left( -1 \right) ^{\sum_{j < i}k_j}(1-k_i)\ket{\cdots k_i+1 \cdots} \\
a_i \ket{\cdots k_i \cdots} & = \left( -1 \right) ^{\sum_{j < i}k_j}k_i\ket{\cdots k_i-1 \cdots}
\end{align}
Thus we can represent the action of $a_i^\dagger$ and $a_i$ by $A_i^+$ and $A_i^-$ respectively and with $\hat{Z}$ on all qubits with index less than $i$,
\begin{align}
a_i^\dagger & \longrightarrow Q_i^+ = Z_{i}^\leftarrow \otimes A_i^+ \\
a_i & \longrightarrow Q_i^- = Z_{i}^\leftarrow \otimes A_i^-
\end{align}
where $Z_{i}^\rightarrow$ is parity operator:
\begin{align*}
Z_{i}^\leftarrow & = Z_0 \otimes Z_1 \otimes \cdots \otimes Z_{i-1}
\end{align*}
and
\begin{align*}
A^+ & = \frac{1}{2}\left( \hat{X} - i\hat{Y} \right) \\
A^- & = \frac{1}{2}\left( \hat{X} + i\hat{Y} \right) 
\end{align*}
and, 
\begin{align*}
A_i^+\ket{\cdots q_i \cdots} & = (1-q_i)\ket{\cdots q_i+1 \cdots} \\
A_i^-\ket{\cdots q_i \cdots} & = q_i\ket{\cdots q_i-1 \cdots}
\end{align*}
The effect of the string of $Z$ gates is to introduce the required phase change of $-1$, if the parity of set of qubits with index less than $i$ is 1 (odd), and do nothing if the parity is 0 (even). This also ensures that qubit operations $Q^+$ and $Q^-$ follow anti-commutation relations.\\

The problem with this method is that the number of $Z$ operators scales linearly with the size of system, i.e., for system of $N$ fermion, a single fermionic operator is $\mathcal{O}(N)$ local.

\begin{example}{}{label}
Prove that the qubit operations $Q^+$ and $Q^-$ follow anti-commutation relations, i.e., 
\begin{align*}
\{Q_m^- , Q_n^- \} & = 0 \\
\{Q_m^+ , Q_n^+ \} & = 0 \\
\{Q_m^+ , Q_n^- \} & = \delta_{mn} 
\end{align*}
\paragraph{Solution:}Consider the qubit state $\ket{q_m,q_n}$, then
\begin{align*}
\{Q_m^- , Q_n^- \}\ket{q_m,q_n} & = Q_m^-Q_n^-\ket{q_m,q_n} + Q_n^-Q_m^-\ket{q_m,q_n} \\
& = \left[(-1)^{S'_m}(-1)^{S'_n}+(-1)^{S_m}(-1)^{S_n}\right]q_mq_n\ket{q_m-1,q_n-1} \\
& = \left[ (\pm 1) + (\mp 1) \right]q_mq_n \ket{q_m-1,q_n-1} \\
& = 0
\end{align*}
Similarly, 
\begin{align*}
\{Q_m^+ , Q_n^+ \}\ket{q_m,q_n} & = Q_m^+Q_n^+\ket{q_m,q_n} + Q_n^+Q_m^+\ket{q_m,q_n} \\
& = \left[(-1)^{S'_m}(-1)^{S'_n}+(-1)^{S_m}(-1)^{S_n}\right](q_m+1)(q_n+1)\ket{q_m+1,q_n+1} \\
& = \left[ (\pm 1) + (\mp 1) \right] (q_m+1)(q_n+1)\ket{q_m+1,q_n+1} \\
& = 0
\end{align*}
Now, 
\begin{align*}
\{Q_m^+ , Q_n^- \} \ket{q_m,q_n} & = Q_m^+Q_n^-\ket{q_m,q_n} + Q_n^-Q_m^+\ket{q_m,q_n} \\
& = \left[(-1)^{S'_m}(-1)^{S'_n}+(-1)^{S_m}(-1)^{S_n}\right] (1-q_m)q_n \ket{q_m+1,q_n-1} \\
& = \begin{cases}
0 & m\neq n \\
1 & m=n
\end{cases}
\end{align*}
where $\displaystyle{(-1)^{S_i}=(-1)^{\sum_{j<i}q_j}}$.
\end{example}

\section{Parity Transformation}
Unlike Jordan-Wigner transformation, application of creation/annihilation operator on qubit $j$ is represented by applying $Z$ on all qubits with index less than $j$ that introduce the required phase change, we can now accomplish this task by applying $Z$ once if qubit $j$ stores the parity of all the filled orbitals upto $j$. \footnote{In Jordan-Wigner transformation, $k_i$ stores the parity of $i$-th qubit only.} That is, qubit $j$ stores $\displaystyle{p_j=\sum_{i=0}^j k_i\ \text{mod}(2)}$. This encoding scheme is called \emph{parity transformation}:
\begin{align*}
\ket{k_0, k_1, \cdots , k_{n-1}} & \longrightarrow \ket{p_0,p_1,\cdots,p_{n-1}}
\end{align*}
This transformation can be represented by the action of matrix on vector $\ket{\textbf{k}}$ in occupation number basis to get corresponding vector $\ket{\textbf{p}}$ in parity basis. Since, parity transformation involves the sum of bits mod 2, then the matrix must have the form:
\begin{align}
\pi_{n\times n} & = \begin{pmatrix}
1 & 0 & \cdots & 0 \\
1 & 1 & \cdots & 0 \\
\vdots & \vdots & \ddots & \vdots \\
1 & 1 & \cdots & 1
\end{pmatrix}
\end{align}
\begin{example}{}{label}
Transform the vector $\ket{11100101}$ in occupation number basis to a parity basis.
\paragraph{Solution:}Since there are 8 qubit, thus size of $\pi$ will be $8\times 8$ and is given by:

\begin{center}
$\displaystyle{\pi_{8\times 8}}$ = \begin{blockarray}{ccccccccc}
& $k_0$ & $k_1$ & $k_2$ & $k_3$ & $k_4$ & $k_5$ & $k_6$ & $k_7$ \\
\begin{block}{c(cccccccc)}
$p_0$ & 1 & 0 & 0 & 0 & 0 & 0 & 0 & 0 \\
$p_1$ & 1 & 1 & 0 & 0 & 0 & 0 & 0 & 0 \\
$p_2$ & 1 & 1 & 1 & 0 & 0 & 0 & 0 & 0 \\
$p_3$ & 1 & 1 & 1 & 1 & 0 & 0 & 0 & 0 \\
$p_4$ & 1 & 1 & 1 & 1 & 1 & 0 & 0 & 0 \\
$p_5$ & 1 & 1 & 1 & 1 & 1 & 1 & 0 & 0 \\
$p_6$ & 1 & 1 & 1 & 1 & 1 & 1 & 1 & 0 \\
$p_7$ & 1 & 1 & 1 & 1 & 1 & 1 & 1 & 1 \\
\end{block}
\end{blockarray}
\end{center}
\begin{align*}
\ket{p} & \equiv \pi_{8\times 8}\begin{pmatrix}
1 & 1 & 1 & 0 & 0 & 1 & 0 & 1
\end{pmatrix}^T \\
& = \begin{pmatrix}
1 & 0 & 1 & 1 & 1 & 0 & 0 & 1
\end{pmatrix}^T \equiv \ket{10111001}
\end{align*}
\end{example}
Now, we cannot represent the creation/annihilation of particle in orbital $j$ by simply operating $A^{\pm}$ on qubit $j$, as we did in Jordan-Wigner transformation. Because, qubit $j$ now stores the parity of all orbitals with index $\leq j$. Operating $A^+$ or $A^-$ on qubit $j$ now depends on qubit $j-1$. If $\ket{q_{j-1}}=\ket{0}$, then $a_j^\dagger \mapsto A_j^+$ and $a_j \mapsto A_j^-$ as usual. But if $\ket{q_{j-1}}=\ket{1}$, then qubit $j$ will have inverted parity, thus $a_j^\dagger\mapsto A_j^-$ and $a_j\mapsto A_j^+$. Therefore, we can define new operations as:
\begin{align}
\mathcal{P}_j^\pm & \equiv \op{0}{0}_{j-1}\otimes A_{j}^\pm - \op{1}{1}_{j-1}\otimes A_j^\mp
\end{align}
\begin{exercise}{}{label}
Prove that:
\begin{align*}
\mathcal{P}_j^\pm & = \frac{1}{2}\left( Z_{j-1}\otimes X_j \mp iY_j \right)
\end{align*}\\
\paragraph{Hint:}Substitute value of $A^\pm$ and expand.
\end{exercise}
In addition to that, operating $P_j^\pm$ changes the parity data of all orbitals with index $>j$. Thus, we must update the comulative sum $p_i$ for $i>j$ by applying $X$ to all qubits $\ket{p_i}$ for $i>j$. The representation of creation and annihilation operators is thus given by:
\begin{align}
a_j^\dagger & = \mathcal{P}_j^+ \otimes X_{j+1}^\rightarrow \\
a_j & = \mathcal{P}_j^- \otimes X_{j+1}^\rightarrow 
\end{align}
where $X_i^\rightarrow$ is the ``update operator'' that updates all the qubits that store partial sum including orbital $i-1$ when the occupation number of orbital changes by applying $P^\pm$.
\begin{align*}
X_i^\rightarrow & = X_i \otimes X_{i+1} \otimes \cdots \otimes X_{n-1}
\end{align*}
We see that the string of $Z$ in Jordan-Wigner Transformation has been replaced by a string of $X$ in parity basis which also scales as $\mathcal{O}(n)$, i.e., as compared to Jordan-Wigner transformation, parity basis has no advantage in terms of complexity.
\section{Bravyi-Kitaev Transformation}
We noticed that two kinds of information was required to simulate a single fermionic operators $a_j^\dagger,a_j$: occupation of $q_j$ and parity of set of orbitals with index $<j$. We see that both Jordan-Wigner and Parity transformations are $\mathcal{O}(N)$ local for system of $N$ fermions. In Jordan-Wigner transformation, occupation information is stored locally whereas, parity information is non-local. On the other hand, in parity transformation, parity information is stored locally and occupation information is non-local. \footnote{In Jordan-Wigner transformation, $q_j$ stores the parity of its own corresponding orbital $k_j$, by applying $a_j/a_j^\dagger$, parity of other qubits is unaffected, thus parity information in this transformation is non-local. On other hand, $q_j$ stores the parity of all the spin orbitals upto $j$, hence applying $a_j/a_j^\dagger$, does affect the parity of the orbitals with index $>j$, hence parity is local in this case.} The Bravyi-Kitaev transformation provides something sort of middle ground, i.e., it balances the locality of both occupation number and parity as well.
\subsection{Bravyi-Kitaev Basis}
The scheme is to use a basis $\{\ket{b_j}\}$ that store the \emph{partial sum} of occupation number. In this scheme, qubit stores the parity of $2^x$ orbitals, where $x\geq 0$. A qubit $q_j$ will always store the parity of occupation number of orbital $j$. For even $j$, this is the only thing it stores. But for odd $j$, it will store the parity of certain sets of adjacent orbitals with index $<j$. This transformation can be represented by the action of matrix on vector $\ket{\textbf{k}}$ in occupation number basis to get the corresponding vector $\ket{\textbf{b}}$ in Bravyi-Kitaev basis:
\begin{align*}
\ket{k_{n-1},\cdots,k_1,k_0}& \longrightarrow \ket{b_{n-1},\cdots,b_1,b_0}
\end{align*}
\begin{align}
\ket{\textbf{b}} & \equiv  \beta_{n\times n} \ket{\textbf{k}} \\
b_i & = \sum_j [\beta_{n\times n}]_{ij} k_j\ \ \text{mod }2
\end{align}
where
\begin{align}
\beta_1 & = \beta_1^{-1} = \begin{pmatrix}
1
\end{pmatrix}_{1\times 1}
\end{align}
\begin{align}\label{normal}
\beta_{2^{x+1}} & = \left(\begin{array}{c|c}
\beta_{2^x} & \begin{array}{c}
\leftarrow 1 \rightarrow \\
\mathcal{O}
\end{array} \\
\hline
\mathcal{O} & \beta_{2^x}
\end{array} \right)
\end{align}
\begin{align}\label{Inverse}
\beta_{2^{x+1}}^{-1} & = \left( \begin{array}{c|c}
\beta_{2^x}^{-1} & \begin{array}{cc}
1 & \\
& \mathcal{O}
\end{array} \\
\hline
\mathcal{O} & \beta_{2^x}^{-1}
\end{array} \right)
\end{align}
In Bravyi-Kitaev transformation, we need to have following three set of qubits:
\begin{enumerate}
\item \emph{Parity Set:} It stores the parity of all the orbitals with index $<j$.
\item \emph{Update Set:} It stores the partial sum (mod 2) of the orbitals upto $j$ (including $j$).
\item \emph{Flip Set:} It determines whether qubit $j$ has the same parity as orbital $j$.
\end{enumerate}\newpage

\begin{example}{}{label}
Construct $\beta_8$ and $\beta_8^{-1}$ using Eq. (\ref{normal}) and Eq. (\ref{Inverse}).
\paragraph{Solution:}Eq. (\ref{normal}) and Eq. (\ref{Inverse}) for $x=0$ are:
\begin{align}
\beta_2 & = \left( \begin{array}{c|c}
\beta_1 & \begin{array}{c}
\leftarrow 1 \rightarrow \\
\mathcal{O}
\end{array} \\
\hline
\mathcal{O} & \beta_1
\end{array}\right) = \begin{pmatrix}
1 & 1 \\
0 & 1
\end{pmatrix}
\end{align}
For $x=1$,
\begin{align}
\beta_4 & = \left( \begin{array}{c|c}
\beta_2 & \begin{array}{c}
\leftarrow 1 \rightarrow \\
\mathcal{O}
\end{array} \\
\hline
\mathcal{O} & \beta_2
\end{array}\right) = \begin{pmatrix}
1 & 1 & 1 & 1 \\
0 & 1 & 0 & 0 \\
0 & 0 & 1 & 1 \\
0 & 0 & 0 & 1
\end{pmatrix}
\end{align}
For $x=2$,
\begin{align}\label{beta8}
\beta_8 & = \left( \begin{array}{c|c}
\beta_4 & \begin{array}{c}
\leftarrow 1 \rightarrow \\
\mathcal{O}
\end{array} \\
\hline
\mathcal{O} & \beta_4
\end{array}\right) = \begin{pmatrix}
1 & 1 & 1 & 1 & 1 & 1 & 1 & 1 \\
0 & 1 & 0 & 0 & 0 & 0 & 0 & 0 \\
0 & 0 & 1 & 1 & 0 & 0 & 0 & 0 \\
0 & 0 & 0 & 1 & 0 & 0 & 0 & 0 \\
0 & 0 & 0 & 0 & 1 & 1 & 1 & 1 \\
0 & 0 & 0 & 0 & 0 & 1 & 0 & 0 \\
0 & 0 & 0 & 0 & 0 & 0 & 1 & 1 \\
0 & 0 & 0 & 0 & 0 & 0 & 0 & 1  
\end{pmatrix}
\end{align}
In the similar fashion, we can create $\beta_8^{-1}$.
\end{example}

\subsubsection{Parity Set}
In Bravyi-Kitaev transformation, we want to know which set of qubits tells us whether the state aquire the phase change of -1 or not under the action of creation/annihilation operator. The parity of this set of qubit has the same parity as the set of orbitals with index $<j$. Thus this set of qubits can be called as ``parity set'' of index $j$, $P(j)$. To determine $P(j)$, we have to transform Bravyi-Kitaev basis to parity basis. We know that 
\begin{align}
\ket{\textbf{p}} & \equiv \pi_{n\times n} \ket{\textbf{k}} \\
\ket{\textbf{b}} & \equiv \beta_{n\times n}\ket{\textbf{k}} \\
\Rightarrow \ket{\textbf{k}} & \equiv \beta_{n\times n}^{-1} \ket{\textbf{b}} \\
\therefore \ket{\textbf{p}} & \equiv \pi_{n\times n}\beta_{n\times n}^{-1} \ket{\textbf{k}}
\end{align}
We can also write this as:
\begin{align}
p_i & = \sum_k \left[\pi\beta^{-1} \right]_{ik} b_k
\end{align}
The matrix $\displaystyle{\pi\beta^{-1}}$ transform Bravyi-Kitaev basis to parity basis. As an example, consider a state $\ket{k_7,\cdots,k_0}=\ket{10100111}$ and the matrix $\pi_8\beta_8^{-1}$ will be thus:

\begin{center}
$\displaystyle{\pi_8\beta_8^{-1}}$ = \begin{blockarray}{ccccccccc}
& 7 & 6 & 5 & 4 & 3 & 2 & 1 & 0 \\
\begin{block}{c(cccccccc)}
7 & 1 & 1 & 1 & 0 & 1 & 0 & 0 & 0 \\
6 & 0 & 1 & 1 & 0 & 1 & 0 & 0 & 0 \\
5 & 0 & 0 & 1 & 1 & 1 & 0 & 0 & 0 \\
4 & 0 & 0 & 0 & 1 & 1 & 0 & 0 & 0 \\
3 & 0 & 0 & 0 & 0 & 1 & 1 & 1 & 0 \\
2 & 0 & 0 & 0 & 0 & 0 & 1 & 1 & 0 \\
1 & 0 & 0 & 0 & 0 & 0 & 0 & 1 & 1 \\
0 & 0 & 0 & 0 & 0 & 0 & 0 & 0 & 1 \\
\end{block}
\end{blockarray}
\end{center}
Thus, the parity sets would be:
\begin{align*}
P(7) & = \{6,5,3 \} \\
P(6) & = \{5,3 \} \\
P(5) & = \{4,3 \} \\
P(4) & = \{3 \} \\
P(3) & = \{2,1 \} \\
P(2) & = \{1 \} \\
P(1) & = \{0 \} \\
P(0) & = \emptyset \\
\end{align*}
\subsubsection{Update Set}
The sets of qubits other than $j$ must be updated after application of creation/annihilation operator on qubit $j$. This new set of qubits can be called as ``update set'', $U(j)$. It must stores the partial sum of the orbitals upto $j$ including $j$. We already know that for even $j$, the occupation of only orbital $j$ is stored in $q_j$, thus the update set will have qubits with odd indices. We can determine the update set from the matrix $\beta$. The columns of this matrix shows that which qubits in Bravyi-Kitaev basis store the particular orbital. Thus, non-zero entries in column $j$, above the main diagonal, determines which qubit (other than $j$) must be updated. Re-writing $\beta_8$ given in Eq. (\ref{beta8}):
\begin{center}
$\displaystyle{\beta_8}$ = \begin{blockarray}{ccccccccc}
& $k_7$ & $k_6$ & $k_5$ & $k_4$ & $k_3$ & $k_2$ & $k_1$ & $k_0$ \\
\begin{block}{c(cccccccc)}
$b_7$ & 1 & 1 & 1 & 1 & 1 & 1 & 1 & 1 \\
$b_6$ & 0 & 1 & 0 & 0 & 0 & 0 & 0 & 0 \\
$b_5$ & 0 & 0 & 1 & 1 & 0 & 0 & 0 & 0 \\
$b_4$ & 0 & 0 & 0 & 1 & 0 & 0 & 0 & 0 \\
$b_3$ & 0 & 0 & 0 & 0 & 1 & 1 & 1 & 1 \\
$b_2$ & 0 & 0 & 0 & 0 & 0 & 1 & 0 & 0 \\
$b_1$ & 0 & 0 & 0 & 0 & 0 & 0 & 1 & 1 \\
$b_0$ & 0 & 0 & 0 & 0 & 0 & 0 & 0 & 1 \\
\end{block}
\end{blockarray}
\end{center}
Thus, the update set would be:
\begin{align*}
U(7) & = \emptyset \\
U(6) & = \{7 \} \\
U(5) & = \{7 \} \\
U(4) & = \{5,7 \} \\
U(3) & = \{7 \} \\
U(2) & = \{3,7 \} \\
U(1) & = \{3,7 \} \\
U(0) & = \{1,3,7 \} 
\end{align*}
\subsubsection{Flip Set}
We want to know whether qubit $j$ has the same parity as orbital $j$ or not. This set of qubits is called ``flip set'', $F(j)$. It stores the parity of occupation number other than $k_j$ in the sum $b_j$ and tells whether $b_j$ flipped the parity with respect to $k_j$ or not. Since $\ket{\textbf{k}}=\beta^{-1}\ket{\textbf{b}}$. For even $j$, qubit $j$ will store the parity of only orbital $j$. Thus we can conclude that flip of set of even indices is always going to be empty set. We can determine the flip set by using $\beta^{-1}$. The non-zero entries in the column $j$, to the right of the main diagonal, determines the set of qubits that together store the same set of orbitals as stored by $b_i$. Matrix $\beta_8^{-1}$ is given by:
\begin{center}
$\displaystyle{\beta_8^{-1}}$ = \begin{blockarray}{ccccccccc}
& $k_7$ & $k_6$ & $k_5$ & $k_4$ & $k_3$ & $k_2$ & $k_1$ & $k_0$ \\
\begin{block}{c(cccccccc)}
$b_7$ & 1 & 1 & 1 & 0 & 1 & 0 & 0 & 0 \\
$b_6$ & 0 & 1 & 0 & 0 & 0 & 0 & 0 & 0 \\
$b_5$ & 0 & 0 & 1 & 1 & 0 & 0 & 0 & 0 \\
$b_4$ & 0 & 0 & 0 & 1 & 0 & 0 & 0 & 0 \\
$b_3$ & 0 & 0 & 0 & 0 & 1 & 1 & 1 & 0 \\
$b_2$ & 0 & 0 & 0 & 0 & 0 & 1 & 0 & 0 \\
$b_1$ & 0 & 0 & 0 & 0 & 0 & 0 & 1 & 1 \\
$b_0$ & 0 & 0 & 0 & 0 & 0 & 0 & 0 & 1 \\
\end{block}
\end{blockarray}
\end{center}
Thus, the flip set would be:
\begin{align*}
F(7) & = \{6,5,3 \} \\
F(6) & = \emptyset \\
F(5) & = \{4 \} \\
F(4) & = \emptyset \\
F(3) & = \{2,1 \} \\
F(2) & = \emptyset \\
F(1) & = \{0 \} \\
F(0) & = \emptyset \\
\end{align*}
Given these three set, a transformation from fermion operators to qubit operators can be derived.
\subsection{Operator transformation}
It is useful to define the operators $\hat{E}_S$ and $\hat{O}_S$, where $S$ is the set of qubits with parity either even or odd. For set of qubits $\ket{\textbf{q}}$ with \emph{even} parity, 
\begin{align*}
E_S\ket{\textbf{q}} & = \ket{\textbf{q}} \\
O_S\ket{\textbf{q}} & = 0
\end{align*}
For set of qubits $\ket{\textbf{q}}$ with \emph{odd} parity, 
\begin{align*}
E_S\ket{\textbf{q}} & = 0 \\
O_S\ket{\textbf{q}} & = \ket{\textbf{q}}
\end{align*}
In term of pauli matrices, $E_S$ and $O_S$ can be written as:
\begin{align*}
E_S & = \frac{1}{2} \left( I + Z \right) \\
O_S & = \frac{1}{2} \left( I - Z \right)
\end{align*}
\begin{center}
\textbf{Case I: Representation of fermionic operators in Bravyi-Kitaev basis for \emph{even} $j$} 
\end{center}
For even $j$, qubit $j$ stores only the occupation of orbital $j$. So, we operate $A^{\pm}$ on qubit $j$, just like we did in Jordan-Wigner transformation, $Z$ on parity set to determine the parity of occupied orbitals with index $<j$, $X$ on qubits with index $>j$, to update the qubits that store partial sum that include occupation number $j$. Hence the fermionic opeartors map in the following way:
\begin{align}
a_j^{\dagger} & \longrightarrow X_{U(j)} \otimes A_j^{+} \otimes Z_{P(j)} \\
a_j & \longrightarrow X_{U(j)} \otimes A_j^{-} \otimes Z_{P(j)}
\end{align}

\begin{center}
\textbf{Case II: Representation of fermionic operators in Bravyi-Kitaev basis for \emph{odd} $j$}
\end{center}
For odd $j$, qubit $j$ stores the partial sum of the occupation of the orbitals upto $j$ including $j$. Thus, $q_j$ can either stores even parity or odd parity of the orbitals other than $j$. Therefore, to represent creation/annihilation operators, we need to replace them by $A^+$ or $A^-$ which depends on the parity of qubits in flip set. If the parity of qubits in $F(j)$ is even, then creation and annihilation of particle will requires us to operate $A^+$ and $A^-$, respectively, as usual. But if the parity of this set is odd, then creation of particle requires us to operate $A^-$ and annihilation requires us to operate $A^+$. Thus, new creation and annihilation operators are defined as:
\begin{align}
\Pi_j^\pm & \equiv A^{\pm}_j \otimes E_{F(j)} - A^{\mp}_j \otimes O_{F(j)}
\end{align}
The updating of qubits work exactly the same way as it does for $j$ even. But for parity operators, we only need to consider those qubits in $R(j)$: qubits that are in $P(j)$ but not in $F(j)$. Because $\Pi^\pm$ already calculates the parity of \emph{subset} of $P(j)$ that is also in $F(j)$.
\begin{align*}
R(j) & \equiv P(j) \backslash F(j)
\end{align*}
Hence the fermionic opeartors map in the following way:
\begin{align}
a_j^{\dagger} & \longrightarrow X_{U)j)} \otimes \Pi_j^+ \otimes Z_{R(j)} \\
a_j & \longrightarrow X_{U)j)} \otimes \Pi_j^- \otimes Z_{R(j)}
\end{align}
We can define:
\begin{align}
\rho(j) & = \begin{cases}
P(j) &\ j\ \text{even} \\
R(j) &\ j\ \text{odd}
\end{cases}
\end{align}
Now, the fermionic operators in Bravyi-Kitaev basis can be represented generally as:
\begin{align*}
a_j^\dagger & \longrightarrow X_{U(j)}\otimes \Pi_j^+ \otimes Z_{\rho(j)} \\
a_j & \longrightarrow X_{U(j)}\otimes \Pi_j^- \otimes Z_{\rho(j)}
\end{align*}
We see that number of qubits in each set scales as $\mathcal{O}(\log N)$ for system of $N$ fermions.

\end{document}